%% file: Thesis.tex
\documentclass[a4paper,12pt]{muthesis}

\usepackage{times}
\usepackage{url}
\usepackage{graphicx}
\usepackage{subfigure}
\usepackage{amsmath}
\usepackage{setspace}
\usepackage{fancyhdr}
\usepackage[margin=1in]{geometry}
\usepackage[pdfborder={0 0 0}]{hyperref}   
\newtheorem{corollaries}{Corollary}[chapter]
\newtheorem{theorems}{Theorem}[chapter]
\newtheorem{definitions}{Definition}[chapter]

\begin{document}
\include{titlePage}

\newpage
\include{abstract}

\newpage
\include{acknowledgements}

\newpage
\tableofcontents

\newpage
\listoffigures

\newpage
\listoftables

\newpage
\include{conventions}

\newpage
\include{acronyms}

\newpage
\include{chapter1}

\newpage
\include{chapter2}

\newpage
\include{chapter3}

\newpage
\include{chapter4}

\newpage
\include{chapter5}

\newpage
\include{chapter6}

\appendix
\include{proofs}

\bibliographystyle{plain}
\bibliography{references}

\end{document}

%% file: titlePage.tex
\begin{titlepage}
	\begin{center}
		\includegraphics[scale=.5]{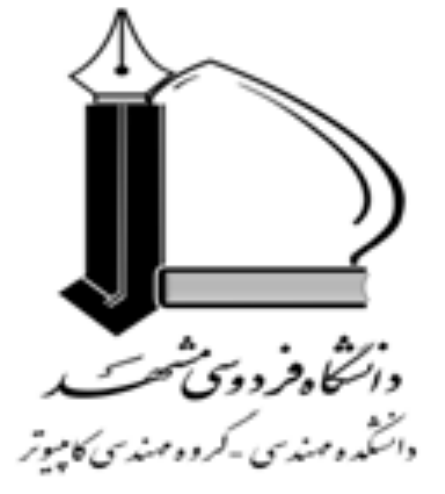}\\
		\noindent {\large \textbf{Ferdowsi University of Mashhad}} \\
		\vspace*{0.3cm}
		\noindent {\large {Department of Computer Engineering}} \\
		
		\vspace*{0.5cm}
		\noindent \Huge \textbf{Master's \ \ T H E S I S} \\
		
		\vspace*{0.3cm}
		\noindent \large {By\\ }
		\noindent \LARGE Hamid Reza Hassanzadeh\\
		
		\vspace*{0.8cm}
		\noindent {\Large \textbf{A New Type-II Fuzzy Logic Based Controller for Non-linear Dynamical Systems with Application to a 3-PSP Parallel Robot }} \\
		
		\vspace*{0.8cm}
		\noindent \Large Advisor: Professor M.-R Akbarzadeh-T \\
		\noindent \Large Co-Advisor: Dr. A. Akbarzadeh-T \\ 
		
		\vspace*{0.5cm}
		\noindent\linebreak\linebreak\linebreak\linebreak\linebreak\linebreak	
		\large \textbf{ A thesis submitted to Ferdowsi University of Mashhad in accordance with the requirements for the degree of Master of Science in the Faculty of Engineering, Department of Computer Engineering}
	\end{center}
\end{titlepage}

\sloppy

%% file: abstract.tex
\section* {\textbf{Abstract}}
The concept of \emph{uncertainty} is posed in almost any complex system including \emph{parallel robots} as an outstanding instance of dynamical robotics systems. As suggested by the name, {uncertainty}, is some missing information that is beyond the knowledge of human thus we may tend to handle it properly to minimize the side-effects through the control process.

Type-II fuzzy logic has shown its superiority over traditional fuzzy logic when dealing with uncertainty. Type-II fuzzy logic controllers are however newer and more promising approaches that have been recently applied to various fields due to their significant contribution especially when noise (as an important instance of uncertainty) emerges. During the design of Type-I fuzzy logic systems, we presume that we are almost certain about the fuzzy membership functions which is not true in many cases. Thus T2FLS as a more realistic approach dealing with practical applications might have a lot to offer. Type-II fuzzy logic takes into account a higher level of uncertainty, in other words, the membership grade for a type-II fuzzy variable is no longer a crisp number but rather is itself a type-I linguistic term \cite{Men02a,Men01,Kar01a,Zad75,Men00,Men99}.

Parallel robots on the other hand, are rather new sort of industrial and scientific tools that are being used in diverse research and industrial academia. The most problematic issues that engineers and designers face when using such robots are the high computational complexity needed for calculation of the inverse dynamics (which should be recalculated continuously at the beginning of each iteration) in the presence of structural uncertainty.

In this thesis the effects of uncertainty in dynamic control of a parallel robot is considered. More specifically, it is intended to incorporate the Type-II Fuzzy Logic paradigm into a model based controller, the so-called \emph{computed torque control}
method, and apply the result to a 3 degrees of freedom parallel manipulator.

One of the most well-known dynamic controllers that relies on the dynamic calculation of parameters of the underlying robot (in the feedback) is called the Computed Torque Control method. The CTC converts the non-linear dynamics of a robot into a linear one provided that the dynamics of the system at hand is completely identified. Having designed a system with a linear dynamic, it is easy for a control engineer to design a PID (or maybe PD) controller for it so that the final motion of the robot would be to follow a predetermined trajectory precisely.  

The problem with the aforementioned method is that even if we manage to determine the foregoing parameters accurately we are yet to recalculate several matrices in each iteration. This imposes a high amount of computational burden. To overcome this demanding task, we should find a closed form formula for each of the dynamic terms so that not to perform intense computations that eventually lead to calculation of those parameters again and again.
\\\\\noindent \textbf {Keywords:} Robot Dynamic Control, Parallel Manipulator, 3PSP Robot, Type-II Fuzzy Logic, Computed Torque Control Method

%% file: acknowledgements.tex
\section*{\textbf{Acknowledgments}}
Firstly, I would like to express my gratitude to my supervisors, Professor Mohammad-R Akbarzadeh-T and Dr. Alireza Akbarzadeh-T for their invaluable helps, comments, suggestions and advices throughout the whole project. Their instrumental perspectives on theoretical aspects of robotics and intelligent systems along with their efforts in providing me with an easy access to the required platforms at the Robotics Laboratory opened my lines of thoughts to the robotics world. Without their contributions I would have never managed to come up with valuable results.

Secondly, I should thank my parents, esp. my mother, who have spent a great deal of their time supporting me on different aspects while I was busy carrying out my investigations.

Last, but not the least, my thanks are given to the guys at Robotics Lab who let me know what is teamwork and what is a real scientific collaboration.

%% file: conventions.tex
\chapter*{CONVENTIONAL NOTATIONS}

\begin{tabular}{p{2.5cm} p{10cm}}
$\int$ &Used to denote the union over elements of continuous fuzzy sets\\
$\sum$ &Used to denote the union over elements of discrete fuzzy sets\\
$/$ & Notation to separate a point in the domain with its membership grade\\
$A$, $B$, $C$ & Widespread alphabets denoting T1 fuzzy sets\\
$\tilde A$, $\tilde B$ &Widespread alphabets denoting T2 fuzzy sets\\
$\mu _{\tilde A} (.,.)$, $\mu _{\tilde A} (x,u)$  &Another way to represent a T2 fuzzy set\\
$\mu _{\tilde A} (x',.)$, $\mu _{\tilde A} (x',u)$  &Secondary membership function of a T2FS at point, 
$x'$\\
$\mu _{\tilde A} (x',u')$, $\mu _{\tilde A} (u=x',u=u')$&Secondary membership grade 
corresponding to domain point, $x'$ with associated primary membership grade $y'$
(note that sometimes the prime sign is omitted for convenience\\
$*$ & T-norm\\
$\vee$ & T-conorm\\
$\wedge$ & Minimum t-norm\\
$\otimes$ & A general binary operation\\
$\Im$ & Used for denoting consecutive t-norms\\
$\prod$& Depending on the context, used for denoting both consecutive products and \emph{meet}
operation\\
$\coprod$& Used for denoting \emph{join} operation 
\end{tabular}

%% file: acronyms.tex
\chapter*{LIST OF ACRONYMS}

\begin{tabular}{p{2.5cm} p{10cm}}
FS & Fuzzy Set\\
FLS & Fuzzy Logic System\\
MF & Membership Function\\
T1 & Type-I\\
T2 & Type-II\\
T2FS & Type-II Fuzzy Set\\
T2FLS & Type-II Fuzzy Logic System\\
IT1FS & Interval Type-I Fuzzy Set\\
IT1FLS & Interval Type-I Fuzzy Logic Set\\
IT2FS & Interval Type-II Fuzzy Set\\
IT2FLS & Interval Type-II Fuzzy Logic Set\\
TR & Type Reduction\\
TG & Trajectory Generation\\
CTC & Computed Torque Control\\
SNR & Signal to noise ratio
\end{tabular}

%% file: chapter1.tex
\chapter{{L}iterature {R}eview, {T}ype-{II} {F}uzzy {L}ogic}
\section{\textbf{Preface}}
The knowledge to build a fuzzy logic system is itself uncertain, hence we expect our designed system 
to consider this uncertainty. This means that the output should no longer be a crisp number.
This is due to the fact that uncertainty has been propagated to the output as a result of uncertain
information and inputs. Therefore the output should somehow represent this uncertainty. 
This is one of the most meaningful reasons why 
ordinary fuzzy logic systems (henceforth called T1FLSs) have given way to Type-II fuzzy logic systems
(T2FLSs). In other words (\cite{His81}): increased fuzziness in a description requires increased ability to handle 
inexact information in a reasonably proper way. Uncertainty may come into existence 
from four main sources as follows \cite{Men02a}:
\begin{itemize}
\item The linguistic words being used in both antecedent and consequents of the fuzzy rules may
mean different to different people. 
\item Some consequents derived from insights of different experts, which may differ.

\item These noise interferences are in almost any real-world application.
\item Also, the measuring devices which provide the inputs to fuzzy systems are themselves not exact and introduce noise to the inputs.

\end{itemize}
Accordingly, in many situations we prefer to opt a Type-II fuzzy logic based approach the most 
appropriate instances of which are illustrated in \cite{Men02a} as follows:
\begin{itemize}
\item "Measurement noise is non-stationary, but the nature of the non-stationarity cannot be 
expressed mathematically ahead of time," e.g., approximating a function under a variable SNR parameter.
\item	"A data-generating mechanism is time-varying, but the nature of the time variations cannot
be expressed mathematically ahead of time," e.g., time variant communication channels.
\item "Features are described by statistical attributes that are non-stationary, but the nature of
the non-stationarity cannot be expressed mathematically ahead of time," e.g., rule-based
classification of video traffic.
\item "Knowledge is mined from experts using IF-THEN questionnaires."
\end{itemize}

The question that strikes the mind of any newcomer to the realm of Type-II fuzzy logic is that why, 
despite the fact that the concept of Type-II fuzzy was introduced by Zadeh \cite{Zad75} in 1975, it was not welcomed and did not receive the attention of researchers, until the late nineties? In fact even now, the T2FL is considered as a state-of-the-art area of knowledge that has been emerging by introducing 
hundreds of valuable applications and supplementary theoretical techniques each year. One reason 
might be that science flourishes progressively. That is why old theories always give way 
to new ones and so forth. After introduction of T1FL by Zadeh in his seminal paper, no one understood or could foretell the future in progress; even in scientific academia some prejudicedly denied to accept it. 
It took some time until new applications of that theory became tangible to human and put it into practice. During the course of time scientists and more importantly engineers developed this logic until it became what it is today. The same history has been repeated for T2FL. That is to say, aside from few researches that considered the mathematical aspects of T2FL \cite{Dub87,Dub79,Gor87}, it had not received much attention until the shortcomings of the predecessor became 
understood in a progressive way. Today we all know that the assertion "Type-I fuzzy logic models uncertainty" that once was an undeniable fact, see for example paradox of Type-I fuzzy sets in \cite{Men03,Kli88} , is no longer believed to be true as they are inherently crisp and not fuzzy.
\pagebreak
\section{\textbf{Type-II Fuzzy Sets}}
\begin{figure}[htp]
  \begin{center}
    \subfigure [An ordinary fuzzy set]{\label{fig1-a}\includegraphics[width=3cm, height=3cm] {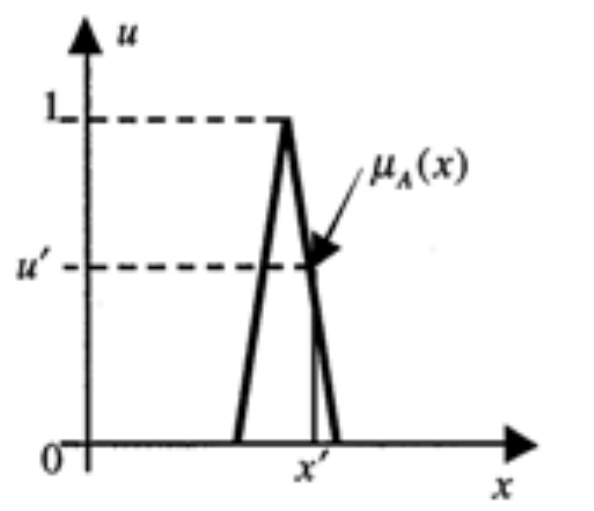}}\\
    \subfigure[having the membership function perturbed] {\label{fig1-b}\includegraphics[width=3cm, height=3cm]{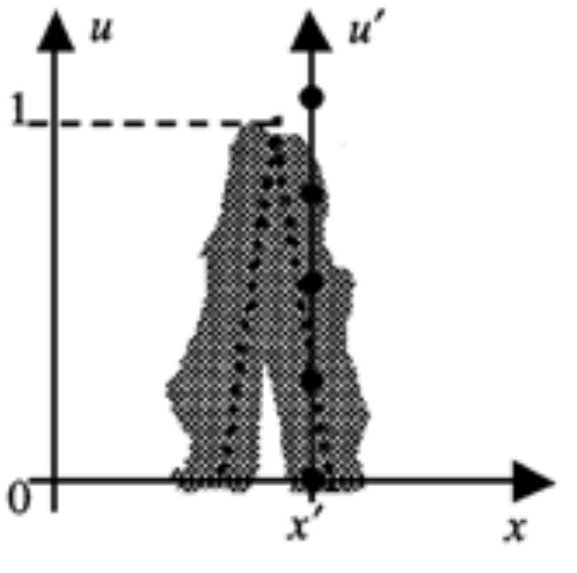}} \\
    
  \end{center}
  \caption{Ordinary vs. perturbed fuzzy set. \cite{Men02b}}
  \label{fig1}
\end{figure}

In this subsection we provide definitions for Type-II fuzzy sets (T2FSs) and the related but important concepts. This helps us to communicate effectively by laying a well-defined foundation for the language we use extensively through the rest of this thesis. Consider an ordinary fuzzy set depicted in Figure \ref{fig1-a}. What if we blur the portrayed membership function by shifting (not necessarily evenly) the points on the triangle up and down? Figure \ref{fig1-b} shows the modified image. In this manner for a specific value in the domain, say $x'$, there will not be any unique membership 
grade attributed to it, rather the membership values now take on a continuous set of real values. This is illustrated in the latter figure. We can also 
ascribe a real value in [0,1] to each element of the aforementioned set to make up the third dimension.
This second membership function is called \emph{secondary membership function}, in literature.

Doing so for all points in the domain, we come up with three-dimensional membership function which we call it a Type-II fuzzy set (T2FS). Among general Type-II fuzzy sets, two main groups are well-known and have been employed successfully, to date, 
namely the \emph{Gaussian} Type-II fuzzy sets and \emph{interval} Type-II fuzzy sets . The former includes Gaussian secondary membership functions while the latter includes interval valued Type-I secondary membership functions.

To communicate easier through the Type-II fuzzy literature we define the following important concepts,
as well \cite{Men02b}:\\
\begin{definitions} A Type-II fuzzy set, denoted by \begin{math}\tilde{A}\end{math} , is characterized by \begin{math}\mu _{\tilde{A}}(x,u)\end{math} , where \begin{math} x \in X\end{math} and \begin{math}u \in J_x \subseteq [0,1]\end{math}. That is to say,
\begin{equation}
\tilde A = \{ (x,u),\mu _{\tilde A} (x,u))|\forall x \in X,\forall u \in J_x  \subseteq [0,1]\}\label{1.1} 
\end{equation}
where \begin{math}\begin{array}{c} 0 \le \mu _{\tilde A} (x,u) \le 1 \end{array}\end{math}. 
${\tilde A}$  can also be re-expressed as 
\begin{equation} \tilde A = \int_{x \in X} {\int_{u \in J_x } {\mu _{\tilde A} (x,u)/(x,u)} } \,\,\,\,\,J_x  \subseteq [0,1] \label{1.2}\end{equation}
where the integral sign represents the union over all admissible $x$, $u$. For discrete T2FSs $ \int $  is replaced by the summation, $ \Sigma $. 
\end{definitions}
\begin{definitions} The domain of the secondary membership is also called the \emph{primary membership} and denoted by $J_x$.\\
\end{definitions}
\begin{definitions} At each value, $x'$, in the domain $X$, the 2D plane whose axes are $u$ and $\mu _{\tilde A} (x',u)$ is called a \emph{vertical slice} of the T2FS. Obviously this is a T1FS which can be stated as 
\begin{equation}
\label{1.3}
\mu _{\tilde A} (x = x',u) \equiv \mu _{\tilde A} (x = x') = \int_{u \in J_{x'} } {f_{x'} (u)/(u)} \,\,\,\,\,J_{x'}  \subseteq [0,1]
\end{equation}
Note that for convenience, it is common to use $\mu _{\tilde A} (x')$ and $\mu _{\tilde A} (x)$ interchangeably. Note also that, as already stated $\mu _{\tilde A} (x')$ is the secondary membership function of a T2FS at $x = x'$.\\
\end{definitions}
\begin{definitions} The amplitude of a secondary membership is called a \emph{secondary grade}. The secondary grade at $x=x'$, $u=u'$   is denoted by $\mu _{\tilde{A}}(x=x',u=u')$.\\
\end{definitions}
\begin{definitions} The union of all primary membership grades create a bounded region that we call the \emph{footprint of uncertainty} (or FOU for short), i.e., 
\begin{equation}
\label{1.4}
FOU{\tilde{A}}=\bigcup\limits_{x \in X} {J_x }.
\end{equation}
The shaded region in Fig. \ref {fig3} is the FOU of the associated T2FS. The term FOU is very useful
in T2~fuzzy literature in that it provides us with a convenient description of entire domain of the 
\emph{support} for all the secondary membership grades in a T2FS.\\
\end{definitions}
\begin{definitions} For a general T2FS, $\tilde{A}$, an \emph{embedded Type-II set}, $\tilde{A}_e$  , is a T2FS such that for every admissible $x\in X$ in $\tilde{A}_e$  there is one and only one element, 
say $u_x$, in domain of $\tilde{A}_e(x)$  such that $\tilde A_e (x,u_x ) = \tilde A(x,u_x )$ . 
For a discrete T2FS,
\begin{equation}
\tilde A = \sum\limits_{i = 1}^N {\sum\limits_{j = 1}^{M_i } {\mu _{\tilde A} (x_i ,u_{ij} )/(x_i ,u_{ij} )} } \,\,\,\,\,J_{x_i }  \subseteq [0,1]
\label{1.5}
\end{equation}
$\tilde A_e$ can be found as any of possible IT2FSs of the following form,
\begin{equation}
\tilde A_e  = \sum\limits_{i = 1}^N {\mu _{\tilde A} (x_i ,u_i )/(x_i ,u_i )} \,\,\,
\label{1.6}
\end{equation}
where $u_i \in M_i$. Accordingly for a discrete T2FS, there are a total of $\prod\limits_{i = 1}^N {M_i}$ different embedded T2FSs.\\
\end{definitions}
\begin{definitions} For a T2FS, a T1 embedded FS is similar to T2 embedded one except that the secondary grade is now omitted so that the resulting set is in the form of a Type-I fuzzy set. For a general T2FS, $\tilde{A}$, this set is denoted as $A_e$ and can be expressed (assuming the discrete case) as,\begin{equation}
A_e = \sum\limits_{i = 1}^N {u_i /x_i \,\,\,u_i  \in J_{x_i } } \,\,\,
\label{1.7}
\end{equation}\\	
\end{definitions}
\begin{definitions} A Type-II fuzzy set whose primary membership variable is discrete but its secondary membership functions are continuous is called a 
\emph{partially discrete} Type-II fuzzy set. In the same manner, a Type-II fuzzy set whose both the primary variable and the associated secondary membership functions are crisp is called a \emph{discrete membership} Type-II fuzzy set.\\
\end{definitions}
\begin{definitions} An \emph{Upper} membership function (henceforth called \emph{MF}) and a \emph{Lower MF} are two Type-I MFs that are the upper and lower limits of FOU, respectively.\\
\end{definitions}
\subsection{\textbf{Gaussian Type-II Fuzzy Sets}}
\begin{figure}[htp]
  \begin{center}
    \includegraphics[width=7cm, height=5cm]{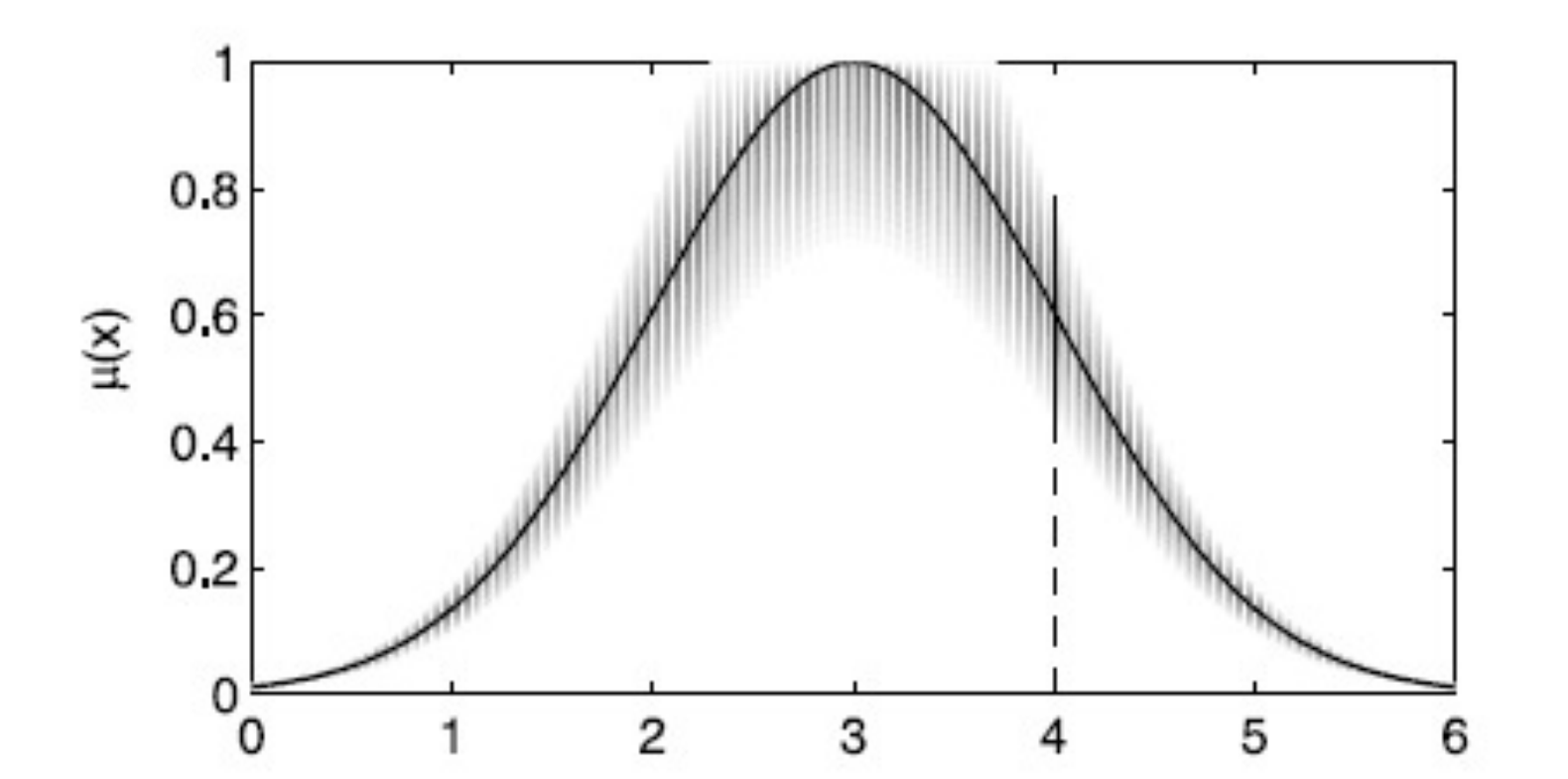}\\
  \end{center}
  \caption{Type-II Gaussian fuzzy set portrayed from above view. \cite{Kar01b}}
  \label{fig2}
\end{figure}

When all the secondary membership functions corresponding to the domain of a Type-II fuzzy set are Type-I Gaussian membership functions, we call such a set a Gaussian Type-II set.\\
Figure \ref{fig2} depicts a 3-dimensional Gaussian Type-II fuzzy set in a 2D picture where the third dimension has been transferred to the image intensity. In the figure, the darker points represent higher secondary membership grades and the solid line shows those points with unity secondary membership grade.

\subsection{\textbf{Interval Type-II Fuzzy Sets}}
An interval Type-II fuzzy set (IT2FS) is a special kind of general T2FSs for which the secondary membership grades equal to 1. An IT2FS is completely defined by the footprint of uncertainty. These sets are the most widely used T2FSs due to several reasons (esp. the implementation issues, demonstrated in subsequent sections.) More specifically, when all the secondary membership functions corresponding to the domain of a Type-II fuzzy set are interval Type-I membership functions, we call such a set an interval Type-II set (IT2FS).\\

\begin{figure}[htp]
  \begin{center} 
    \subfigure [Gaussian primary IT2FS with uncertain mean]{\label{fig3-a}\includegraphics[width=7cm, height=5cm] {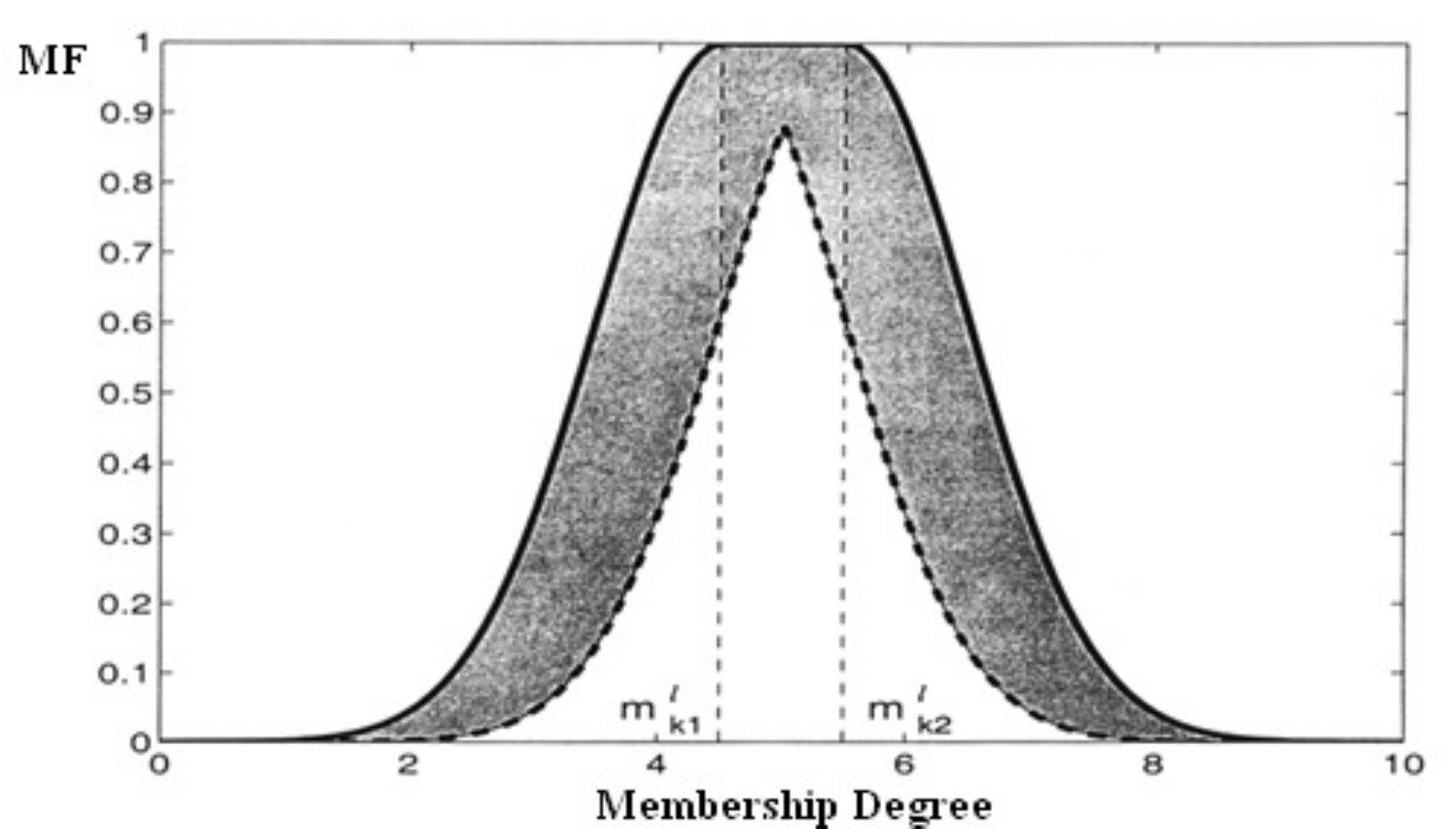}}\\
    \subfigure [Gaussian primary IT2FS with uncertain sigma]{\label{fig3-b}\includegraphics[width=7cm, height=5cm] {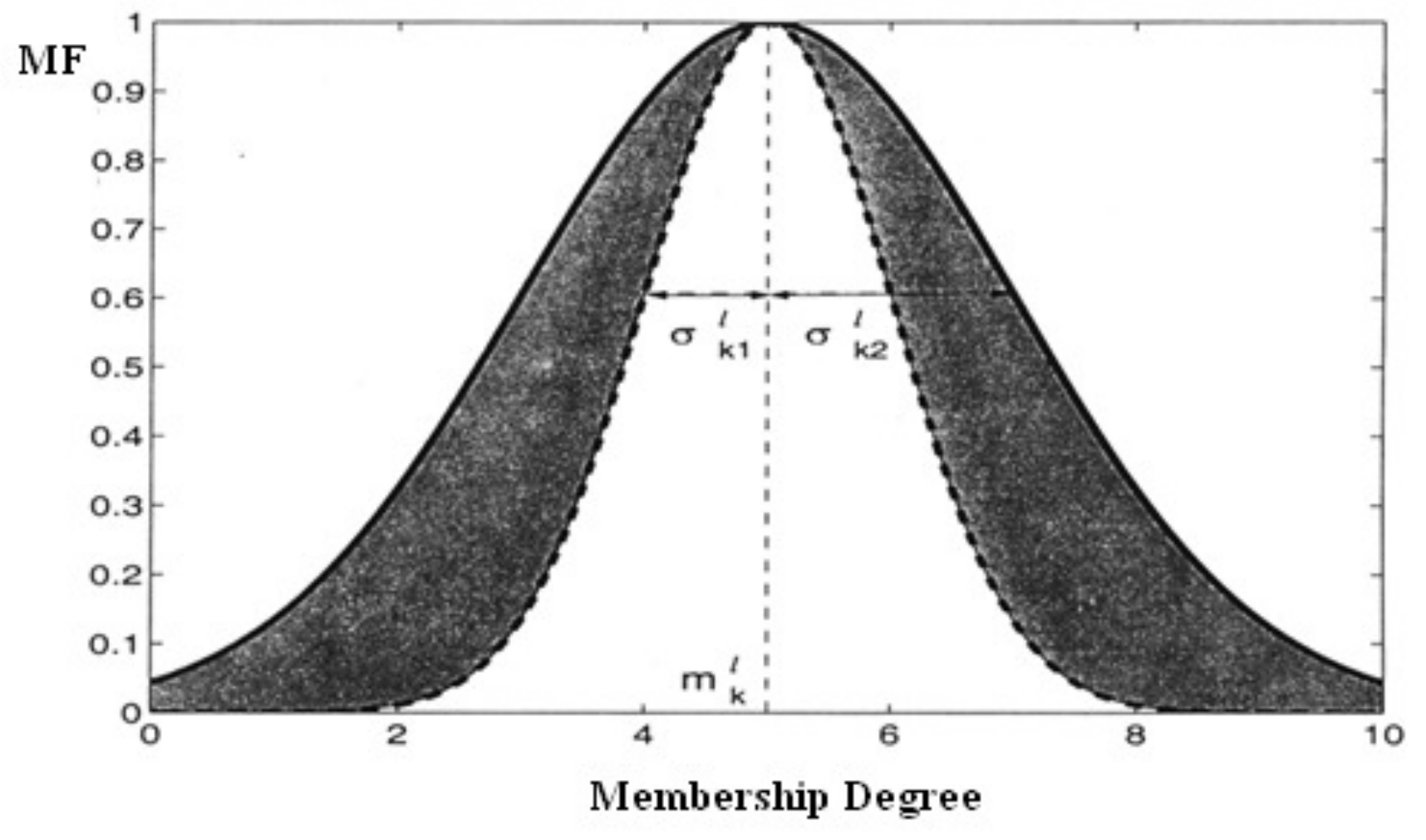}}\\
  \end{center}
  \caption{Two types of Type-II Gaussian primary membership functions \cite{Lia00}}
  \label{fig3}
\end{figure}

An important class of interval Type-II fuzzy sets (henceforth called IT2FSs) is Gaussian primary Type-II sets. Two different types of Gaussian primary T2FSs are: Gaussian IT2FSs with uncertain mean and Gaussian IT2FSs with uncertain standard deviation. Figure \ref{fig3} illustrates these types of sets.

\section{\textbf{Extension Principle}} 
This principle extends the classical operations defined in crisp mathematics to the realm of fuzzy 
logic. In fact this is the basis of all T2FL analysis either explicitly or implicitly. The principle 
states that if an operation $\otimes$, is defined on a set of numbers, then its counterpart in fuzzy 
domain is obtained by,
\begin{equation} 
F \otimes G = \int_u {\int_w {(f(u) * g(w))} /(u \otimes w)}
\label{1.8a}
\end{equation}
where the $\int$ sign and the $*$ denote the union and t-norm, respectively. In other words, 
if a relation $f(x)$, is defined on crisp variable $x\in X$ then we extend it to operate on fuzzy 
set, $A$, as follows: 
\begin{equation}
\mu_B(y)=max_{x=f^{-1}(y)}\mu_A(x)
\label{1.8b}
\end{equation}
\section{\textbf{Operations in T2FLSs \cite{Kar01b}}}
A T1FLS is a system that operates on T1FSs. Similarly a T2FLS is a system which manipulates T2FSs possibly along with T1FSs. These manipulations are mostly set-theoretic operations; however as is in this thesis, some systems perform algebraic operations, too. The algebraic operations in T2FLSs are mainly performed on T1FSs which come out of the type reduction unit, as explained in upcoming sub-sections. Of course we can define algebraic operations on T2FSs, however normally they do not play a role in T2FLSs. 
For more details on such operations see \cite{Ble07}. Thus it is necessary to have a rather thorough discussion on both types of fuzzy operations.\\
\subsection{\textbf{Set-Theoretic Operations on Type-II Sets}}

Based on extension principle \cite{Zad75} the set-theoretic operations on T2FSs are declared as follows,\\

Union:\begin{equation}
\tilde A \cup \tilde B \Leftrightarrow \mu _{\tilde A \cup \tilde B} (x) = \mu _{\tilde A} (x)\coprod \mu _{\tilde B} (x) = \int_u {\int_w {(f_x (u) * g_x (w) )}/(u \vee w)} 
\label{1.9}
\end{equation}\\
Intersection:\begin{equation}
\tilde A \cap \tilde B \Leftrightarrow \mu _{\tilde A \cap \tilde B} (x) = \mu _{\tilde A} (x)\prod \mu _{\tilde B} (x) = \int_u {\int_w {(f_x (u) * g_x (w) )}/(u * w)} 
\label{1.10}
\end{equation}\\ 
Complement:\begin{equation}
\overline{ \tilde A} \Leftrightarrow \mu _{\overline{ \tilde A}} (x) = \neg \mu _{\tilde A} (x) = \int_u {f_x (u)} /(1 - u)
\label{1.11}
\end{equation}\\
where $\vee$ and $*$ represent any t-conorm and t-norm, respectively and the operations $\prod$ , $\coprod$  and $\neg $ are called \emph{meet, joint} (defined shortly) and negation, respectively. If the domains are discrete, the integral notation (which denotes the union) is replaced by summation. Notice that if two different u,w yield the same point in the domain of the resulting set followed by a set-theoretic operation, then we keep the one with higher membership grade (this conform to t-conorm operation implied by integral and summation signs). 
To define the $\prod$ and $\coprod$  recall that the corresponding operations in T1FSs are the t-norm and t-conorm respectively, 
which take two crisp numbers and return another crisp number. Similarly we can extend the concept of t-norm and t-conorm for T2FSs. Hence, join and meet operations should take two T1FSs and return another T1FS. Thus, using the extension principle and substituting $\vee$  and $*$ operations for $\otimes$, in order, we define, $\prod$ and $\coprod$ as follows \cite{Wan08}:
\begin{equation}
F\prod G = \int_u {\int_w {(f(u) * g(w)} )/(u * w)} 
\label{1.12}
\end{equation}\\
\begin{equation}
F\coprod G = \int_u {\int_w {(f(u) * g(w)} )/(u \vee w)} 
\label{1.12}
\end{equation}\\

\begin{figure}[htp]
  \begin{center}
    \subfigure [ ]{\label{fig4-a}\includegraphics[width=7cm, height=3cm] {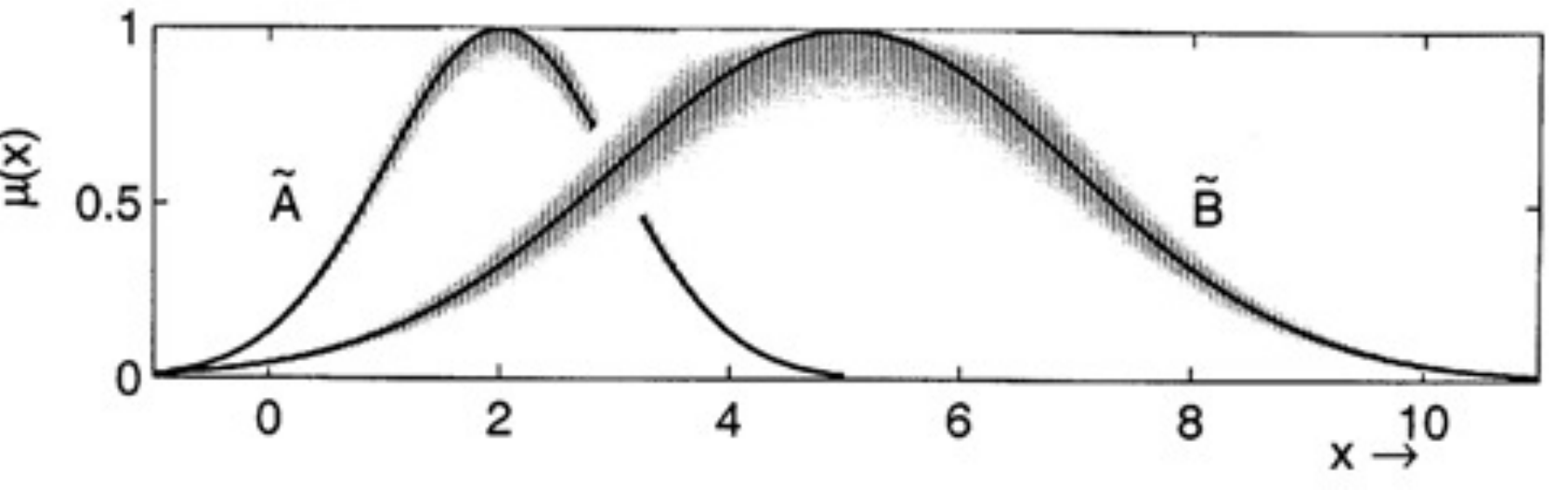}}\\
    \subfigure [ ]{\label{fig4-b}\includegraphics[width=7cm, height=3cm]{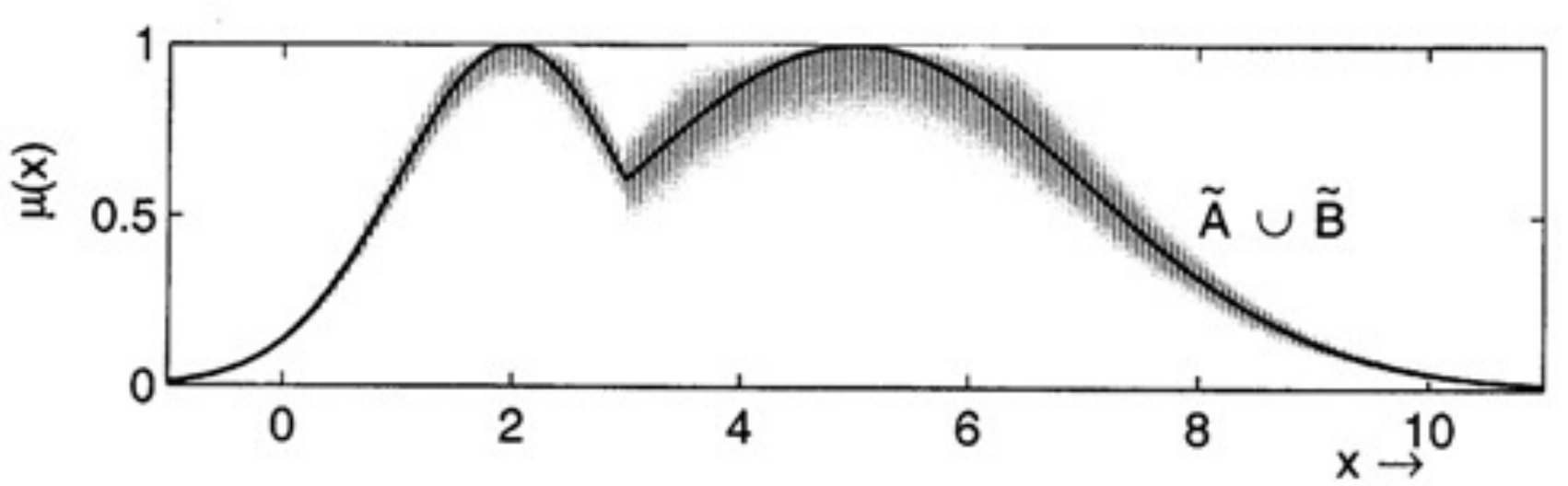}} \\
    \subfigure [ ]{\label{fig4-c}\includegraphics[width=7cm, height=3cm]{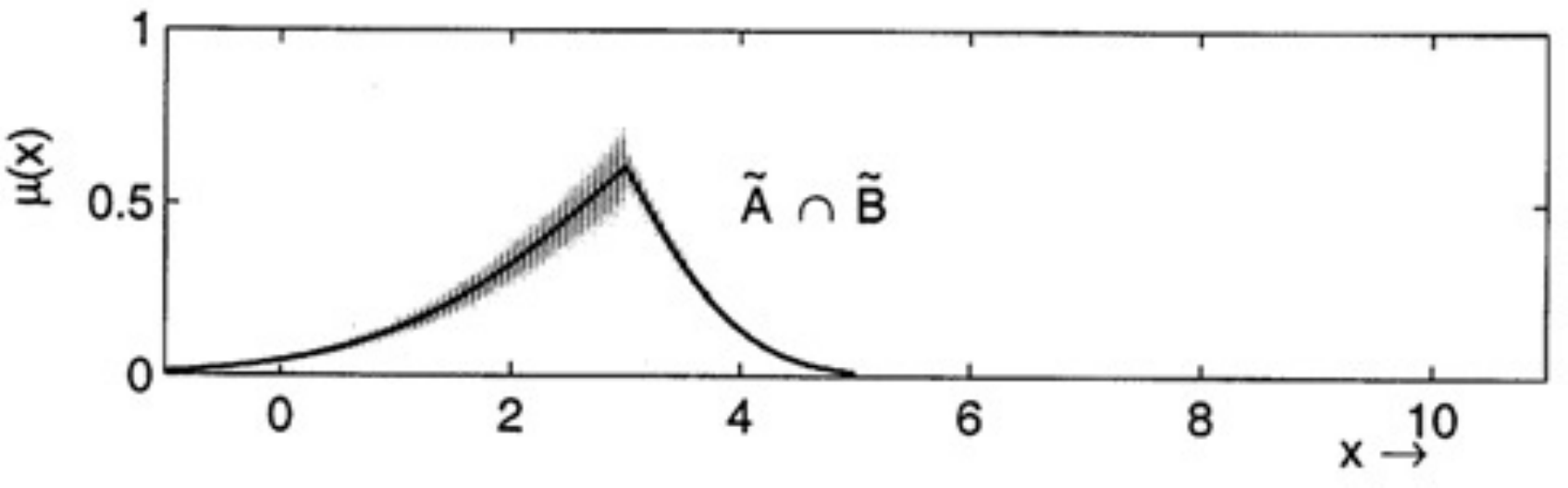}} \\
    
  \end{center}
  \caption{Pictorial representation of union and intersection performed with join and meet operations. \cite{Kar01b}}
  \label{fig4}
\end{figure}

Figure \ref{fig4} depicts pictorial representation of the union and intersection of two general Type-II fuzzy sets. Darker regions in this figure correspond to higher values for secondary membership grades. The black solid curves are thus denoting the principle membership function of the T2FS. \\

The following useful theorems hold for some special sorts of T2FSs (namely IT1FSs and Gaussian T1FSs). 
Some of these appear useful when handling T2FLSs and may mitigate the computational complexity 
of required operations considerably. Others may be necessary as a future extension of the work 
presented in this thesis.

\begin{theorems}{(\cite{Kar01b})} 
Given $n$ convex, normal Type-I fuzzy sets $F_1,\dots,F_n$ characterized by membership functions
$f_1,\dots,f_n$ respectively and a set of real numbers $v_1,\dots,v_n$ such that 
$v_1\leq\dots\leq v_n$ and $f_1(v_1)=\dots=f_n(v_n)=1$ and for \emph{max} t-conorm and \emph{min}
t-norm we have,

\begin{equation}
\mu _{\coprod\nolimits_{i = 1}^n {F_i } } (\theta ) = \left\{
\begin{array}{c l}
\wedge_{i = 1}^n f_i (\theta ) & \theta  < v_1  \hfill \\
\wedge_{i = k + 1}^n f_i (\theta ) & v_k  \le \theta  < v_{k + 1} \,,\,1 \le k \le n - 1 \hfill\\
\vee_{i = 1}^n f_i (\theta )&v_n  \le \theta  \hfill 
\end{array} \right. 
\label{1.14}
\end{equation}

\begin{equation}
\mu _{\prod_{i = 1}^n {F_i } } (\theta ) = \left\{ 
\begin{array}{c l}
  \vee _{i = 1}^n f_i (\theta ) & \theta  < v_1  \hfill\\
  \wedge _{i = 1}^k f_i (\theta ) & v_k  \le \theta  < v_{k + 1} \,,\,1 \le k \le n - 1 \hfill\\
  \wedge _{i = 1}^n f_i (\theta ) & v_k  \le \theta  \hfill
\end{array} \right.
\label{1.15}
\end{equation}
\label{thr1.1}
The proof for this theorem is presented in the Appendix \ref{prf1.1}.
\end{theorems}

A direct application of previous theorem can be derived when the employed fuzzy sets are similar in shape. Formally speaking, if there are $n$ convex, normal Type-I fuzzy sets $F_1,\dots F_n$ characterized by membership functions $f_1,\dots,f_n$ , respectively, such that $f_i(\theta)=f_1(\theta-k_i)$ and $ 0 \leq k_1 \leq k_2\leq k_3 \leq \dots \leq k_n$ , then using max t-conrom and min t-norm we have $\prod\limits_{i = 1}^n {F_i }  = F_1 $ and $\coprod\limits_{i = 1}^n {F_i }  = F_n $.\\
The other useful by-product of previous theorem is when we substitute \emph{product} t-norm in place of \emph{min} t-norm in which case for join operation, 
the same results are acquired with the only modification that $\wedge$ is replaced by products as follows:
\begin{equation}
\mu _{\coprod_{i = 1}^n {F_i } } (\theta ) = \left\{ 
\begin{array}{c l}
  \prod\limits_{i = 1}^n {f_i (\theta )} & \theta  < v_1  \hfill \\
  \prod\limits_{i = k + 1}^n {f_i (\theta )} & v_k  \le \theta  < v_{k + 1} \,,\,1 \le k \le n - 1 \hfill \\ 
  \mathop  \vee _{i = 1}^n f_i (\theta ) & v_n  \le \theta \hfill \\
\end{array}
\right.
\label{1.28}
\end{equation}
where $\prod$ indicates multiplication of several numbers and should not be mistaken for \emph{meet}
operation. Notice that the consequence alluded above does not hold for meet operation, a fact which makes our 
future calculations demanding and inexact until the fuzzy sets own some special features (an instance of which is presented in the next theorem). We will soon present other approaches that make approximations when the \emph{product} t-norm is used.\\
\begin{theorems}{(\cite{Lia00})}
Let $F_1,\dots,F_n$ be n interval Type-I fuzzy sets with domains $[l_1,r_1],\dots,[l_n,r_n]$ , then join and meet operations for these fuzzy sets are:
\begin{equation}
F_1 \prod ...\prod F_n  = \int_{u \in [l_1  *  \cdots  * l_n ,r_1  *  \cdots  * r_n ]} {1/u}   
\label{1.29a}
\end{equation}
\begin{equation}
F_1 \coprod ...\coprod F_n  = \int_{u \in [l_1  \vee  \cdots  \vee l_n ,r_1  \vee  \cdots  \vee r_n ]} {1/u}
\label{1.29b}
\end{equation}
Using the induction principle, the proof is simple and left to the reader. According to this theorem we do not need to compute $F_1(\theta_1)*\dots*F_n(\theta_n)$ and $\theta_1*\dots * \theta_n$ or $\theta_1 \vee \dots \vee \theta_n$ for all possible combinations of $\theta_1, \dots, \theta_n$ which 
leads to an enormous saving in computational resources. This makes T2FLS easier to implement. Unfortunately there is not any exact closed form formula for the case of general T2FSs (unless they are arranged in a specific ordered way or under certain conditions, see \eqref{1.29a},\eqref{1.29b}) and the whole computations should be performed by a T2FLS.
\end{theorems}
\begin{theorems}{(\cite{Kar01b})}
\label{thr1.3}
Given $n$ Gaussian fuzzy numbers $F_1,\dots,F_n$ with respective means and standard deviations $m_1,\dots,m_n$ and $\sigma_1,\dots,\sigma_n$, then, under product t-norm, 
\begin{equation}
\mu _{F_1 \prod F_2 \prod  \cdots \prod F_n } (\theta ) \approx \exp \left\{ { - {1 \over 2}((\theta  - m_1 m_2  \cdots m_n )/\bar \sigma )^2 } \quad i=1,2,\dots,n \right\}
\label{1.30}
\end{equation}
where $\bar \sigma  = \sqrt {\sigma _1 ^2 \prod\limits_{i,i \ne 1} {m_i ^2 }  +  \cdots  + \sigma _j ^2 \prod\limits_{i,i \ne j} {m_i ^2 }  +  \cdots  + \sigma _n ^2 \prod\limits_{i,i \ne n} {m_i ^2 } } 
$.
\end{theorems}
The proof for this theorem is included in Appendix \ref{prf1.5}.
\subsection{\textbf{Algebraic Operations on Type-I Sets}}
Just as the t-norm and t-conorm operations have been generalized from crisp membership grades 
to T1FSs by means of extension principle, the algebraic operations can also be extended to the class 
of T1FSs. Consider the binary operation $\otimes$ defined on crisp numbers, we can extend 
this operator to work on class of fuzzy sets $F_1  = \int\limits_v {f_1 (v)} /v$
and $F_2  = \int\limits_w {f_2 (w)} /w$, as follows:
\begin{equation}
F \otimes G = \int_v {\int_w {(f(v) * g(w)} )/(v \otimes w)} 
\label{1.43}
\end{equation}
where $*$ indicates the t-norm and $\otimes$ is the predefined operator. Similarly and n-ary operation $f(\theta _1 ,..,\theta _n )$ on n Type-I fuzzy sets can be defined as follows:
\begin{equation}
f(F_1 ,..,F_n ) = \int_{\theta _1 } { \cdots \int_{\theta _n } {\mu _{F_1 } (\theta _1 ) *  \cdots  * \mu _{F_n } (\theta _n )} /f(\theta _1 ,..,\theta _n )} 
\label{1.44}
\end{equation}
Using algebraic operations we can perform several important operations on \emph{fuzzy numbers}. We take advantage of this bless from extension principle in later chapters where we develop our newly designed model. More specifically, much the same way we do operations on crisp numbers we can perform on fuzzy numbers. 
The definition for fuzzy addition is given bellow:
\begin{equation}
F + G = \int_u {\int_w {(f(u) * g(w)} )/(u + w)}
\label{1.45}
\end{equation}
We can also extend the notion of fuzzy additions to more than one fuzzy numbers as we already did with join and meet operations. Due to the significant importance of algebraic operations in T2FLSs, we express some theorems to cover this topic in more depth and we will make use of them in subsequent sections and chapters.
\begin{theorems}{\cite{Kar01b}} 
\label{affineIT1comb}
Given $n$ IT1 fuzzy numbers $F_1 , \cdots ,F_n $, with means $
m_1 , \cdots ,m_n 
$ and spreads $s_1 , \cdots ,s_n 
$, their affine combination defined as $
\sum\limits_{i = 1}^n {\alpha _i F_i  + \beta }$ where the related constants, $\alpha_i,  i=1,2,\cdots,n$ and $\beta$ are crisp numbers, is also an IT1FS whose mean and spread can be find by $
m = \sum\limits_{i = 1}^n {\alpha _i m_i  + \beta } 
$ and $s = \sum\limits_{i = 1}^n {|\alpha _i |s_i } 
$ respectively.\\
\emph{Proof}: Let $
F_i  = [m_i  - s_i ,m_i  + s_i ]
$. According to the extension principle \cite{Zad75}, we can multiply $F_i$ by an IT1FS defined as $G_i  = \{ 1/\alpha _i \}$ which gives,
\begin{equation}
F_i  \times G_i  = \int_v {1/(\alpha _i v)} \,\,\,v \in [m_i  - s_i ,m_i  + s_i ]
\label{1.46}
\end{equation}
Similarly we can add the result with $H  = \{ 1/\beta \}$ which yields,
\begin{equation}
F_i  \times G_i  + H_i  = \alpha _i  \times G_i  + \beta  = \int_v {1/(\alpha _i v + \beta )} \quad v \in [m_i  - s_i ,m_i  + s_i ]
\label{1.47}
\end{equation}
Substituting $w$ for $
\alpha _i v + \beta 
$ we come to,
\begin{equation}
\alpha _i  \times G_i  + \beta  = \int_w {1/w} \,\,\,w \in [\alpha _i m_i  - |\alpha _i |s_i  + \beta ,\alpha _i m_i  + |\alpha _i |s_i  + \beta ]
\label{1.48}
\end{equation}
which brings about the result of this theorem. 
\end{theorems}
\begin{theorems}{\cite{Kar01b}}
\label{thr1.5}
Given $n$ Gaussian fuzzy numbers $
F_1 , \cdots ,F_n 
$, with means $
m_1 , \cdots ,m_n 
$ and standard deviation $s_1 , \cdots ,s_n $, their affine combination defined as $
\sum\limits_{i = 1}^n {\alpha _i F_i  + \beta } 
$ where the related constants, $\alpha_i, \quad i=1,2,\dots, n$ and $\beta$ are crisp numbers, is also a Gaussian Type-I FSs whose mean and standard deviation can be found by $
m = \sum\limits_{i = 1}^n {\alpha _i m_i  + \beta } 
$ and $
\Sigma ' = \left\{{ \begin{array}{c l}
   \sqrt {\sum\limits_{i = 1}^n {\alpha _i ^2 \sigma _i ^2 }} & \hbox{for product t-nrom}\\ 
   \sum\limits_{i = 1}^n {|\alpha _i |\sigma _i} & \hbox{for min t-norm}\\   
	\end{array}}
   \right.$ respectively.
\end{theorems}
The proof for this theorem is included in Appendix \ref{prf1.5}.
\section{\textbf{Type-II Fuzzy Relations, Compositions and Cartesian Products}}
Let $X_1 ,X_2 , \cdots ,X_n $ be $n$ universes of discourses. A crisp relation in $X_1  \times X_2  \times  \cdots  \times X_n $ is defined as a crisp subset of this product space. Similarly a (Type-I/Type-II) fuzzy relation on $X_1  \times X_2  \times  \cdots  \times X_n $ is defined as a (Type-I/Type-II) fuzzy subset of the aforementioned product space. In this section our focus is on T2 fuzzy relations. For a thorough discussion of T1 fuzzy relations, refer to \cite{Wan97}.
To perform union and intersection of two T2 fuzzy relations we follow the same approach as we did for the union and intersection of two T2 fuzzy sets. More specifically, let $\tilde{R}(U,V)$ and $\tilde{S}(U,V)$ be two fuzzy relations on the product space $U\times V$, their union and intersection are thus determined as, $\mu _{\tilde R \cup \tilde S} (u,v) = \mu _{\tilde R} (u,v)\coprod \mu _{\tilde S} (u,v)$ and $\mu _{\tilde R \cap \tilde S} (u,v) = \mu _{\tilde R} (u,v)\prod \mu _{\tilde S} (u,v)$.
Now, consider two different product spaces, $U\times V$ and $V\times W$ . Let let $R(U,V)$ and $S(U,V)$ be two crisp relations on these spaces, respectively. Obviously, the composition of these relations is
defined as another relation, say $T(U,W)$, such that for every pair of $(u,w)\in T$ , we necessarily have $\exists v \in V \ni (u,v) \in \Re,(v,w) \in S$; conversely if $(u,v) \in \Re,(v,w) \in S$ then $(u,w) \in T$  . This can be expressed in the following sense (also called sup-star composition),
\begin{equation}
\mu _{R \circ S} (u,w) = \sup _{v \in V} [\mu _R (u,v).\mu _S (v,w)]
\label{1.61}
\end{equation}
Where $*$ is any t-norm. Note that in the above formula, we are manipulating crisp sets meaning that the membership grades are either 0 or 1 and consequently the result of product is 1 if $(u,w) \in S$ and $(v,w) \in \Re$ and 0 otherwise. We simply extend \eqref{1.61} to the class of (T1/T2) fuzzy sets but with the only consideration that this time the membership functions are either crisp numbers in the domain [0,1] or a fuzzy numbers in the same domain. Therefore the previous t-norm can not do the job any longer; rather it should be replaced by its counterparts associated to each type (either of T1 or T2) namely, the t-norm or meet respectively. Accordingly, we define the respective composition of T1 and T2 sets as follows,
\begin{equation}
\mu _{R \circ S} (u,w) = \sup _{v \in V} [\mu _R (u,v) * \mu _S (v,w)]
\label{1.62a}
\end{equation}
\begin{equation}
\mu _{\tilde R \circ \tilde S} (u,w) = \coprod _{v \in V} [\mu _{\tilde R} (u,v)\prod \mu _{\tilde S} (v,w)]
\label{1.62b}
\end{equation}
Interested readers can find the proof for \eqref{1.62a}, \eqref{1.62b} in \cite{Kar01b}.
\section{\textbf{Type-II Fuzzy Logic Systems}}
\begin{figure}[htp]
  \begin{center}
    \includegraphics[width=10cm, height=5cm]{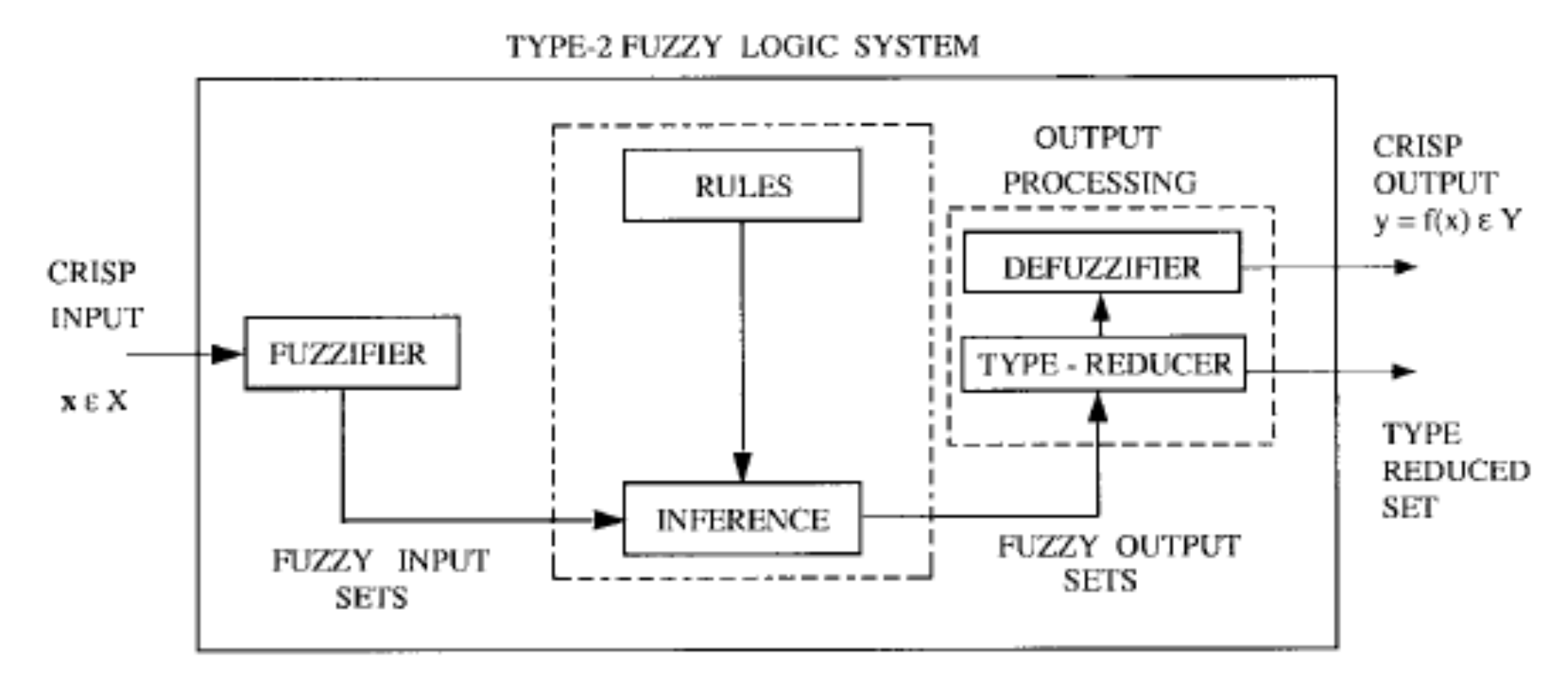}
  \end{center}
  \caption{Architecture of a T2FLS \cite{Kar99c}}
  \label{fig6}
\end{figure}

To date, Type-II fuzzy sets and FLSs have been employed in a vast variety of applications such as clustering \cite{Hwa04,Rhe01}, control of mobile robots \cite{Hag04,Wu96}, decision making \cite{Oze04,Cha87,Yag80}, solving fuzzy relation equations \cite{Wag88}, survey processing \cite{Kar99a}, time series forecasting \cite{Kar99b,Men00}, function approximation \cite{Men98}, and preprocessing of data \cite{Joh00}. There are also other areas where we may take advantage of T2FLSs \cite{Kar99c} including mobile communication, communication networks, pattern recognition and robust control due to existence of uncertain information. Figure \ref{fig6} shows the architecture of a Type-II fuzzy logic system. As easily understood from this block diagram, it takes after the structure of a T1FLS except that the defuzzification block is now replaced by the so-called \emph{output processing} unit, explained shortly. 
In the fuzzification unit the crisp inputs are converted into Type-II fuzzy sets much in the same way a Type-I fuzzifier converts the inputs into Type-I fuzzy sets. The fuzzification in T2FLSs can be classified into two main categories, namely the singelton and non-singleton fuzzification.  The former converts the crisp input into a Type-II fuzzy set having its domain confined to a single point with a singleton Type-I fuzzy set as its fuzzy membership grade and the latter differs in that its domain is an interval instead of a single point with each point in the interval associated to a general Type-I membership function.
The rule base on the other hand, generally contains a set of "IF-THEN" rules having the following form coined as canonical form:
\begin{equation}
\begin{array}{l l}
\hbox{Rule}^1 & \hbox{: if }x_1\hbox{ is }\tilde F_1^1\hbox{ and }x_2\hbox{ is }\tilde F_2^1\hbox{ and } \cdots\hbox{, and }x_p\hbox{ is }\tilde F_p^1\hbox{ then y is }\tilde G^1   \\
\hbox{Rule}^2 & \hbox{: if }x_1\hbox{ is }\tilde F_1^2\hbox{ and }x_2\hbox{ is }\tilde F_2^2\hbox{ and } \cdots\hbox{, and }x_p\hbox{ is }\tilde F_p^2\hbox{ then y is }\tilde G^2    
\\ & \vdots    \\
\hbox{Rule}^M & \hbox{: if }x_1\hbox{ is }\tilde F_1^M\hbox{ and }x_2\hbox{ is }\tilde F_2^M\hbox{ and } \cdots\hbox{, and }x_p\hbox{ is }\tilde F_p^M\hbox{ then y is }\tilde G^M   
\end{array}
\label{1.63}
\end{equation}
where $x_i$s are inputs, $\tilde{F}_i^l$s are the antecedent fuzzy sets, $y$ is the output set and $\tilde{G}$s are the consequent sets. Accordingly, in relation to its Type-I predecessor, the structure of the rule base remains unchanged while the membership functions have been evolved to higher level degrees. Notice that, to call a FLS a T2FLS it is not required for all the antecedents and
consequents to be of Type-II kind, rather a single Type-II fuzzy set in the rule base puts the system in the Type-II category.

As explained in \cite{Wan97}, the inference unit in a T1FLS maps a T1 fuzzy set into another Type-I fuzzy set. Similarly, in the T2FLS, the inference unit maps a T2 fuzzy set into another T2 set. The inference process in the former is established by means of composition of the fuzzy input with the fuzzy relations obtained from each fuzzy rule and later applying sup-star operation. If there are more than one rule, then the output corresponding to each will be merged together using t-conorm operation. In the T2 case however, the so-called join and meet operations are substituted for t-conorm and t-norm 
respectively. Consider a Type-II FLS having   inputs, $x_1 \in X_1 ,\,\,x_2  \in X_2 , \cdots ,x_p  \in X_p$ and one output $y \in Y$.
Assuming that there are M rules available with the $l$th rule having the form
\begin{equation}
\begin{array}{l l}
\hbox{Rule}^l & \hbox{: if }x_1\hbox{ is }\tilde F_1^l\hbox{ and }x_2\hbox{ is }\tilde F_2^l\hbox{ and } \cdots\hbox{, and }x_p\hbox{ is }\tilde F_p^l\hbox{ then y is }\tilde G^l
\end{array}
\label{1.64}
\end{equation}
Similar to T1 rules, this rule represents a Type-II fuzzy relation between the input and the output spaces $X_1  \times X_2  \times  \cdots  \times X_p$ and $Y$, respectively. 
We denote this relation by $\mu _{\tilde F_1^l  \times \tilde F_2^l  \times  \cdots  \times \tilde F_p^l  \to \tilde G^l } (\bar x,y)$ where $\tilde F_1^l  \times \tilde F_2^l  \times  \cdots  \times \tilde F_p^l $ is the Cartesian product of $\tilde F_1^l ,\tilde F_2^l , \cdots ,\tilde F_p^l 
$ and $\bar x = \{ x_1 ,x_2 , \cdots ,x_p \} $. When the input $x'$ is fed to the system, the composition of $\tilde{X}'$ to which $x'$ belongs and the $l$th fuzzy relation is found by using the extended version of the sup-star composition, as follows:
\begin{equation}
\mu _{\tilde X'\, \circ \,\tilde F_1^l  \times \tilde F_2^l  \times  \cdots  \times \tilde F_p^l  \to \tilde G^l } (\bar x,y) = \coprod _{\bar x \in \tilde X'} [\mu _{\tilde X'} (\bar x)\prod \mu _{\tilde F_1^l  \times \tilde F_2^l  \times  \cdots  \times \tilde F_p^l  \to \tilde G^l } (\bar x,y)]
\label{1.65}
\end{equation} 
We denote the output set, $\tilde X'\, \circ \,\tilde F_1^l  \times \tilde F_2^l  \times  \cdots  \times \tilde F_p^l  \to \tilde G^l $ as $\tilde{B}^l$. When the singleton fuzzification is used, $\tilde A^l$, reduces to,
\begin{equation}
\mu _{\tilde X' \circ \,\tilde F_1^l  \times \tilde F_2^l  \times  \cdots  \times \tilde F_p^l  \to \tilde G^l } (\bar x,y) = \mu _{\tilde F_1^l  \times \tilde F_2^l  \times  \cdots  \times \tilde F_p^l  \to \tilde G^l } (\bar x',y)
\label{1.66}
\end{equation} 
where $\bar{x}'$  is the input applied to the system. To compute the membership grade associated to Cartesian product of Type-II fuzzy sets, we perform meet operations between the corresponding membership functions. Hence, \eqref{1.66} can be rewritten as,
\begin{equation}
\begin{array}{l}
\mu _{\tilde B^l } (y) = \mu _{\tilde F_1^l \times  \cdots  \times \tilde F_p^l } (\bar x')\prod \mu _{\tilde G^l } (y)  =\mu _{\tilde F_1^l } (x'_1 )\prod \mu _{\tilde F_2^l } (x'_2 )\prod  \cdots \prod \mu _{\tilde F_p^l } (x'_p )\prod \mu _{\tilde G^l } (y) \\= \mu _{\tilde G^l } (y)\prod [\prod\limits_{i = 1}^p {\mu _{\tilde F_i^l } (x'_i )]} 
\end{array}
\label{1.67}
\end{equation}
where we used the commutative property of the meet operation.\\
In a Type-I FLS, using individual rule of inference mechanism, the output corresponding to each rule is a Type-I fuzzy set. Since there might be many fired rules the defuzzifier firstly combines them all to convert these fuzzy sets into a single one representing the fuzzy output and then extracts a crisp value from this single set. Just as a Type-I defuzzifier combines all the Type-I output sets, a type reducer joins the Type-II fuzzy sets corresponding to the fired rules and then reduces the resulting set into a Type-I fuzzy set. There are several type reducers introduced to date, among which \emph{centroid, center of sets} and \emph{height} are most commonly used.
\subsection{\textbf{Type Reduction (TR)}}
\subsubsection{Centroid Type Reduction}
The \emph{centroid} type reducer is among the well-known TR procedures.
Consider a Type-I fuzzy set, $A$, with continuous domain. The centroid defuzzifier (which converts a Type-I fuzzy set into a Type-0 or crisp number) for this set is given as follows:
\begin{equation}
c_A^c  = {{\int\limits_{x \in A} {x.\mu _A (x)dx} } \over {\int\limits_{x \in A} {\mu _A (x)dx} }}
\label{1.68}
\end{equation}
Note that in digital world applications, however, we have to discretize the domain so that we can calculate the centroid numerically. This yields a number which is computed by
\begin{equation}
c_A^d  = {{\sum\limits_{i = 1}^N {x_i .\mu _A (x_i )} } \over {\sum\limits_{i = 1}^N {\mu _A (x_i )} }}
\label{1.69}
\end{equation}
In the same way we can define the centroid type reduction for a Type-II fuzzy set whose primary domain has been discretized into N points, using extension principle \cite{Zad75}. Assuming $D_i  = \mu _{\tilde A} (x_i )$ follows that
\begin{equation}
C_{\tilde A}  = \mathop{\int\!\!\!\int\!\!\!\int}\limits_{\kern-5.5pt {\theta _1 , \cdots ,\theta _N }} 
 {{{[\mu _{D_1 } (\theta _1 ) *  \cdots  * \mu _{D_N } (\theta _N )]} \mathord{\left/
 {\vphantom {{[\mu _{D_1 } (\theta _1 ) *  \cdots  * \mu _{D_N } (\theta _N )]} {{{\sum\limits_{i = 1}^N {x_i .\theta _i } } \over {\sum\limits_{i = 1}^N {\theta _i } }}}}} \right.
 \kern-\nulldelimiterspace} {{{\sum\limits_{i = 1}^N {x_i .\theta _i } } \over {\sum\limits_{i = 1}^N {\theta _i } }}}}} 
\label{1.70}
\end{equation}
where $\theta_i \in D_i$. Notice that the integral sign is the common notation prevalent in fuzzy literature which denotes the continuous fuzzy sets. Note also that the secondary domain is still continuous; however, this does not cause much trouble for us if the secondary membership functions are interval T1 fuzzy sets, as explained shortly. In other words, for IT2 fuzzy sets it is not required to discretize the secondary domain. Another point is that although not apparent, the supremum is indirectly available in the context of the previous formula. Since according to definition of fuzzy sets, for a domain point with different membership grades the one with highest membership values is considered and the rest are omitted. That is to say, \eqref{1.70} can be restated as 
\begin{equation}
C_{\tilde A}  = Y_C (x) = \int\limits_{x\{ \theta _1 , \cdots ,\theta _N \} } {\sup \,\,{{[\mu _{D_1 } (\theta _1 ) *  \cdots  * \mu _{D_N } (\theta _N )]} \mathord{\left/
 {\vphantom {{[\mu _{D_1 } (\theta _1 ) *  \cdots  * \mu _{D_N } (\theta _N )]} x}} \right.
 \kern-\nulldelimiterspace} x}} 
\label{1.71}
\end{equation}
where $\{ \theta _1 ,\theta _2 , \cdots ,\theta _N \}$ are such that $x = {{\sum\limits_{i = 1}^N {x_i .\theta _i } } \over {\sum\limits_{i = 1}^N {\theta _i } }}$ and $*$ represents the t-norm operator. We can think of the subsequent terms in
$\{ \theta _1 ,\theta _2 , \cdots ,\theta _N \}$ to form a membership function of some Type-I set, say $A'$  having the same domain as $\tilde A$ . We call such set an \emph{embedded} Type-I set since it is embedded inside a higher type fuzzy set disregarding its secondary membership grades. Keeping this in mind, the centroid  is a fuzzy set whose elements are the centroid of all the embedded Type-I fuzzy sets, these sets can be thought to represent the output of a an associated T1FLS, we will talk about this line of thought soon. Also the membership assigned to each embedded set centroid is the t-norm of the memberships associated with each combination  $\{ \theta _1 ,\theta _2 , \cdots ,\theta _N \}$ that makes up the embedded set \cite{Kar99c}. Note that we should select the foregoing t-norm operation carefully. Usually the minimum t-norm is used for the centroid calculation and the product t-norm is used elsewhere. To give you a sense of why this is necessary to perform, consider the case where $N$
tends to infinity. If we select the product t-norm operation, then the membership value approaches zero (unless all except some countable number of them are 1). Assuming that the secondary membership functions are normal fuzzy sets, the resulting centroid would eventually converge to the principal membership function of the underlying Type-II fuzzy set, which is an undesired outcome since in that case our Type-II FLS has been reduced to a single Type-I FLS incurring severely imposed computational overheads. Note that this problem does not occur in IT2FS as the secondary membership grades are either 1 or 0 (refer to \cite{Kar01a} for a further explanations). The reader should observe that if we discretize domain of each $D_i$ into $M$ points, the number of resulting embedded fuzzy sets amounts to $M^N$ which is considerably large even for small $M$ and $N$. This makes it evident why the general Type-II FLSs cannot be realized for real-time applications. In theory the amount of computations required for a continuous general T2FLS is infinitely high due to an uncountable number of existing embedded sets. However, if we know in advance that the related T2 fuzzy set follows some regularities then the computation overhead may be relieved substantially by computing an approximation or an exact centroid set that makes it suitable for implementations. 
The sequence of computations required to obtain $C_{\tilde A}$ is as follows:
\begin{enumerate}
\item Combine all the rules' outputs using join or meet operation to find the final T2 output set, as follows:
\begin{equation}
\mu _{\tilde B} (y) = \coprod\limits_{l = 1}^M {\mu _{\tilde B^l } (y)} 
\label{1.72}
\end{equation}
Note that we adopt the join policy in this thesis for rules' output combination.
\item Discretize the outputs space $Y$ into $N$ points,  $y_1 ,y_2 , \cdots ,y_N $
\item Discretize the secondary domain associated with each sample, $y_i$, into $M_i$ which amount to $M = \prod\limits_{j = 1}^N {M_j }$ number of embedded sets.
\item Compute the type reduced set using \eqref{1.70}
\end{enumerate}

\begin{figure}[htp]
  \begin{center}
    \includegraphics[width=10cm, height=5cm]{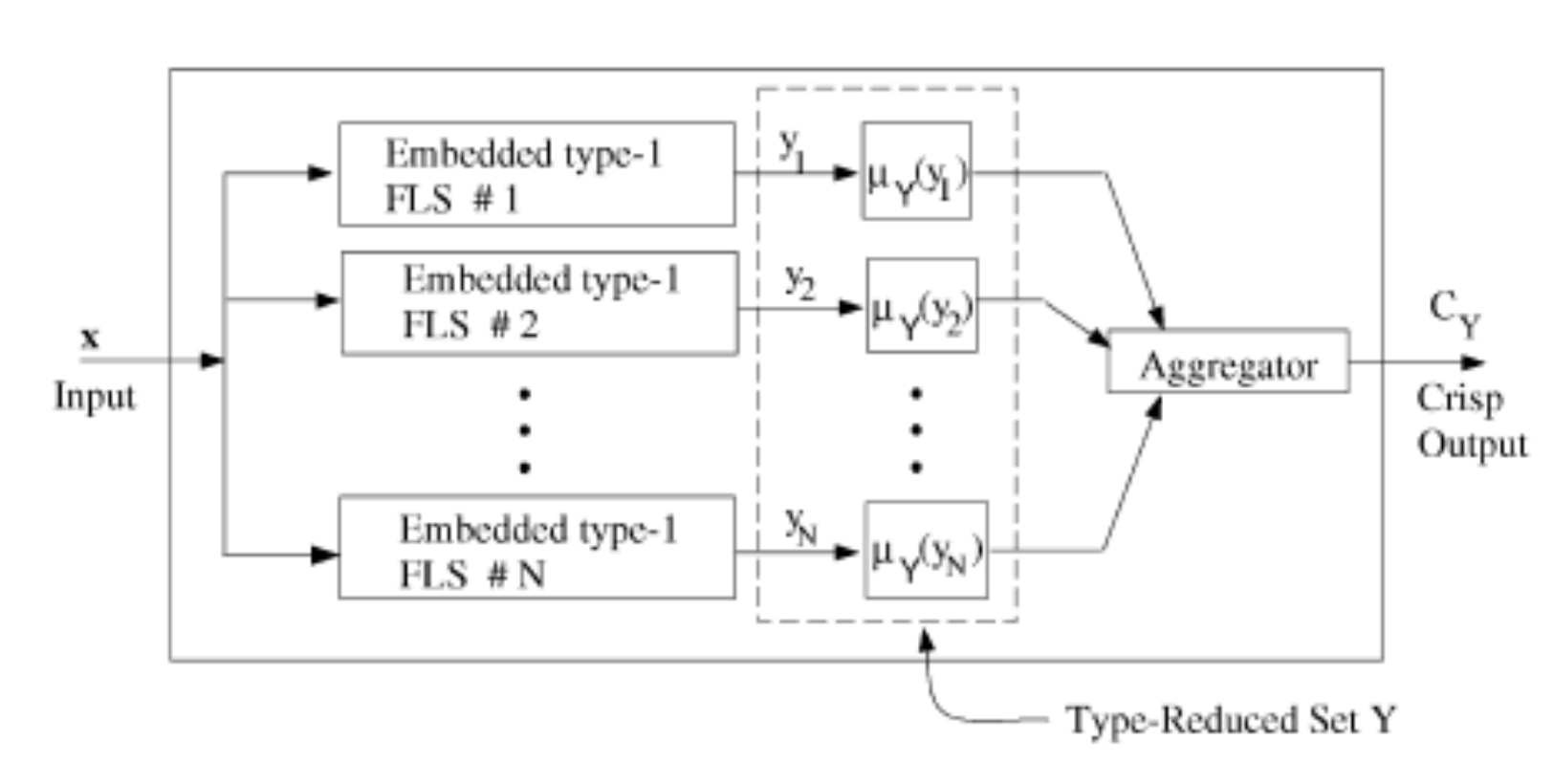}
  \end{center}
  \caption{A general T2FLS is a combination of many T1FLSs \cite{Men00}}
  \label{fig7}
\end{figure}

As discussed yet and later, the type reduced output set of a T2FLS is the centroid of some Type-II fuzzy set (T2FS), and as explained earlier, this set is comprised of the centroids of embedded Type-I fuzzy sets each of which can be thought of as an output for a T1FLS. That is why it is sometimes stated that a T2FLS is in fact a collection of numerous T1FLSs \cite{Men00}. This also makes evident why the class of T2FLSs are more powerful than the T1 FLSs but at cost of much more computational complexity. Figure \ref{fig7} depicts this idea graphically.

Unfortunately to date, only special classes (Interval, Gaussian, ... T2FLS) of general T2FLSs are shown to be realizable in real-time applications due to the high computational burden. The computation bottleneck of the related computations in such systems reside in the already introduced type reduction operation. For the case of an interval Type-II fuzzy logic system (IT2FLS), as will be discussed later, this amount of computations could be reduced to only two T1FLS, each corresponding to one of the output bounds (more on this later.) This drastically decreases the computation burden making it feasible for real-time applications. The related procedures are addressed extensively in \cite{Kar01a,Men02a}. The most significant reason for this faster execution comes from the fact that regardless of the type of fuzzification (singleton or non-singleton) , type reduction (center of sets, centroid, height and etc.) and even inference mechanism the type reduced set is an IT1 set which can be easily described via upper and lower limits. The length of this fuzzy set may be considered as a measure of uncertainty present in the output \cite{Men02a} and works much the same way as a confidence interval when uncertainty is available. This is of the most significant contributions of T2FL in the T2FLS area, which is not a characteristic of their traditional Type-I counterparts.
\subsubsection{Height Type Reduction}
The \emph{height defuzzifier} \cite{Dri96} puts a singleton in the \emph{combined set} for each output set in place with associated maximum membership value and then calculates the centroid of the resulting set. That is,
\begin{equation}
y_h (x) = {{\sum\limits_{l = 1}^M {\bar y^l .\mu _{B^l } (\bar y^l )} } \over {\sum\limits_{l = 1}^M {\mu _{B^l } (\bar y^l )} }}
\label{1.73}
\end{equation}
where $\bar y^l$ is the point with maximum membership value in the $l$th output set (if there is more than one such point then the average value may be considered) for which the corresponding membership value, $\mu _{B^l } (\bar y^l )$  is given as,
\begin{equation}
\mu _{B^l } (\bar y^l ) = \mu _{G^l } (\bar y^l ) * \Im _{i = 1}^p \mu _{F_i^l } (x_i )
\label{1.74}
\end{equation}
where $*$, $\Im$  indicate the selected t-norms.\\ 
Similarly, the \emph{height} TR, replaces each T2 output set by a T2 singleton (i.e. a fuzzy set whose domain is a single point associated to which is a T1 FS in $[0,1]$) situated at $\bar y^l$ where the principle membership function has the maximum membership value. This value can be found by,
\begin{equation}
\mu _{\tilde B^l } (\bar y^l ) = \mu _{\tilde G^l } (\bar y^l )\,\Pi \,[\Pi _{i = 1}^p \mu _{\tilde F_i^l } (x_i )]
\label{1.75}
\end{equation}
Letting $D^l  = \mu _{\tilde B^l } (\bar y^l )$ the extended version of \eqref{1.73} can be expressed as,
\begin{equation}
Y_h (x) = \mathop{\int\!\!\!\int\!\!\!\int}\limits_{\kern-5.5pt {\theta _1 , \cdots ,\theta _M }} 
 {{{[\mu _{D^1 } (\theta _1 ) *  \cdots  * \mu _{D^M } (\theta _M )]} \mathord{\left/
 {\vphantom {{[\mu _{D^1 } (\theta _1 ) *  \cdots  * \mu _{D^M } (\theta _M )]} {{{\sum\limits_{l = 1}^M {\bar y^l .\theta _l } } \over {\sum\limits_{l = 1}^M {\theta _l } }}}}} \right.
 \kern-\nulldelimiterspace} {{{\sum\limits_{l = 1}^M {\bar y^l .\theta _l } } \over {\sum\limits_{l = 1}^M {\theta _l } }}}}} 
\label{1.76}
\end{equation}
where $\theta_l \in D_l$ for $l=1,2,\dots,M$.
The sequence of computations to find $Y_h(x)$ is therefore as follows:
\begin{enumerate}
\item Find the eligible $\bar y^l$ for each output set and compute the associated $
\mu _{\tilde B^l } (\bar y^l )$ according to \eqref{1.74}
\item Discretize the domain for each $\bar y^l $ into required number of points, $N_j\quad (j=1,2,\dots,M)$, similar to the procedure followed in centroid TR.
\item Compute $Y_h(x)$ according to \eqref{1.76}. 
\end{enumerate}
It's worth noting that the t-norm operation in height TR can include the product t-norm as well as other types of t-norms. This is as opposed to centroid TR where due to already presented explanations excludes the employment of product t-nrom. Briefly speaking, the number of terms included in the t-norm operation is a constant while in the former it may tend to infinity. Another point to note is that centroid based T2FLSs and height based T2FLSs should perform the weighted sum computation $\prod\limits_{j = 1}^N {M_j } $ and $\prod\limits_{j = 1}^M {M_j } $ times, respectively. Generally $M<<N$ which means the amount of computation needed for the former is exponentially greater than the latter.
The main drawback with the height type reducer arises when for some special input, 
only one rule fires in which case the output does not represent a confidence interval. This is certainly undesirable as will be 
explained shortly. In the next subsection, we introduce a new type of type reduction which overcomes this shortcoming. 

Consider an input, $x$, for which the only fired rule is the $l=l'$th one. Accordingly, \eqref{1.76} yields,
\begin{equation}
Y_h (x) = \int\limits_{\theta _{l'} } {{{\mu _{D^{_{l'} } } (\theta _{_{l'} } )} \mathord{\left/
 {\vphantom {{\mu _{D^{_{l'} } } (\theta _{_{l'} } )} {{{\bar y^{l'} .\theta _{l'} } \over {\theta _{l'} }}}}} \right.
 \kern-\nulldelimiterspace} {{{\bar y^{l'} .\theta _{l'} } \over {\theta _{l'} }}}}}  = [\begin{array}{c}\\
   \sup \,\,\mu _{D^{_{l'} } } (\theta _{_{l'} } )\\
   {l'} 
\end{array}
]/\bar y^{l'}  = 1/\bar y^{l'}  = \bar y^{l'} 
\label{1.77}
\end{equation}
Equation \eqref{1.77} reveals the fact that in height TR, the result may reduce to a single point or equivalently we have lost the confidence interval which is one of the most significant contributions of T2FLSs. 
The next topic resolves problem with height TR.
\subsubsection{Center of Sets Type Reduction}
The \emph{center of sets} defuzzifier replaces each consequent set with a singleton located at its centroid point and then combines together the resulting sets and computes the centroid of the outcome. The expression for the output is thus given by,
\begin{equation}
y_{\cos } (x) = {{\sum\limits_{l = 1}^M {c^l \Im _{i = 1}^p \mu _{F_i^l } (x_i )} } \over {\sum\limits_{l = 1}^M {\Gamma _{i = 1}^p \mu _{F_i^l } (x_i )} }}
\label{1.78}
\end{equation}
where $\Gamma$ indicates the opted t-norm and $c^l$ is centroid of the $l$th consequent set. Note that for a special case where consequent sets are normal, convex and symmetric, $Y_{cos}(x)=Y_h(x)$ .\\
The center of sets type reducer on the other hand, computes the centroid of each consequent set (notice that this centroid is itself a T1FS) and then finds the weighted average 
of these centroid sets using Extension Principle \cite{Zad75}. The related weights are in fact the firing strengths that are calculated as follows,
\begin{equation}
E_l  = \Pi _{i = 1}^p \mu _{\tilde F_i^l } (x_i )
\label{1.79}
\end{equation}
Put it together, the type reduced set in the aforementioned TR procedure is computed as follows:
\begin{equation}
Y_{\cos } (x) = \mathop{\int\!\!\!\int\!\!\!\int}\limits_{\kern-5.5pt {d_1 ,d_2 , \ldots ,d_M }} 
 {\,\,\mathop{\int\!\!\!\int\!\!\!\int}\limits_{\kern-5.5pt {e_1 ,e_2 , \ldots ,e_M }} 
 {{{\Im _{l = 1}^M \mu _{C^l } (d_l ) * \Im _{l = 1}^M \mu _{E^l } (e_l )} \mathord{\left/
 {\vphantom {{\Im _{l = 1}^M \mu _{C^l } (d_l ) * \Im _{l = 1}^M \mu _{E^l } (e_l )} {{{\sum\limits_{l = 1}^M {d_l e_l } } \over {\sum\limits_{l = 1}^M {e_l } }}}}} \right.
 \kern-\nulldelimiterspace} {{{\sum\limits_{l = 1}^M {d_l e_l } } \over {\sum\limits_{l = 1}^M {e_l } }}}}} } 
\label{1.80}
\end{equation}
where $\Gamma$, $*$ are the selected t-norm operators and $d_l  \in C_l  = C_{\tilde G^l } ,e_l  \in E_l$ are the associated centroids and firing strengths, respectively. \\
It should be noted that in this type of TR, we are no longer constrained by the previous pitfall when only one rule is fired. To see this, let $l=l'$ be the only fired rule; in that case type reduced set becomes:
\begin{equation}
\begin{array}{l}
Y_{\cos } (x) = \int\limits_{d^{l'} } {\int\limits_{e^{l'} } {{{\mu _{C^{l'} } (d_{l'} ) * 
\mu _{E^{l'} } (e_{l'} )} \left/ {{d_{l'}e_{l'}} \over {e_{l'}}}\right.}}}=
{\int\limits_{d^{l'} } {\mu _{C^{l'} } (d_{l'} ) * \int\limits_{e^{l'} } {\mu _{E^{l'} } (e_{l'} )} } } \mathord{\left/
 {\vphantom {{\int\limits_{d^{l'} } {\mu _{C^{l'} } (d_{l'} ) * \int\limits_{e^{l'} } {\mu _{E^{l'} } (e_{l'} )} } } {d_{l'} }}} \right.
 \kern-\nulldelimiterspace} {d_{l'}} \\
 = {\int\limits_{d^{l'} } {\mu _{C^{l'} } (d_{l'} ) * [\begin{array}{l}
   \\
	\sup \,\,\mu _{E^{l'} } (e_{l'} )\\
   e_{l'}
	\end{array} 
 ]}} \mathord{\left/
 d_{l'}\right.}
\end{array}
\label{1.81}
\end{equation}
Assuming that the firing strengths are normal FSs (in fact if each $\mu _{\tilde F_i^l } (x_i )$ is normal, then $E_l$ will also be normal), as a consequence we get,
\begin{equation}
Y_{\cos } (x) = {{\int\limits_{d^{l'} } {\mu _{C^{l'} } (d_{l'} )} } \mathord{\left/
 {\vphantom {{\int\limits_{d^{l'} } {\mu _{C^{l'} } (d_{l'} )} } {d_{l'} }}} \right.
 \kern-\nulldelimiterspace} {d_{l'} }} = C^{l'} 
\label{1.82}
\end{equation}
The only drawback with this type of TR is that in the special case already mentioned, antecedents do not take part in the type reduced set, while this is not the case in centroid TR.

To sum up, the following steps should be performed sequentially to find the type reduced set in a T2FLS,
\begin{enumerate}
\item Discretize the output space $Y$ into relevant number of points and compute $C_l$, the consequent centroid set, for each rule according to \eqref{1.70}. As these sets are independent of future computations, we can find and store them in advance for future uses.
\item Compute the degrees of firing $E_l  = \Pi _{i = 1}^p \mu _{\tilde F_i^l } (x_i )
$ associated with the rule consequents.
\item Discretize the domain of each $C_l$ into suitable number of points say $M_l\quad(l = 1,2,...,M)$ 
\item Discretize the domain of each $E_l$ into suitable number of points say $M_l\quad(l = 1,2,...,M)$.  
\item Enumerate all possible combinations $
\{ c_1 ,c_2 , \cdots ,c_M ,e_1 ,e_2 , \cdots ,e_M \} 
$ where $d_l  \in C_l  = C_{\tilde G^l } ,e_l  \in E_l $ , hence a total number of $\Pi _{j = 1}^M M_j N_j $.
\item Finally, compute the type reduced set using \eqref{1.80} 
\end{enumerate}
The final thing to note is that as explained earlier, a T2FLS can be thought of as a collection of T1FLSs 
operating in parallel \cite{Mel04}. This reminds the notion of parallel processing for FLSs. Therefore the time complexity of a T2FLS can be drastically decreased if we decompose it into relevant T1FLSs that work simultaneously, 
the gain is the considerable speed up and the pain is the implementation cost.

Up to now, we have discussed the most employed type reduction techniques namely the height TR, the centroid TR and the center of sets TR. Lets restate these three once more,
\begin{equation}
\begin{array}{l}
Y_c  = \mathop{\int\!\!\!\int\!\!\!\int}\limits_{\kern-5.5pt {\theta _1 , \cdots ,\theta _N }} 
 {{{[\mu _{D_1 } (\theta _1 ) *  \cdots  * \mu _{D_N } (\theta _N )]} \mathord{\left/
 {\vphantom {{[\mu _{D_1 } (\theta _1 ) *  \cdots  * \mu _{D_N } (\theta _N )]} {{{\sum\limits_{i = 1}^N {x_i .\theta _i } } \over {\sum\limits_{i = 1}^N {\theta _i } }}}}} \right.
 \kern-\nulldelimiterspace} {{{\sum\limits_{i = 1}^N {x_i .\theta _i } } \over {\sum\limits_{i = 1}^N {\theta _i } }}}}} \\
Y_h (x) = \mathop{\int\!\!\!\int\!\!\!\int}\limits_{\kern-5.5pt {\theta _1 , \cdots ,\theta _M }} 
 {{{[\mu _{D^1 } (\theta _1 ) *  \cdots  * \mu _{D^M } (\theta _M )]} \mathord{\left/
 {\vphantom {{[\mu _{D^1 } (\theta _1 ) *  \cdots  * \mu _{D^M } (\theta _M )]} {{{\sum\limits_{l = 1}^M {\bar y^l .\theta _l } } \over {\sum\limits_{l = 1}^M {\theta _l } }}}}} \right.
 \kern-\nulldelimiterspace} {{{\sum\limits_{l = 1}^M {\bar y^l .\theta _l } } \over {\sum\limits_{l = 1}^M {\theta _l } }}}}} \\
Y_{\cos } (x) = \mathop{\int\!\!\!\int\!\!\!\int}\limits_{\kern-5.5pt {d_1 ,d_2 , \ldots ,d_M }} 
 {\,\,\mathop{\int\!\!\!\int\!\!\!\int}\limits_{\kern-5.5pt {e_1 ,e_2 , \ldots ,e_M }} 
 {{{\Im _{l = 1}^M \mu _{C^l } (d_l ) * \Im _{l = 1}^M \mu _{E^l } (e_l )} \mathord{\left/
 {\vphantom {{\Im _{l = 1}^M \mu _{C^l } (d_l ) * \Im _{l = 1}^M \mu _{E^l } (e_l )} {{{\sum\limits_{l = 1}^M {d_l e_l } } \over {\sum\limits_{l = 1}^M {e_l } }}}}} \right.
 \kern-\nulldelimiterspace} {{{\sum\limits_{l = 1}^M {d_l e_l } } \over {\sum\limits_{l = 1}^M {e_l } }}}}} } 
\end{array}
\label{1.83}
\end{equation}
Considering \eqref{1.83}, observe how each of them appears in a similar form. That is to say, regardless of the indexes $M$ and $N$, the first two are the same and last can be thought of as a more general form of the above two. In other words, the only difference between the center of sets TR and the previous ones is that in the former, the coefficients, $d_l$ , are fuzzy sets as well, 
while for the latter cases (the centroid and height TRs ) the associated coefficients are crisp number. Hence, since any crisp number can be considered as a fuzzy set whose domain includes only one element equal to related crisp number (with unity degree of membership), we can treat them all in the same framework. Accordingly, we develop new techniques for the last formula keeping in mind that the results are applicable to the other two, as well. For convenience we henceforth call the last formula, the generalized centroid, due the explanations given above.
\subsubsection{Different TR Methods Based on a Realization Perspective}
The generalized centroid developed earlier is not amenable to computer (or even hardware) implementation which motivates us to search for new ways and techniques that address the implementation issues in their context. Unfortunately, except for the very special (but important) case of interval Type-II fuzzy logic systems (IT2FLSs), there is not any technique proposed yet that results in an exact calculation of the generalized centroid TR, however few approximation methods have been also 
developed in \cite{Men02a,Kar01a} which is illustrated in this thesis.
\subsubsection{An Approximate Method for Generalized Centroid}
Karnik et. al. \cite{Kar01a}, have presented a new method for calculation of generalized centroid provided that the amount of Type-II uncertainty is small, as follows,
\begin{theorems} \label{approxgc}
Consider the \emph{generalized centroid} bellow,
\begin{equation}
Y_{gc}  = \int_{z_1  \in Z_1 } { \cdots \int_{z_N  \in Z_N } {\int_{w_1  \in W_1 } { \cdots \int_{w_N  \in W_N } {{{\Im _{l = 1}^N \mu _{Z^l } (z_l ) * \Im _{l = 1}^N \mu _{W^l } (w_l )} \mathord{\left/
 {\vphantom {{\Im _{l = 1}^N \mu _{Z^l } (z_l ) * \Im _{l = 1}^N \mu _{W^l } (w_l )} {{{\sum\limits_{l = 1}^N {w_l z_l } } \over {\sum\limits_{l = 1}^N {w_l } }}}}} \right.
 \kern-\nulldelimiterspace} {{{\sum\limits_{l = 1}^N {w_l z_l } } \over {\sum\limits_{l = 1}^N {w_l } }}}}} } } } \,
\label{1.84}
\end{equation}
where $\Im$ and $*$ indicate the t-norm operations. Presuming that each $Z_l$ is a Type-I fuzzy set with support $[c_l  - s_l ,c_l  + s_l ]$ and each $W_l$ is also a Type-I fuzzy set with support $[h_l  - \Delta _l ,h_l  + \Delta _l ]$ , then we have  
\begin{equation}
Y_{gc}  \approx \sum\limits_{l = 1}^N {\left[ {\underbar {Z}_l \left( {{{h_l } \over {\sum\limits_{l = 1}^N {h_l } }}} \right) + \underbar {W}_l \left( {{{c_l  - \xi } \over {\sum\limits_{l = 1}^N {h_l } }})} \right)} \right]}  + \xi 
\label{1.85}\
\end{equation}
where
\begin{equation}
  \underbar{ Z}_l  = Z_l  - c_l 
\label{1.86a}	
\end{equation}
\begin{equation}
  \underbar{W}_l  = W_l  - h_l 
\label{1.86b}
\end{equation}
and 
\begin{equation}
\xi  = {{\sum\limits _{l = 1}^N {h_l c_l } } \over {h_l }}
\label{1.87}
\end{equation}
provided that 
\begin{equation}
{\sum\limits_{l = 1}^N {\Delta _l }  \over {\sum\limits_{l = 1}^N {h_l }}} \ll 1
\label{1.88}
\end{equation}
The approximation gets more rigorous as the above metric grows smaller and is exact whenever $
\sum\nolimits_{l = 1}^N {\Delta _l }  = 0$ which implies that $\Delta_l=0$  for $l=1,2,�,N$. Note that we have not imposed any 
restriction on $Z_l$ and $W_l$ except that they are arbitrary fuzzy numbers. The significance of this theorem becomes more evident if we know that the right hand side of \eqref{1.85} is much easier to compute than RHS of \eqref{1.84} and hence it relieves the computational burden considerably.\\ 
\emph{Proof}: Defining slack variables $\gamma _l  = z_l  - c_l $ and $\delta _l  = w_l  - h_l $ for $l=1,2,\dots,N$ (1.83) becomes,
\begin{equation}
\begin{array}{l}
Y_{gc}  = \int_{\gamma _1 } { \cdots \int_{\gamma _N } {\int_{\delta _1 } { \cdots \int_{\delta _N } {{{\Im _{l = 1}^N \mu _{Z^l } (c_l  + \gamma _l ) * \Im _{l = 1}^N \mu _{W^l } (h_l  + \delta _l )}}}}}} \\
{{{{{ \mathord{\left/
 {\vphantom {{\Im _{l = 1}^N \mu _{Z^l } (c_l  + \gamma _l ) * \Im _{l = 1}^N \mu _{W^l } (h_l  + \delta _l )} {{{\sum\limits_{l = 1}^N {(h_l  + \delta _l )} (c_l  + \gamma _l )} \over {\sum\limits_{l = 1}^N {(h_l  + \delta _l )} }}}}} \right.
 \kern-\nulldelimiterspace} {{{\sum\limits_{l = 1}^N {(h_l  + \delta _l )} (c_l  + \gamma _l )} \over {\sum\limits_{l = 1}^N {(h_l  + \delta _l )} }}}}} } } } \,
\end{array}
\label{1.89}
\end{equation}
where each $\gamma_l$ takes values in $[-s_l,s_l]$ and each $\delta_l$ takes values inside $[-\Delta_l \Delta_l]$. The term to the right of the slash in \eqref{1.89} can therefore be rephrased as,
\begin{equation}
{{\sum\limits_{l = 1}^N {z_l w_l } } \over {\sum\limits_{l = 1}^N {z_l } }} = {{\sum\nolimits_l {h_l c_l }  + \sum\nolimits_l {h_l \gamma _l }  + \sum\nolimits_l {\delta _l c_l  + \sum\nolimits_l {\delta _l \gamma _l } } } \over {\sum\nolimits_l {h_l }  + \sum\nolimits_l {\delta _l } }}
\label{1.90}
\end{equation}  
Let expand the denominator of \eqref{1.90} and replace ${1 \over {1 + \left( {{{\sum\nolimits_l {\delta _l } } \over {\sum\nolimits_l {h_l } }}} \right)}}
$ with its first order approximation as follows,
\begin{equation}
{1 \over {\sum\nolimits_l {h_l }  + \sum\nolimits_l {\delta _l } }} = {1 \over {\sum\nolimits_l {h_l } }}\left( {{1 \over {1 + \left( {{{\sum\nolimits_l {\delta _l } } \over {\sum\nolimits_l {h_l } }}} \right)}}} \right) \approx {1 \over {\sum\nolimits_l {h_l } }}\left( {1 - \left( {{{\sum\nolimits_l {\delta _l } } \over {\sum\nolimits_l {h_l } }}} \right)} \right)
\label{1.91}
\end{equation}
which is true for ${{|\sum\nolimits_l {\delta _l } |} \over {\sum\nolimits_l {h_l } }} \ll 1
$ or ${{\sum\nolimits_l {\Delta _l } } \over {\sum\nolimits_l {h_l } }} \ll 1
$. Putting the result of \eqref{1.91} into \eqref{1.90} yields,
\begin{equation}
{{\sum\limits_{l = 1}^N {w_l z_l } } \over {\sum\limits_{l = 1}^N {w_l } }} = {{\sum\nolimits_l {h_l c_l }  + \sum\nolimits_l {h_l \gamma _l }  + \sum\nolimits_l {\delta _l c_l  + \sum\nolimits_l {\delta _l \gamma _l } } } \over {\sum\nolimits_l {h_l } }}\left( {1 - {{\sum\nolimits_l {\delta _l } } \over {\sum\nolimits_l {h_l } }}} \right)
\label{1.92}
\end{equation}
As a second approximation, we neglect the terms containing powers of $\left( {{{\sum\nolimits_l {\delta _l } } \over {\sum\nolimits_l {h_l } }}} \right)$ higher than 1, to get,
\begin{equation}
\begin{array}{l}
{{\sum\limits_{l = 1}^N {w_l z_l } } \over {\sum\limits_{l = 1}^N {w_l } }} = {{\sum\nolimits_l {h_l c_l } } \over {\sum\nolimits_l {h_l } }} - {{\sum\nolimits_l {\delta _l } } \over {\sum\nolimits_l {h_l } }}\left( {{{\sum\nolimits_l {h_l c_l } } \over {\sum\nolimits_l {h_l } }}} \right) + \,{{\sum\nolimits_l {h_l \gamma _l } } \over {\sum\nolimits_l {h_l } }} - {{\sum\nolimits_l {\delta _l } } \over {\sum\nolimits_l {h_l } }}\left( {{{\sum\nolimits_l {h_l \gamma _l } } \over {\sum\nolimits_l {h_l } }}} \right) 
\\
+ {{\sum\nolimits_l {\delta _l c_l } } \over {\sum\nolimits_l {h_l } }} + \,{{\sum\nolimits_l {\delta _l \gamma _l } } \over {\sum\nolimits_l {h_l } }}
\end{array}
\label{1.93}
\end{equation}
Let $\xi  = {{\sum\nolimits_{l = 1}^N {h_l c_l } } \over {h_l }}
$ ; then \eqref{1.93} can be approximated as,
\begin{equation}
{{\sum\limits_{l = 1}^N {w_l z_l } } \over {\sum\limits_{l = 1}^N {w_l } }} \approx \xi  - \xi {{\sum\nolimits_l {\delta _l } } \over {\sum\nolimits_l {h_l } }} + {{\sum\nolimits_l {\delta _l c_l } } \over {\sum\nolimits_l {h_l } }} + \,\,{{\sum\nolimits_l {h_l \gamma _l } } \over {\sum\nolimits_l {h_l } }} - {{\sum\nolimits_l {\delta _l } } \over {\sum\nolimits_l {h_l } }}\left( {{{\sum\nolimits_l {h_l \gamma _l } } \over {\sum\nolimits_l {h_l } }}} \right) + \,{{\sum\nolimits_l {\delta _l \gamma _l } } \over {\sum\nolimits_l {h_l } }}
\label{1.94}
\end{equation}
Considering the last two terms in \eqref{1.93}, we have,
\begin{equation}
\left| {{{\sum\nolimits_l {\delta _l \gamma _l } } \over {\sum\nolimits_l {h_l } }}} \right| \le \left| {\mathop {\max }\limits_l \gamma _l \left( {{{\sum\nolimits_l {\delta _l } } \over {\sum\nolimits_l {h_l } }}} \right)} \right| \le \mathop {\max }\limits_l s_l \left| {\left( {{{\sum\nolimits_l {\delta _l } } \over {\sum\nolimits_l {h_l } }}} \right)} \right|
\label{1.95}
\end{equation}
since $\left| {\gamma _l } \right| \le s_l $;
Similarly, ${{\sum\nolimits_l {h_l \gamma _l } } \over {\sum\nolimits_l {h_l } }} \le \mathop {\max }\limits_l s_l
\label{1.96}
$.
Therefore the result of summation over the last two terms in \eqref{1.91} can be bounded from above as follows,
begin
\begin{equation}
\begin{array}{l}
\left| { - {{\sum\nolimits_l {\delta _l } } \over {\sum\nolimits_l {h_l } }}\left( {{{\sum\nolimits_l {h_l \gamma _l } } \over {\sum\nolimits_l {h_l } }}} \right) + \,{{\sum\nolimits_l {\delta _l \gamma _l } } \over {\sum\nolimits_l {h_l } }}} \right| \le \left| {{{\sum\nolimits_l {\delta _l } } \over {\sum\nolimits_l {h_l } }}\left( {{{\sum\nolimits_l {h_l \gamma _l } } \over {\sum\nolimits_l {h_l } }}} \right)} \right| + \left| {{{\sum\nolimits_l {\delta _l \gamma _l } } \over {\sum\nolimits_l {h_l } }}} \right| 
\\
\le 2\mathop {\max }\limits_l s_l \left| {{{\sum\nolimits_l {\delta _l } } \over {\sum\nolimits_l {h_l } }}} \right|
\end{array}
\label{1.97}
\end{equation}
Since we already presumed that $\left| {{{\sum\nolimits_l {\delta _l } } \over {\sum\nolimits_l {h_l } }}} \right| \ll 1$ we can ignore the last two terms in \eqref{1.94} in comparison with other terms, which yields, 
\begin{equation}
{{\sum\limits_{l = 1}^N {w_l z_l } } \over {\sum\limits_{l = 1}^N {w_l } }} \approx \sum\limits_{l = 1}^N {\left[ {\gamma _l \left( {{{h_l } \over {\sum\limits_L {h_l } }}} \right) + \delta _l {{\left( {c_l  - \xi } \right)} \over {\sum\limits_l {h_l } }}} \right]}  + \xi 
\label{1.98}
\end{equation}
Finally using \eqref{1.89} in conjunction with \eqref{1.98} the result can be rewritten as,
\begin{equation}
\begin{array}{l}
Y_{gc}  = \int_{\gamma _1 } { \cdots \int_{\gamma _N } {\int_{\delta _1 } { \cdots \int_{\delta _N } {{{\Im _{l = 1}^N \mu _{Z^l } (c_l  + \gamma _l ) * \Im _{l = 1}^N \mu _{W^l } (h_l  + \delta _l )} \mathord{\left/
 {\vphantom {{\Im _{l = 1}^N \mu _{Z^l } (c_l  + \gamma _l ) * \Im _{l = 1}^N \mu _{W^l } (h_l  + \delta _l )} {\sum\limits_{l = 1}^N {\left[ {\gamma _l \left( {{{h_l } \over {\sum\limits_L {h_l } }}} \right) + \delta _l \left( {{{c_l  - \xi } \over {\sum\limits_l {h_l } }}} \right)} \right]}  + \xi }}} \right.}}}}}}\\
{{{{{{
 \kern-\nulldelimiterspace} {\sum\limits_{l = 1}^N {\left[ {\gamma _l \left( {{{h_l } \over {\sum\limits_L {h_l } }}} \right) + \delta _l \left( {{{c_l  - \xi } \over {\sum\limits_l {h_l } }}} \right)} \right]}  + \xi }}} } } } \,
\end{array}
\label{1.99}
\end{equation}
Now let define fuzzy sets $\underbar {Z}_l  = Z_l  - c_l $ and $\underbar {W}_l  = W_l  - h_l $ 
with supports $[-s_l ,s_l ]$ and $-\Delta_l , \Delta_l$ respectively. In other words, 
\begin{equation}
\underbar Z_l  = \int_{z_l } {\mu _{Z_l } (z_l )/(z_l }  - c_l ) = \int_{z_l } {\mu _{Z_l } (\gamma _l  + c_l )/} \gamma _l  \Rightarrow \mu _{Z_l } (\gamma _l  + c_l ) = \mu _{\underbar Z_l } (\gamma _l )
\label{1.100}
\end{equation}
Note that $z_l  - c_l  = \gamma _l $.
In the same sense, $\mu _{W_l } (h_l  + \delta _l ) = \mu _{\underbar W_l } (\delta _l )
$. Hence we can rewrite the RHS of \eqref{1.98} as,
\begin{eqnarray}
\sum\limits_{l = 1}^N {\left[ {\underbar{Z}_l \left( {{{h_l } \over {\sum\limits_L {h_l } }}} \right) + \underbar W_l \left( {{{c_l  - \xi } \over {\sum\limits_l {h_l } }}} \right)} \right]}  + \xi 
\label{1.101}
\end{eqnarray}
As a special case of \eqref{1.101} one can easily observe that if $W_l$ are crisp numbers (recall that a crisp number can be considered as a fuzzy set, too), i.e. $\Delta _l  = 0$ for $l=1,2,�,N$ then \eqref{1.101} reduces to $\sum\limits_{l = 1}^N {\left[ {\underbar Z_l \left( {{{h_l } \over {\sum\limits_L {h_l } }}} \right)} \right]}  + \xi 
$ which follows,
\begin{equation}
\begin{array}{l}
Y(Z_1 ,...,Z_N ,h_1 ,...,h_N ) = \int_{\gamma _1 } { \cdots \int_{\gamma _N } {{{\Im _{l = 1}^N \mu _{Z^l } (c_l  + \gamma _l )} \mathord{\left/
 {\vphantom {{\Im _{l = 1}^N \mu _{Z^l } (c_l  + \gamma _l )} {{{\sum\limits_{l = 1}^N {h_l z_l } } \over {\sum\limits_{l = 1}^N {h_l } }}}}} \right.
 \kern-\nulldelimiterspace} {{{\sum\limits_{l = 1}^N {h_l z_l } } \over {\sum\limits_{l = 1}^N {h_l } }}}}} } \, =  \cdots  =   \\
\int_{\gamma _1 } { \cdots \int_{\gamma _N } {{{\Im _{l = 1}^N \mu _{Z^l } (c_l  + \gamma _l )} \mathord{\left/
 {\vphantom {{\Im _{l = 1}^N \mu _{Z^l } (c_l  + \gamma _l )} {\sum\limits_{l = 1}^N {\left[ {_l \left( {{{h_l } \over {\sum\limits_L {h_l } }}} \right)} \right]}  + \xi }}} \right.
 \kern-\nulldelimiterspace} {\sum\limits_{l = 1}^N {\left[ {_l \left( {{{h_l } \over {\sum\limits_L {h_l } }}} \right)} \right]}  + \xi }}} }  \\
\end{array} 
\label{1.102}
\end{equation}
\end{theorems}
The reader should observe that the result of the theorem in this case is exact since the source of inexactness in the proof has been removed.
\subsubsection{Approximate Generalized Centroid for IT2FSs}
As a direct application of previous theorem, one can derive a new formula when the fuzzy sets $Z_l$ and $W_l$ are interval Type-I sets. 
\begin{corollaries}{(\cite{Kar01a})}
If $Z_l$ and $W_l$ in Theorem \ref{approxgc} are IT1 sets, the result of generalized centroid TR is approximately an IT1 set with mean $\xi$, defined earlier, and spread $\kappa $ given by,
\begin{equation}
\kappa  = {{\sum\limits_{l = 1}^N {[(h_l s_l ) + \left| {c_l  - \xi } \right|\Delta _l ]} } \over {\sum\nolimits_{l = 1}^N {h_l } }}
\label{1.103}
\end{equation}
provided that,
\begin{equation}
{{\sum\nolimits_{l = 1}^N {\Delta _l } } \over {\sum\nolimits_{l = 1}^N {h_l } }} \ll 1
\label{1.104}
\end{equation}
\emph{Proof}: The proof to this corollary is straight forward by applying Theorem \ref{affineIT1comb} to \eqref{1.84} and keeping in mind the fact that the mean for both $\underbar W_l$ and $\underbar{Z}_l$ are zero and the respective 
spreads are $s_i$ and $\Delta_i$.\\ 
Now consider the centroid TR introduced in \eqref{1.70} as a special case of generalized centroid TR; to show the power of the introduced theorem for approximate TR, lets consider the result of the previous corollary as an important application for centroid TR. For an IT2FS, \eqref{1.70} reduces to, 
\begin{equation}
C_{\tilde A}  = \int_{\theta _1  \in J_{x_1 } } { \cdots \int_{\theta _N  \in J_{x_N } } {1/{{\sum\limits_{i = 1}^N {x_i .\theta _i } } \over {\sum\limits_{i = 1}^N {\theta _i } }}} }  = [c'_l ,c'_r ]
\label{1.105}
\end{equation}
Note $x_i$s in \eqref{1.105} represent $c_l$ in \eqref{1.103} and $s_i=0$. If we define $
J_{x_i }  \buildrel \Delta \over = [L_i ,R_i ]$ entails that $h_l  = {{L_l  + R_l } \over 2}
$ and $\Delta _l  = {{R_l  - L_l } \over 2}$. These altogether give us an IT1FS with support $ [\xi  - \kappa ,\xi  + \kappa ] $ as an approximate centroid where,
\begin{equation}
\kappa  = {{\sum\limits_{l = 1}^N {\left| {c_l  - \xi } \right|\Delta _l } } \over {\sum\nolimits_{l = 1}^N {h_l } }}
\label{1.106}
\end{equation}
and $\xi $ as defined earlier. We take advantage of this important result in future chapters where we develop our model to control a model for a 3PSP parallel robot.
\end{corollaries}
\subsubsection{Approximate Generalized Centroid for Gaussian FSs}
The second direct application of the previous theorem which offers a significant contribution to the field of Type-II fuzzy logic is that it provides an approximation for generalized centroid of Gaussian T2 fuzzy sets which are among the most promising and useful T2 fuzzy sets. Accordingly, the following corollary paves the way towards development of T2FLSs based on Gaussian fuzzy sets.
\begin{corollaries}{(\cite{Kar01a})}
Consider the generalized centroid introduced in \eqref{1.84}. If each of $Z_l$ and $W_l$ is a Gaussian Type-I set with respective means and standard deviations $m_l,\sigma_l$ and $h_l,\Delta_l$ then the generalized centroid is approximately a T1 Gaussian set with mean $\varsigma$ and standard deviation $\Sigma$ , where  
\begin{equation}
\varsigma = {{\sum\nolimits_{l = 1}^N {h_l m_l } } \over {\sum\nolimits_{l = 1}^N {h_l } }}
\label{1.107}
\end{equation}
and
\begin{equation}
\sum  = \left\{ \begin{array}{c l}
  {{\sqrt {\sum\nolimits_{l = 1}^N {\left[ {(h_l \sigma _l )^2 [m_l  -\varsigma ]^2 \Delta _l ^2 } \right]} } } \over {\sum\nolimits_{l = 1}^N {h_l } }} & \hbox{if product t-norm used} \hfill \\ 
  {{k\sum\nolimits_{l = 1}^N {\left[ {(h_l \sigma _l ) + \left| {m_l  - \varsigma} \right|\Delta _l } \right]} } \over {\sum\nolimits_{l = 1}^N {h_l } }} & \hbox{if min t-norm is used} \hfill \\ \end{array} \right.
\label{1.108}
\end{equation}
provided that 
\begin{equation}
{{k\sum\nolimits_{l = 1}^N {\Delta _l } } \over {\sum\nolimits_{l = 1}^N {h_l } }} \ll 1
\label{1.108}
\end{equation}
where $k$ is a number generally chosen to be 2 or 3. Roughly speaking, for the condition above to be satisfied, $W_l$'s should be narrow enough. In the limit case when $W_l$'s approach crisp numbers the results get exact. Notice however that there are not any limitations on $Z_l$'s which of course make sense. \\
\emph{Proof}: Let's separate the terms in \eqref{1.85} into three parts as follows,
\begin{equation}
Y_{gc}  \approx \sum\limits_{l = 1}^N {\left[ {\underline Z _l \left( {{{h_l } \over {\sum\limits_{l = 1}^N {h_l } }}} \right) + \underline W _l \left( {{{c_l  - \xi } \over {\sum\limits_{l = 1}^N {h_l } }}} \right)} \right]}  + \xi  = \overbrace {\sum\limits_{l = 1}^N {\underline Z _l \left( {{{h_l } \over {\sum\limits_{l = 1}^N {h_l } }}} \right)} }^{Y_{gc_1 } } + \overbrace {\sum\limits_{l = 1}^N {\underline W _l \left( {{{c_l  - \xi } \over {\sum\limits_{l = 1}^N {h_l } }}} \right)} }^{Y_{gc_2 } } + \overbrace \xi ^{Y_{gc_3 } }
\label{1.110}
\end{equation}
Based on Theorem \ref{affineIT1comb}, the first two terms in the right hand side of \eqref{1.110} are T1 Gaussian fuzzy sets whose mean and standard deviation can be found as follows (note that the mean for both $\underbar{Z}_l$ and $\underbar{W}_l$ are zero) 
\\
\begin{equation}
 \left\{
 \begin{array}{l}
  m_{Y_{gc_1 } }  = 0 \hfill \\
  m_{Y_{gc2} }  = 0 \hfill \\ \end{array}  \right. \Rightarrow m_{Y_{gc} }  = \varsigma = \xi  = {{\sum\nolimits_{l = 1}^N {h_l m_l } } \over {\sum\nolimits_{l = 1}^N {h_l } }}
\label{1.111a}
\end{equation}
\\
\begin{equation}
	{ \begin{array}{l}
		{ \left\{
			{ \begin{array}{l}
  				\sigma _{Y_{gc_1 } }  = 
					{ \left\{
						{\begin{array}{l l}
   						{\sqrt {\sum\limits_{i = 1}^N {\left( {{{h_i } \over {\sum\limits_{l = 1}^N {h_l} }}}\right)^2 \sigma _i ^2 } }} & \hbox {for product t-nrom}  \\ 
   						{\sum\limits_{i = 1}^N {\left| {{{h_i } \over {\sum\limits_{l = 1}^N {h_l } }}} \right|\sigma}} & \hbox{for min t-norm} \end{array} 
						}	\right. 
					} 
				\hfill \\ 
		  		\sigma _{Y_{gc_{_2 } } }  = 
				{\left\{ 
					{\begin{array}{l l}
   					{\sqrt {\sum\limits_{i = 1}^N {\left( {{{m_i  - \xi } \over {\sum\limits_{l = 1}^N {h_l } }}} \right)^2 \Delta _i ^2 } }} & \hbox{for product t-nrom}  \\ 
   					{\sum\limits_{i = 1}^N {\left| {{{m_i  - \xi } \over {\sum\limits_{l = 1}^N {h_i } }}} \right|\Delta _i }}& \hbox{for min t-norm} \\ \end{array} 
					} \right. 
				}\end{array}
		} \right. 
	} \\ \\
	\Rightarrow 
	
	\sigma _{Y_{gc} }  = \sum  = 
	{\left\{ 
		{\begin{array}{l l}
	 		{{\sqrt {\sum\limits_{i = 1}^N {(h_i \sigma _i )^2  + (m_i  - \xi )^2 \Delta _i ^2 } } } \over {\sum\limits_{l = 1}^N {h_l } }} & \hbox{for product t-norm} \hfill \\ 
 			{{\sum\limits_{i = 1}^N {\left( {h_i \sigma _i  + \left| {m_i  - \xi } \right|\Delta _i } \right)} } \over {\sum\limits_{l = 1}^N {h_l } }} & \hbox{for min t-norm} \hfill \\ \end{array} 
		} \right.
	} \end{array} 	
}
\label{1.111b}
\end{equation}
\end{corollaries}

\subsubsection{Approximate Centroid for Gaussian FSs (Under Minimum T-Norm)}
Recall from our previous discussion about centroid TR that we should not select product for 
t-norm due to undesired results. Hence, for a special case of generalized centroid 
(that is the centroid TR,) we present an approximation in the subsequent corollary. The reader 
can show the validity of the next corollary in a straightforward fashion.
\begin{corollaries}
The centroid for a Gaussian T2FS (GT2FS) is an approximate Gaussian T1FS, with mean and standard 
deviations obtained by,
\begin{equation}
\varsigma = {{\sum\nolimits_{l = 1}^N {x_l m(x_l )} } \over {\sum\nolimits_{l = 1}^N {m(x_l )} }}
\label{1.112}
\end{equation}
and
\begin{equation}
\sum  = {{\sum\limits_{i = 1}^N {\left| {x_i  - \varsigma } \right|\Delta _i } } \over {\sum\limits_{l = 1}^N {m(x_l )} }}
\label{1.113}
\end{equation}
provided that,
\begin{equation}
{{k\sum\nolimits_{l = 1}^N {\sigma (x_l )} } \over {\sum\nolimits_{l = 1}^N m (x_l )}} \ll 1
\label{1.114}
\end{equation}
\end{corollaries}
\subsubsection{Exact Generalized Centroid TR for IT2FSs}
Up to now, we have talked about approximate TR in general T2FLSs. However for the important class of IT2FLSs, there is still an exact and efficient way to handle the intensive computational costs. Thus, the gain is the significant computational speed up and the pain is restricting ourselves to IT2FLSs. 
As explored earlier, the three different types of TR procedures we have seen thus far, can be re-expressed by the following general expression,
\begin{equation}
\begin{array}{l}
Y(Z_1 ,Z_2 , \cdots ,Z_M ,W_1 ,W_2 , \cdots ,W_M ) =\\ \mathop{\int\dots\int}\limits_{\kern-5.5pt {z_1 , \ldots ,z_M }} 
 {\,\,\mathop{\int\dots\int}\limits_{\kern-5.5pt {w_1 ,\ldots ,w_M }} 
 {{{\Im _{l = 1}^M \mu _{Z_l } (z_l ) * \Im _{l = 1}^M \mu _{W_l } (w_l )} \mathord{\left/
 {\vphantom {{\Im _{l = 1}^M \mu _{Z_l } (z_l ) * \Im _{l = 1}^M \mu _{W_l } (w_l )} {{{\sum\limits_{l = 1}^M {w_l z_l } } \over {\sum\limits_{l = 1}^M {w_l } }}}}} \right.
 \kern-\nulldelimiterspace} {{{\sum\limits_{l = 1}^M {w_l z_l } } \over {\sum\limits_{l = 1}^M {w_l } }}}}} } 
\end{array}
\label{1.115}
\end{equation}
Where $*$, $\Im$ indicate t-norm operators and $w_l  \in W_l ,\,z_l  \in Z_l $  for $l=1,2,\dots, M$. Note that $W_l, Z_l$ are secondary membership functions and of Type-I characteristic. 
We show how things get much simplified if we use IT2FSs, introduced in future sections, according to \cite{Kar99c}. In the presented method, the computation cost imposed by the associated complex iterative process in \eqref{1.115} is vastly mitigated by confining the exhaustive exploration of the search space constructed over $z_l$'s and $w_l$'s . \\
As explained for IT2FLSs each $W_l$ and $Z_l$ $(l=1,2,�,M)$ is an interval Type-I set therefore $\mu {}_{W_l }(w_l ) = \mu {}_{Z_l }(z_l ) = 1$. Hence, \eqref{1.115} can be re-expressed as,
\begin{equation}
Y(Z_1 ,Z_2 , \cdots ,Z_M ,W_1 ,W_2 , \cdots ,W_M ) = \mathop{\int\dots \int}\limits_{\kern-5.5pt {z_1 ,z_2 , \ldots ,z_M }} 
 {\,\,\mathop{\int\dots \int}\limits_{\kern-5.5pt {w_1 ,w_2 , \ldots ,w_M }} 
 {{1 \mathord{\left/
 {\vphantom {1 {{{\sum\limits_{l = 1}^M {w_l z_l } } \over {\sum\limits_{l = 1}^M {w_l } }}}}} \right.
 \kern-\nulldelimiterspace} {{{\sum\limits_{l = 1}^M {w_l z_l } } \over {\sum\limits_{l = 1}^M {w_l } }}}}} } 
\label{1.116}
\end{equation}
which emerges in the form of an IT1FS. Accordingly, it suffices to find the minimum and maximum 
bounds of this fuzzy set which completely represent it. That is to say, an IT1FS can be rigorously characterized by its domain interval or simply by two real numbers which serve as the upper and lower limits of the support of the fuzzy set. In other words, we may describe the domain interval by the left and right ends, $[l,r]$, or by its center and spread factors as $[c-s,c+s]$ where $c$, $s$ equal $(l+r)/2$ and $(r-l)/2$, respectively. We prefer the second notation in this thesis. Notice that $Y$ in the previous formula is a function of several IT1FSs (i.e. $Z_1 ,Z_2 , \cdots ,Z_M ,W_1 ,W_2 , \cdots ,W_M$). Let $c_l$, $s_l$ be the center and spread of $Z_l$ ,respectively and $h_l, \Delta_l$be the corresponding terms for $W_M $. Obviously $h_l\ge \Delta_l$  since the firing strengths are not defined for negative values.
We only present the step by step sequence of actions to compute $c$, $s$; the interested reader can refer to \cite{Kar01a} for the proof.\\
Let $y(w_1 , \cdots ,w_N ) = {{\sum\nolimits_{l = 1}^N {(c_l  + s_l )w_l } } \over {\sum\nolimits_{l = 1}^N {w_l } }}$ follow the next steps in order,
\begin{enumerate}
\item Set  for $l=1,2,..,N$ and compute $y' = y(h_1 , \cdots ,h_N )$
\item	Find $k$, $1 \le k \le N - 1$, such that $z_k  \le y' \le z_{k + 1} $
\item Set $w_l  = h_l  - \Delta _l $ for $l \leq k$ and $w_l  = h_l  + \Delta _l $ for $l \geq k+1$ and compute  $y'' = y(h_1  - \Delta _1 , \cdots ,h_k  - \Delta _k ,h_{k + 1}  + \Delta _{k + 1} , \cdots ,h_N  + \Delta _N )$
\item Check whether $y'' = y'$ , if yes we have reached the maximum at $y=y'$ , else, go to the next stage.
\item Set $y'=y''$ and jump to step 2.     
\end{enumerate}
It is easy to show that this iterative procedure stops at a maximum number of iterations equal to $N$. See \cite{Kar01a} for details.
\subsection{\textbf{Defuzzification}}
The last step to be done in T2FLSs is the \emph{defuzzification}. Defuzzification is also the last step to be done in the output processing unit. The defuzzification unit produces a crisp number so that it can be exploited to control the system that the controller was already designed for. In this thesis however, we develop a new strategy that does not make use of this unit until the last minute. Consider a system which has been designed to cope with uncertainty, for example. As discussed earlier, the output of a TR unit, is a T1FS which represents the uncertainty present in the output, namely, the confidence interval. In this approach we skip the defuzzification unit and work with T1 fuzzy numbers instead.
We benefit from T1 fuzzy algebra which is derived from Extension Principle and postpone the defuzzification only to the last stage where we eventually defuzzify the output FS to get to the final crisp number.
Among different defuzzifiers that can be used in a T2FLS, centroid is the most popular one. The expression for centroid defuzzifier  is already defined in \eqref{1.69}. Applying the centroid defuzzifier is like a weighted average of the outputs of all embedded FLSs.
We should clarify about an important misconception in the defuzzification of the type reduced sets. If there is a single point associated to which a unit membership value, then one might think that for simplicity it can be opted as the defuzzified value which is equivalent to the height defuzzifier.
 However this deteriorates the resulting output in the following sense; Consider the TR introduced in \eqref{1.70} as an example;
 for the type reduced set to have a unit membership grade at some point, there should be some combination of $\{ \theta _1 , \cdots ,\theta _N \}$ (note that as explained in previous sections, this combination represents a wavy slice in the T2FS) such that the term $\mu _{D_1 } (\theta _1 ) *  \cdots  * \mu _{D_N } (\theta _N )$ amounts to 1. This means that all the secondary membership values at the corresponding points must be 1, too. In that case $\{ \theta _1 , \cdots ,\theta _N \} $ makes a principle membership function of the underlying T2FS! Thus,
 the selected point in the type reduced set, ${{\sum\limits_{i = 1}^N {x_i .\theta _i } } \over {\sum\limits_{i = 1}^N {\theta _i } }}$, 
 would be the centroid of the principle membership function and as explained previously this process is eventually equivalent to discarding all Type-II fuzzy analyses. Consequently use of height defuzzification means wasting a great part of our computational resources for nothing except a simple analysis of a T1FLS.
\subsection{\textbf{Type-II Fuzzy Logic Systems (T2FLS)}}
In summary, we merge all the previously mentioned concepts, theorems and corollaries to come to a general T2FLS. The results can be further simplified for special T2FLSs such as IT2FLSs. 
Having the rules defined in the form of canonical representation,
\begin{equation}
\begin{array}{l l}
\hbox{Rule}^1 & \hbox{: if }x_1\hbox{ is }\tilde F_1^1\hbox{ and }x_2\hbox{ is }\tilde F_2^1\hbox{ and } \cdots\hbox{, and }x_p\hbox{ is }\tilde F_p^1\hbox{ then y is }\tilde G^1   \\
\hbox{Rule}^2 & \hbox{: if }x_1\hbox{ is }\tilde F_1^2\hbox{ and }x_2\hbox{ is }\tilde F_2^2\hbox{ and } \cdots\hbox{, and }x_p\hbox{ is }\tilde F_p^2\hbox{ then y is }\tilde G^2    
\\ & \vdots    \\
\hbox{Rule}^M & \hbox{: if }x_1\hbox{ is }\tilde F_1^M\hbox{ and }x_2\hbox{ is }\tilde F_2^M\hbox{ and } \cdots\hbox{, and }x_p\hbox{ is }\tilde F_p^M\hbox{ then y is }\tilde G^M   
\end{array}
\label{1.117}
\end{equation}
a T1 fuzzy output set can be derived according to the choice of type reduction methods. Here we consider T2FLSs built based on center of sets and centroid TR. To realize other types of TRs, one should proceed in rather same way with slight considerations. We assume that the fuzzification block is a non-singleton one. The procedure needed to follow for singleton fuzzifiers are quite the same as singleton FSs and can be considered a special case of non-singleton FSs with less computations.
As stated earlier each rule in the above rule set represents a T2 relation between the input $X_1  \times  \cdots  \times X_p $ and output space, $Y$ . Based on the discussion given in previous sections, we can derive a T2 implication, as follows: Let $\tilde A^l  = \tilde F_1 ^l  \times  \cdots  \times \tilde F_p ^l $ ; the $l$th rule in \eqref{1.117} which has the form,
\begin{equation}
R^l :\,\tilde F_1 ^l  \times  \cdots  \times \tilde F_p ^l  \to \tilde Y^l  \Rightarrow \tilde A^l  \to \tilde Y^l\hbox{ l=1,2,...,M} 
\label{1.118}
\end{equation}
and can be restated as:
\begin{equation}
\mu _{R^l } (\bar x,y) = \mu _{\tilde A^l  \to \tilde Y^l } (\bar x,y) = [\prod _{i = 1}^p \mu _{\tilde F^l _i } (x_i )]\prod \mu _{\tilde G^l } (y)
\label{1.119}
\end{equation}
where by $\mu _{R^l } (\bar x,y)$ we mean $\mu _{R^l } (x_1 , \cdots ,x_p ,y)$ . Usually, the \emph{p}-dimensional input $x_1 , \cdots ,x_p $ is given by a T2FS, $\tilde A_{\bar x} $, which is defined as: $\mu _{\tilde A_{\bar x} } (\bar x) = \prod _{i = 1}^p \mu _{\tilde X_i (x_i )}$ . Therefore, the output of the $l$th rule, $\tilde B_l$ can be found using the composition of T2 relations, as follows:
\begin{equation}
\mu _{\tilde B^l } (y) = \mu _{\tilde A_{\bar x}  \circ R^l } (y) = \coprod _{x \in X} [\mu _{\tilde A_{\bar x} } \prod \mu _{R^l } (\bar x,y)]\,\quad l=1,2,�,M   
\label{1.120}
\end{equation}
or
\begin{equation}
\begin{array}{l}
\mu _{\tilde B^l } (y) = \mu _{\tilde Y^l } (y)\prod \\ \left\{ {[\coprod _{x_1  \in X_1 } \mu _{\tilde X_1 } (x_1 )\prod \mu _{\tilde F^l _1 } (x_1 )]\prod  \cdots 
\prod [\coprod _{x_p  \in X_p } \mu _{\tilde X_p } (x_p )\prod \mu _{\tilde F^l _p } (x_p )]} \right\}\quad \forall y \in Y
\end{array}
\label{1.121}
\end{equation}
The term enclosed by the bracket on the right hand side of \eqref{1.121} is called the Firing Set which signifies to what extent the lth rule has been fired. For singleton T2FLSs, \eqref{1.121} reduces to,
\begin{equation}
\mu _{\tilde B^l } (y) = \mu _{\tilde Y^l } (y)\prod \left\{ {\mu _{\tilde F^l _1 } (x_1 )\prod  \cdots \prod \mu _{\tilde F^l _p } (x_p )} \right\}\,\,\,\,\,\,\forall y \in Y
\label{1.122}
\end{equation}
which simplifies the computations considerably and consequently the firing set for this type of T2FLSs can be expressed as:
\begin{equation}
\hbox{Firing Set = }\prod _{i = 1}^p \mu _{\tilde F^l _i } (x_i )
\label{1.123}
\end{equation}
Now, depending on the selected TR, the required actions differ as follows:
\begin{itemize}
\item Centroid TR\\
In this case we must aggregate all the output T2FSs into a single T2FS. This can be done using join operation. In other words:
\begin{equation}
\mu _{\tilde B} (y) = \coprod _{i = 1}^M \mu _{\tilde B^l } (y)
\label{1.124}
\end{equation}
and finally using the Centroid TR, one can compute the type reduced output which is defuzzified subsequently to produce the final crisp number.
\item Center of Set (CoS) TR\\
In this case the procedure is straight forward; the type reduced set can be obtained by first computing the centroid type reduced set of each of the consequent sets and treat them as  $Z_l$'s in \eqref{1.84} and to consider firing sets as the weights, $W_l$'s 
and to proceed with the center of sets computations.
\end{itemize}

By this section, we come to the end of a rather thorough exploration of T2FLSs and T2FSs. We will take
advantage of the materials presented in this chapter to develope a new fuzzy logic based system for 
control of complex dynamic systems.

In the next chapters, we organize the robotics foundations that lead us to a basic understanding of
dynamic equations pertaining to every robotic system.

%% file: chapter2.tex
\chapter{Mechanical Foundations of Parallel Robots}
\section{\textbf{Preface}}
\begin{figure}[htp]
  \begin{center}
    \includegraphics[width=10cm, height=5cm]{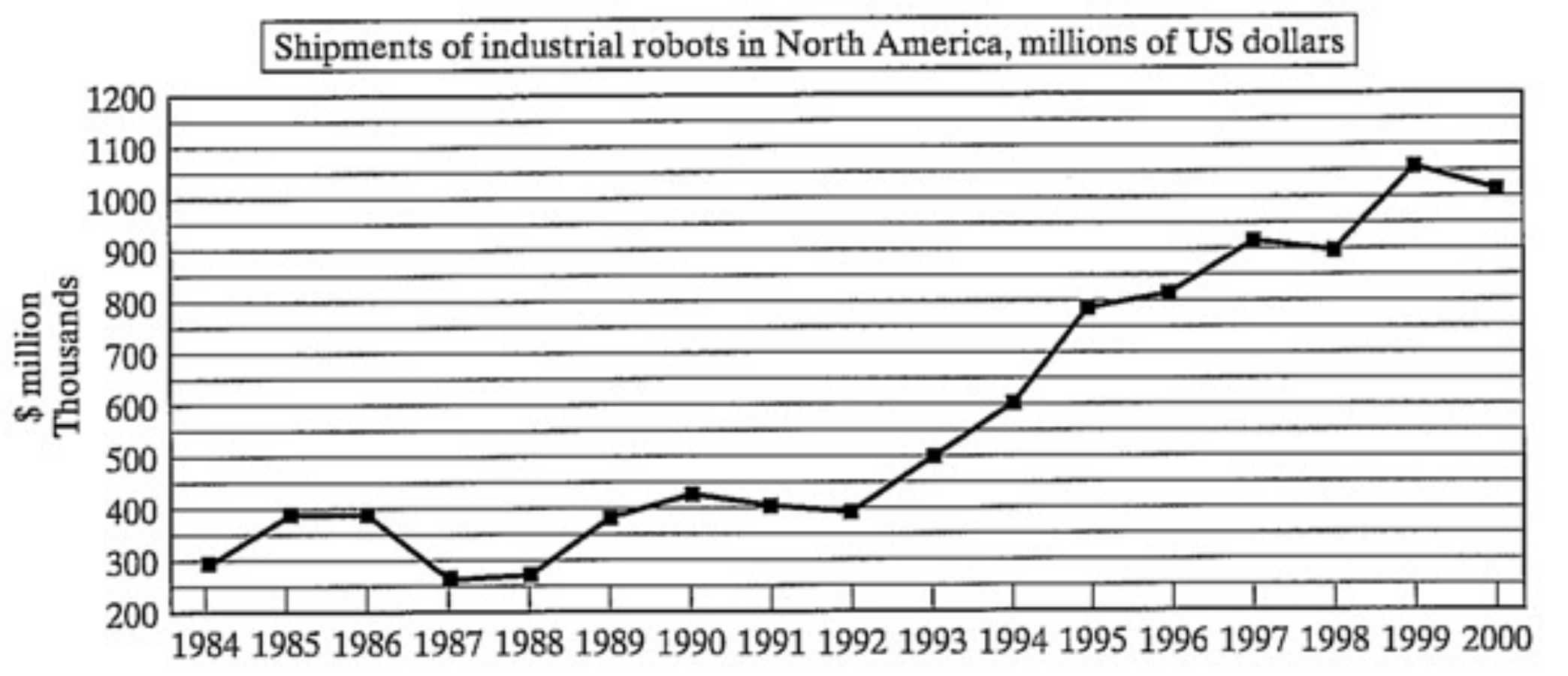}
  \end{center}
  \caption{Shipment of industrial robots in North America, \cite{Cra04}}
  \label{fig8}
\end{figure}

Industrial automation has always been described by rapidly changing eras in popular methods \cite{Cra04}. These changes have undoubtedly had close relation to the world economics and the industrial trends. Today, by emergence of computer aided design systems and tools (CAD tools) along with computer aided manufacturing systems (CAM) the field of robotics has flourished towards its unknown promising future. The shipment of industrial robots has increased significantly during the past two decades; see Figure \ref{fig8}. The story does not end to the price and shipment rates, the quality and effectiveness of industrial robots are also increasing continuously; meaning that these tools are becoming faster, more accurate and flexible when comparing to their predecessors. If we could consider this factor into account then the real prices drop down even in a greater pace \cite{Cra04}. This is while the labor price has not been decreased during the time and in fact this itself is another incentive for industry to automate most part of the factories. These all together are not the only reasons which have encouraged human on developing automation systems but other influential reasons forced him to do so including working in dangerous situations or tasks impossible for him, to name two.

Nowadays, millions of robots of different types are working in industry and scientific academia; each designed keeping many considerations (workspace, size, speed, quality, stiffness, the structure , load capability and etc.) in mind. Due to importance of industrial robots and how to come up with programs that robustly and reliably control their different parts, in this thesis, we have focused on an open issue of interest in control of a 3PSP robot (to be introduced later) that tries to handle efficiently one of outstanding matters of concern, the uncertainty.

\pagebreak

\section{\textbf{Mechanics Terminologies}}
\subsection{\textbf{Frame}}
In the context of robotics we are constantly concerned with the current position (and probably the orientation) of all the segments and the end, the so-called \emph{tool-tip} or \emph{end-effector}, of the underlying robot during the time. Hence, at the very basic step, we have to define and adhere coordinate systems, or frames to each part so that they could be easily described in terms of the corresponding coordinate systems and later in term of a global frame. In fact the global frame is the one that we are seeking to express all the system parameters in connection with it, which serves as a global reference. Among the most well-known frames are the tool frame and base frame which are stuck to the tool-tip and the non-moving base of robots (especially the manipulators) respectively.
\subsection{\textbf{Kinematics}}
\emph{Kinematics} is the science of motion [1] which considers motion disregarding the forces that are the cause of them. Therefore kinematics is a framework that describes the motions in term of geometrical time variant component of the motion.
\subsection{\textbf{Link, Joint}}
A link is a nearly rigid connection between two joints. That is to say, two links are connected to each other via a joint which lets them have a relative movement. This movement could be a rotary one (the \emph{revolute} joints) or either a sliding one (\emph{prismatic} joints). In case of the prismatic joint, the link displacement is called joint\emph{angles} and in the latter case it is coined as joint \emph{offset}.
\subsection{\textbf{Degree of Freedom}}
\begin{figure}[htp]
  \begin{center}
    \includegraphics[width=10cm, height=5cm]{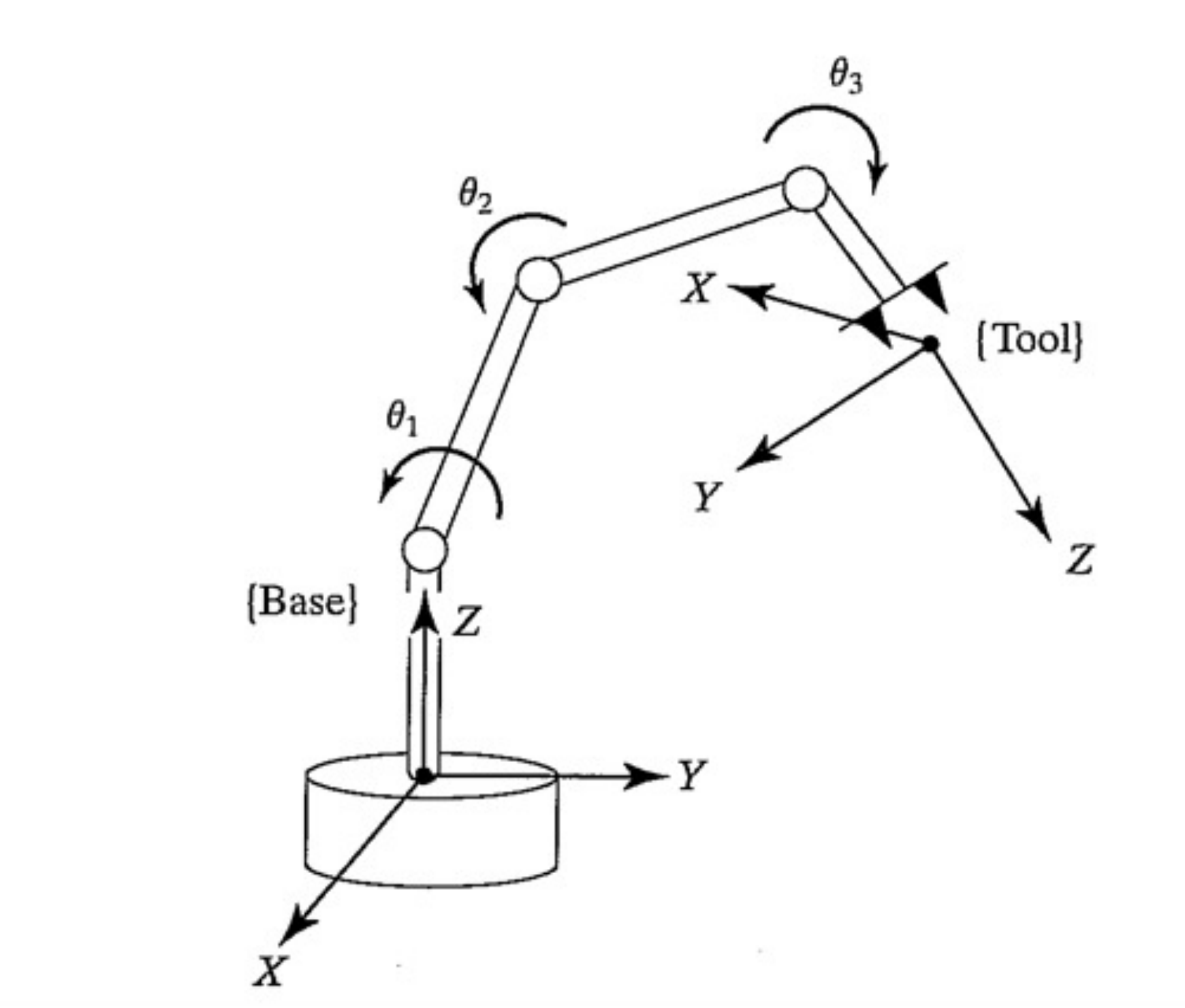}
  \end{center}
  \caption{Equations describe the tool frame relative to the base frame, \cite{Cra04}}
  \label{fig9}
\end{figure}

The number of independent variables that completely describe the system mechanism is called the \emph{degree of freedom} or \emph{D.O.F} for short. Note that different set of variables could be selected as a completely describing set some of which might be more or less comprehensive and intuitively plausible. For example, the robot manipulator depicted in Figure \ref{fig9}, illustrates a 3 DOF robot which are denoted by $\Theta_1$, $\Theta_2$ and $\Theta_3$.
\subsection{\textbf{Forward Kinematics}}
The first problem that comes up during the course of mechanical analysis of a robot is the \emph{forward kinematics}. In other works given the set of joint angles (offsets) it is desired to know the exact position and orientation of the end-effector. The forward kinematics is sometimes attributed as a change in representation space namely from joint space to Cartesian space. While this is a fundamental step in study of an underlying robot, it's not of a practical value while we are mostly in quest of the reverse.
\subsection{\textbf{Inverse Kinematics}}
The problem of inverse kinematics is what we practically willing to find and of course it is the most challenging (and maybe the most computationally intensive) part of robotics. When dealing with such problem several hurdles my come to existence that we didn't use to see beforehand, to name an example, the solution to the foregoing is not necessarily unique and one which itself is contingent upon other parameters; meaning that depending on the status of the robot the number of solutions will differ. Also it might be the case that some of the solutions are not plausible or valid and among those that are valid one might be of more practical interests due to fulfilling different constraints (a multi-objective problem). A final end of robotics in such circumstances is that given a trajectory find a feasible solution in the joint space such that starting from the initial conditions in the Cartesian space our method converges to the final desired one. To sum up, the inverse kinematics has the following limitations:
\begin{itemize}
\item It imposes a great amount of computational burden
\item The solutions are not unique
\item The number of solutions are dependent upon current state of the robot
\item The solution is a non-linear one which consist of a great deal of trigonometric functions
\item Due to non-linearity of the kinematic equations, generally, there is not a closed form solution available
\item Also there might not be any feasible solution at all in different circumstances, in that case it is said that the robot cannot obtain the required configuration or simply it resides outside the robot \emph{workspace}
\end{itemize}
\subsection{\textbf{Jacobian, Singular Points}}
\begin{figure}[htp]
  \begin{center}
    \includegraphics[width=10cm, height=5cm]{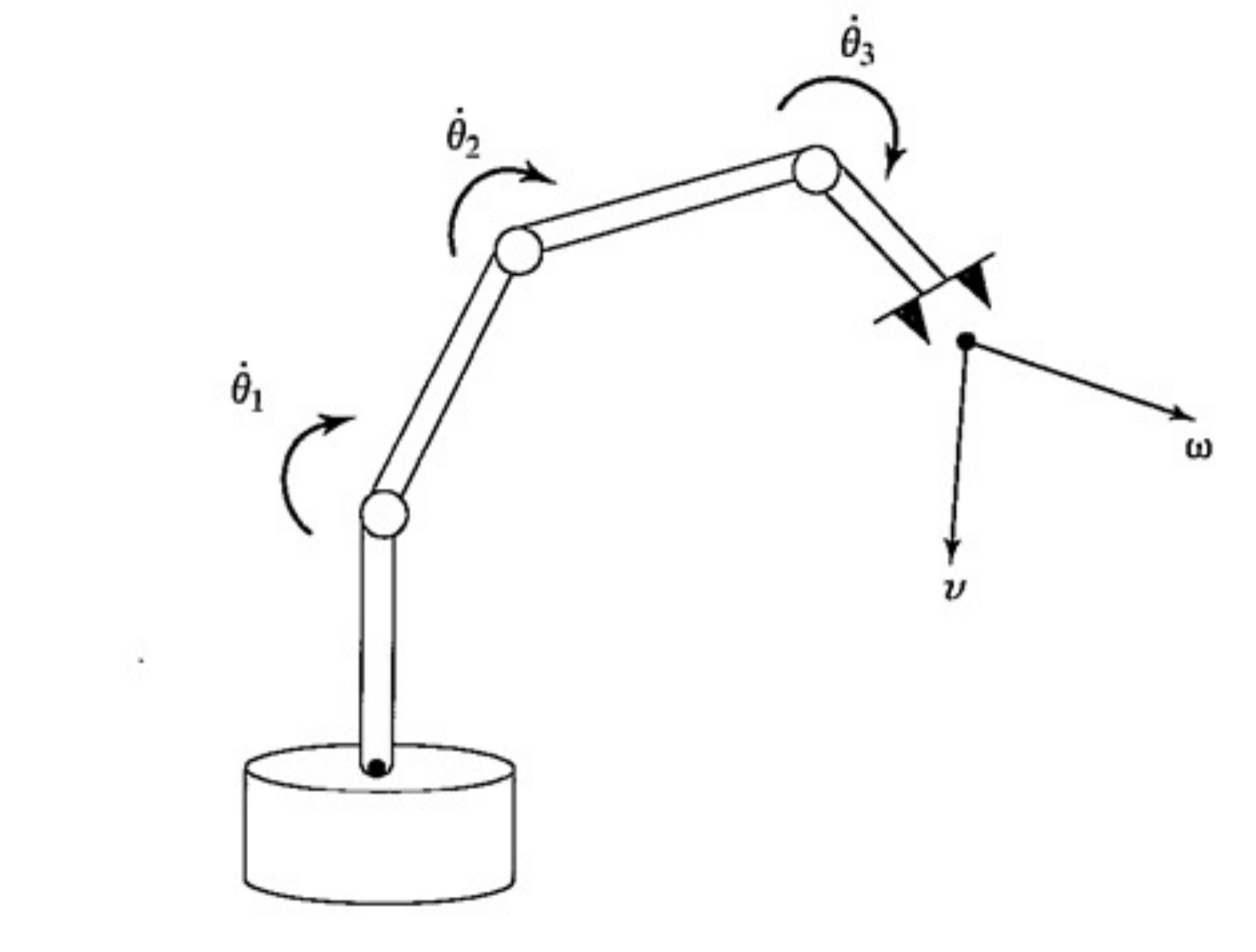}
  \end{center}
  \caption{geometrical relation between joint rates and velocity of the end-effector can be described in a matrix called Jacobian, \cite{Cra04}}
  \label{fig10}
\end{figure}

\begin{figure}[htp]
  \begin{center}
    \includegraphics[width=10cm, height=5cm]{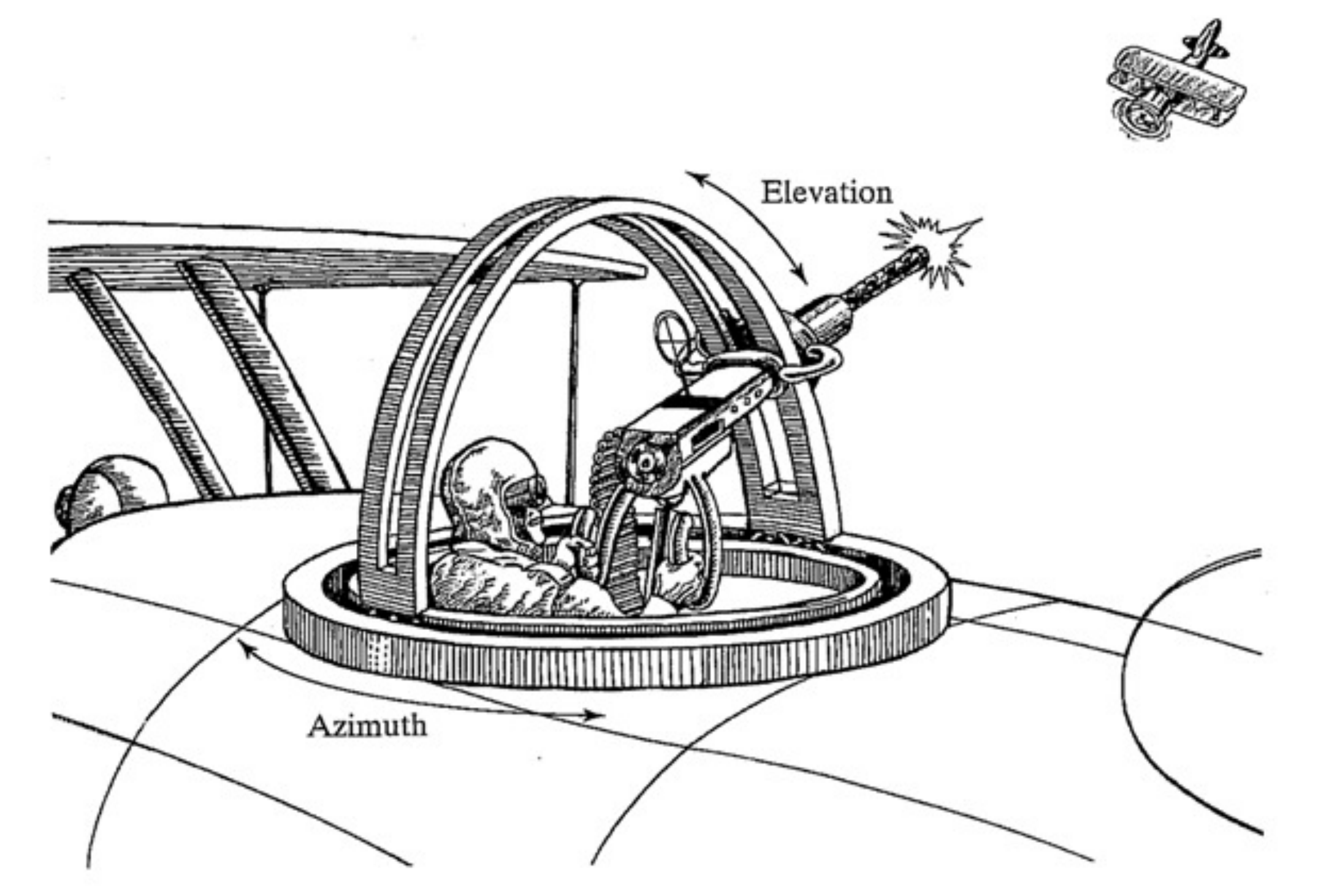}
  \end{center}
  \caption{A World War I biplane with a pilot and a rear gunner subject to singular positions, \cite{Cra04}}
  \label{fig11}
\end{figure}

Consider a robot in motion, just as the position in joint space is easily converted into the Cartesian space, the velocity in the joint space could be mapped into the Cartesian space via multiplication by a matrix called the \emph{Jacobian}. However unlike the position, there might come across situations in which the transfer from joint space into the Cartesian space is not reversible. In those specific points we say that the robot is trapped in a singular point. That is to say, it is undesirable to approach the singular points through the workspace. For this reason, finding singular points so that not to come near to them is an important task in design of a controller. Figure \ref{fig10} depicts the joint and Cartesian speed.
A tangible example of singularity is maybe the biplane fighters that had been used during World War I. The fighters were equipped with a 2 DOF gun which could easily rotate about two axes called Azimuth and Elevation. The realized mechanical system used to work well unless the target passes almost over the gunner's head! In that case tracking became a difficult task. In theory that means if the target flies exactly above the gunner's head then he should rotate the gun with an infinite speed! Due to the singular points that exist in that direction (see Figure \ref{fig11}. In fact this is an inherent limitation of any 2 DOF orienting mechanism that has exactly two rotational joints \cite{Cra04}. 
In fact when a robot reaches a singular point it loses some degrees of freedom. In that case it could not move around (along) the aforementioned direction any longer as it has already lost it! In literature we call the mechanism \emph{locally degenerate}.  Although singular points do not prevent the robot from reaching a specific point in the workspace, however, it makes problem for the motion when approaching those certain points.
Notice that considering our platform model singularities is something beyond the scope of this thesis
and thus, is not covered in this work.
\subsection{\textbf{Dynamics}}
\emph{Dynamics} is a field of research in robotics which takes into account the forces needed to make a robot follow a trajectory with a predefined velocity and acceleration. Consider a robot arm that takes an object then starts moving from initial rest and increases its speed (acceleration) to a specific amount then continues with a constant pace and finally stops with a damping velocity and puts the object on the destination point. Through the whole motion the required actuators' torques should be calculated correctly such that the smoothed motion occurs. 
\subsection{\textbf{Spatial Description}}
To describe an object in space it should be ascribed with attributes indicating where and to what direction that object is placed. By objects we mostly mean the parts or tools. We can also specify a \emph{frame} using both of these attributes. More specifically, if incorporate both position vector and orientation matrix (to be stated shortly) we come to a more general term, \emph{frame}. Description is relative concept meaning that it should be performed in relation to a \emph{universal coordinate system}. To specify position in 3D space we need to determine 3 elements which together form a \emph{position vector}. For the orientation, we attach rigidly a coordinate system to the object and denote three (although two suffices) unit axes as elements of the corresponding vector. Notice that each of the orientation elements themselves are a combination of three elements which denotes its components along the X, Y and Z axes of the universal coordinate system. That means unlike the position, the orientation is described by a $3\times 3$ matrix. We call the foregoing matrix the rotation matrix (for we see shortly). Accordingly an orientation matrix is expressed as follows:
\begin{equation}
{}_B^A R = [{}^A\mathord{\buildrel{\lower3pt\hbox{$\scriptscriptstyle\frown$}}\over 
 X} _B \,{}^A\mathord{\buildrel{\lower3pt\hbox{$\scriptscriptstyle\frown$}}\over 
 Y} _B \,{}^A\mathord{\buildrel{\lower3pt\hbox{$\scriptscriptstyle\frown$}}\over 
 Z} _B ]\, = \,\left[ {\begin{array}{l}
   {{}^B\mathord{\buildrel{\lower3pt\hbox{$\scriptscriptstyle\frown$}}\over 
 X} _A^T }  \\
   {{}^B\mathord{\buildrel{\lower3pt\hbox{$\scriptscriptstyle\frown$}}\over 
 Y} _A^T }  \\ 
   {{}^B\mathord{\buildrel{\lower3pt\hbox{$\scriptscriptstyle\frown$}}\over 
 Z} _A^T }  \\ 

 \end{array}} \right]
\label{2.1}
\end{equation}
where the leading superscript, A, is reference coordinate system and the B is the attached one. The trailing subscript, A or B, denotes the related component of A or B respectively. Note that it's not hard to verify the last equality. One can easily show that $_B^AR$ is in fact the description of frame {A} relative to {B} assuming that they coincide. In other words given a position vector,  in the former, one can easily find its equivalent in the latter simply by calculating ${}_A^B R \times {}^AP$. As stated earlier a frame is completely declared using 4 vectors (3 for orientation and one for position) which describes in detail a coordinate system relative to a reference. Based on what has stated yet, we can convert a vector description from a source frame to a destination frame in a more general fashion, that is when the frames are not coincident. In that case we have:
\begin{equation}
{}^AP = {}_B^A R \times {}^BP\, + \,{}^AP_{BORG} 
\label{2.2}
\end{equation}
where ${}^AP_{BORG} $ is the description of $\{B\}$'s origin in the $\{A\}$ frame. There is a more convenient notation that states \eqref{2.2} in a rather better closed form by introducing a slack constant into the position vectors, as follows:
\begin{equation}
{}^AP' = {}_B^A T \times {}^BP'
\label{2.3}
\end{equation}
where ${}^AP' = \left[ {\begin{array}{c}   {{}^AP}  \\   1  \\\end{array} } \right]_{4 \times 1} $ and ${}_B^A T = \left[ {\begin{array}{c c}
   {{}_B^A R} & {{}^AP_{BORG} }  \\ 
   \bf{0} & 1  \\ 
	\end{array}
 	}  \right]_{4 \times 4} $ is called a \emph{homogeneous} transform indicating the description of frame {B} relative to {A}.
\subsection{\textbf{Operators}}
A little thought reveals that equations \eqref{1.2} and \eqref{2.3} are also applicable to the case where a vector is rotated, translated or both inside the same frame. In other words we have,
\begin{equation}
{}^AP' = T \times {}^AP''
\label{2.4}
\end{equation}
where the leading subscript and superscript of the homogenous transform matrix are omitted for that the operation is done inside the same frame. As an example, a transform matrix which describes a frame that has rotated 30 degrees about the Z axis of a reference frame is the same as the transform matrix which exemplifies a 30 degree (and not a -30 degree) rotation of a vector inside a specific frame.
\subsection{\textbf{Manipulators}}
In this section we try to present a rather lucid background of parallel robots to ease understanding of materials presented in subsequent chapters. Firstly, several elementary and yet important concepts and notations are introduced which help us to develop further concepts as much as needed in future sections/chapters. We present several robot architectures then and finally introduce the newly designed type of parallel structure at Robotics Lab, FUM. This section serves as the foundation for robotics and more specifically parallel robots. The introduced subjects and topics are mostly covered by Merlet( \cite{Mer06}) in details.
Mechanical systems whose aims are to move a rigid-solid (henceforth called end-effector) object in space are appreciated considerably in industry and academia. Every rigid body can be moved (rotated or translated) in various ways. This makes up the robot's degree of freedom (DOF). Each body can have at most 6 DOFs (3 translational freedoms and 3 rotational freedoms about the main axes; the former determines the position and the latter defines the orientation of that body). In literature the combination of position and orientation of a rigid body is called its \emph{pose}.
\subsection{\textbf{A Short Study of Classical Robots' Characteristics}}
\begin{figure}[htp]
  \begin{center}
    \includegraphics[width=10cm, height=5cm]{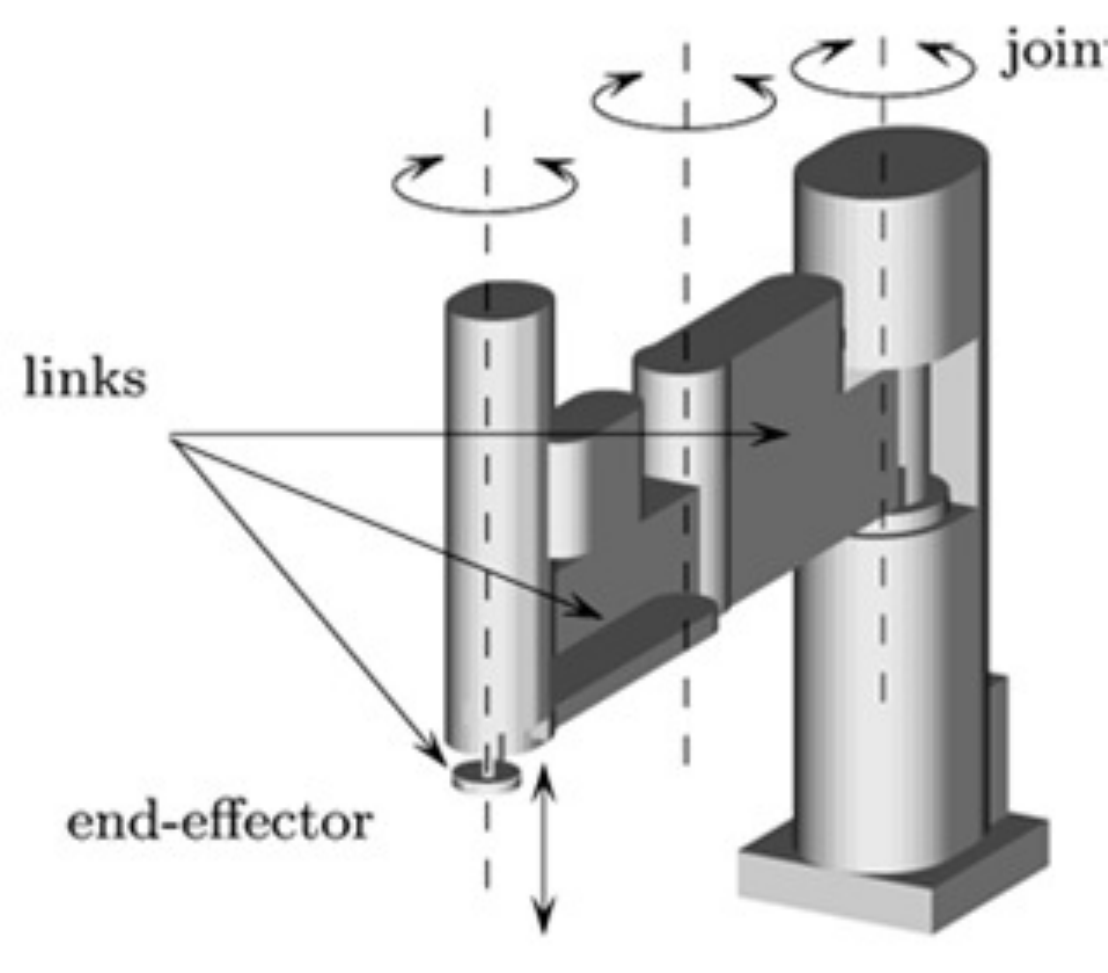}
  \end{center}
  \caption{SCARA robot, \cite{Mer06}}
  \label{fig12}
\end{figure}

\begin{figure}[htp]
  \begin{center}
    \includegraphics[width=10cm, height=5cm]{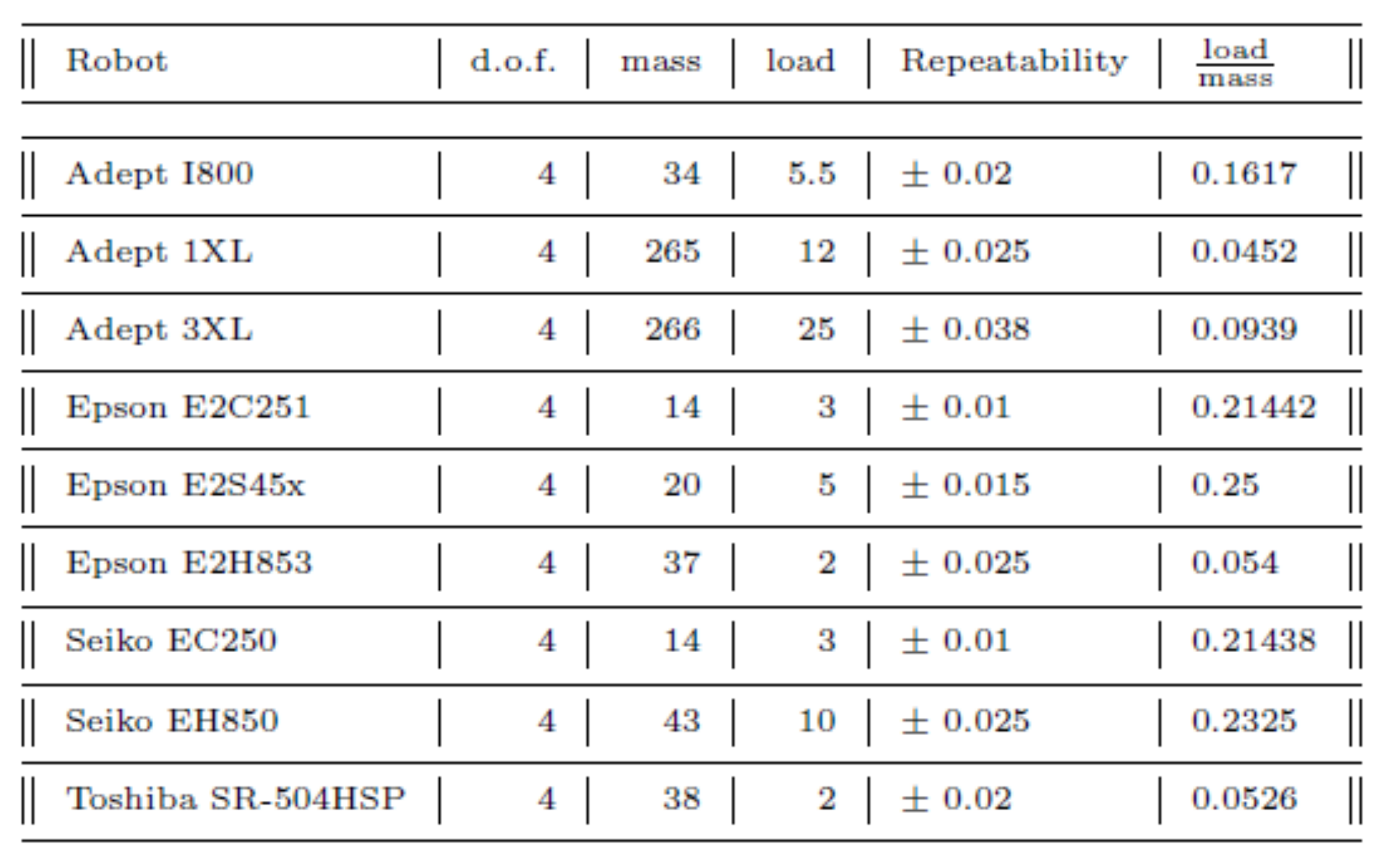}
  \end{center}
  \caption{Characteristics of industrial SCARA type robots (Repeatability in mm), \cite{Mer06}}
  \label{fig13}
\end{figure}

Presently, most existing manipulators are human like (anthropomorphic) robots such as those taking after a human arm. That is they are made up of a succession of rigid bodies (called \emph{links}) linked to each other through a 1 DOF joint (links with higher DOFs can be decoupled to a number of imaginary joints each possessing only 1 DOF). These architectures are called \emph{serial} robots due to serial connection of the robots elements. An example of this mechanism is the \emph{spherical} robot, where a succession of segments goes from the base to the tool-tip that are linked via revolute joints. \emph{SCARA} is a good pedagogical instance of this type of robots which is depicted in Figure \ref{fig12}. It has 4 actuated revolute joints totally representing 4 degrees of freedom for the tool tip. Figure \ref{fig13} represents characteristics of industrial SCARA robots. 
\emph{Repeatability}, in this figure, is a measure that defines the maximum error between two consecutive movements started from the same pose. This may be considered as a metric for positioning accuracy. Uncertainty (originated from different sources) is the reason for this inexactness; as discussed later. Several factors may be the cause of this uncertainty the most important of which are (\cite{Mer06}): Errors imposed by actuator sensors (such as pulse counters), clearance in drives, flexure of the links, and quality of geometric realization. The noteworthy point about serial robots is that generally, the absolute accuracy in a serial robot is poor. Apart from the positioning accuracy the amount of load mass that a serial robot can handle is relatively low with respect to its mass, too (consider the load to mass ration in the last column). In a serial architecture each joint and link has to withstand the weight of the subsequent links and joints along with the load. This itself requires large amounts of torques and highly stiff segments which ultimately increases the weight of a serial robot to high above an endurable amount. The heavy segments also imply higher inertia which confines the movement and control of the robot especially when high speed is needed as centrifugal and Coriolis forces become significant. Notice that in robots of serial architectures the error becomes magnified as it propagates to the tool tip. This severs the situation while trying to impose a threshold on the error.
To sum up, serial robots can not serve as a good selection when manipulation of heavy objects is needed or when the accuracy is a matter of concern.
\subsection{\textbf{Parallel Manipulators}}
\begin{figure}[htp]
  \begin{center}
    \includegraphics[width=10cm, height=7cm]{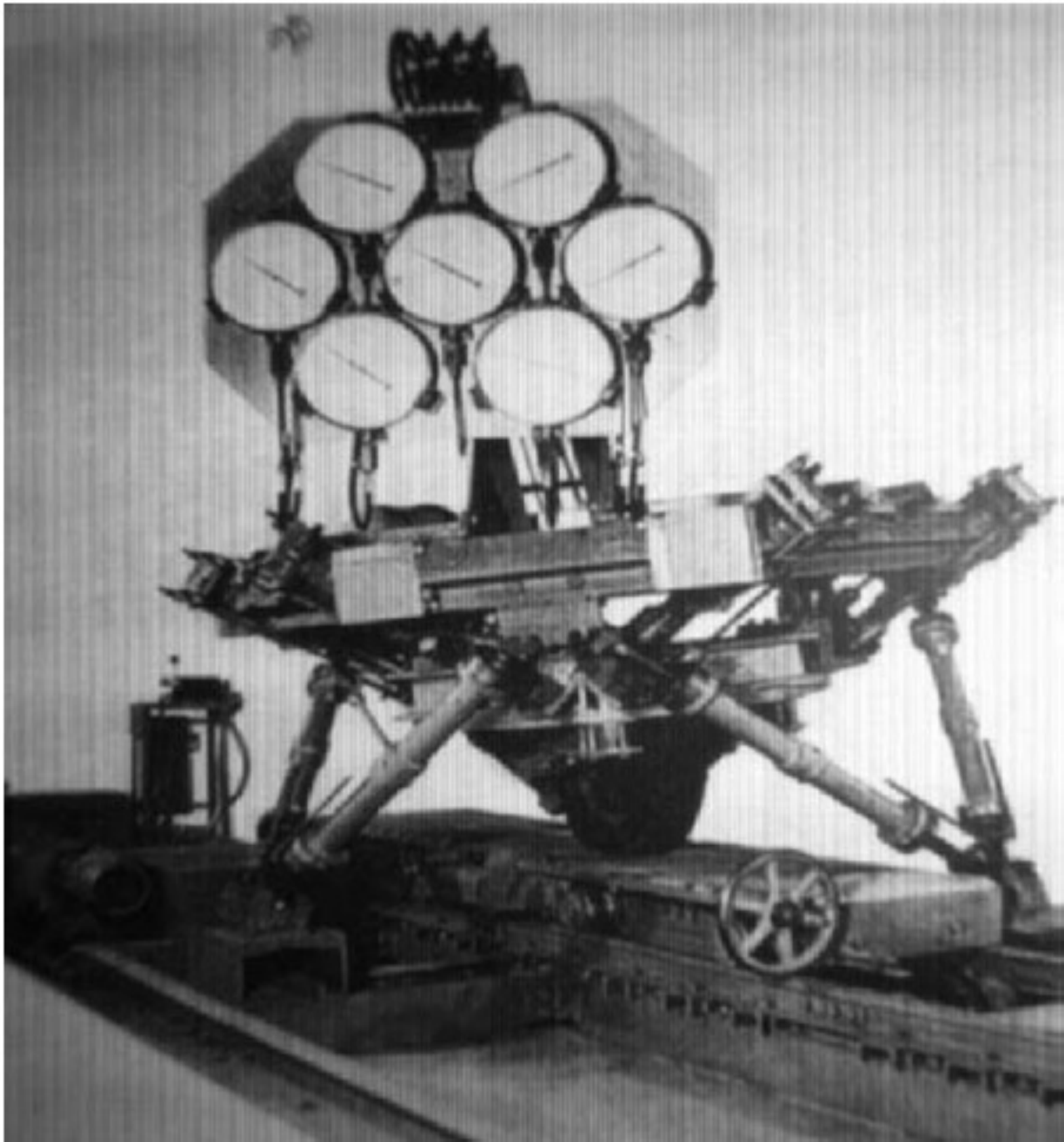}
  \end{center}
  \caption{Gough platform, a closed loop kinematic structure implemented in 1955, \cite{Mer06}}
  \label{fig14}
\end{figure}

\begin{figure}[htp]
  \begin{center}
    \includegraphics[width=10cm, height=7cm]{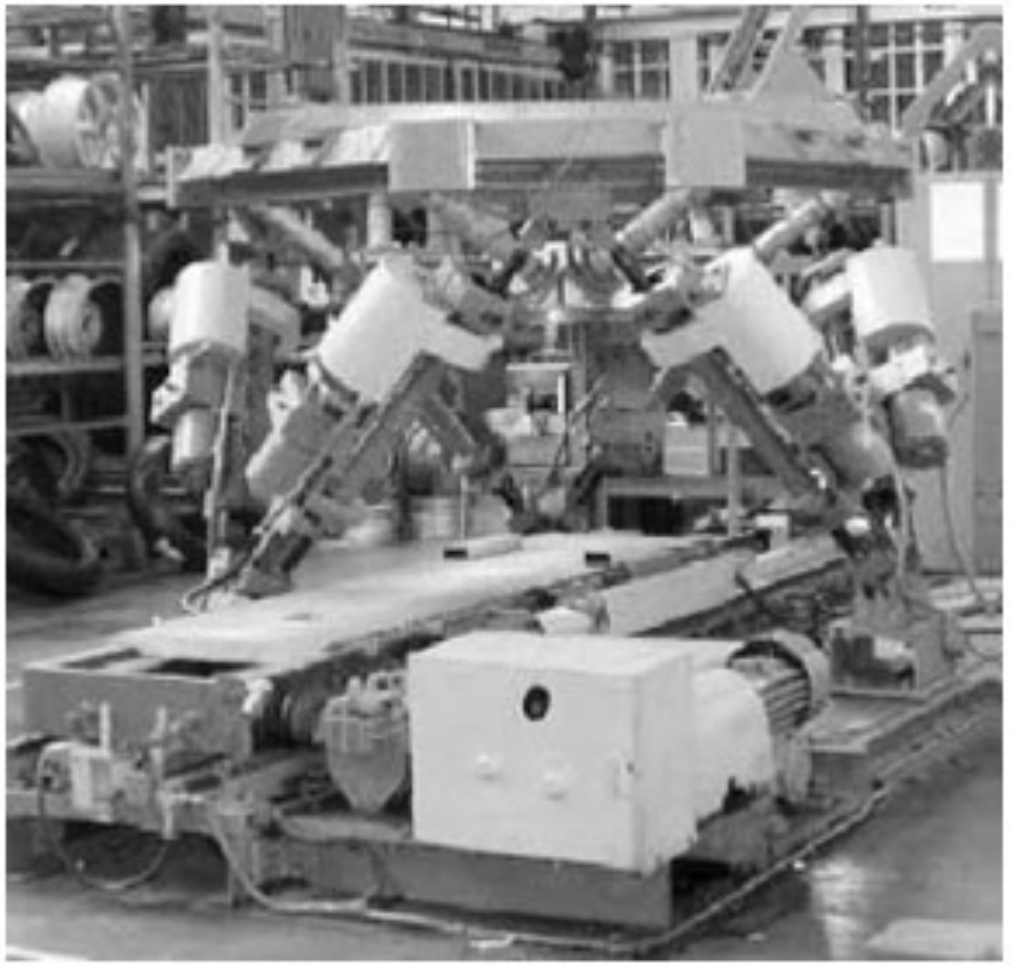}
  \end{center}
  \caption{A more modern platform based on the design introduced by Gough, \cite{Mer06}}
  \label{fig15}
\end{figure}

In order to introduce parallel robots some formal notions should be defined first:
\begin{definitions} The number of independent degrees of freedom pertaining to an end-effector is called \emph{Mobility}.
\end{definitions}
\begin{definitions}
For a link, \emph{connection degree} is defined as the number of rigid bodies attached to a joint.
\end{definitions}
\begin{definitions}
If each element possesses a connection degree of 2 or less then it can be called a \emph{simple kinematic chain} like a serial arm. Since a serial manipulator includes links whose connection degrees are either 2 or 1 (which is associated to the base and the end-effector). A serial manipulator is also called an \emph{open-loop} kinematic chain.
\end{definitions}
\begin{definitions}
Contrary to \emph{open-loop} kinematic chain, a \emph{closed-loop} chain is obtained if the connection degree associated to one of the links (but not the base) is equal or more than 3.
\end{definitions}
\begin{definitions} 
\emph{Generalized Parallel Manipulator} as stated in \cite{Mer06} is an open definition. It for instance includes redundant mechanisms with the number of actuators more than the degree of freedom associated with the end effector or it can also be defined as manipulators working in parallel. In this thesis we only consider the parallel manipulators of the following characteristics:
\begin{itemize}
\item At least two chains support the end-effector where each chain must contain at least one simple actuator and of course an appropriate sensor to measure the associated variables (corresponding to either of rotational movements or linear motion)
\item There is not any redundant actuator. The number of DOF of the end-effector is exactly the same as the number of actuators.
\item The mobility of the manipulator is zero when all the actuators are locked.
\end{itemize}
This makes the problem at hand interesting due to the following reasons:
\begin{enumerate}
\item The load mass distributes over each chain which makes it suitable for handling heavy masses
\item The number of actuators becomes minimum which makes is suitable from different aspects including complexity issues
\item The number of required sensors are also minimum
\item When the actuators are locked the end-effector (and other elements) remains at a certain position. This is especially important when safety issues are a matter of concern such circumstances appear in different robots including medical robots.
\end{enumerate}
\end{definitions}
\begin{definitions}
A robot is considered \emph{parallel} when comprised of an end-effector with $n$ degrees of freedom and a fixed base which are connected together by at least two independent kinematic chains. In this robot the actuation takes place using $n$ simple actuators.
\end{definitions}
\begin{definitions}
A parallel robot for which the number of DOF of the end-effector is strictly equal to the number of chains is called a \emph{fully parallel} manipulator. 
\end{definitions}
In 1947 Gough, \cite{Gou56}, designed a mechanism for with a closed loop kinematic structure \ref{fig14} which caused positioning and orienting a moving platform to test the wear and tear issues of a tire. He later made the prototype of his design in 1955. There are 6 links each of which is connected to the base via a universal joint and via a ball-and-socket joint to the moving plate. The mechanism is actuated using 6 linear motors. The platform was such a successful one that had worked till 2000. Figure \ref{fig15} depicts the rather modern prototype of his design used in the Dunlop Tyres company (courtesy of Mike Beeson from this company). It should be noted however that although Gough was the first to propose a functional prototype of a parallel structure (of hexapod form), parallel robots of this type (hexapod) were already recognized. Today this structure can be seen in diverse range of flight simulators to produce the required shakes and movements.

\begin{figure}[htp]
  \begin{center}
    \includegraphics[width=10cm, height=7cm]{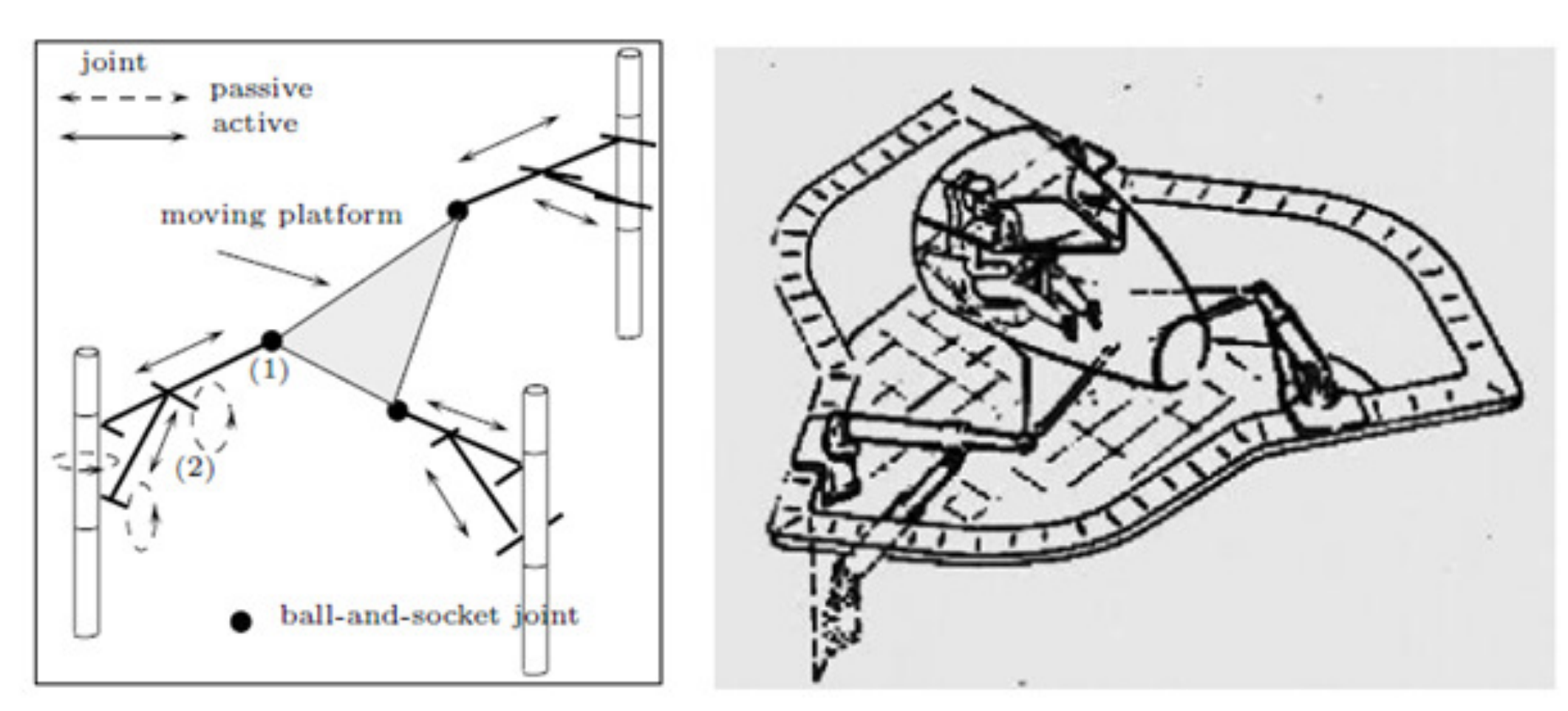}
  \end{center}
  \caption{Flight simulator model proposed by Stewart, \cite{Mer06}}
  \label{fig16}
\end{figure}

Let's consider the aforementioned design in more details so that to underline the benefits of a parallel robot in relation to a serial manipulator. It is clear that the load to mass ratio must be remarkably higher when considering a serial robot in the same size since each of the links only bears 1/6 of the total load as opposed to the former in which each link should withstand both the complete load and the succeeding links. Hence, the essential power suited for handling loads in parallel structures are considerably lower. In fact it has been shown that for a 3 DOF parallel structure, the average energy usage is 26\% of a serial manipulator of similar size. The choice of linear actuators employed in this structure is also interesting since such actuators widely exist in variety of specifications such as speed, acceleration, movement amplitude and mass with high performances. One may consider that the positioning accuracy is also good for two main reasons. Firstly, the uncertain deformation of the links, which serial robots widely undergo due to flexure, is eliminated or reduced considerably. Second, the errors in internal sensors merely translate into a slight error in the platform position and it is not magnified.
In 1965, Stewart showed that a flight simulator should follow the structure depicted in Figure \ref{fig16}. According to the proposed structure, the triangular moving platform is connected to the associated succession of two links via its vertices using ball-and-socket joints (see Fig. \ref{fig16}, left). Each of the aforementioned links is actuated via its own jack (linear actuator) which is connected to a vertical axes rotating pillar via a revolute joint on one end. The other end of one of the jacks is as already explained attached to the foregoing ball-and-socket joint and the other end of the latter is attached to the first jack via another revolute joint. The right hand side of Fig. \ref{fig16} portrays the blueprint of the Stewart design serving a as flight simulator.
\subsubsection{Architectures of Parallel Manipulators}
\begin{figure}[htp]
  \begin{center}
    \includegraphics[width=10cm, height=7cm]{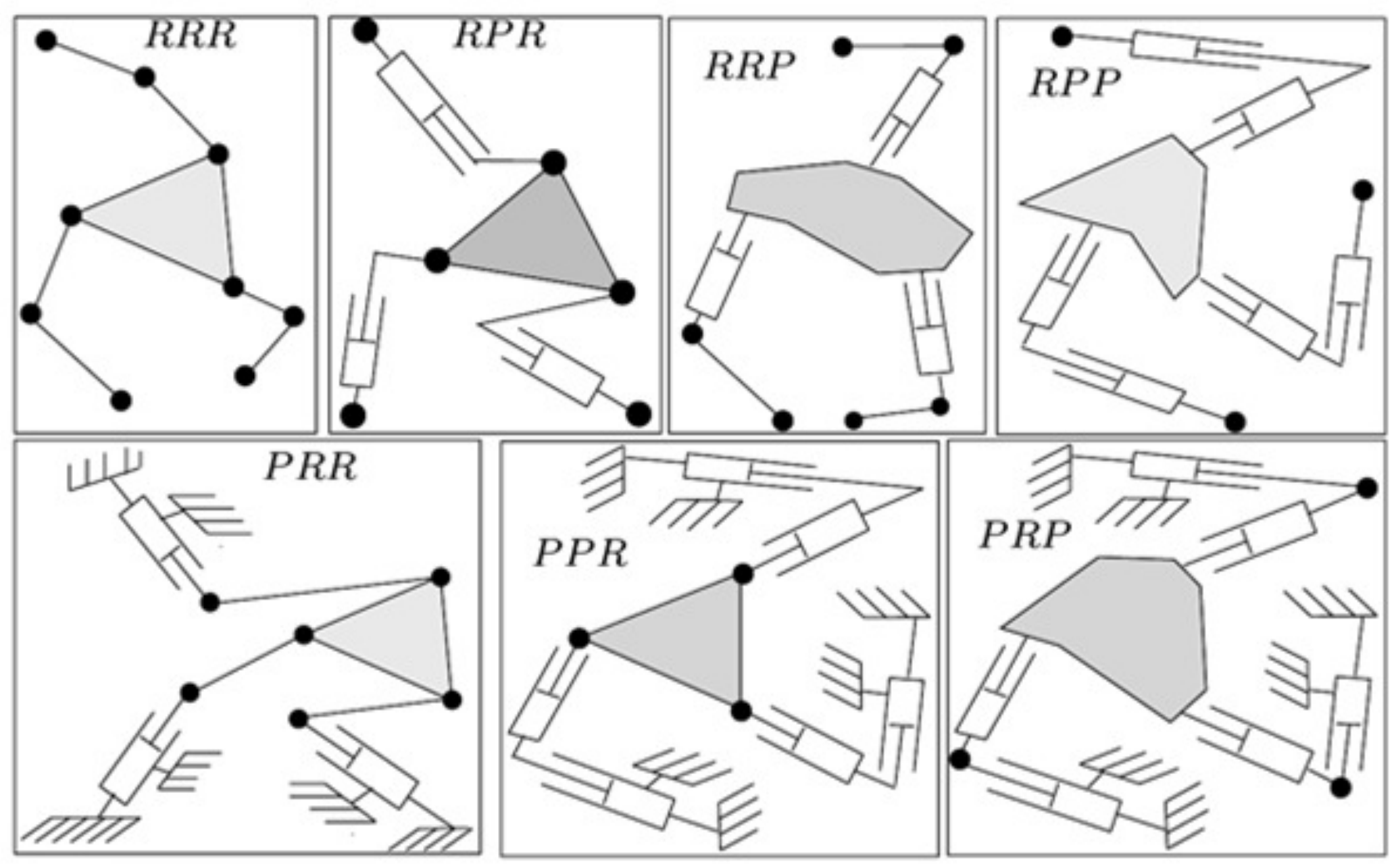}
  \end{center}
  \caption{possible sequence of joints in a kinematic chain associated with a planar parallel robot, \cite{Mer06}}
  \label{fig17}
\end{figure}

\begin{figure}[htp]
  \begin{center}
    \includegraphics[width=12cm, height=9cm]{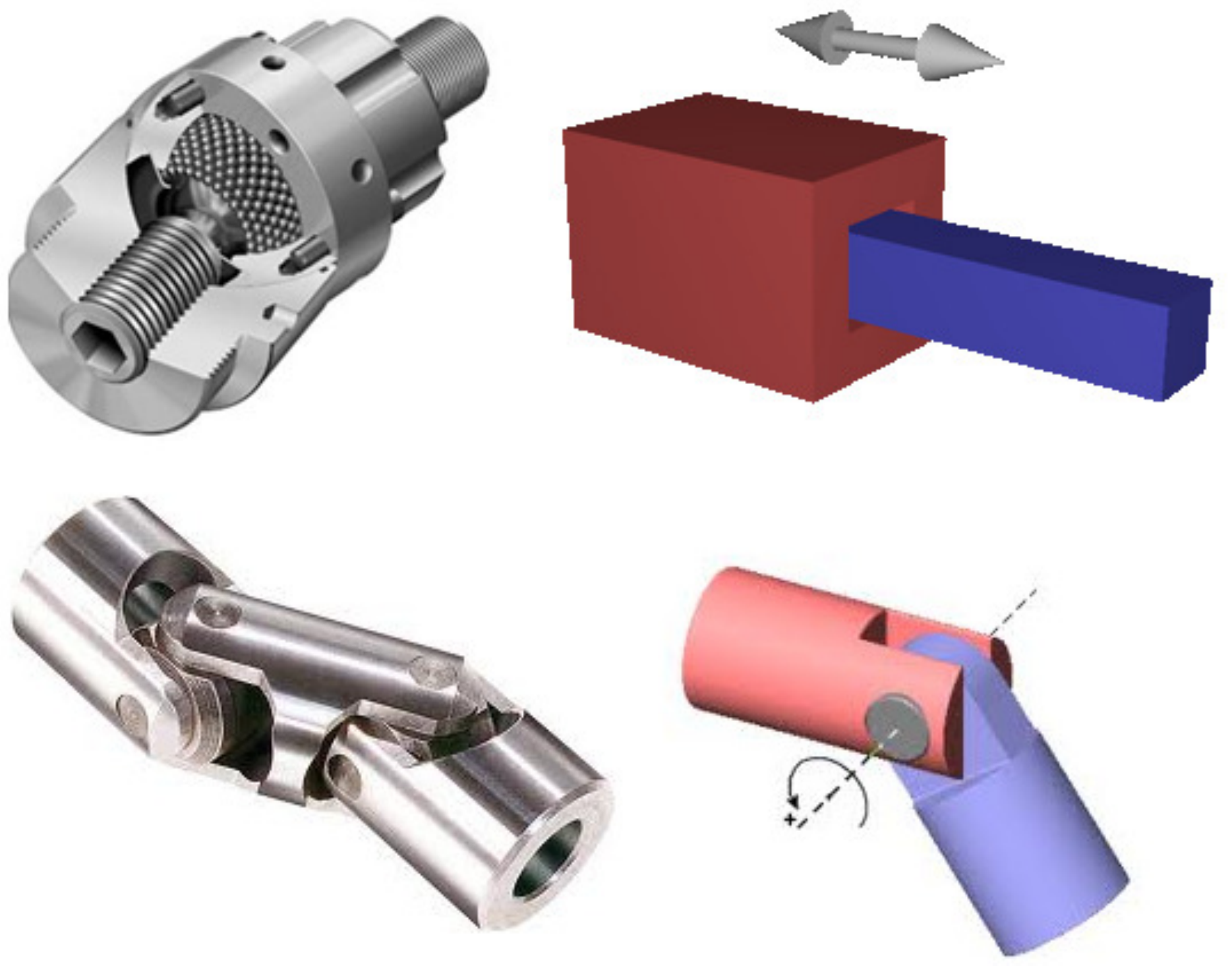}
  \end{center}
  \caption{Types of joints (top-left) spherical \cite{Mer06}, (top-right) prismatic \cite{Web1}, (bottom-left) universal \cite{Web2}, (bottom-right) revolute \cite{Web3}}
  \label{fig18}
\end{figure}

This section briefly discusses different architectures of parallel robots and present a taxonomy of such architectures so that to ease the identification and characterization of our underlying robot. As a general agreement, we use the arrowed vector for actuated joints and the dashed arrow for passive ones as a notation. We present diverse applications of different parallel robots from \cite{Mer06} to emphasize the applicability and importance of such parallel manipulators.
Due to inefficacious architecture of serial manipulators, few researchers tried to develop new ones differing substantially from the formers including Minsky \cite{Min72} and Hunt \cite{Hun78}. However those structures where not designed according to a systematic approach rather they where more relied on intuitive and initiative ideas. In contrast, \emph{structure (or type) synthesis}, is a formal approach towards systematic design of any desired parallel structure with a desired kinematic performance. The structure synthesis is however beyond the scope of this thesis and mostly includes one of the following three synthesis approaches: graph theory, group theory and screw theory. The interested reader may refer to \cite{Mer06} for detailed knowledge of structural synthesis. 
Unlike serial manipulators for which a well-defined literature has been developed, for robots of parallel architectures there hasn't been such a terminology available yet. This is very unfortunate of course since having a uniform notation bank is considered as a prerequisite for further analyses especially if the analyses should be automated by means of CAD tools. For example for the Gough architecture introduced earlier, different names (including 6-6 robot or 6-UPS or 3T-3R) may imply the same architecture each of which specifying a special attribute and none of which may express the complete characteristics. Nevertheless, this by no means implies that we cannot analyze the robots' architectures at all. In the subsequent discussion we consider several parallel manipulators of planar type. For notational convenience, we denote prismatic and revolute joints by P and R respectively and the actuated joints are underlined. 
Consider a 3 DOF robot which can move freely in the $x-y$ two dimensional plane and which can rotate around the z axis (the axis perpendicular to the foregoing plane). We are in quest of a fully parallel robot based on the definition given earlier. In other words the robot should possess three independent kinematic chains with three actuators. As each of these chains has to link the moving platform to the ground, there should be three attachments on the ground and three on the moving plate. This indirectly suggests that moving plate should have the shape of a triangle. Synthetic analysis (see chapter 1 of \cite{Mer06}) shows that under these constraints each chain should be constituted of two rigid bodies connected to each other by a joint which means that each chain should have exactly three joints. In other words there can be different permutation of joint types if counted from base upward. These sequences are as follows: RRR, RPR, RRP, RPP, PRR, PPR, PRP and PPP which are depicted in Figure \ref{fig17}. Note that the last set of joints (i.e. PPP) is omitted since the succession joints reduce to a single prismatic joint. 
It is also worthy of noting that by exchanging the base and moving plate the RRP and PRR chains and also the PPR and RPP chains become similar. As already stated, the actuator can be placed on each of the joints; however, it is more desired not to place the actuator on the end-effector to due its weight overhead. Therefore, the first joint is the ideal case to be actuated. Another possibility is that each chain can hold one of the above sequences without considering the sequence attributed to the other two chains. 
In the architectures presented above we only considered the revolute and prismatic joints. There are also two important types of joints namely the \emph{universal} and \emph{spherical} joints with associated DOFs of 2 and 3 respectively. Figure \ref{fig18} depicts all four types of joints alluded by now.
Parallel manipulators are usually classified based on their corresponding degrees of freedom, an important question that may arise at the first time a model is posed. In this thesis however, we only deal with robots which possess 3 DOFs. One of the main motivations for this discrimination is that the underlying robot that we will present in later sections is also a 3-DOF robot with a new architecture and structure. 

\subsubsection{3 D.O.F Manipulators}
\begin{figure}[htp]
  \begin{center}
    \includegraphics[width=10cm, height=7cm]{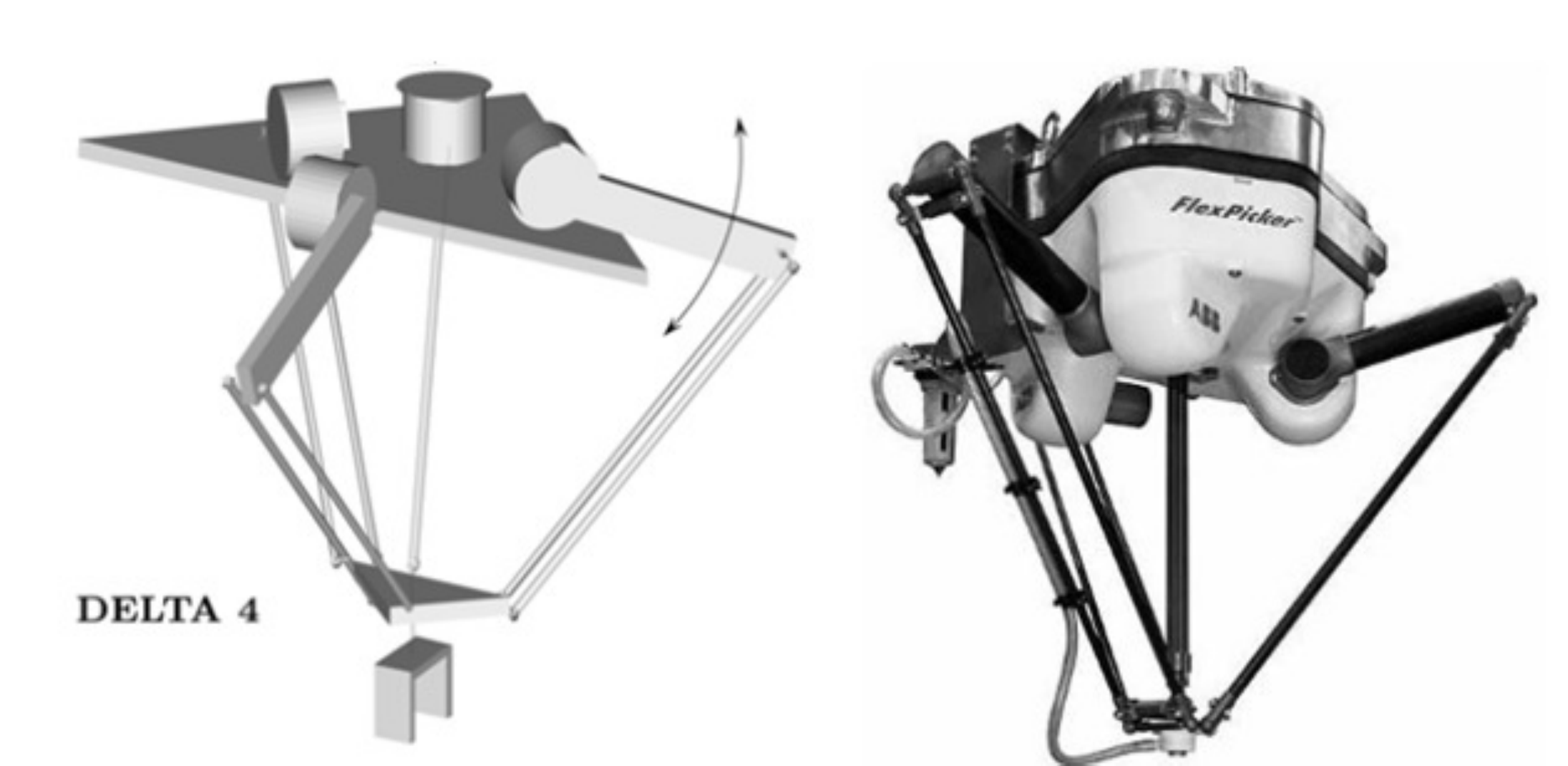}
  \end{center}
  \caption{Delta robot, \cite{Mer06}}
  \label{fig19}
\end{figure}

\begin{figure}[htp]
  \begin{center}
    \includegraphics[width=7cm, height=5cm]{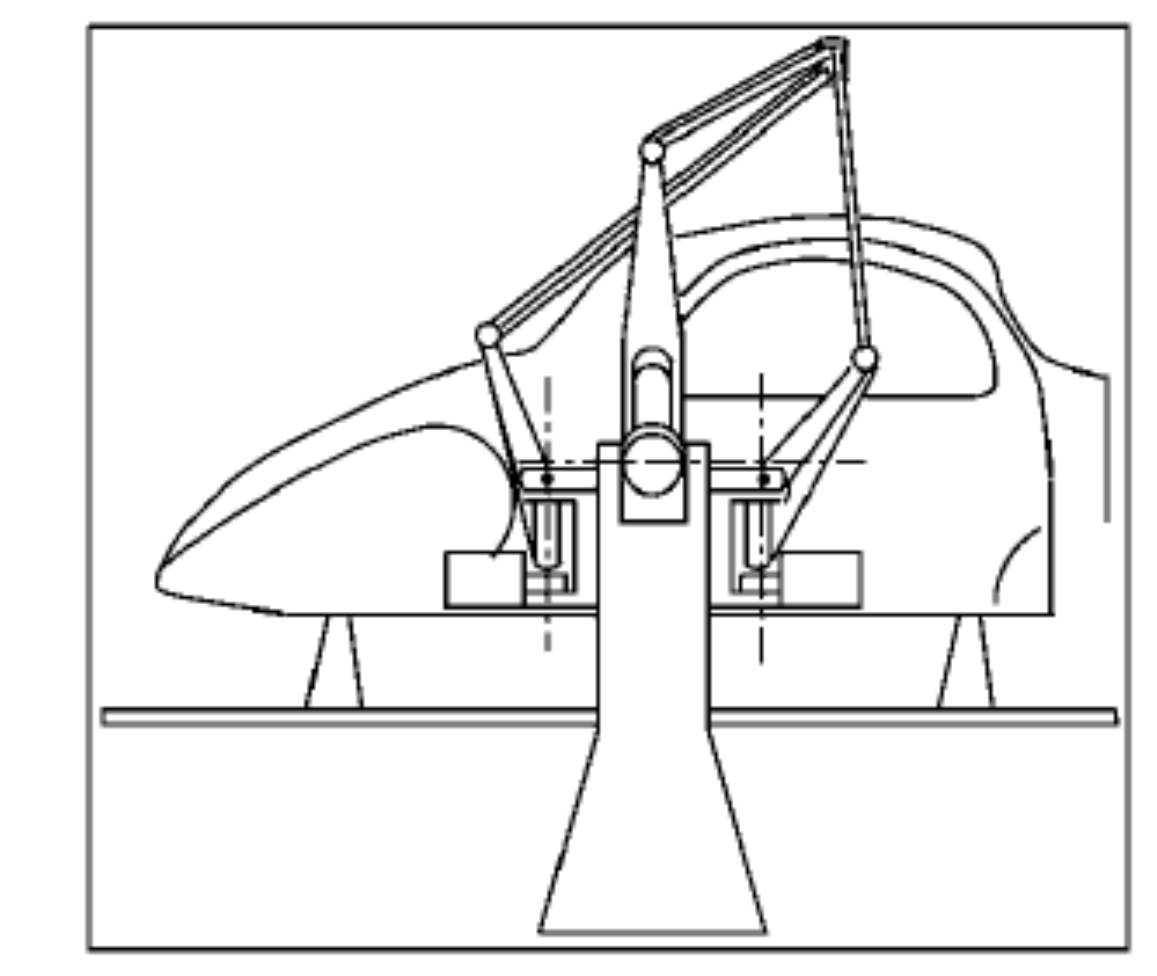}
  \end{center}
  \caption{Pollard mechanism, \cite{Mer06}}
  \label{fig20}
\end{figure}

\begin{figure}[htp]
  \begin{center}
    \includegraphics[width=10cm, height=7cm]{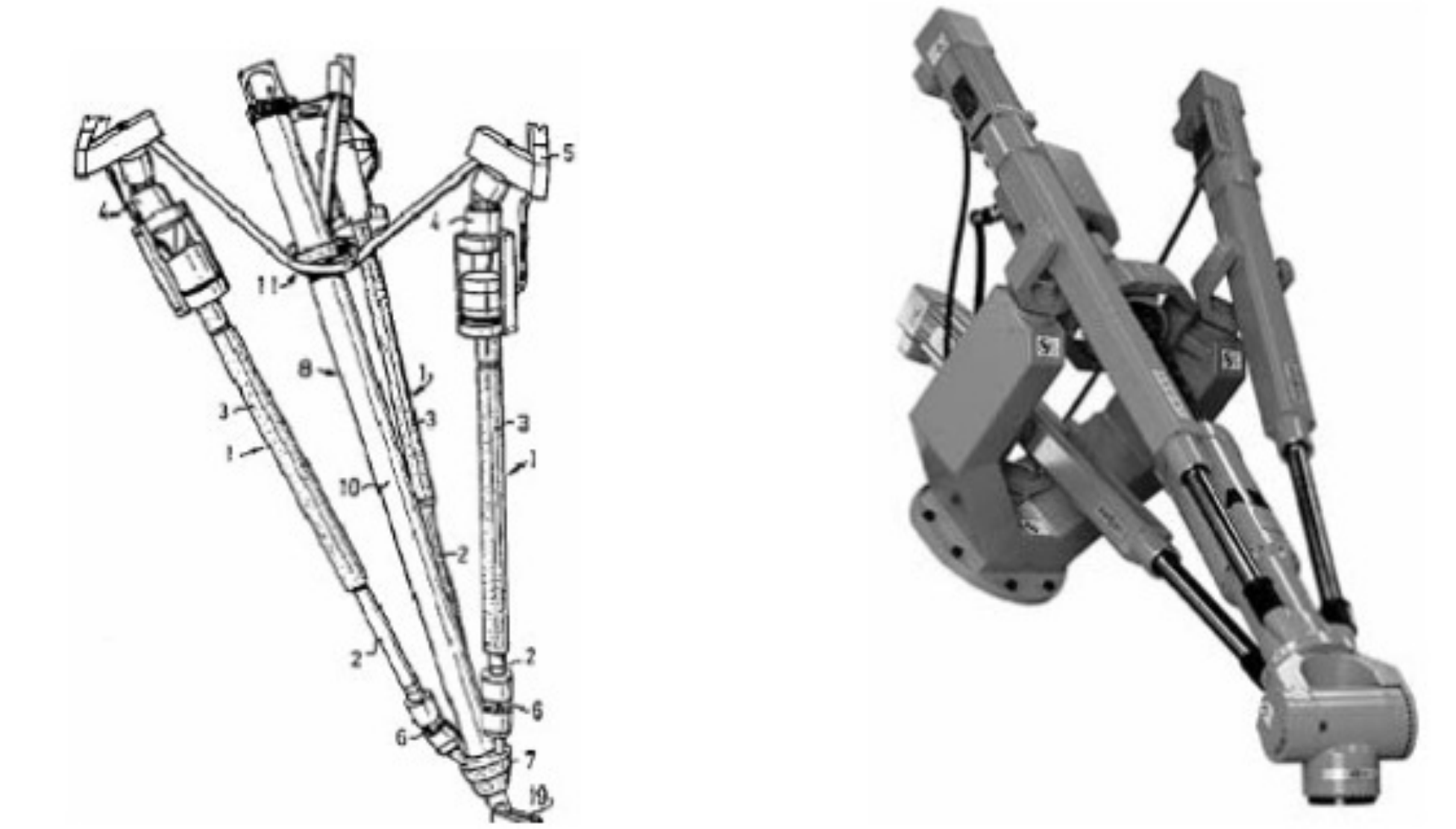}
  \end{center}
  \caption{(On the left) the Neumann patent, (on the right) IRB940, from ABB, \cite{Mer06}}
  \label{fig21}
\end{figure}

\begin{figure}[htp]
  \begin{center}
    \includegraphics[width=7cm, height=5cm]{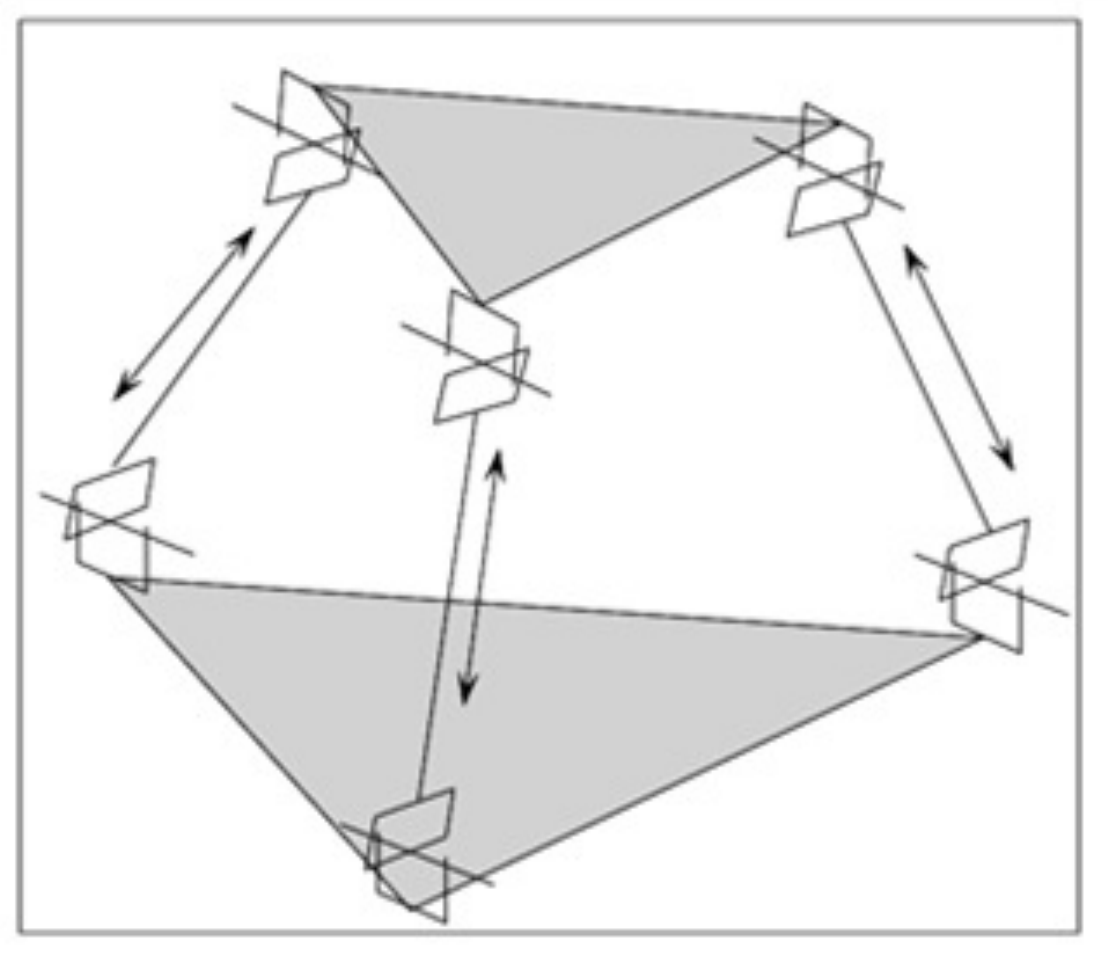}
  \end{center}
  \caption{Abstract model (architecture) of a 3-UPU manipulator, \cite{Mer06}}
  \label{fig22}
\end{figure}

\begin{figure}[htp]
  \begin{center}
    \includegraphics[width=10cm, height=5.5cm]{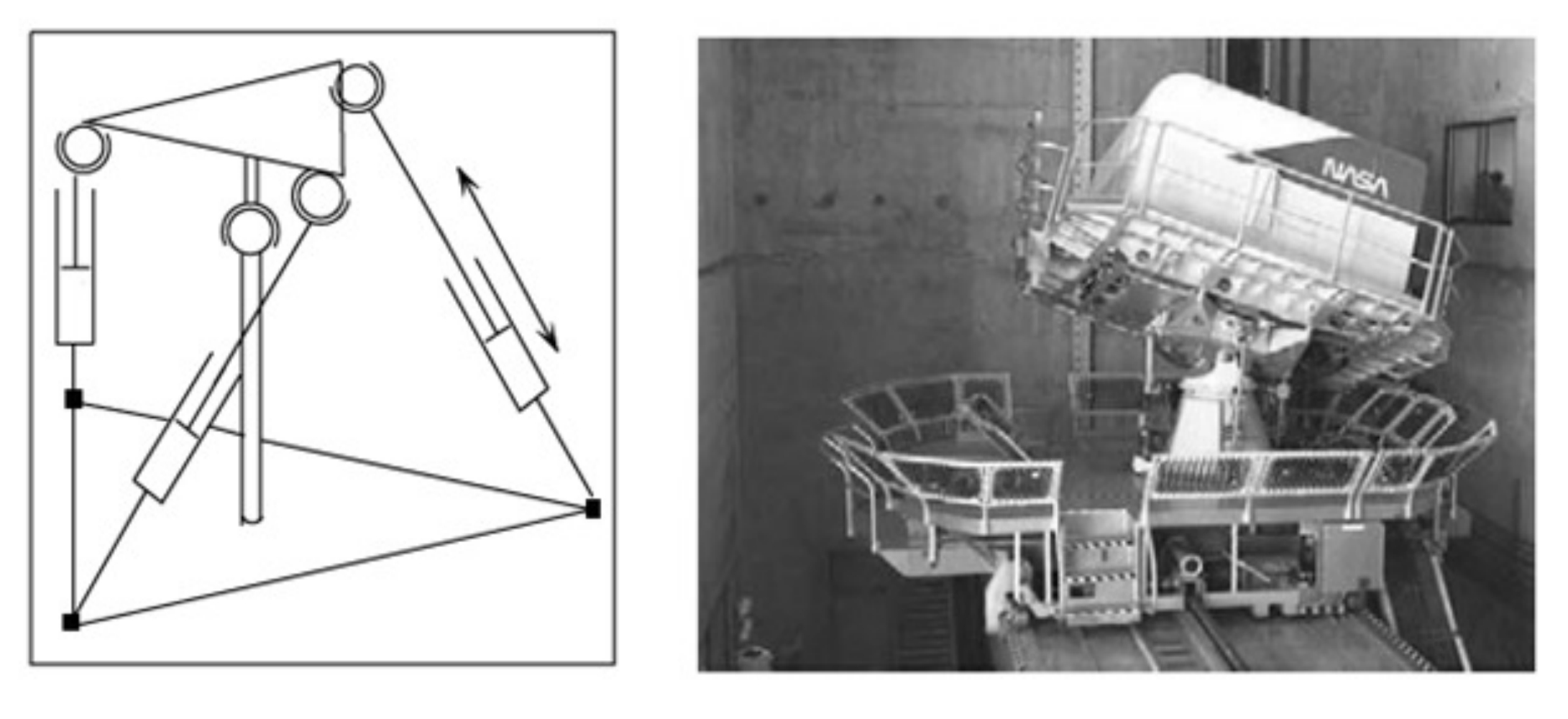}
  \end{center}
  \caption{A parallel wrist (left), a realized version of the model, \cite{Mer06}}
  \label{fig23}
\end{figure}

\begin{figure}[htp]
  \begin{center}
    \includegraphics[width=10cm, height=7cm]{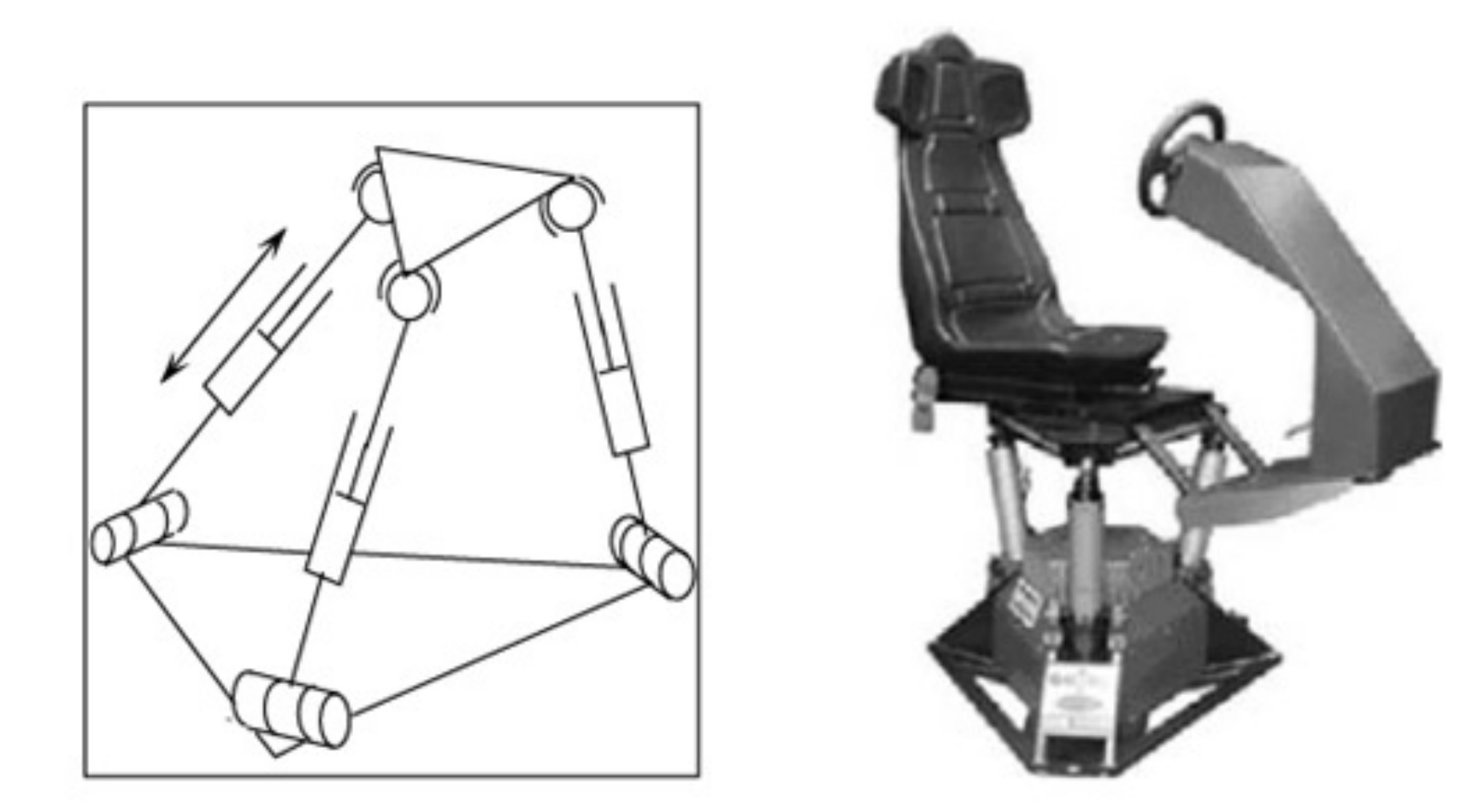}
  \end{center}
  \caption{A three degree of freedom manipulator proposed by Hunt (left), an entertainment tool as a realized version of this model (right), \cite{Mer06}}
  \label{fig24}
\end{figure}

\begin{figure}[htp]
  \begin{center} 
    \subfigure [ ]{\label{fig25-a}\includegraphics[width=6cm, height=5cm]{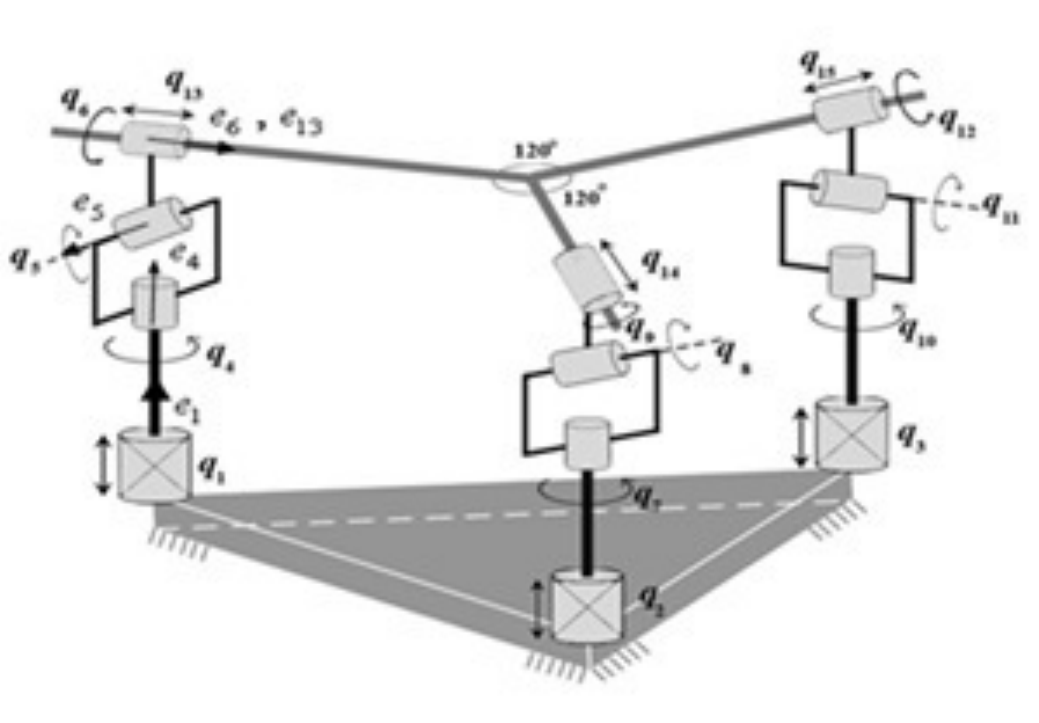}}
    \subfigure [ ]{\label{fig25-b}\includegraphics[width=6cm, height=6cm]{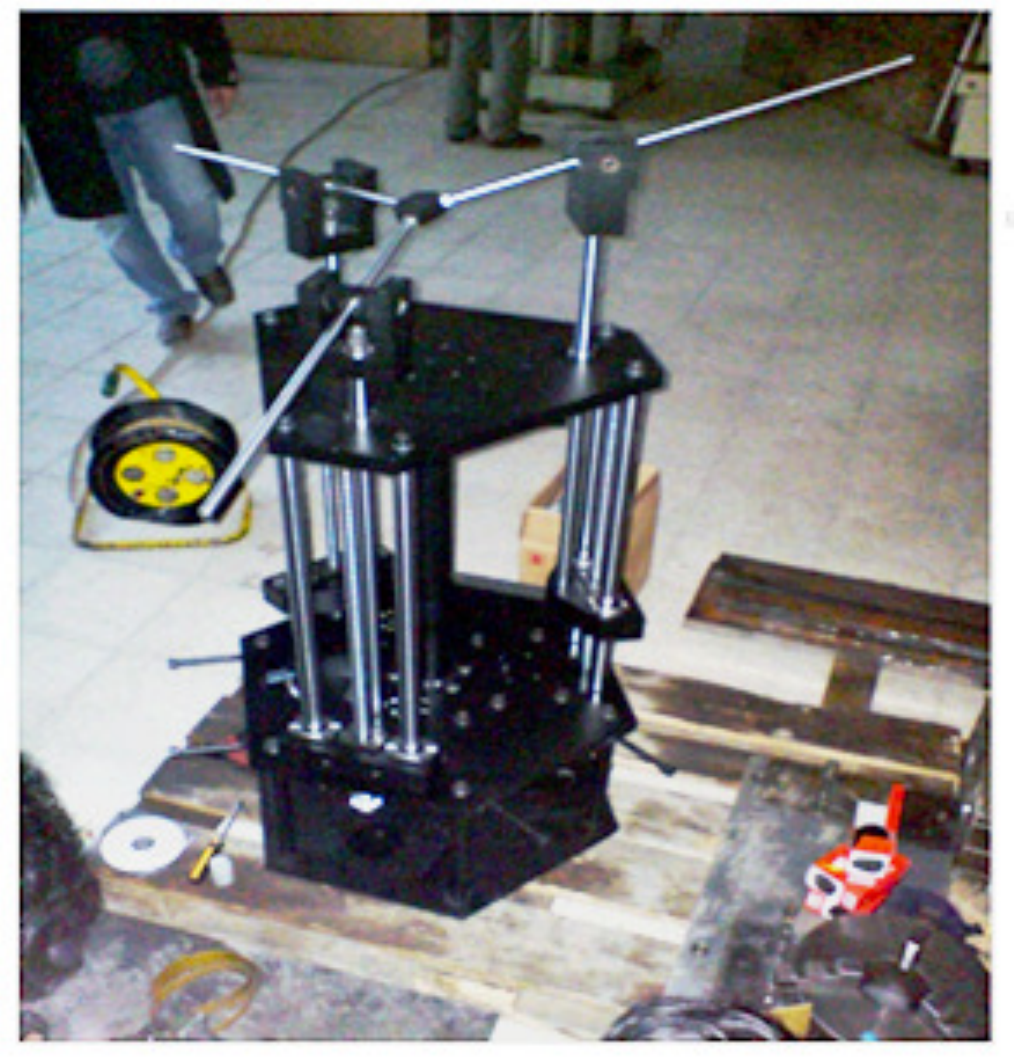}}\\
    \subfigure [ ]{\label{fig25-c}\includegraphics[width=7cm, height=7cm]{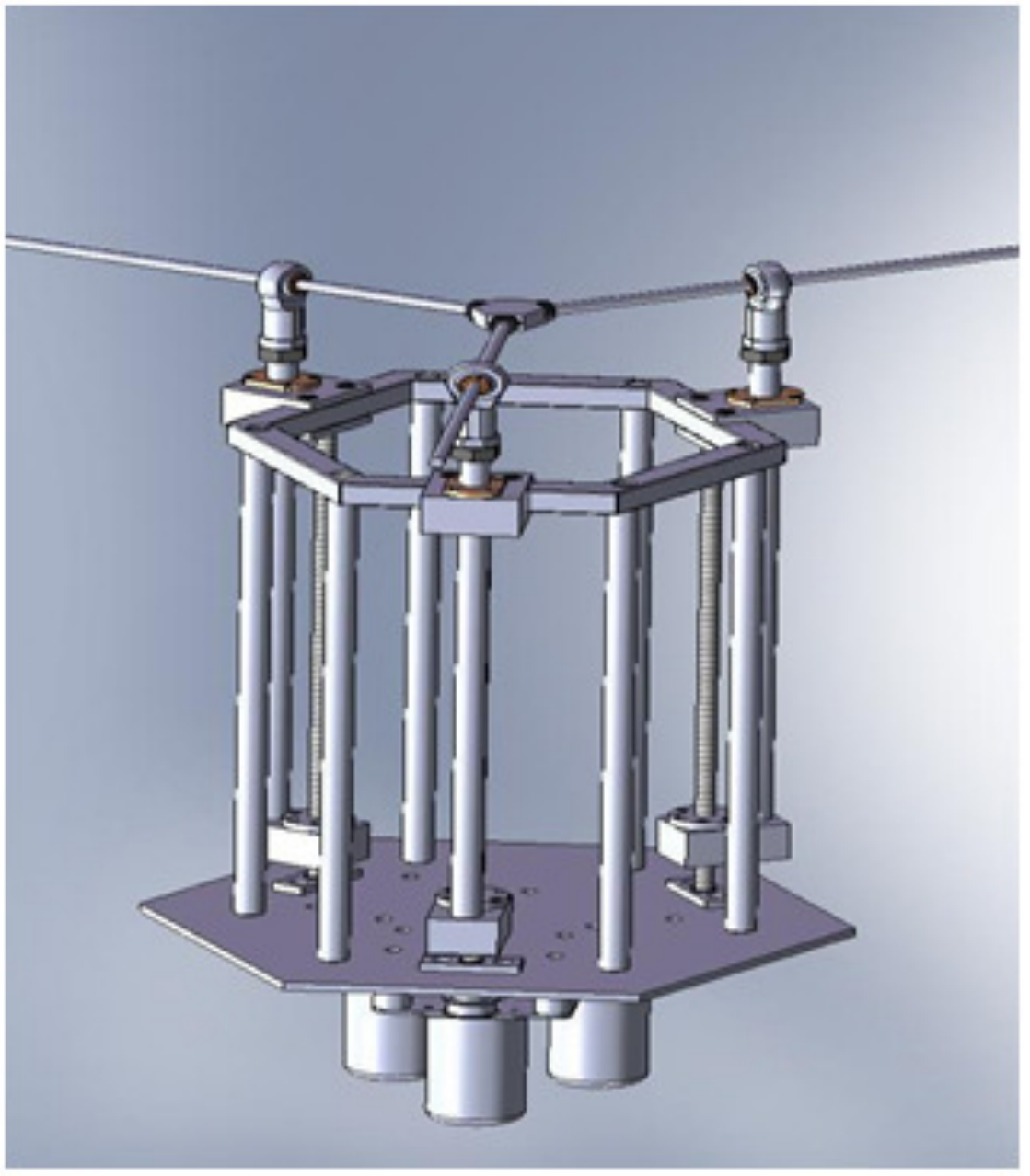}}\\
  \end{center}
  \caption{(a) A 3PSP architecture developed at robotics lab, FUM (b) the fabricated architecture, 
  (c) the abstract model in \emph{Solid Works}, \cite{Mer06}}
  \label{fig25}
\end{figure}

In this section we review several 3 D.O.F structures so that to become familiar with diverse applications of these robots in industrial world. 3 D.O.F manipulators are categorized into three main classes namely \emph{Translation}, \emph{Orientation} and \emph{Mixed Degrees of freedom} Manipulators. The first is suitable for applications that need to perform moving actions; the second type is specialized for arranging the end-effector in a desired orientation and the last is a mixture of both. Each of these methods has its own applications. We finally present our new design and realization of a 3-PSP parallel manipulator which is a hybrid of both translation and orientation mechanisms (falls in the last category).
\paragraph{Translation Manipulators}
Manipulators with 3 D.O.F are particularly applicable to the pick-and-place and machining tasks \cite{Mer06}. Of this type the most well-known robot is the Delta robot \cite{Cla88}, developed at EPFL. This robot can move its end-effector along each of the Cartesian coordinate system axes, see Figure \ref{fig19}. Each of the robot's chains is made up of a RRPR joint sequence. All the revolute joints share the same rotation axis, say $w$. The actuator operates on the first revolute joint. The first revolute joint is attached to the second revolute joint via a link after which comes the parallelogram which is permitted to move along an axis parallel to $w$. This joint then meets the last revolute joint which is attached to the end-effector. This is a very useful architecture and the implemented robot has been employed widely in industry since its first introduction to the market. A word under notation is that the ancestor of Delta is a mechanism described by Pollard which was intended to be used for car painting applications, see Figure \ref{fig20}. To date several variants of this architecture has been developed yet each of which is specialized for a specific application.
The next industrial 3 D.O.F robot is the Tricept made according to a patent by Neumann \cite{Neu88}. Each kinematic chain in this robot is constituted of a joint sequence of RRPS type that acts on the end-effector. Figure \ref{fig21} depicts this robot and its associated patent. This architecture has been also employed to realize a huge hybrid of serial and parallel manipulators made for assembly issues (named Tetrabot) but never introduced to the market.
In academic studies, the most investigated 3 D.O.F parallel robot is the 3-UPU (a special family instance of the family of 3RRPRR) robot suggested by Tsai \cite{Tsa96}, for the first time. Figure \ref{fig22} depicts this architecture of such model.
Two of the main motivations for academia to consider this manipulator is that is high sensitivity  to construction tolerances and also the fact that it is located in a singular situation at its nominal position.
\paragraph{Orientation Manipulators}
Manipulators allowing three rotations about one point are called orientation manipulators. What is most important in such platforms is not the position of robot's end-effector but rather the orientation of the tool-tip. The need for such robots arises especially in tracking radar systems and missile launching bases. Figure \ref{fig23} depicts the so-called parallel wrist (on the left) and a realized version of the model from NASA (on the right). The manipulator is comprised of three RRPS (or UPS) kinematic chains.
Note that in Figure \ref{fig23}, the solid squares denote the universal joints.
\paragraph{Mixed Degrees of Freedom Manipulators}
The important characteristic of manipulators of this type is that their freedom is comprised of both movement and orientation. Figure \ref{fig24}(left) depicts a 3 D.O.F manipulator proposed by Hunt \cite{Hun83} which has been studied by several authors to date. Several variants of this robot have been developed yet. 
\subparagraph{The Underlying 3 D.O.F PSP Robot}
A mixed degree of freedom robot with two rotational and one translational movement has been developed (and yet under upgrades) at Robotics Lab, Ferdowsi University of Mashhad. The robot is a parallel manipulator with 3 PSP kinematic chains. Figure \ref{fig25} depicts this robot. The manipulator is a fully parallel robot with the bottom most joints (the prismatic ones) actuated.

%% file: chapter3.tex
\chapter{Dynamics of the 3-PSP Parallel Robot}
\section {\textbf{Preface}}
\begin{figure}[htp]
  \begin{center} 
	\includegraphics[width=10cm, height=5cm]{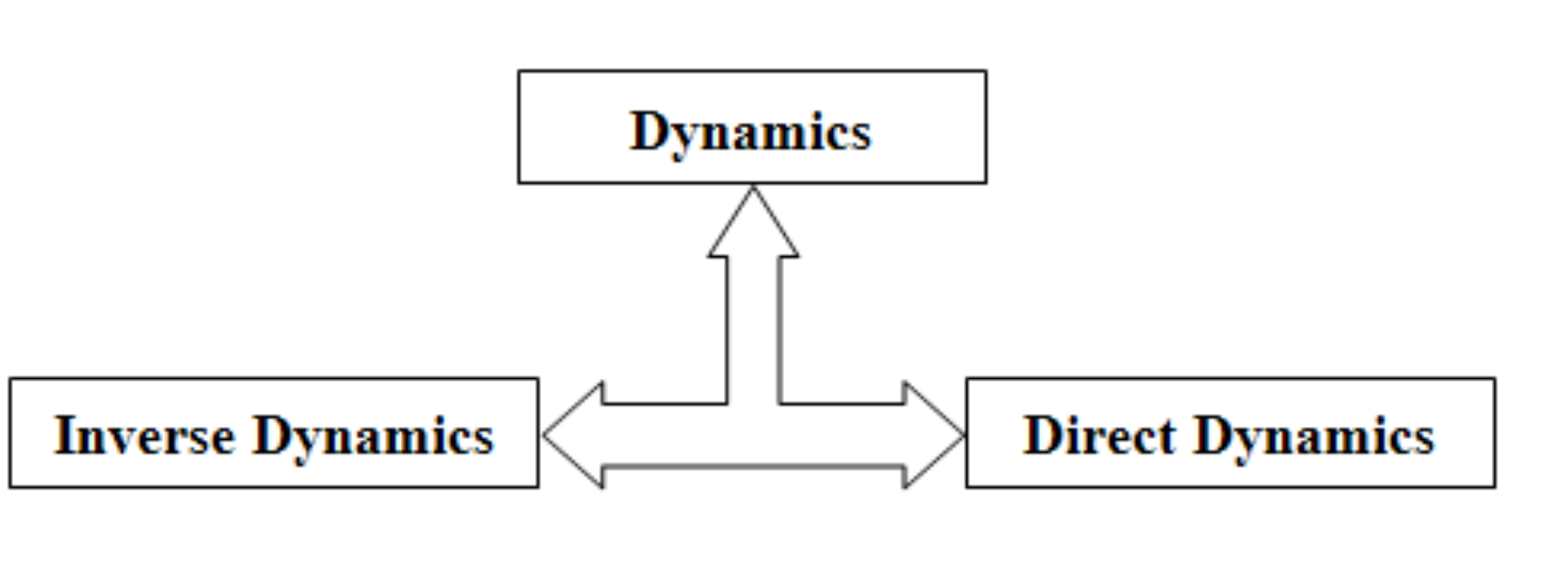}
  \end{center}
  \caption {Direct/Inverse dynamics}
  \label{fig26}
\end{figure}

In this chapter, we develop the foundation of dynamic model of the underlying robot. To date, a variety of methods have been introduced for analysis of robots dynamics' including \emph{Newton-Euler}, \emph{Lagrange} and \emph{virtual work} methods. In this work however we take advantage of the \emph{natural orthogonal complement} method, or NOC for short, for dynamics analysis of a special kind of parallel robots, called 3PSP. NOC, introduced by Ou. Ma. et. al.  \cite{Ma91}, models dynamics of parallel robots which opens new ways towards modeling the dynamics of 3PSP robots by specializing the general method to reduce the results to the case of 3PSP parallel robots. Based on this method, one can determine position, velocity and acceleration of each of the joints and combining the results with that of Lagrange method the dynamic equations of the underlying robot with respect to the position of moving and non-moving joints can be described. Accordingly, the dynamic model of the robot in the form of Lagrange equations is found without having introduced the kinematic constraints and by merely solving the foregoing equations. By finding the solution to a robot dynamic problem we mostly mean finding differential equations
which precisely describe the robots movement in the following form:
\begin{equation}
M\ddot q^a  + C\dot q^a  + G = \tau ^a 
\label{3.1}
\end{equation}
where $\ddot q^a$, $\dot q^a$ in this work are vectors proportional to speed and acceleration of the three actuators in joint space, respectively and $\tau ^a $ is the torque vector applied to the actuators. These torques generate a linear acceleration in each of the ball-screws derived by the actuators. Also, we will present a straight forward method for calculation of $M$, $C$, $G$ which play an important role in \eqref{3.1} using the NOC method. We will show that these matrices can be found according to the following equations: 
\begin{equation}
M = M(q) = T^T M_{total} T
\label{3.2}
\end{equation}
\begin{equation}
C = C(q,\dot q) = T^T M_{total} \dot T + T^T \Omega M_{total} T
\label{3.3}
\end{equation}
\begin{equation}
G = G(q) =  - T^T w^g 
\label{3.4}
\end{equation}
where $M$ is the inertia positive definite matrix which depends on $M_{total}$ (that itself 
is contingent upon the mass and inertia parameters of robot links); $G$ is a matrix that is 
associated to gravity such that the $w^g$ in it indicates the link weights; $C$ is a 
matrix pertaining to Coriolis and centripetal forces and $\Omega$ indicates the 
angular velocities pertaining to the links and finally $T$, $\dot T$ represent the NOC matrix and its derivative with respect to time. The former includes the dimensions and constraint equations associated to the underlying robot. Now depending on the characteristics of inputs of the problem we can solve the equations in two different ways, as follows:
\begin{enumerate}
\item Inverse Dynamics
\item Direct Dynamics
\end{enumerate}
This is portrayed in Figure \ref{fig26} in a more descriptive manner which we discuss in details. Note however in thesis, we only consider the inverse dynamics of the system. 

\pagebreak

\section{\textbf{Inverse Dynamics}}
\begin{figure}[htp]
  \begin{center} 
	\includegraphics[width=13cm, height=5cm]{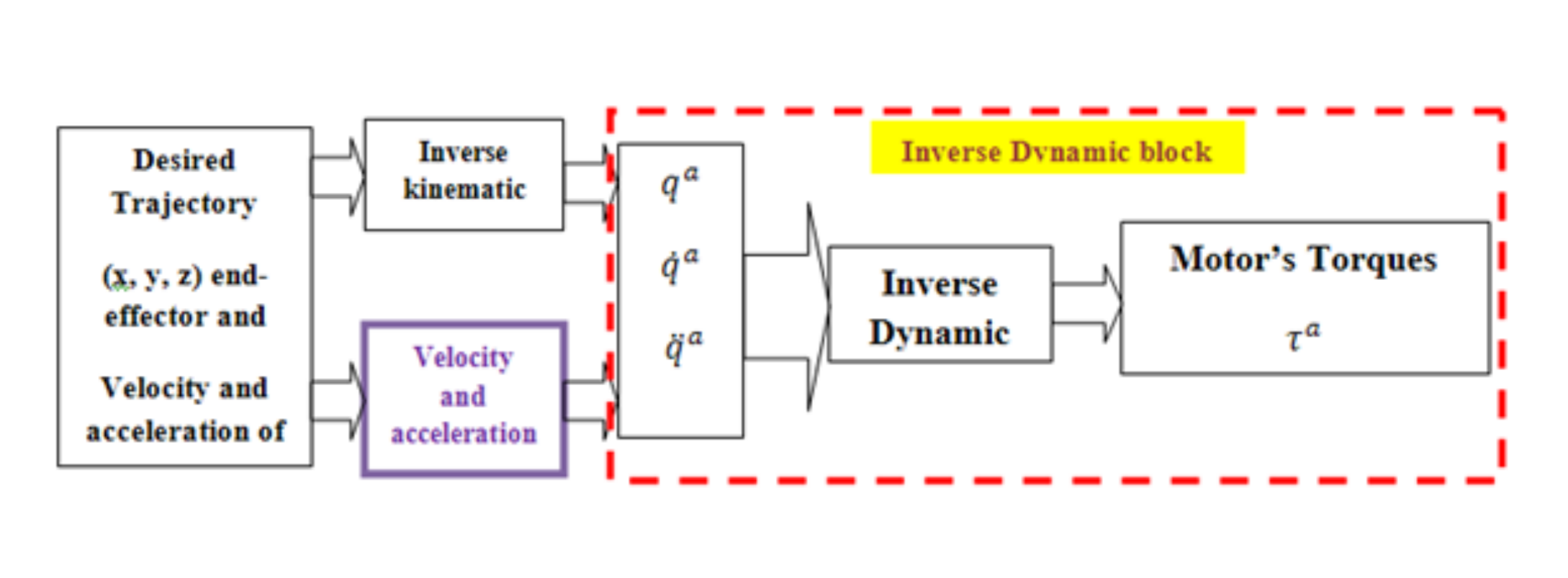}
  \end{center}
  \caption {Steps to solving inverse dynamics}
  \label{fig27}
\end{figure}

\begin{figure}[htp]
  \begin{center} 
	\includegraphics[width=10cm, height=2.1cm]{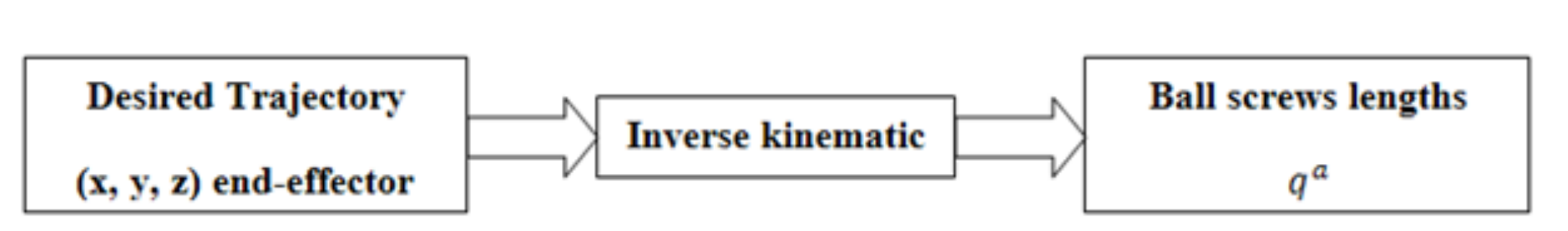}
  \end{center}
  \caption {Inverse kinematics}
  \label{fig28}
\end{figure}

In inverse dynamic modeling, given trajectories, velocities and accelerations associated with robot tool tip we are in quest of the torques related to each joint's actuator such that when applied to the actuators the desired trajectory come into existence. This is illustrated in Fig. \ref{fig27}. \\
According to the figure, during the procedure for solving inverse dynamics, one should first find the solution to the inverse kinematic problem so that he/she can transfer the desired trajectory of the tool-tip from Cartesian space into joint space variables, or equivalently the ball screw length, denoted by $q^a$. This procedure is individually depicted in Fig. \ref{fig28}.
Having the desired trajectory in the joint space, one can take the first and second derivatives with respect to time to compute $\dot q^a$, $\ddot q^a$. To find the solution of the inverse dynamics of our 3PSP robot we should first calculate the $M$, $C$, $G$ matrices that play important roles inside \eqref{3.1}. The ingredients of each of these matrices already given in \eqref{3.2}-\eqref{3.4}.
Next we give the required steps to be done for computation of inverse dynamics of a 3PSP robot. These step can also extended to the more general classes of parallel and serial robots.
\subsection{\textbf{Specifying Robot Constant Parameters}}
\begin{figure}[htp]
  \begin{center} 
    \subfigure [Manipulator's links]{\label{fig29-a}\includegraphics[width=5.5cm, height=5cm]{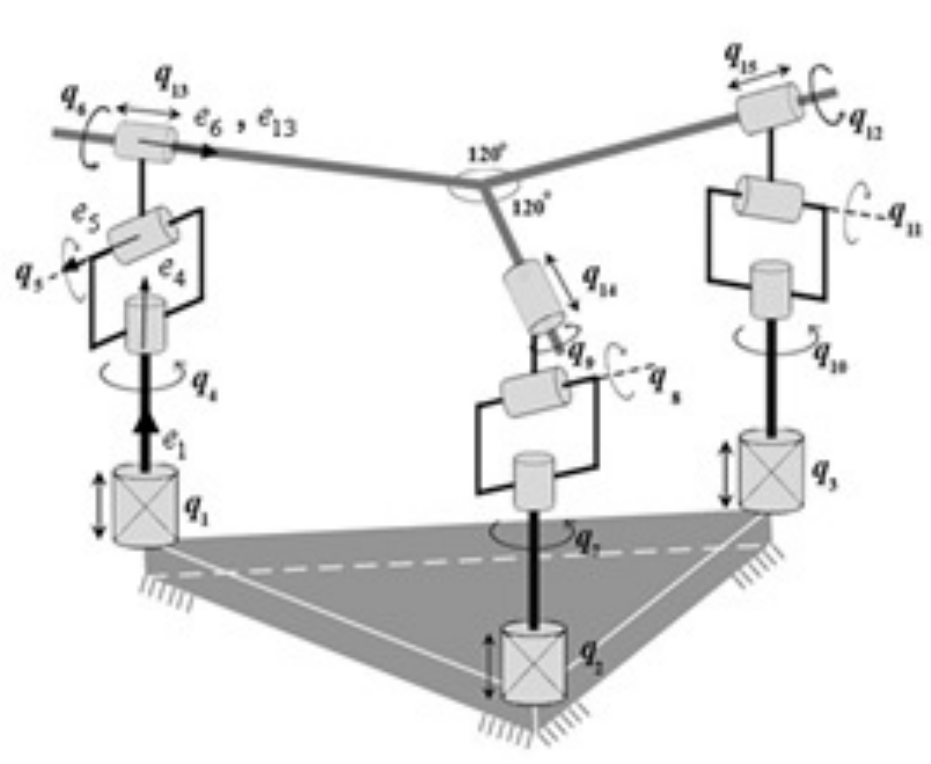}} 
    \subfigure [Manipulator's joints]{\label{fig29-b}\includegraphics[width=5.5cm, height=5cm]{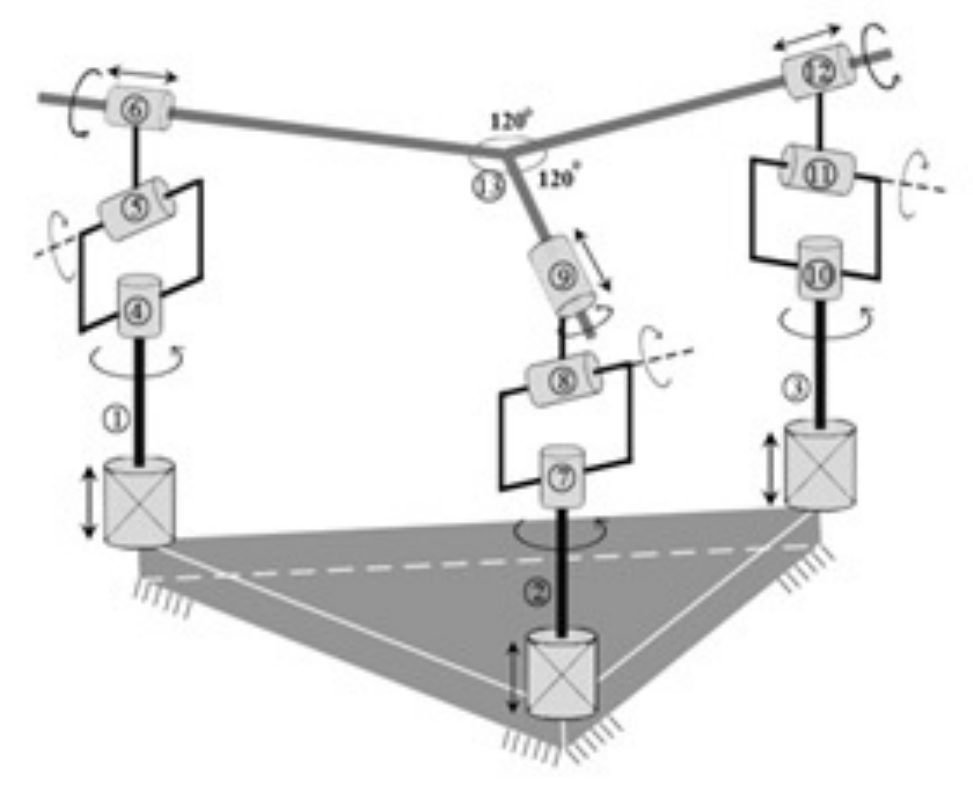}}
  \end{center}
  \caption{Links (a) and joints (b) }
  \label{fig29}
\end{figure}

The robot under discussion has 3 degrees of freedom (DOF). This robot has as many as $r$=13 rigid elements which are connected with each other through 15 ($m=15$) one D.O.F joints. Figure \ref{fig29} depicts all these links and joints.
Now let's introduce some constant parameters and the amounts of masses and inertias pertaining to the robot. The values associated to these parameters are available in table \ref{tbl3.1}. \\
The first constant to be defined is the parameter, $d$, which denotes the distance between the center of the stable triangular plate and robot's vertical axes. 
The second parameter is the size of the robot tool tip which is denoted by $h$. Then we need to define the mass associated to each \emph{nut}
and finally the inertia of each moving element should be rigorously defined.
\begin{table}[htp]
\caption{The value associated with each of the defined metrics}
\begin{tabular}{p{3.2cm} p{2.5cm} c c}
\hline\hline
Distance from center of non-moving plate to each axis & d &181 & $mm$\\
Size of tool-tip & h & 70 & $mm$\\
Mass of each nut & $m_1=m_2=m_3$ &7.175 & $kg$\\
Mass of each spherical joint & $m_4=m_5=\dots=m_{12} $ & 0.357 & $kg$\\
Mass of Star & $m_{13}$ & 1.758 & $kg$\\
Inertia of each nut & $I_1=I_2=I_3$ &	$\left[ \begin{array}{c c c }
0.003 &0 & 0\\
0 &0.39&0\\
0 & 0 & 0.3294\\
\end{array} \right] $ & $kg/m^2$\\
The inertia associated with each spherical joint & $I_4=I_5=\dots=I_{12}$ & $\left[ \begin{array}{c c c }
0.094 &0 & 0\\
0 &0.094&0\\
0 & 0 & 0.1117\\
\end{array} \right]\times 10^{-3} $ & $kg/m^2$\\
Star inertia & $I_{13}$	& $\left[ \begin{array}{c c c }
0.6451 &0 & 0\\
0 &1.2901&0\\
0 & 0 & 0.6451\\
\end{array} \right] $ & $kg/m^2$\\
\hline \hline
\label{tbl3.1}
\end{tabular}
\end{table}
Having determined the aforementioned parameters, the problem of direct kinematics and inverse dynamics can be solved.
\subsection{\textbf{Finding Constraint Equations Associated with Robot Joints}}
To find constraint equations, the following steps should be carried out in order:
\begin{itemize}
\item Determining general coordinates corresponding to each joints depicted in Fig. \ref{fig29}.
\item Finding the difference between robot D.O.F and the number of general coordinates. This difference is key item that determines the number of constraint equations required for solving the dynamic system.
\item Finding similar arms and set each of them apart from others (this makes some serial robots whose number equals to the number of arms.)
\item Assign two coordinate systems one to the center of the non-moving plate at the bottom of the robot and the other to the end of the tool tip connected to the star.
\item Create D.H table (refer to \cite{Cra04} for details) for each of the aforementioned 3 arms (serial chains) individually.
\item For each of the foregoing arms, find the transfer matrix corresponding to the tool tip end with respect to the coordinate system already defined for the non-moving plate. Observe that we can find the same number of transfer matrices as the number of separated arms.
\item As stated in the previous chapter, a homogeneous transformation includes a rotation and a translation, thus, we can equates the translation and rotation terms individually to find position constraint equations. Notice that in order to avoid redundant constraints, one should consider the number of independent kinematic loops. By equating the 
Euler angles (which don not include redundant data) associated to the rotations as well as the 3 
rotation terms, we find the rest of constraint equations.
\end{itemize}
\subsection{\textbf{Introducing General Coordinates for Robot}}
First, the joint coordinates should be determined. As already expressed, this robot posseses 15 joints, the first 3 of which are the actuated joints and the rest are un-actuated 
(passive) joints. Based on Fig. \ref{fig29}, we can express these variables as follows:
\begin{equation}
q = \left\{ {a_1 ,\,a_2 ,\,a_3 ,\,\varphi _1 ,\,\theta _1 ,\,\lambda _1 ,\,\varphi _2 ,\,\theta _2 ,\,\lambda _2 ,\,\varphi _3 ,\,\theta _3 ,\,\lambda _3 ,\,b_1 ,\,b_2 ,\,b_3 } \right\}^T  = \left\{ {\begin{array}{l}
   {q^a }  \\
   {q^u }  \\ 

\end{array}
 } \right\}
\label{3.5a}
\end{equation}
\begin{equation}
q^a  = \left\{ {a_1 ,\,a_2 ,\,a_3 } \right\}^T 
\label{3.5b}
\end{equation}
\begin{equation}
q^u  = \left\{ {\,\varphi _1 ,\,\theta _1 ,\,\lambda _1 ,\,\varphi _2 ,\,\theta _2 ,\,\lambda _2 ,\,\varphi _3 ,\,\theta _3 ,\,\lambda _3 ,\,b_1 ,\,b_2 ,\,b_3 } \right\}^T 
\label{3.5c}
\end{equation}
where superscripts $a$, $u$ indicate the status of being actuated or un-actuated, respectively. In fact the actuated joints are the independent ones, too, and the un-actuated ones are those that are not dependent but rather contingent upon these three.
\subsection{\textbf{Assigning Frames to Robot and Generating Transformation Matrices Based on D.H. Table}}
\begin{figure}[htp]
  \begin{center} 
    \subfigure [First arm]{\label{fig30-a}\includegraphics[width=5.5cm, height=5cm]{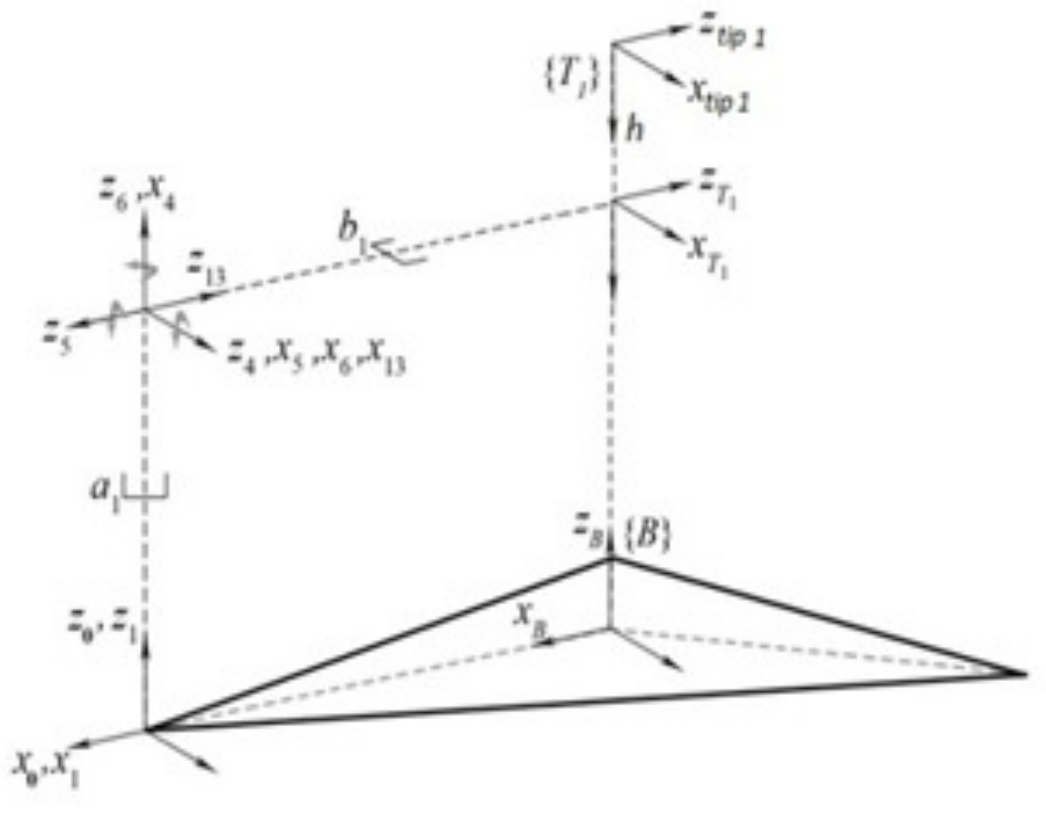}} 
    \subfigure [Second arm]{\label{fig30-b}\includegraphics[width=5.5cm, height=5cm]{Images/robotarmframesa}}
    \subfigure [Third arm]{\label{fig30-c}\includegraphics[width=5.5cm, height=5cm]{Images/robotarmframesa}}
  \end{center}
  \caption{Assigning frames to each robot arm}
  \label{fig30}
\end{figure}

After having the general coordinates on each of the joints determined and the difference between these coordinates and the DOF known, the required number of constraint equations will be determined consequently. As stated earlier, there are totally 15 general coordinates and 3 actuated joints, meaning that one should find 12 constraints to specify the robot setting precisely. 
Now, we define two important coordinate systems, namely {B},{T}. The former is assigned to non-moving plate and the latter to tool tip end. The homogeneous transform matrix relating the fixed coordinate of non-moving plate to the coordinate associated to moving plate is then obtained according to the paths pertaining to each arm. Let's call the set of all constraints, $\varphi$ . In other words we need the solution to:
\begin{equation}
\varphi(\bf{q})=\bf{0}
\label{3.6}
\end{equation}
This is depicted in a more descriptive manner in Fig. \ref{fig30} for each robot arm and will be discussed in details shortly.

\subsubsection{First Arm}
The transform matrices for the first arm, Fig. \ref{fig30-a}, can be obtained as follows:
\begin{equation}
\begin{array}{c c}
	{{}_{{\rm{01}}}^{\rm{B}} {\rm{T }} = \left[ 
	{\begin{array}{c c c c}
	   1 & 0 & 0 & d  \\
	   0 & 1 & 0 & 0  \\ 
	   0 & 0 & 1 & 0  \\ 
	   0 & 0 & 0 & 1  \\ 
	\end{array}
 	} \right] } 
	& 	
	{{}_1^{01} {\rm{T }} = \left[ 
	{\begin{array}{c c c c}
	   1 & 0 & 0 & 0  \\ 
	   0 & 1 & 0 & 0  \\ 
	   0 & 0 & 1 & {a_1}  \\ 
	   0 & 0 & 0 & 1  \\ 
	\end{array}
	} \right] } \\
	
	{{}_{\rm{4}}^{\rm{1}} {\rm{T}} = \left[ {\begin{array}{c c c c}
   {{\rm{sin(}}\varphi _{\rm{1}} {\rm{)}}} & {{\rm{cos(}}\varphi _{\rm{1}} {\rm{)}}} & {\rm{0}} & {\rm{0}}  \\ 
   {\rm{0}} & {\rm{0}} & {\rm{1}} & {\rm{0}}  \\
   { {\rm{cos(}}\varphi _{\rm{1}} {\rm{)}}} & {{\rm{ - sin(}}\varphi _{\rm{1}} {\rm{)}}} & {\rm{0}} & {\rm{0}}  \\ 
   {\rm{0}} & {\rm{0}} & {\rm{0}} & {\rm{1}}  \\
	\end{array}
 	} \right] } &
	{ {}_{\rm{5}}^{\rm{4}} {\rm{T }} = \left[ {\begin{array}{c c c c}
   {{\rm{sin(}}\theta _1 {\rm{)}}} & {{\rm{cos(}}\theta _1 {\rm{)}}} & {\rm{0}} & {\rm{0}}  \\
   {\rm{0}} & {\rm{0}} & {\rm{1}} & {\rm{0}}  \\ 
   {{\rm{cos(}}\theta _1 {\rm{)}}} & {{\rm{ - sin(}}\theta _1 {\rm{)}}} & {\rm{0}} & {\rm{0}}  \\ 
   {\rm{0}} & {\rm{0}} & {\rm{0}} & {\rm{1}}  \\
	\end{array} }   \right]}
	\\
	{
	{}_{\rm{6}}^{\rm{5}} {\rm{T }} = \left[ {\begin{array}{c c c c }
   {{\rm{cos(}}\lambda _1 {\rm{)}}} & {{\rm{ - sin(}}\lambda _1 {\rm{)}}} & {\rm{0}} & {\rm{0}}  \\
   {\rm{0}} & {\rm{0}} & {\rm{1}} & {\rm{0}}  \\ 
   {{\rm{ - sin(}}\lambda _1 {\rm{)}}} & {{\rm{ - cos(}}\lambda _1 {\rm{)}}} & {\rm{0}} & {\rm{0}}  \\
   {\rm{0}} & {\rm{0}} & {\rm{0}} & {\rm{1}}  \\
	\end{array}
 	}  \right]
	} & {{}_{{\rm{13}}}^{\rm{6}} {\rm{T }} = \left[ {\begin{array}{c c c c}
   1 & 0 & 0 & 0  \\
   0 & 0 & 1 & {b_1 }  \\
   0 & { - 1} & 0 & 0  \\ 
   0 & 0 & 0 & 1  \\
	\end{array}
  } \right]
	}\\
	{{}_{{\rm{tip1}}}^{{\rm{13}}} {\rm{T }} = \left[ {\begin{array}{c c c c }
   1 & 0 & 0 & 0  \\ 
   0 & 1 & 0 & { - h}  \\ 
   0 & 0 & 1 & 0  \\
   0 & 0 & 0 & 1  \\ 
	\end{array}
	}  \right] } & {}\\
\label{3.7}
\end{array}
\end{equation}
Consequently, to find the foregoing homogeneous transform matrix, one should multiply all these matrices in order, which follows:
\begin{equation}
{}_{{\rm{13}}}^{\rm{B}} {\rm{T}}_{\rm{1}}  = {}_{01}^B {\rm{T}} \times {}_1^{01} {\rm{T}} \times {}_4^1 {\rm{T}} \times {}_5^4 {\rm{T}} \times {}_6^5 {\rm{T}} \times {}_{13}^6 {\rm{T}} \times {}_{tip1}^{13} {\rm{T}}
\label{3.8}
\end{equation}

\subsubsection{Second Arm}
In a rather similar approach, the transform matrix for the second arm can be found as:
\begin{equation}
\begin{array}{c c }
{_{{\rm{02 - 1}}}^{\rm{B}} {\rm{T}} = \left[ {\begin{array}{c c c c}
   1 & 0 & 0 & { - d/2}  \\ 
   0 & 1 & 0 & {\sqrt 3 /2*d}  \\ 
   0 & 0 & 1 & 0  \\ 
   0 & 0 & 0 & 1  \\ 
	\end{array}
  } \right]
} & {{}_{{\rm{02 - 2}}}^{\rm{B}} {\rm{T }} = \left[ {\begin{array}{c c c c}
   {\cos \left( {2\pi /3} \right)} & { - {\rm{sin}}\left( {2\pi /3} \right)} & 0 & 0  \\ 
   {{\rm{sin}}\left( {2\pi /3} \right)} & {\cos \left( {2\pi /3} \right)} & 0 & 0  \\ 
   0 & 0 & 1 & 0  \\ 
   0 & 0 & 0 & 1  \\ 
\end{array}
 } \right]
} \\
{{}_{\rm{2}}^{{\rm{02}}} {\rm{T }} = \left[ {\begin{array}{c c c c}
   1 & 0 & 0 & 0  \\ 
   0 & 1 & 0 & 0  \\ 
   0 & 0 & 1 & {a_2 }  \\ 
   0 & 0 & 0 & 1  \\ 
	\end{array}
 } \right]
} &
{_{\rm{7}}^{\rm{2}} {\rm{T }} = \left[ {\begin{array}{c c c c}
   {{\rm{sin(}}\varphi _{\rm{2}} {\rm{)}}} & {{\rm{cos(}}\varphi _{\rm{2}} {\rm{)}}} & {\rm{0}} & {\rm{0}}  \\ 
   {\rm{0}} & {\rm{0}} & {\rm{1}} & {\rm{0}}  \\ 
   {{\rm{cos(}}\varphi _{\rm{2}} {\rm{)}}} & {{\rm{ - sin(}}\varphi _{\rm{2}} {\rm{)}}} & {\rm{0}} & {\rm{0}}  \\ 
   {\rm{0}} & {\rm{0}} & {\rm{0}} & {\rm{1}}  \\ 
\end{array}
 } \right]
} \\
{_9^8 {\rm{T }} = \left[ {\begin{array}{c c c c}
   {{\rm{cos(}}\lambda _2 {\rm{)}}} & {{\rm{ - sin(}}\lambda _2 {\rm{)}}} & {\rm{0}} & {\rm{0}}  \\ 
   {\rm{0}} & {\rm{0}} & {\rm{1}} & {\rm{0}}  \\ 
   {{\rm{ - sin(}}\lambda _2 {\rm{)}}} & {{\rm{ - cos(}}\lambda _2 {\rm{)}}} & {\rm{0}} & {\rm{0}}  \\ 
   {\rm{0}} & {\rm{0}} & {\rm{0}} & {\rm{1}}  \\ 
	\end{array}
 } \right]
} &
{_{\rm{8}}^{\rm{7}} {\rm{T }} = \left[ {\begin{array}{c c c c}
   {{\rm{sin(}}\theta _2 {\rm{)}}} & {{\rm{cos(}}\theta _2 {\rm{)}} } & {\rm{0}} & {\rm{0}}  \\ 
   {\rm{0}} & {\rm{0}} & {\rm{1}} & {\rm{0}}  \\ 
   {{\rm{cos(}}\theta _2 {\rm{)}}} & {{\rm{ - sin(}}\theta _2 {\rm{)}}} & {\rm{0}} & {\rm{0}}  \\ 
   {\rm{0}} & {\rm{0}} & {\rm{0}} & {\rm{1}}  \\ 
	\end{array}
 } \right]
} \\{{}_{{\rm{14}}}^{\rm{9}} {\rm{T }} = \left[ {\begin{array}{c c c c}
   1 & 0 & 0 & 0  \\ 
   0 & 0 & 1 & {b_2 }  \\ 
   0 & { - 1} & 0 & 0  \\ 
   0 & 0 & 0 & 1  \\ 
	\end{array}
 } \right]
} 
& { {}_{{\rm{T}}_{\rm{2}} }^{T_1 } {\rm{T }} = \left[ {\begin{array}{c c c c}
   {\cos \left( {2\pi /3} \right)} & 0 & {\sin \left( {2\pi /3} \right)} & 0  \\ 
   0 & 1 & 0 & 0  \\ 
   { - \sin \left( {2\pi /3} \right)} & 0 & {\cos \left( {2\pi /3} \right)} & 0  \\ 
   0 & 0 & 0 & 1  \\ 
	\end{array}
} \right]
}\\
{{}_{{\rm{tip2}}}^{{\rm{T2}}} {\rm{T }} = \left[ {\begin{array}{c c  c c}
   1 & 0 & 0 & 0  \\ 
   0 & 1 & 0 & { - h}  \\ 
   0 & 0 & 1 & 0  \\ 
   0 & 0 & 0 & 1  \\ 
\end{array}
} \right]
}
&{}
\end{array}
\label{3.9}
\end{equation}
and the final transform matrix appears in the following form:
\begin{equation}
{\rm{T}}_{\rm{2}}  = {}_{02 - 1}^B {\rm{T}} \times {}_{02 - 2}^B {\rm{T}} \times {}_2^{02} {\rm{T}} \times {}_7^2 {\rm{T}} \times {}_8^7 {\rm{T}} \times {}_9^8 {\rm{T}} \times {}_{14}^9 {\rm{T}} \times {}_{T_2 }^{T_1 } {\rm{T}} \times {}_{{\rm{tip2}}}^{{\rm{T2}}} {\rm{T}}
\label{3.10}
\end{equation}

\subsubsection{Third Arm}
The transform matrices can be found as:
\begin{equation}
\begin{array}{l l}
{{}_{{\rm{03 - 1}}}^{\rm{B}} {\rm{T}} = \left[ {\begin{array}{c c c c}
   1 & 0 & 0 & { - d/2}  \\ 
   0 & 1 & 0 & { - \sqrt 3 /2*d}  \\ 
   0 & 0 & 1 & 0  \\ 
   0 & 0 & 0 & 1  \\ 
\end{array}
} \right]
} &{{}_{{\rm{03 - 2}}}^{\rm{B}} {\rm{T }} = \left[ {\begin{array}{c c c c}
   {\cos \left( {4\pi /3} \right)} & { - \sin \left( {4\pi /3} \right)} & 0 & 0  \\ 
   {\sin \left( {4\pi /3} \right)} & {\cos \left( {4\pi /3} \right)} & 0 & 0  \\ 
   0 & 0 & 1 & 0  \\ 
   0 & 0 & 0 & 1  \\ 
\end{array}
} \right]
}\\
{{}_{\rm{3}}^{{\rm{03}}} {\rm{T}}_{\rm{ }}  = \left[ {\begin{array}{c c c c}
   1 & 0 & 0 & 0  \\ 
   0 & 1 & 0 & 0  \\ 
   0 & 0 & 1 & {a_3 }  \\ 
   0 & 0 & 0 & 1  \\ 
\end{array}
 } \right]
} &{{}_{{\rm{10}}}^{\rm{3}} {\rm{T }} = \left[ {\begin{array}{c c c c}
   {{\rm{sin(}}\varphi _{\rm{3}} {\rm{)}}} & {{\rm{cos(}}\varphi _{\rm{3}} {\rm{)}}} & {\rm{0}} & {\rm{0}}  \\ 
   {\rm{0}} & {\rm{0}} & {\rm{1}} & {\rm{0}}  \\ 
   {{\rm{cos(}}\varphi _{\rm{3}} {\rm{)}}} & {{\rm{ - sin(}}\varphi _{\rm{3}} {\rm{)}}} & {\rm{0}} & {\rm{0}}  \\ 
   {\rm{0}} & {\rm{0}} & {\rm{0}} & {\rm{1}}  \\ 
\end{array}
 } \right]
}\\
{
{}_{{\rm{11}}}^{{\rm{10}}} {\rm{T }} = \left[ {\begin{array}{c c c c}
   {{\rm{sin(}}\theta _3 {\rm{)}}} & {{\rm{cos(}}\theta _3 {\rm{)}}} & {\rm{0}} & {\rm{0}}  \\ 
   {\rm{0}} & {\rm{0}} & {\rm{1}} & {\rm{0}}  \\ 
   {{\rm{cos(}}\theta _3 {\rm{)}}} & {{\rm{ - sin(}}\theta _3 {\rm{)}}} & {\rm{0}} & {\rm{0}}  \\ 
   {\rm{0}} & {\rm{0}} & {\rm{0}} & {\rm{1}}  \\ 
\end{array}
 } \right]
} &{{}_{{\rm{12}}}^{{\rm{11}}} {\rm{T }} = \left[ {\begin{array}{c c c c}
   {{\rm{cos(}}\lambda _{\rm{3}} {\rm{)}}} & {{\rm{ - sin(}}\lambda _{\rm{3}} {\rm{)}}} & {\rm{0}} & {\rm{0}}  \\ 
   {\rm{0}} & {\rm{0}} & {\rm{1}} & {\rm{0}}  \\ 
   {{\rm{ - sin(}}\lambda _{\rm{3}} {\rm{)}}} & {{\rm{ - cos(}}\lambda _{\rm{3}} {\rm{)}}} & {\rm{0}} & {\rm{0}}  \\ 
   {\rm{0}} & {\rm{0}} & {\rm{0}} & {\rm{1}}  \\ 
\end{array}
 } \right]
}\\
{{}_{{\rm{15}}}^{{\rm{12}}} {\rm{T }} = \left[ {\begin{array}{c c c c}
   1 & 0 & 0 & 0  \\ 
   0 & 0 & 1 & {b_3 }  \\ 
   0 & { - 1} & 0 & 0  \\ 
   0 & 0 & 0 & 1  \\ 
\end{array} 
} \right]
} & {{}_{{\rm{T}}_{\rm{3}} }^{T_1 } {\rm{T }} = \left[ {\begin{array}{c c c c}
   {\cos \left( {4\pi /3} \right)} & 0 & {\sin \left( {4\pi /3} \right)} & 0  \\ 
   0 & 1 & 0 & 0  \\ 
   { - \sin \left( {4\pi /3} \right)} & 0 & {\cos \left( {4\pi /3} \right)} & 0  \\ 
   0 & 0 & 0 & 1  \\ 
\end{array}
 } \right]
}\\
{{}_{{\rm{tip3}}}^{{\rm{T3}}} {\rm{T }} = \left[ {\begin{array}{c c c c}
   1 & 0 & 0 & 0  \\ 
   0 & 1 & 0 & { - h}  \\ 
   0 & 0 & 1 & 0  \\ 
   0 & 0 & 0 & 1  \\ 
\end{array}
} \right]
} &{}\\
\end{array}
\label{3.11}
\end{equation}
and the overall transform matrix can be calculated as follows:
\begin{equation}
{\rm{T}}_{\rm{3}}  = {}_{03 - 1}^B {\rm{T}} \times {}_{03 - 2}^B {\rm{T}} \times {}_3^{03} {\rm{T}} \times {}_{10}^3 {\rm{T}} \times {}_{11}^{10} {\rm{T}} \times {}_{12}^{11} {\rm{T}} \times {}_{15}^{12} {\rm{T}} \times {}_{T_3 }^{T_1 } {\rm{T}} \times {}_{{\rm{tip3}}}^{{\rm{T3}}} {\rm{T}}
\label{3.12}
\end{equation}
Now, equating the translation and rotation parameters, we come to a series of constraint equations as follows:
\begin{equation}
{\rm{S}}_{\rm{i}}  = {\rm{(I}}_{{\rm{3}} \times {\rm{3}}}  + {\rm{R}}_{\rm{i}} {\rm{)}}^{{\rm{ - 1}}}  \times {\rm{(R}}_{\rm{i}} {\rm{ - I}}_{{\rm{3}} \times {\rm{3}}} {\rm{)    i}} = \left\{ {{\rm{1,2,3}}} \right\}
\label{3.13}
\end{equation}
which is equivalent to,
\begin{equation}
\begin{array}{l}
{\rm{Constraint(1)}} = {\rm{T}}_{\rm{2}} {\rm{(1,4) - T}}_{\rm{1}} {\rm{(1,4),}}  \\ 
{\rm{Constraint(2)}} = {\rm{T}}_{\rm{2}} {\rm{(1,4) - T}}_{\rm{3}} {\rm{(1,4),}}  \\ 
{\rm{Constraint(3)}} = {\rm{T}}_{\rm{2}} {\rm{(2,4) - T}}_{\rm{1}} {\rm{(2,4),}}  \\ 
{\rm{Constraint(4)}} = {\rm{T}}_{\rm{2}} {\rm{(2,4) - T}}_{\rm{3}} {\rm{(2,4),}}  \\ 
{\rm{Constraint(5)}} = {\rm{T}}_{\rm{2}} {\rm{(3,4) - T}}_{\rm{1}} {\rm{(3,4),}}  \\ 
{\rm{Constraint(6)}} = {\rm{T}}_{\rm{2}} {\rm{(3,4) - T}}_{\rm{3}} {\rm{(3,4),}}  \\ 
{\rm{Constraint(7)}} = {\rm{S}}_{\rm{1}} {\rm{(1,2) - S}}_{\rm{2}} {\rm{(1,2),}}  \\ 
{\rm{Constraint(8)}} = {\rm{S}}_{\rm{1}} {\rm{(1,2) - S}}_{\rm{3}} {\rm{(1,2),}}  \\ 
{\rm{Constraint(9)}} = {\rm{S}}_{\rm{1}} {\rm{(1,3) - S}}_{\rm{2}} {\rm{(1,3),}}  \\ 
{\rm{Constraint(10)}} = {\rm{S}}_{\rm{1}} {\rm{(1,3) - S}}_{\rm{3}} {\rm{(1,3),}}  \\ 
{\rm{Constraint(11)}} = {\rm{S}}_{\rm{1}} {\rm{(2,3) - S}}_{\rm{2}} {\rm{(2,3),}}  \\ 
{\rm{Constraint(12)}} = {\rm{S}}_{\rm{1}} {\rm{(2,3) - S}}_{\rm{3}} {\rm{(2,3)}}{\rm{.}} \\
\end{array}
\label{3.14}
\end{equation}
As already mentioned, in order to avoid redundant equations one should consider only the independent kinematic loops, and for the case of this 3PSP robot there exist two independent loops. Thus we select two pairs among three possible combinations of $T_1$, $T_2$, $T_3$ and equate them to find two matrix relations. Each matrix relation generates six different equations which add up to 12 different constraint equations mentioned in \eqref{3.14}.
\subsection{\textbf{Finding Un-Actuated Variables with Respect to Actuated Ones by Solving the Inverse Kinematics Equations}}
As already stated, the input to the inverse dynamics problem is the desired trajectory for the tool tip end. As is obvious from \eqref{3.1}, given the required torques to the actuators, the equation of direct dynamics finds the actuated variables, namely $q^a, \dot q^a, \ddot q^a$. Most of the time, we have a known trajectory and we are looking for the sequence of torques that if applied to the actuators leads to tracking the desired trajectory. Hence, we are interested in inverse dynamics more than the direct one. Thus, as a first step we should find $q^a, \dot q^a, \ddot q^a$ to feed into our inverse dynamic system which is easily obtained by taking the first and second derivatives of $q^a$ . We also noted that the values associated with un-actuated variables are found by solving 12 constraint equations (in our platform). Thus in this step, one should perform numerical analysis (generally there is not any analytic solution for the inverse kinematic problem) based on an initial condition which in this work presumed to be the 
zero status (also called \emph{home}).
\begin{equation}
{\rm{q}}_{\rm{0}}^{\rm{u}}  = \left[ {\begin{array}{ l l l l l l l l l l l l}
   0 & 0 & 0 & 0 & 0 & 0 & 0 & 0 & 0 & d & d & d  \\ 
\end {array}
 } \right]
\label{3.15}
\end{equation}
\subsection{\textbf{Finding Joint Orthogonal Complement (L) and Natural Orthogonal Complement (T) Matrices}}
In this step we introduce orthogonal complement matrices that are acquired via the NOC method.
\subsubsection{Joint Orthogonal Complement (L)}
In the previous section we discussed how we can relate the position (in joint space) of the actuated variables to the un-actuated ones. In this section we briefly show that we can do the same for joint speeds via matrix L; that is:
\begin{equation}
{\bf{\dot q}} = {\bf{L\dot q}}^a 
\label{3.16}
\end{equation}
where $L_{m\times n}$ can be defined as follows:
\begin{equation}
{\bf{L}} = \left[ \begin{array}{l}
  {\bf{I}}_{n \times n}  \hfill \\ 
   - ({\bf{\Phi }}^u )^{ - 1} {\bf{\Phi }}^a  \hfill \\
	\end{array}
  \right]_{m \times n} 
\label{3.17}
\end{equation}
$\bf{I_{n\times n}}$ is the identity matrix where in this case study, $n$, the robot's D.O.F 
is 3 and $m$, the number of robot elements is 15. $ \bf{\Phi}^u$, $\bf{\Phi^a}$ can be found via constraint equations as will be explained shortly.
Taking derivative of \eqref{3.6}, the relation for joint speeds is calculated as:
\begin{equation}
{\bf{\Phi \dot q}} = 0
\label{3.18}
\end{equation}
where $\bf{\Phi_{12\times 15}}$ is joint Jacobean matrix and is obtained by taking derivative of 
$\varphi(\bf{q})$ with respect to the vector, $\bf{q}$, as follows:
\begin{equation}
{\bf{\Phi }} = {{\partial \phi ({\bf{q}})} \over {\partial {\bf{q}}}}
\label{3.19}
\end{equation}
Let 
\begin{equation}
{\bf{\Phi }}^a  = {{\partial \phi ({\bf{q}})} \over {\partial {\bf{q}}^a }}
\label{3.20}
\end{equation}
and 
\begin{equation}
{\bf{\Phi }}^u  = {{\partial \phi ({\bf{q}})} \over {\partial {\bf{q}}^u }}
\label{3.21}
\end{equation}
where $\bf{\Phi^a}$ is a 12 by 3 matrix and $\bf{\Phi^u}$ is a 12 by 12 one. It is easy to show that,
\begin{equation}
{\bf{\Phi }} = [{\bf{\Phi }}^a \,\,{\bf{\Phi }}^u ]_{12 \times 15} 
\label{3.22}
\end{equation}
Substituting \eqref{3.22} into \eqref{3.18} yields,
\begin{equation}
{\bf{\dot q}}^u  =  - ({\bf{\Phi }}^u )^{ - 1} {\bf{\Phi }}^a {\bf{\dot q}}^a 
\label{3.23}
\end{equation}
As discussed earlier by solving direct kinematics equations we can find the values associated to un-actuated joints with respect to the independent (actuated) ones and hence we have the matrices $\bf{\Phi^u}$ and $\bf{\Phi^a}$ available. $\bf{\dot q^a}$ can also be calculated by taking the derivative from $\bf{q^a}$ . Thus, $\bf{\dot q_{15\times 1}}$ can be specified easily using 
\eqref{3.16}. This means that having the joint speed of actuated joints all other dependent variables can be computed based on numerical analysis.
\subsubsection {Natural Orthogonal Complement (T)}
When considering the locomotion of a rigid body, we should take into account its position, velocity and acceleration. The position associated with the body is precisely determined via a translation vector and a rotation matrix with respect to a universal point. In the same sense, speed is determined by means of twist vector. The twist vector pertaining to the ith element of a robot can be specified by means of a 6 by 1 vector, as follows:
\begin{equation}
{\bf{t}}_i  = \left[ \begin{array}{l}
  \bf{\omega _i}  \hfill \\
  {\bf{v}}_i  \hfill \\
\end{array}  \right]
\label{3.24}
\end{equation}
Taking the derivative of the twist vector with respect to time follows,
\begin{equation}
{\bf{\dot t}} \equiv \left[ \begin{array}{l}
  \bf{\dot \omega _i}  \hfill \\ 
  {\bf{\dot v}}_i  \hfill \\ \end{array}  \right]
\label{3.25}
\end{equation}
where $\bf{\omega_i}$ and $\bf{v_i}$ are angular and linear speed, respectively, and their derivatives  $\bf{\dot \omega _i}$ and $\bf{\dot v_i}$ are the associated angular and linear acceleration of the $i$th element. 
For a system comprised of 13 elements, the twist vector and its derivative w.r.t. time is denoted as follows:
\begin{equation}
{\bf{t}} \equiv \left[ \begin{array}{l}
  {\bf{t}}_1  \hfill \\ 
  {\bf{t}}_2  \hfill \\ 
   \vdots  \hfill \\ 
  {\bf{t}}_{r = 13}  \hfill \\ \end{array}  \right]
\label{3.26a}
\end{equation}
,
\begin{equation}
{\bf{\dot t}} \equiv \left[ \begin{array}{l}
  {\bf{\dot t}}_1  \hfill \\
  {\bf{\dot t}}_2  \hfill \\ 
   \vdots  \hfill \\
  {\bf{\dot t}}_{r = 13}  \hfill \\ \end{array}  \right]
\label{3.26b}
\end{equation}
It is possible to write the twist vector for each element as an affine combination of all the joints as follows:
\begin{equation}
{\bf{t}}_i  = {\bf{K}}_i {\bf{\dot q}}
\label{3.27}
\end{equation}
where in the previous equation, ${\bf{K}}_i$ is a $6\times m$  matrix (where in our case $m$, the number of robot elements, is 15) in the following form:
\begin{equation}
{\bf{K}}_i  = \left[ \begin{array}{l}
  {\bf{A}}_i  \hfill \\
  {\bf{B}}_i  \hfill \\ \end{array}  \right]
\label{3.28}
\end{equation}
If the $j$th joint is prismatic and the remainders are rotational then we have
\begin{equation}
\bf{ \left[ e_1\; e_2 \; \dots \; e_{j-1} \; 0 \; e_{j+1} \; \dots \; e_m  \right]}
\label{3.29a}
\end{equation}
\begin{equation}
\bf{ \left[ e_1\times r_1\; e_2\times r_2 \; \dots \; e_{j-1}\times r_{j-1} \; 0 \; e_{j+1}\times r_{j+1} \; \dots \; e_m\times r_m  \right]}
\label{3.29b}
\end{equation}
where $\bf{r_i}$ in the above matrix is a vector that connects the center of $i$th system to 
the operation point, and $\bf{e_i}$ is a unit vector associated to the $i$th element along its twist (alongside the axis that corresponds to body's inertia).
Now, let define $\bf{K}^a$, $\bf{K}^u$ and $\bf{K}$ as follows:
\begin{equation}
{\bf{K}}^a  = \left[ {\begin{array}{l}
   {{\bf{K}}_1 }  \\ 
    \vdots   \\ 
   {{\bf{K}}_{n = 3} }  \\ 
	\end{array}
 } \right]
\label{3.30a}
\end{equation}
\begin{equation}
{\bf{K}}^u  = \left[ {\begin{array}{l}
   {{\bf{K}}_{n = 3} }  \\ 
    \vdots   \\ 
   {{\bf{K}}_{m = 15} }  \\ 
	\end{array}
 } \right]
\label{3.30b}
\end{equation}
\begin{equation}
{\bf{K}} = \left[ {\begin{array}{l}
   {{\bf{K}}^a }  \\ 
   {{\bf{K}}^u }  \\ 
	\end{array}
 } \right]
\label{3.30c}
\end{equation}
The process through which ${\bf{K}}^a ,{\bf{K}}^u ,{\bf{K}}$ are calculated is something beyond the scope of this thesis and interested reader can refer to \cite{Ma91} for details. The role of natural orthogonal complement matrix, $\bf{T}$, is to find the twist vector, $\bf{t}$, as a linear function of $\bf{\dot q^a}$, as follows,
\begin{equation}
{\bf{t}} = {\bf{T\dot q}}^a 
\label{3.31}
\end{equation}
where $\bf{T_{6r\times n}}$ is the natural orthogonal complement matrix (in our platform $r=13$, $n=3$).
Considering \eqref{3.16} and \eqref{3.27} $\bf{T}$ and $\bf{L}$ are related to each other via the following equations,
\begin{equation}
{\bf{T}} = {\bf{KL}}
\label{3.32}
\end{equation}
\begin{equation}
{\bf{T}} = {\bf{K}}^a  - {\bf{K}}^u ({\bf{\Phi }}^u )^{ - 1} {\bf{\Phi }}^a 
\label{3.33}
\end{equation}
where we have already attained the values associated to each of ${\bf{\Phi }}^u ,{\bf{\Phi }}^a ,{\bf{K}}^a ,{\bf{K}}^u $ in the previous subsections. Now, taking derivative from \eqref{3.31} yields,
\begin{equation}
{\bf{\dot t}} = {\bf{T\ddot q}}^a  + {\bf{\dot T\dot q}}^a 
\label{3.34}
\end{equation}
In order to find $\bf{\dot T}$ one should take derivative of \eqref{3.33} which follows:
\begin{equation}
{\bf{\dot T}} = {\bf{\dot K}}^a  - {\bf{\dot K}}^u ({\bf{\Phi }}^u )^{ - 1} {\bf{\Phi }}^a  + {\bf{K}}^u ({\bf{\Phi }}^u )^{ - 1} {\bf{\dot \Phi }}^a ({\bf{\Phi }}^u )^{ - 1} {\bf{\Phi }}^a  - {\bf{K}}^u ({\bf{\Phi }}^u )^{ - 1} {\bf{\dot \Phi }}^a 
\label{3.35}
\end{equation}
where ${\bf{\dot \Phi }}^u ,{\bf{\dot \Phi }}^a ,{\bf{\dot K}}^a ,{\bf{\dot K}}^u 
$ are the derivatives of ${\bf{\Phi }}^u ,{\bf{\Phi }}^a ,{\bf{K}}^a ,{\bf{K}}^u 
$ with respect to time.
\subsection{\textbf{Finding the remaining parameters $M_{total},\,\Omega,\,w^g$}}
Up to now, we have calculated $\bf{\dot T, T}$ in \eqref{3.2}, \eqref{3.3}, \eqref{3.4} given by NOC. In the next paragraphs we focus on three more matrices needed to complete the dynamic model of a parallel robot, namely $M_{total}$, $\Omega$, $w^g$.
$M_{total}$ includes terms associated to mass and inertia of the robot links, $\Omega$ includes angular velocities pertaining to each link and the weight vector, $w^g$ , includes the links' weights. These matrices are obtained as follows:
\begin{equation}
\bf{w^g}=\left[\bf{0}\; \bf{m_1}g\;0\;\bf{m_2}g\;0\;\dots\;0\;\bf{m_{13}}g \right]^T
\label{3.36a}
\end{equation}
\begin{equation}
\bf{M_{total}}=\left[ \begin{array}{c c c c} 
\bf{M_1} & & & \\
&\bf{M_2}& \ &\\\
\ & \ &\ddots & \\\
\ & \ & \ & \bf{M_{13}}  \\
\end{array} \right]
\label{3.36b}
\end{equation}
\begin{equation}
\bf{\Omega}=\left[ \begin{array}{c c c c} 
\bf{\Omega _1} &&&\\
&\bf{\Omega_2}&  &\\
 &  &\ddots & \\
 &  &  & \bf{\Omega_{13}}  \\
\end{array} \right]
\label{3.36c}
\end{equation}
where $\bf{0}=[0\;0\;0]^T$, $\bf{m_i}g=[0\;0\;m_ig]^T$ and $\bf{M_i}=\left[ \begin{array}{l l}
\bf{I_i}\times \bf{1} & \bf{0}\\
\bf{0} & \bf{m_i} \times \bf{1}\\
\end{array} \right]$.

\subsection {\textbf{Dynamic Equations of the Underlying Robot}}
\begin{figure}[htp]
  \begin{center} 
	\includegraphics[width=13cm, height=9cm]{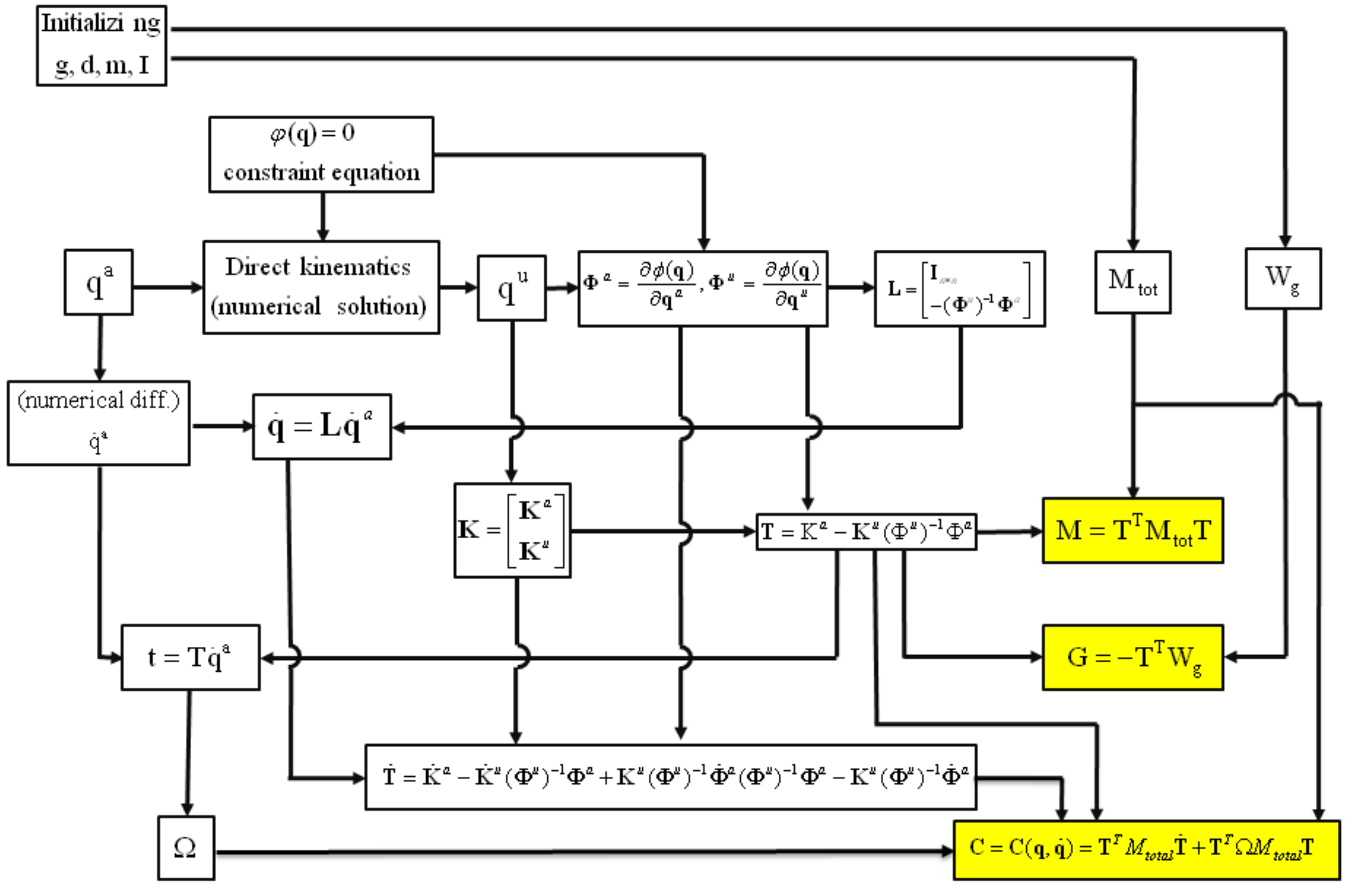}
  \end{center}
  \caption {Solution procedure for inverse dynamics problem}
  \label{fig31}
\end{figure}

As already stated, the ultimate goal of solving a robot's dynamics is to find robot differential 
equations in the form of $M\ddot q^a  + C\dot q^a  + G = \tau ^a $, where $\tau ^a $ is the torque vector applied to motors (actuators) and $\dot q^a ,\ddot q^a $ are linear speed and acceleration of the ball screws in our 3PSP platform. After deriving the $M$, $G$, $C$, the steps towards solving the inverse dynamic is complete. Figure \ref{fig31} depicts the block diagrams as well as the order of computations from beginning to the end. Using the NOC approach discussed in this chapter we managed to solve the inverse dynamic system and hence given a trajectory we can claim that an open loop controller can be designed to make our dynamic system to track it. This is only possible if there is not any source of uncertainty in our model. However, as explained in the previous chapter, ample of uncertainty sources exist when the design structures are implemented physically, hence inverse dynamic process exposes to the uncertainty, too. As alluded in the same chapter, finding a maximum value for parasitic resonances in the robots' workspaces has still remained an open problem. Thus if we can't find such values then there should be a way to handle the inverse dynamics in the presence of this uncertainty. Uncertainty in robots are not limited to realization process rather other factors such as numerical solutions influence our developed model, too. A significant uncertainty of this type exists in dynamic control programs unless there is a closed form formula available according to which the required torques and forces can be calculated. Hence, an open loop controller may not be desirable at all but rather it should be controlled with the contribution of feedback and controlling techniques. In later chapters we develop a promising technique towards control of such model based on a linearization perspective.

%% file: chapter4.tex
\chapter{Linear and Non-linear Control}
\section{\textbf{Preface}}
Based on the subjects presented in previous chapters, we are now able to easily find the position of the end-effector and hence to record and to analyze the traversed trajectories. Thus we are now ready to talk about how to make the robot to follow the trajectories given to us, namely the \emph{desired} trajectories. 

The control methods that we will talk about in this chapter fall into two main classes, the \emph{linear} control and the non-linear one. Rigorously speaking, the linear control can be applied only if the underlying system can be modeled by a set of linear differential equations. Of course one can employ a \emph{linear} for the case of \emph{non-linear} systems, but the performance degrades depending on the degree of non-linearity. Similarly the \emph{non-linear} control is the selected choice when the system at hand is non-linear. As presented in the previous chapter, the dynamics of most of the manipulators are highly non-linear; this inspires the significance of non-linear systems and non-linear control as well. Thus we should not exploit a linear control method for manipulators unless it can be approximated with a linear model provided that the incurred errors are negligible. A word under notation is that although mechanical systems in industry are non-linear in most of the cases; linear control approaches are considered as most prevalent due to their simplicity. 

Last but not the least, consideration of linear methods can serve as a good pedagogical basic for introducing and applying the non-linear approaches as we will see later.

\newpage
\section{\textbf{Linear Control}}
\subsection{\textbf{Feedback and Closed Loop Control}}
\begin{figure}[htp]
  \begin{center} 
	\includegraphics[width=13cm, height=5cm]{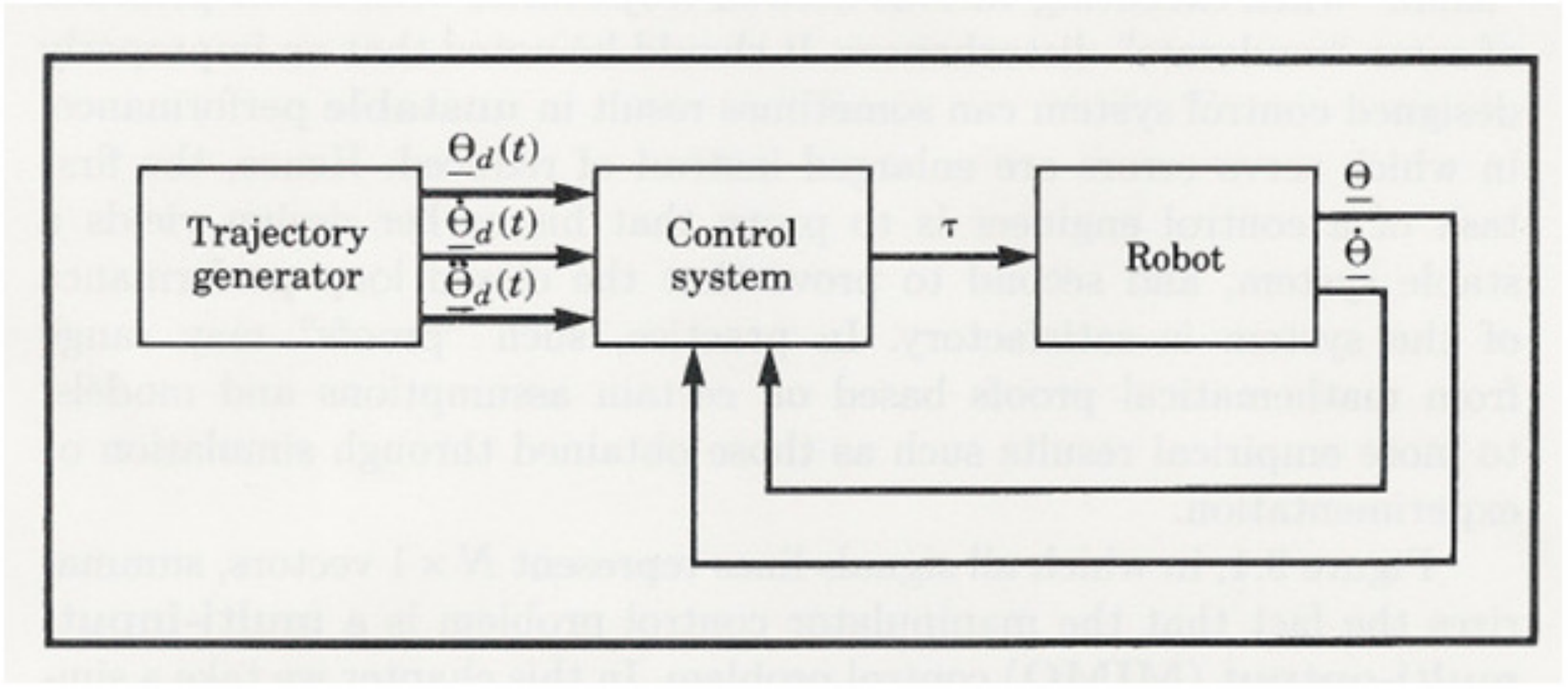}
  \end{center}
  \caption {Abstract model for controlling a robot \cite{Cra04}}
  \label{fig32}
\end{figure}

Before starting to introduce feedback and closed loop control lets define what we mean by a manipulator. Briefly speaking, a manipulator is a set of joints (either actuated or un-actuated) and links connected to each other in some way to make a mechanism. Some of the joints are actuated in this mechanism with each actuator equipped with a sensor. The sensors are most of the times position sensors however other types of sensors may also exist such as speedometers. We also assume that there is a maximum of 1 actuator associated with each joint.
Usually (and especially in industrial robotics) the actuators do not accept positioning commands rather they receive torque values. Therefore it is the role of controller (the programmer in a higher level perspective) to control the actuators (servo motors, jacks and etc.) such that the mechanism ultimately follows the desired trajectory. We call this type of control program a \emph{control system}. Almost always the toque values associated to each (say) servo motor is calculated with the knowledge of previous output of the system by means of a \emph{feedback} from the joint variables (sensor outputs). Figure \ref{fig32} depicts an abstract model of the just mentioned process. After a trajectory is generated by a trajectory generator routine (TG routine), the aim of the control system would be to keep the position of end-effector (which is the focus of the designers) as much close as possible to the desired one (the one generated by TG unit). Note also that the outputs corresponding to the TG unit is in joint space therefore one should follow a direct kinematic process to find the tool-tip position in Cartesian space.

Given the system feedback and the desired positions the control system produces the torque vector in some way. The toques are applied to the physical robot which is the final goal of all these calculations and the process proceeds iteratively. Let's write down the dynamic equation of a mechanical system, as described in earlier chapters;
\begin{equation}
M(\Theta _d )\ddot \Theta _d  + V(\Theta _d ,\dot \Theta _d ) + G(\Theta _d ) = \tau 
\label{4.1}
\end{equation}

The first idea that comes to mind is to find the required torque values suitable for controlling the manipulator by means \eqref{4.1} without a need to closed-loop controller system. However this is only possible if the dynamic model was exact and also no noise or any other disturbances would affect the process but this rarely, if not never, come to truth. Notice that \eqref{4.1} is a function of $\Theta_d$ and its derivatives and not dependent on $\Theta$ or any of its derivatives. This type of controller scheme is also called an open-loop controller since no feedback is needed.

Accordingly, the only way to make an efficient controller is via the use of feedbacks acquired from the sensors, as easily inferred from Figure \ref{fig32}. More specifically the feedback is used to find the \emph{servo} errors 
(the differences between the desired and the actual sensor values) and likewise their derivatives as follows:
\begin{equation}
E = \Theta _d  - \Theta 
\label{4.2a}
\end{equation}
\begin{equation}
\dot E = \dot \Theta _d  - \dot \Theta  \\  
\label{4.2b}
\end{equation}
\subsection{\textbf {A Second Order Linear System}}
\begin{figure}[htp]
  \begin{center} 
	\includegraphics[width=13cm, height=5cm]{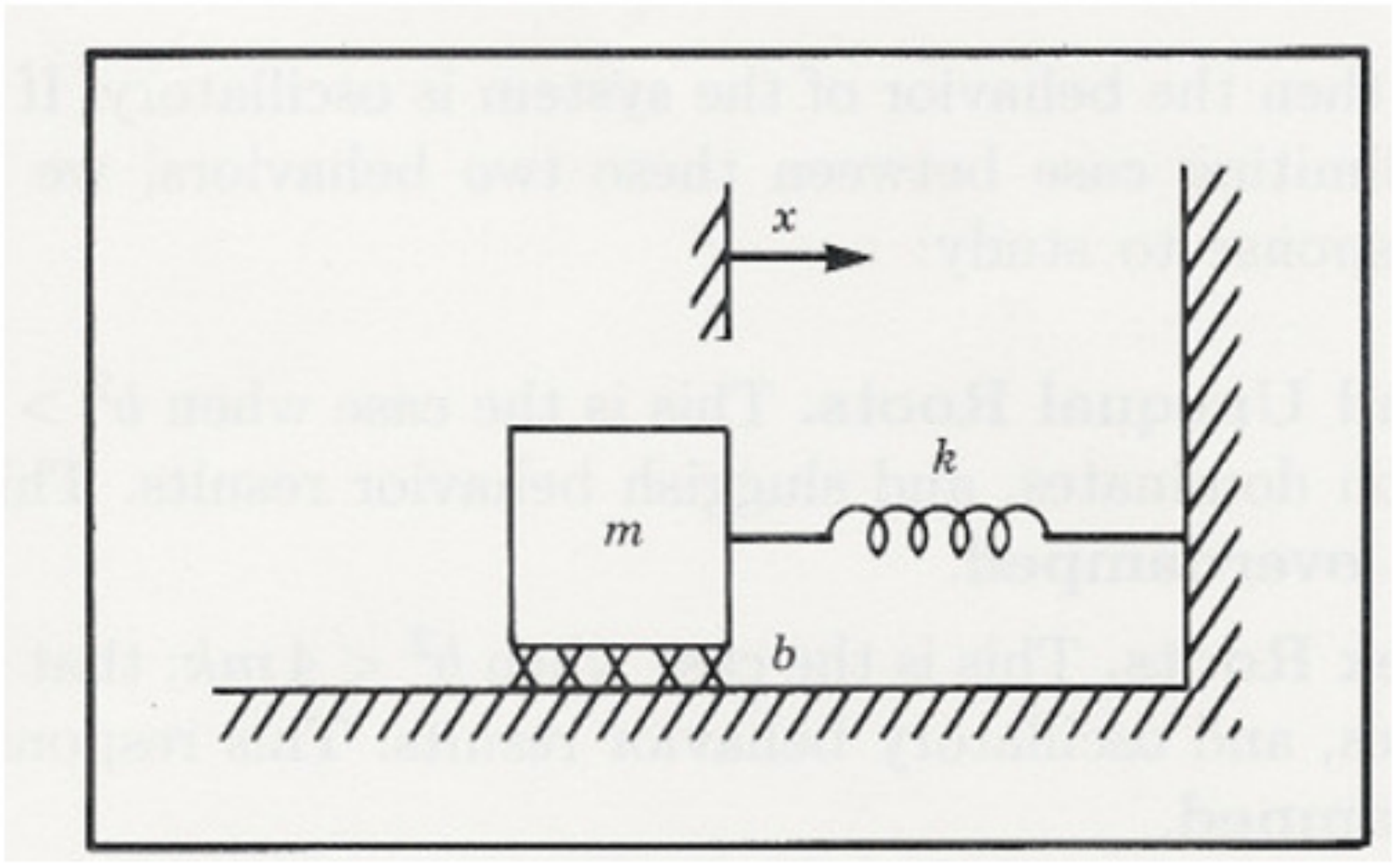}
  \end{center}
  \caption {A mechanical system with dynamics described by second order differential equations \cite{Cra04}}
  \label{fig33}
\end{figure}

\begin{figure}[htp]
  \begin{center} 
	\includegraphics[width=13cm, height=5cm]{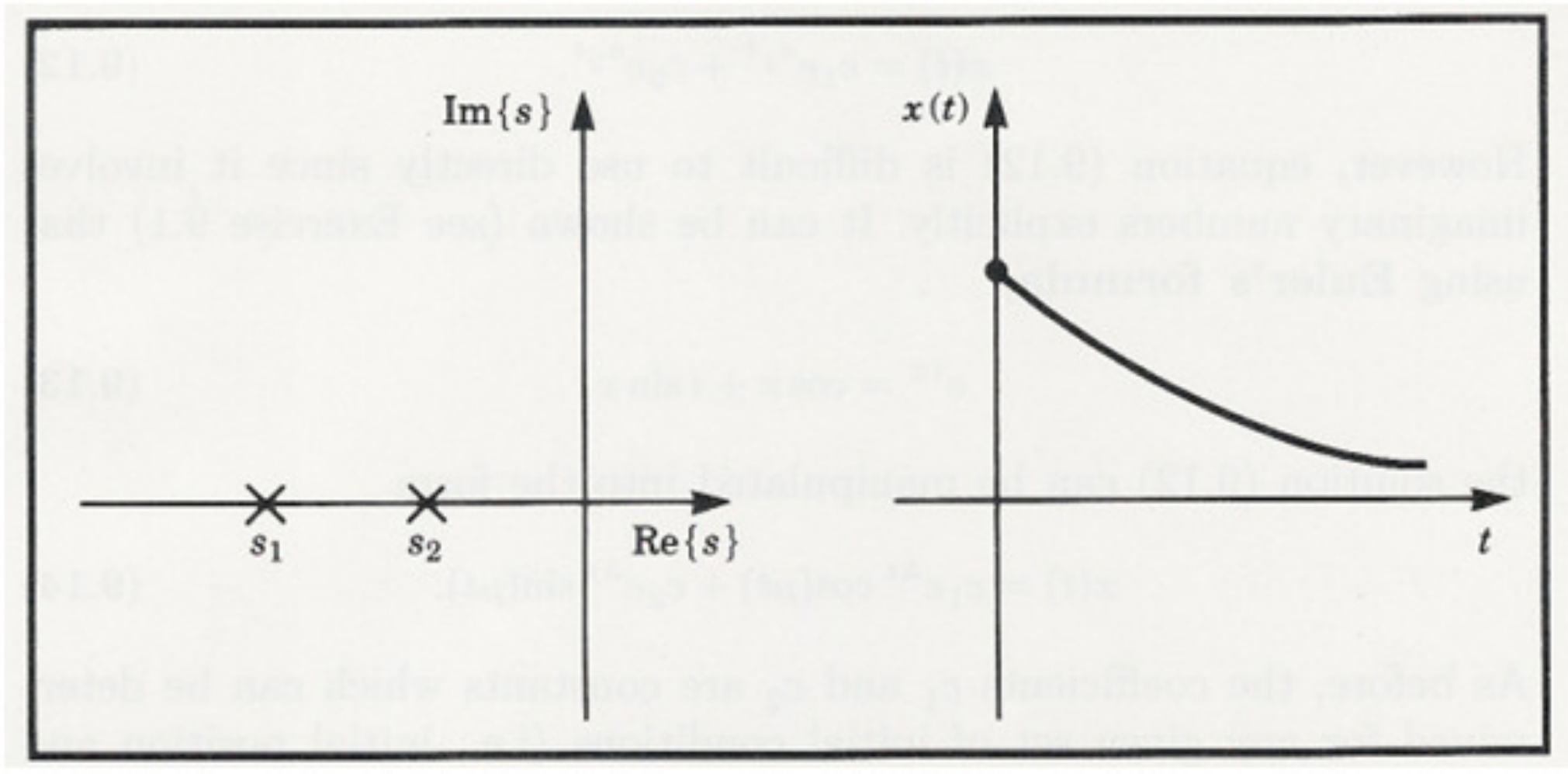}
  \end{center}
  \caption {The response of a system with real and non-equal poles, released at $x=-1$ \cite{Cra04}}
  \label{fig34}
\end{figure}

\begin{figure}[htp]
  \begin{center} 
	\includegraphics[width=13cm, height=5cm]{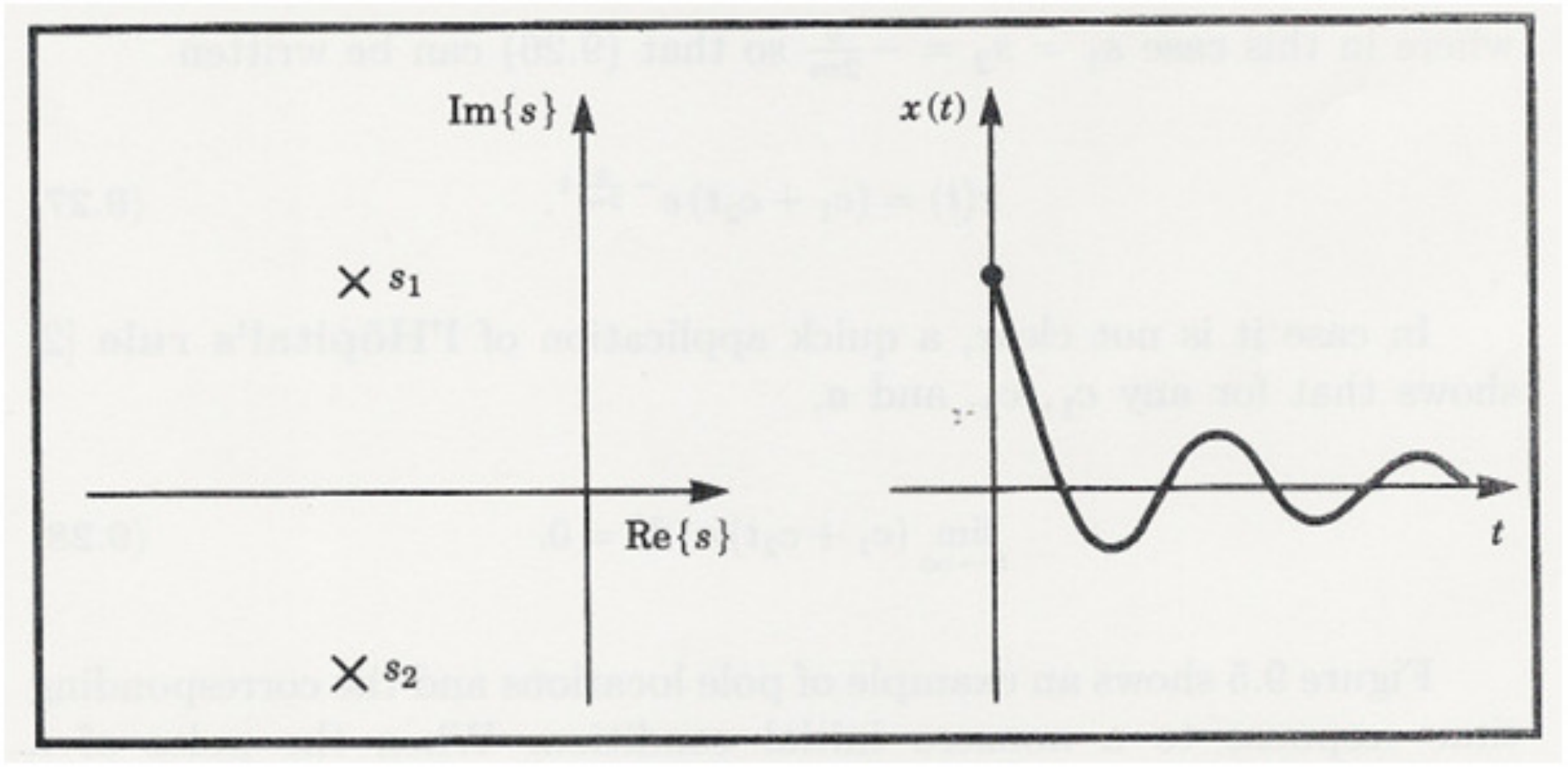}
  \end{center}
  \caption {The response of a system with complex and non-equal poles, released at $x=-1$ \cite{Cra04}}
  \label{fig35}
\end{figure}

\begin{figure}[htp]
  \begin{center} 
	\includegraphics[width=13cm, height=5cm]{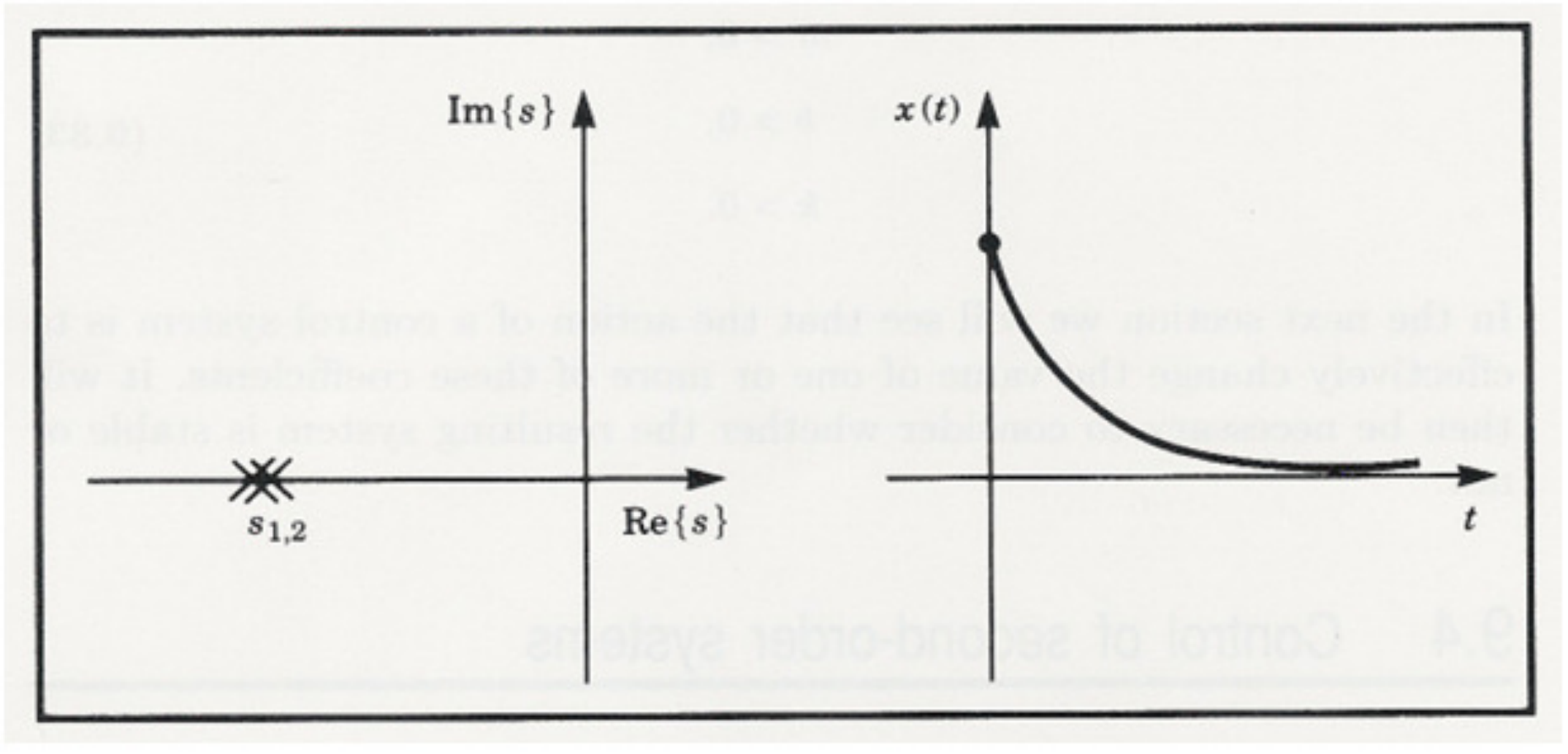}
  \end{center}
  \caption {The response of a system with real and equal poles, released at $x=-1$ \cite{Cra04}}
  \label{fig36}
\end{figure}

In order to get ready to analyze a manipulator system, we present a simple second order linear system
, the spring system, as an example. Figure \ref{fig33} shows a block of mass attached to a spring of 
stiffness $k$ and subject to friction of coefficient $b$. The figure assumed to be in zero condition and the positive sense is indicated with vector $x$. Assuming that the frictional force is linearly proportional to the position of the box, the dynamic equation of the system follows:
\begin{equation}
m\ddot x + b\dot x + kx = 0
\label{4.3}
\end{equation}
Notice that the there is no external force applied to this 1 D.O.F system. The solution to \eqref{4.3}, $x(t)$, is the position of the box which can be found given the initial conditions of the system, that is the initial position and velocity of the box.
Intuitively, we expect to observe different reactions based on the system configuration as soon as perturbation imposed to the system. For example, if the stiffness of the spring is high and the friction is low then we expect to see more oscillatory movements while for a spring with low stiffness and a ground with high friction the movements would converge soon without any overshoot.
From study of differential equations, we know that the form of the solutions of a second order differential equation is contingent upon its \emph{characteristic equation},
\begin{equation}
ms^2  + bs + k = 0
\label{4.4}
\end{equation}
which has the roots,
\begin{equation}
s_1  =  - {b \over {2m}} - {{\sqrt {b^2  - 4mk} } \over {2m}}
\label{4.5a}
\end{equation}
\begin{equation}
s_2  =  - {b \over {2m}} + {{\sqrt {b^2  - 4mk} } \over {2m}}
\label{4.5b}
\end{equation}
In fact it is these \emph{poles} of the system that characterizes the motion of a system. If each of these terms includes an imaginary part then the resulting motion will inherently oscillatory but it in contrast, if they are both real then the system exhibits a damping response. Accordingly, we may observe three different responses, as follows:
\begin{itemize}
\item Real and Unequal Roots.\\
This occurs when $b^2-4ac>0$, the response has a sluggish characteristic as the friction overcomes. It is also called an \emph{overdamped} response. 
\item Complex Roots.\\
This happens when the spring stiffness dominates the friction. It arises when $b^2-4ac>0$ and consequently the system exhibits an oscillatory motion. We call the natural response of this type an \emph{underdamped} response.
\item Real and Equal Roots\\
Finally, this is a compromise between stiffness and friction and is the best answer which yields the fastest possible non-oscillatory solution. The response is called \emph{critically damped}. The requisite to produce a critically damped response is the condition, $b^2-4ac=0$ . This is the most desirable solution we usually in quest of.
\end{itemize}
Figures \ref{fig34}, \ref{fig35}, \ref{fig36} depict the locations of the poles and the response of the systems corresponding to the first, second and third case, respectively, when the system is released from $x=-1$ and at initial rest. Notice how the last response is more desirable than the others.
It's noteworthy that for passive mechanical systems (like the one just described) that they are always stable. Such systems always have the properties that $m>0$, $b>0$, $k>0$ . In the next section we will see that the role of a controller is in fact to change these coefficients and hence might yield an unstable system. Therefore \emph{stability} has always been considered as a mater of concern in controlling a dynamic system.
\subsection{\textbf{Control of Second-Order Systems}}
\begin{figure}[htp]
  \begin{center} 
	\includegraphics[width=13cm, height=5cm]{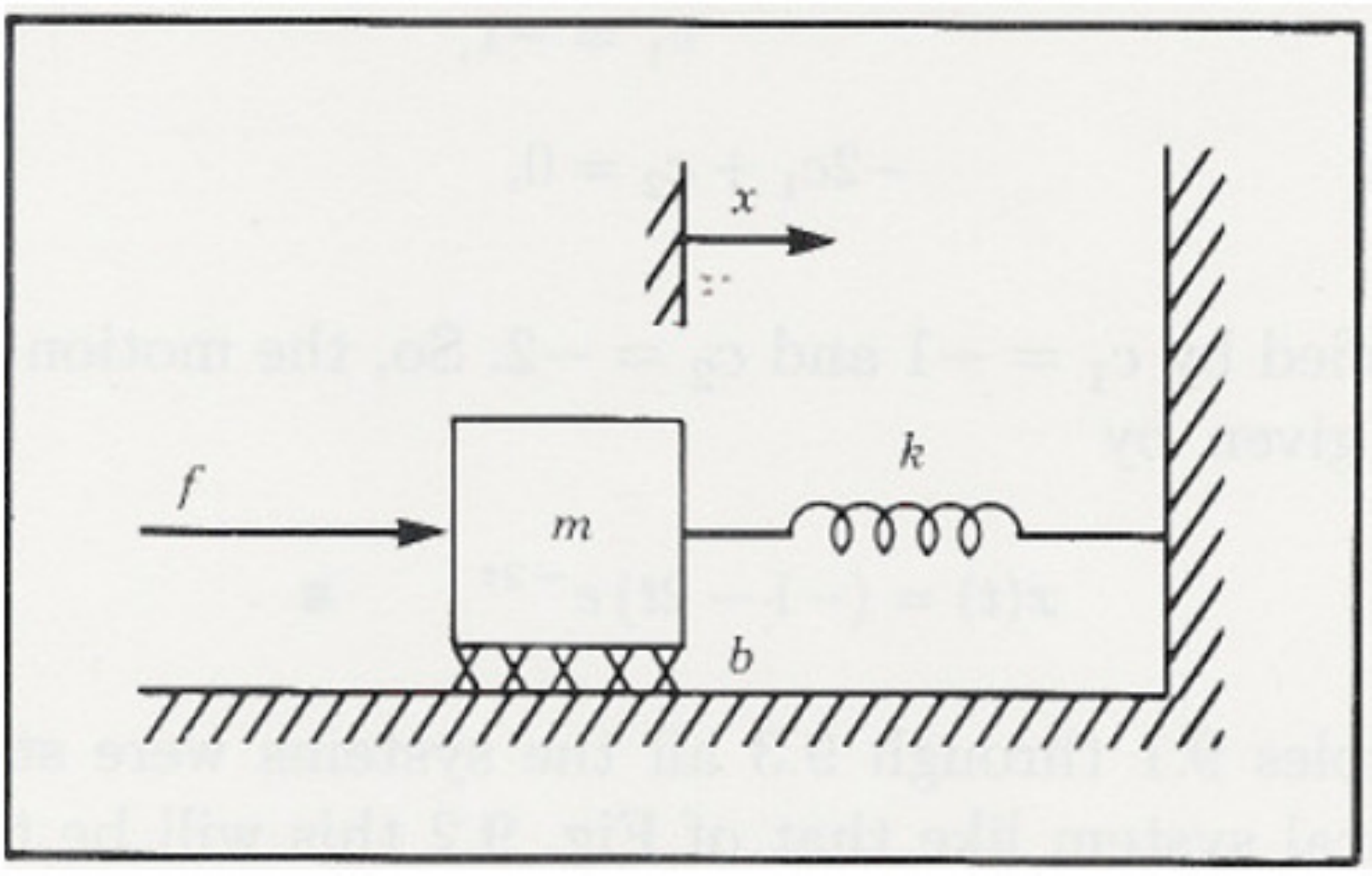}
  \end{center}
  \caption {A damped-spring mass with an actuator \cite{Cra04}}
  \label{fig37}
\end{figure}

\begin{figure}[htp]
  \begin{center} 
	\includegraphics[width=13cm, height=5cm]{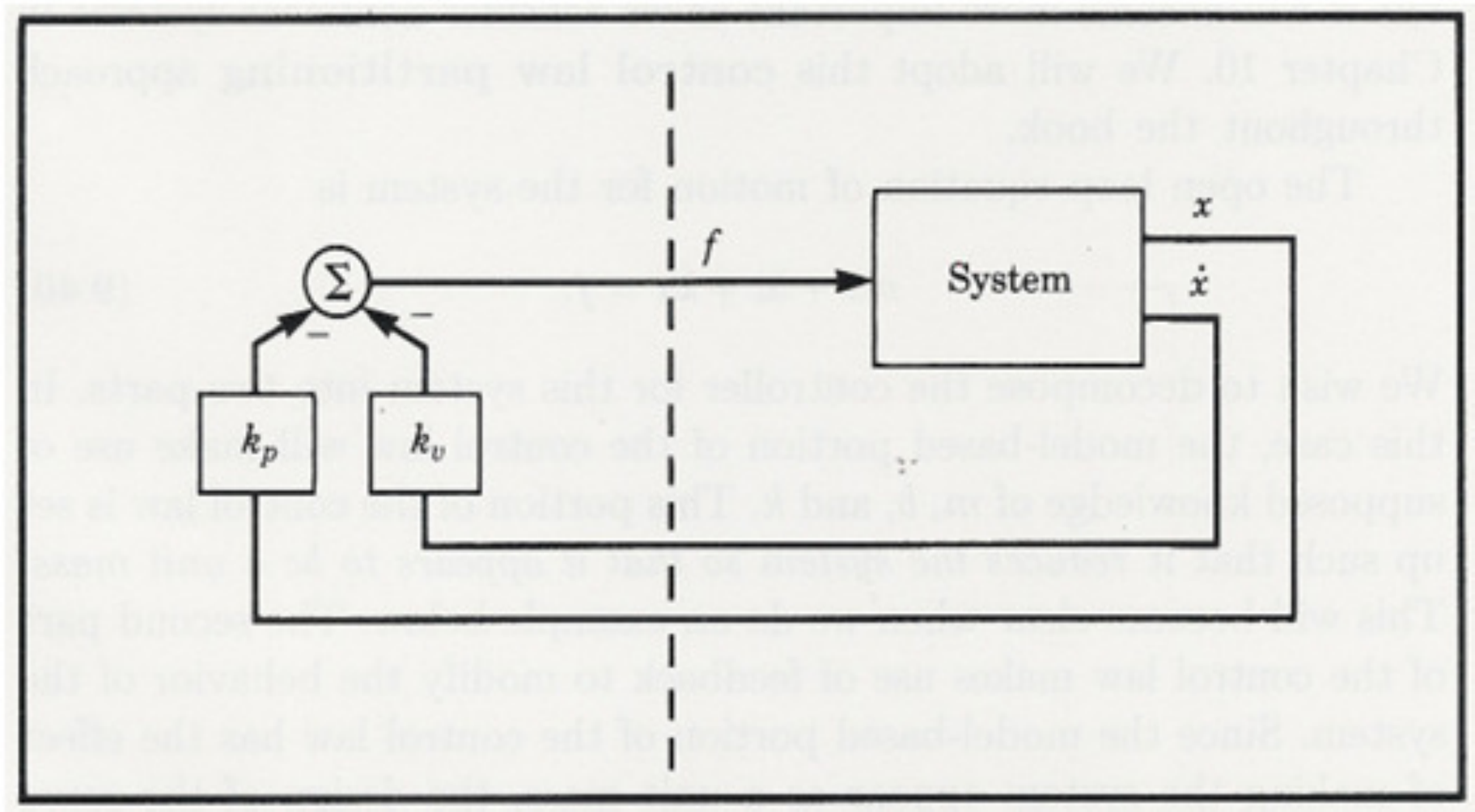}
  \end{center}
  \caption  {A regulation system, for generating every second order response \cite{Cra04}}
  \label{fig38}
\end{figure}

Let's assume that the natural response of an underlying system is not as we desire to be; meaning that it does not behave in a critically damped manner and we would like to make it perform in a critically damped manner or may be the system altogether behaves as if there is not any spring available (or $k=0$ thus it would converge to an undesired final position) and we have to fix it in some way.
Consider Figure \ref{fig37} which differs from \ref{fig33} in that there is additionally an external force applying to the box now. This force is under our control and we can attenuate or amplify its amount by means of an actuator as we wish. Let this amount achieved by following relation:
\begin{equation}
f =  - k_p x - k_v \dot x
\label{4.6}
\end{equation}
Figure \ref{fig38} depicts the block diagram of the closed loop system. Usually the part at the left of the dashed line is implemented in computer side while the one located in the right hand is the physical system. The control system shown in this figure is a \emph{regulation} system and not a \emph{trajectory-following} one in the sense that it simply tries to maintain the position of the box at a desired location. This is more illustrated if we equate and reform \eqref{4.4} and \eqref{4.6} as follows;
\begin{equation}
m\ddot x + b\dot x + kx = f \Rightarrow m\ddot x + b\dot x + kx =  - k_p x - k_v \dot x \Rightarrow m\ddot x + (k_v  + b)\dot x + (k_p  + k)x = 0
\label{4.7}
\end{equation}
or 
\begin{equation}
m\ddot x + b'\dot x + k'x = 0
\label{4.8}
\end{equation}
where $b'=b+k_v$ and $k'=k+k_p$. According to the discussion presented above it is clear from \eqref{4.7} that by giving the proper values to the \emph{control gains}, $k_p$ and $k_v$ , one can produce every second order system behavior that he wishes. Most of the times, a critically damped response is desired hence the gains should be set such that $b'=2\sqrt{mk}$. Moreover if a desired stiffness for system altogether is searched for then the gains would be completely determined. Notice that if each of the $k'$ or $b'$ is negative, then instability happens. This can be verified by finding the solution to the second order differential equations. In case of instability the error is magnified instead of attenuation. The instability may give rise to very undesired situation and should be strictly prohibited.

\subsection{\textbf{Control Law Partitioning}} 
Let's consider Figure \ref{fig37} again. This time we try to control the system with a controller of a different structure. In this method we partition the controller
into two main parts, namely, the \emph{model-based} portion and the \emph{servo} portion such that the system parameters ($m$, $b$, $k$) appears in the model-based part and leave the servo part completely independent of them. This distinction may not appear effective at the moment but its significance becomes evident in later section, on non-linear systems. The open loop equation of the motion, as described earlier, is:
\begin{equation}
m\ddot x + b\dot x + kx = f
\label{4.9}
\end{equation}
Now, we are going to break this system into the aforementioned portions. The model-based system should be set up such that the system appears in the form of a unit mass system. This is done by tuning $m$, $b$ and $k$. The second part of control law then tries to modify the behavior of the system. Since the model-based system has already compensated the effects of friction and stiffness by modeling a system with only a unit mass, the servo portion's job would be very simple, that is to control a system composed of a unit mass. We clarify this procedure in a more illustrative manner in our next discussion.

The model-based portion of the control appears in the form,
\begin{equation}
f = \alpha f' + \beta 
\label{4.10}
\end{equation}
where $\alpha$ and $\beta$ might be functions or constants such that $f'$ is considered as the input to the new unit mass system. Substituting \eqref{4.10} into \eqref{4.9} yields,
\begin{equation}
m\ddot x + b\dot x + kx = \alpha f' + \beta 
\label{4.11}
\end{equation}
Obviously, to convert the system into a unit mass system, the foregoing parameters should be defined as follows:
\begin{equation}
\alpha=m
\label{4.12a}
\end{equation}
\begin{equation}
\beta  = b\dot x + kx
\label{4.12b}
\end{equation}
which follows then,
\begin{equation}
\ddot x=f'
\label{4.13}
\end{equation}
Having a unit mass system, it is the servo portion task to control it. That is we proceed as if \eqref{4.13} is the equation associated to an open-loop system to be controlled in some way. We perform this task just as we did before:
\begin{equation}
f' =  - k_p x - k_v \dot x
\label{4.14}
\end{equation}
Substituting \eqref{4.13} into the previous equation, follows:
\begin{equation}
\ddot x + k_v \dot x + k_p x = 0
\label{4.15}
\end{equation}
Now, we reached to a point where we can set the control gains, independent of system parameters to achieve a desired response. That is for a critical damping reaction, we must set $k_v=2\sqrt{k_p}$.
As implied from the previous figure, the servo portion of the design is in fact a PD controller (a type of classical based controller with proportional and derivative gains) which performs 
promisingly when the system is linear.
\subsection{\textbf{Trajectory-Following Control}}
Instead of maintaining the box at a desired place, we can easily generalize the introduced controller to follow an arbitrary trajectory. The trajectory is generated as a function of time which specifies where the box should be at every instance of time. We also assume that the trajectory is smooth enough so that the first and the second derivative exist. Let's define the servo error as the difference between the desired position and the actual one. The servo control law is accordingly changes to the following: 
\begin{equation}
\ddot x_d  + k_v \dot e + k_p e = f'
\label{4.16}
\end{equation}
and hence we have
\begin{equation}
\ddot x_d  + k_v \dot e + k_p e = \ddot x
\label{4.17}
\end{equation}
or equivalently
\begin{equation}
\ddot e + k_v \dot e + k_p e = 0
\label{4.18}
\end{equation}
\begin{figure}[htp]
  \begin{center} 
	\includegraphics[width=13cm, height=5cm]{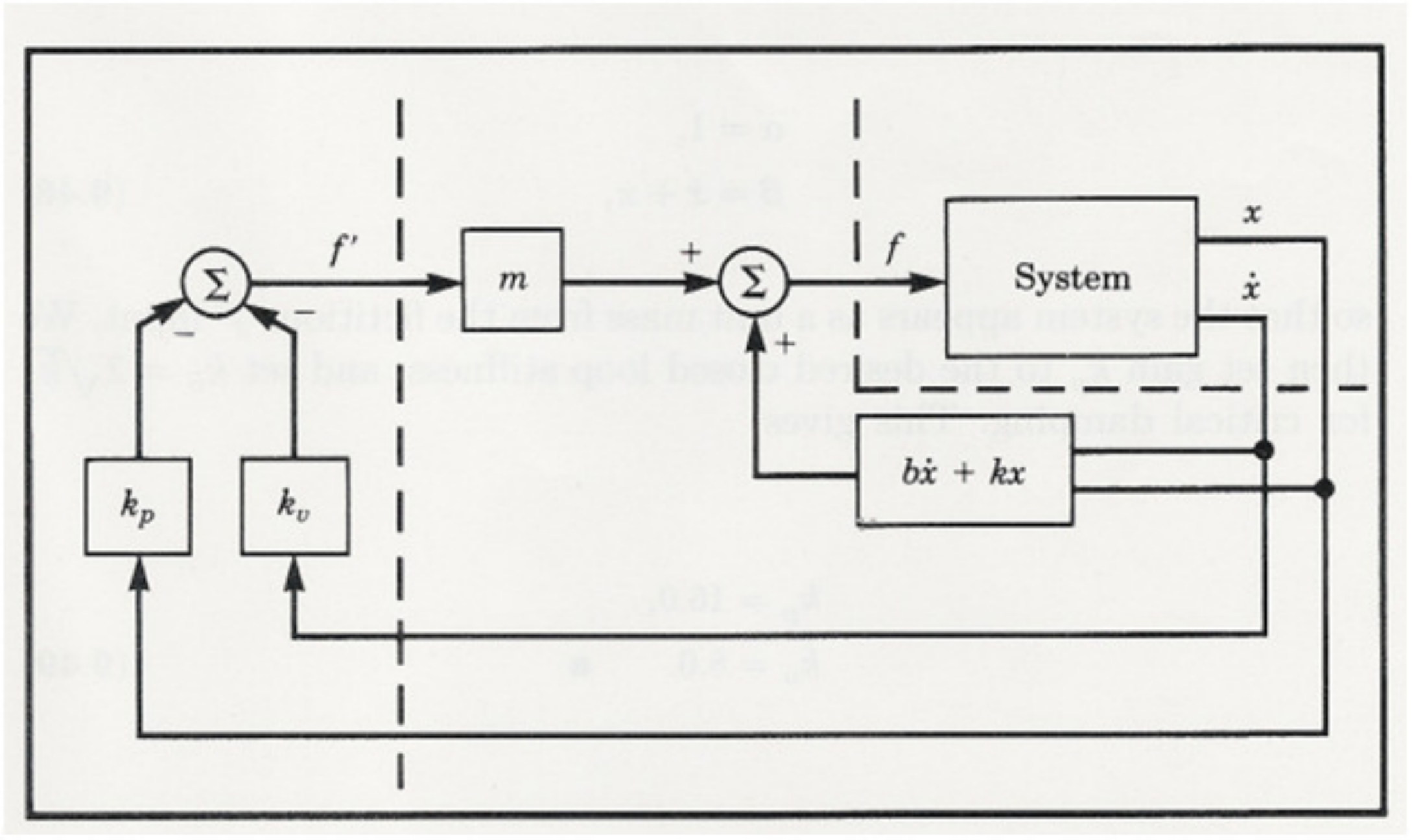}
  \end{center}
  \caption  {The overall closed loop controller based on the partitioning method \cite{Cra04}}
  \label{fig39}
\end{figure} 
This is a second order differential equation in the so-called \emph{error space} as it describes the evolution of error is relative to the desired trajectory. Note that like the previous controller, we can change its behavior by changing the control gains.
The last and perhaps the most important note is that if our model is a perfect one, that is if our knowledge of $b$, $k$ and $m$ is complete (which does not hold in real world applications), and if there is no noise or initial error available in the system, then the box will follow the desired trajectory exactly. If there is an initial error, it will be suppressed accordingly and the system follows the trajectory exactly.\\ 
Finally, Figure \ref{fig40} shows the overall trajectory-tracking controller based on the partitioning law.
\begin{figure}[htp]
  \begin{center} 
	\includegraphics[width=13cm, height=5cm]{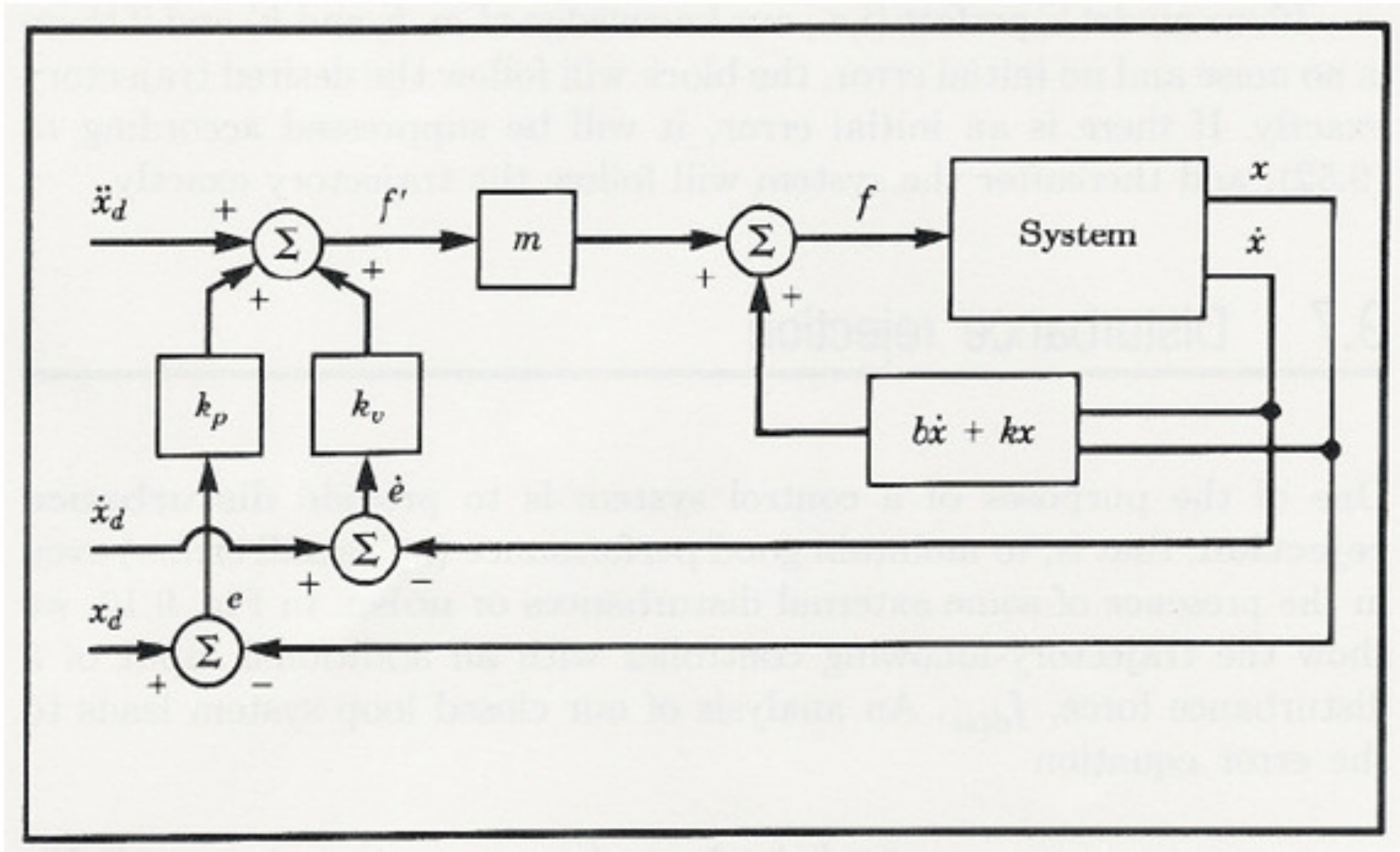}
  \end{center}
  \caption  {The overall trajectory-tracking controller based on the partitioning law \cite{Cra04}}
  \label{fig40}
\end{figure} 
\section{\textbf{Computed Torque Control}}
As the field of non-linear control is a wide area, we must restrict ourselves to a single but important class of controllers for non-linear plants which is called the \emph{Computed Torque Control} (hence force called CTC). 
The CTC, in fact, behaves much similar to the approach just presented but for a broader range of dynamic systems. 
In other words, the parameters are not any longer limited to $m$, $b$ and $k$. A simple instance arises when the spring behaves in a non-linear fassion (e.x. $f(x)=kx^3$). Thus,
the estimated model-based parameters ($\alpha$ and $\beta$) must contain the non-linear terms, too.

A word under notation is that in practical situations the performance implied by \eqref{4.16} is not achievable since:
\begin{itemize}
\item The computer systems whose aims are to execute the control program are inherently digital as opposed to the continuous analogue system. Besides the actuators cannot handle commands received in a frequency higher than a predetermined threshold.
\item Inaccuracies always exist in the manipulator model.
\end{itemize}
\begin{figure}[htp]
  \begin{center} 
	\includegraphics[width=13cm, height=5cm]{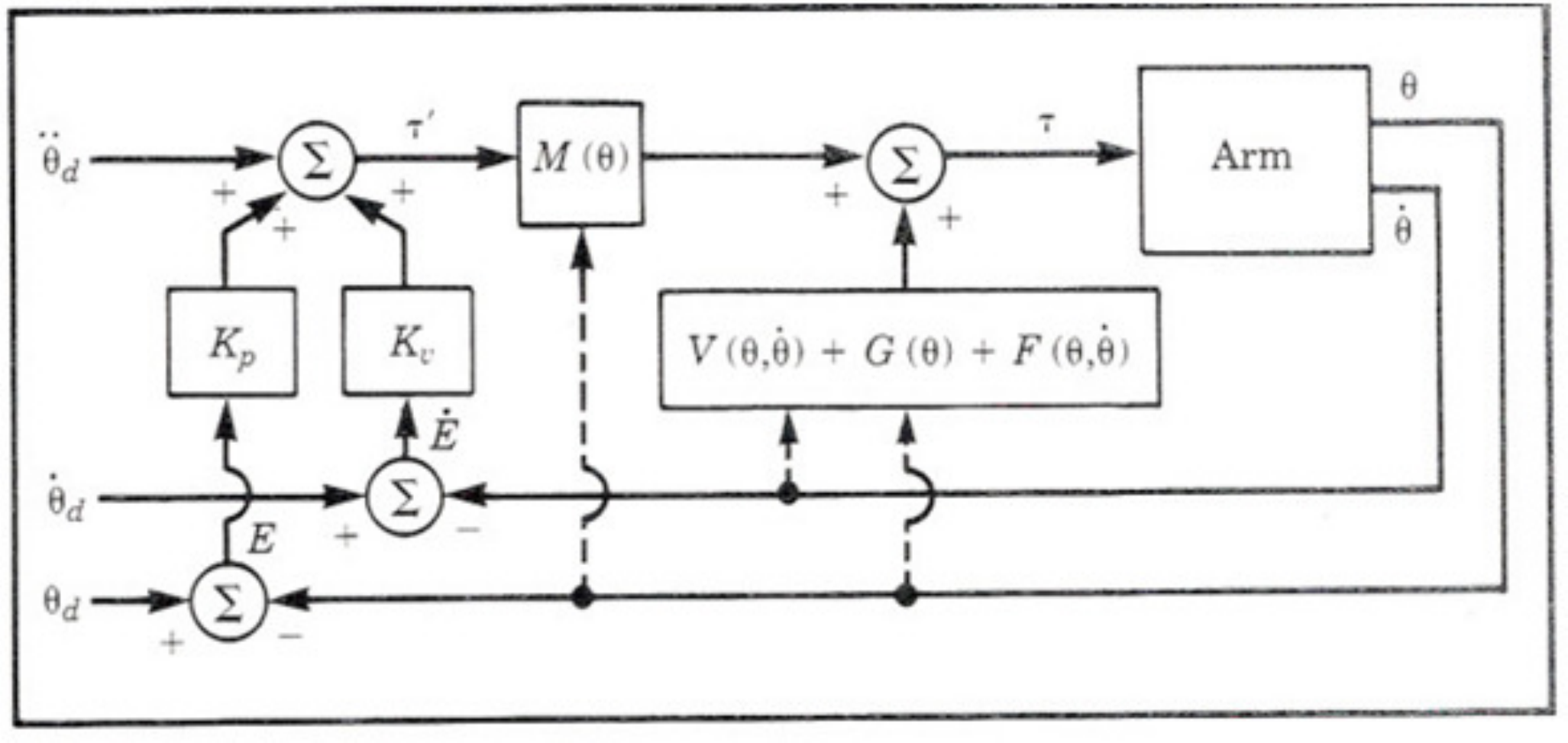}
  \end{center}
  \caption  {Manipulator control via a partitioning controlling scheme \cite{Cra04}}
  \label{fig43}
\end{figure}  
\subsection{\textbf{Practical Considerations}}
Let's discuss the aforementioned items in more details from a practical standpoint.
\subsubsection{Time Required to Compute the Model}
In the method we have presented above, we implicitly assumed that the time taken to perform the required computations is zero and that the overall control process is done in a continuous fashion. Although the computation time can be decreased if more powerful processors and memories are employed in the control process, however this entails the higher prices too, which may not be economically desired. Hence faster methods in control of manipulators are welcomed although there is a compromise between the performance and the speed.	On the other hand, as the computer power is increasing continuously as opposed to the prices, the control laws become more practical too. Another point to note is that the sampling rates of the sensors and also the highest command frequency of the actuators are also matters of concerns when implementing a controller for a real platform.
\subsubsection {Lack of Knowledge of Parameters}
The second main difficulty encountered when implementing the computed torque control method is that the manipulator's dynamics is not precisely known, this holds especially for modeling the friction. In fact it is terribly hard to model the structure of friction as it's highly dependent on several other parameters. Besides the dynamics of manipulators undergo changes due to wear and tear issues. Thus the dynamics parameters do not remain constant during the time.
Another important factor that might influence the dynamics of a system is that since most of the manipulators are designed to pick up tools and parts which might have considerable weights and inertias relative to the weight and inertia of other manipulator's elements, the dynamics of the system changes when the manipulator is performing its role, although in many industrial applications the characteristics of such tools are already known (an thus they may be accounted for in modeled portion of the control law).
Therefore the best way to control a robot which acts in unknown circumstances is to benefit some handy tools which have better performances when the system exposed to the noise (or any other kind of uncertainty). Based on the discussions given in earlier chapters, Type-II fuzzy logic is a valuable and promising technique that can meet our expectations. In later chapter we develop our model keeping all the subjects presented till now in mind.

%% file: chapter5.tex
\chapter{The Proposed Method}
\section{\textbf{Preface}}
This chapter models the uncertainty in dynamical model of our 3 DOF parallel manipulator. More specifically, it is intended to incorporate the Type-II Fuzzy Logic into a model based controller, the previously introduced computed torque method.

Based on previous chapters, Type-II fuzzy logic has proved its superiority over traditional fuzzy logic (or fuzzy logic for short) when dealing with uncertainty. Type-II fuzzy logic controllers are newer and more promising approaches that have been recently applied to various fields due to their significant improvements especially when the noise (as an important instance of uncertainty) emerges. During the design of a Type-I fuzzy logic system, we presume that we are almost certain about the fuzzy membership functions which is not true in all cases. Thus T2FLS as a more realistic approach for dealing with practical applications might offer new interesting results. Type-II fuzzy logic takes into account a higher level of uncertainty. The membership grades for a Type-II fuzzy variable are no longer crisp numbers rather they are themselves Type-I linguistic terms.

Parallel robots on the other hand, are rather new sort of industrial and scientific tools that are being used in many research and industrial academia. The most problematic issues that engineers and designers face when using such robots are the high computational complexity needed for calculation of the inverse dynamics which should be recalculated after each movement step as well as the structural uncertainty present in the underlying robot. Generally, and as discussed earlier, the dynamic equation for a given robot reduces to the following closed form:
\begin{equation}
M(q)\ddot q + C(q,\dot q)\dot q + G(q) = \tau 
\label{5.1}
\end{equation}
where $q$, $M(q)$, $C(q,\dot q)$ and $G(q)$ are trajectory positions (such as the rotation degree of servos), the mass matrix, centrifugal-Coriolis matrix and gravity matrix, respectively. Theoretically, given a trajectory, to calculate the required torque to servos one should solve the inverse dynamic equations and find the amount of torque that he/she must feed into the actuators (servos). Ideally, in that circumstance one is able to design a feed forward (also called open-loop) controller which computes the required torques for each step so that the robot operates as desired. In this model no feedback from the current state of the robot is needed since everything should work fine as all the equations are specified rigorously. But the problem with this perfect model is that we are not capable of estimating exact dynamic parameters of the robot dynamics (and even 
if we do, our computations in digital world (e.x. computers) still incur \emph{truncation} errors), hence, a closed-loop controller is often preferred.

One of the most well-known dynamic controllers that relies on the dynamic parameters of the underlying robot is the \emph{Computed Torque Control} (or CTC for short) Method. The CTC, as introduced in previous chapter, converts the non-linear dynamics of a robot into a linear one provided that the last two matrix terms in \eqref{5.1}, namely $C(q,\dot q)$, $G(q)$  are known. Having designed a system with linear dynamics, it is easy for a control engineer to design a PID (usually PD) controller for it such that the final motion of the robot would be to follow a trajectory precisely.  
\newpage
\section{\textbf{The Computed Torque Control of the Underlying 3PSP Manipulator}}
\begin{figure}[htp]
  \begin{center} 
	\includegraphics[width=13cm, height=6.5cm]{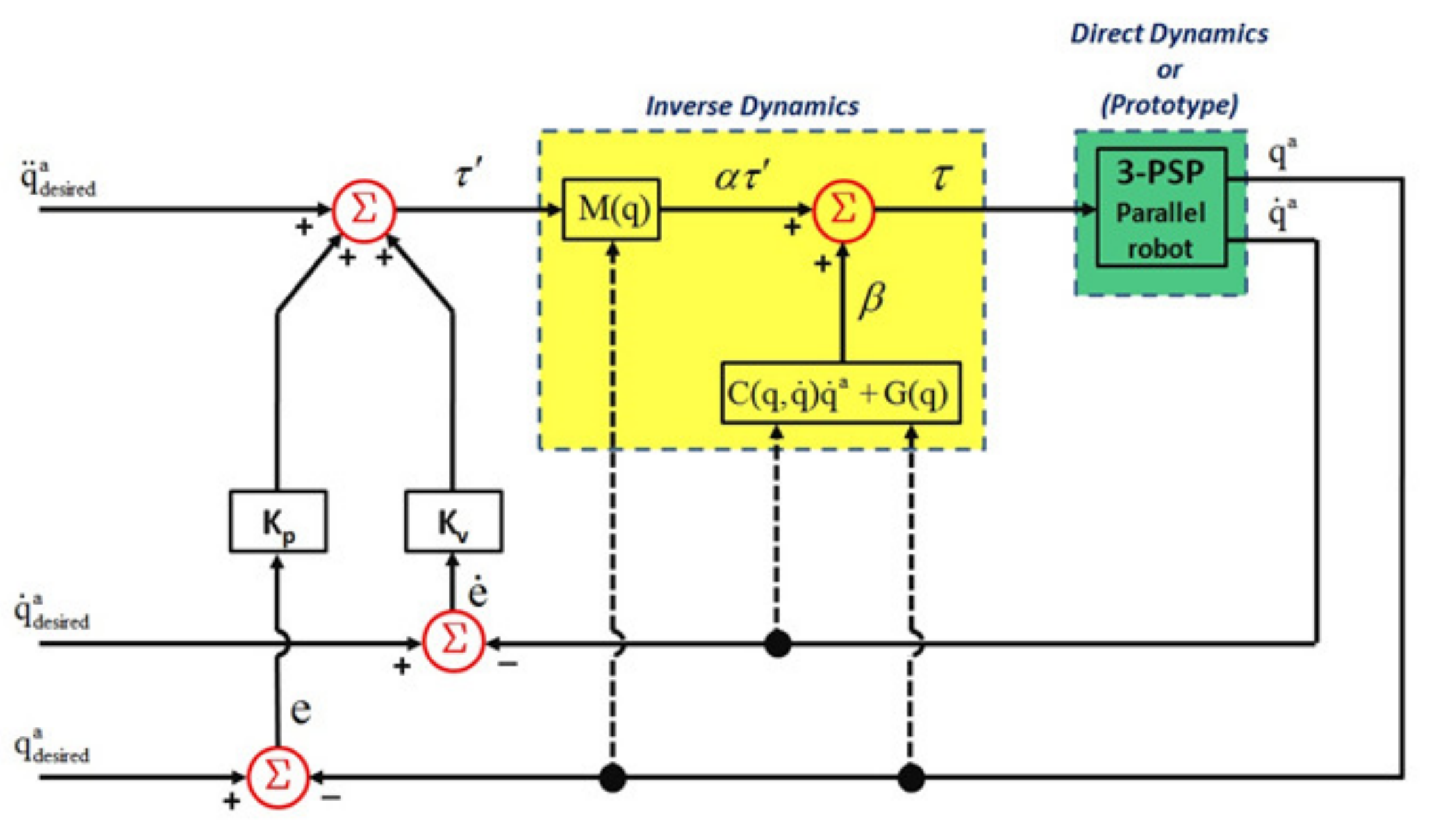}
  \end{center}
  \caption  {Block diagram for computed torque method}
  \label{fig44}
\end{figure}

\begin{figure}[htp]
  \begin{center} 
	\includegraphics[width=13cm, height=7cm]{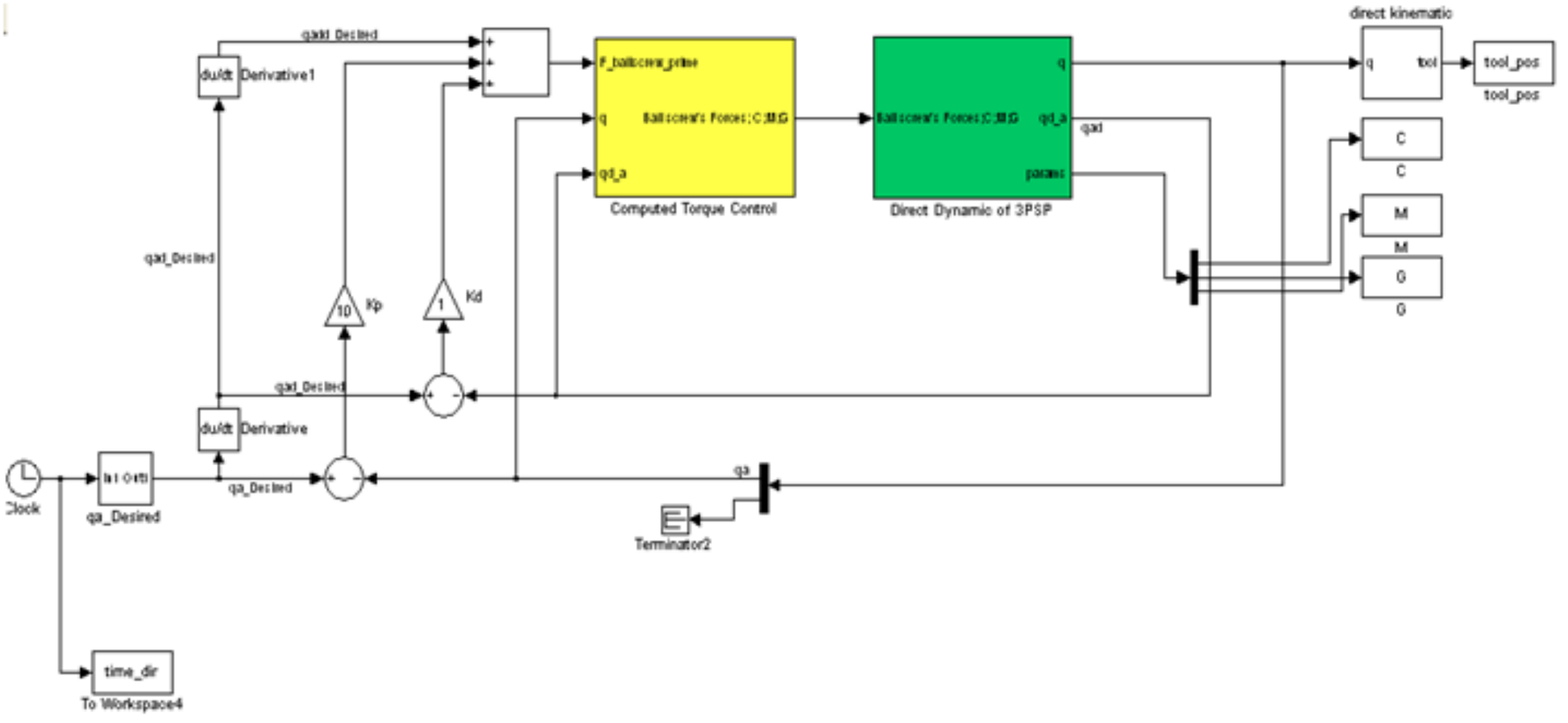}
  \end{center}
  \caption  {Simulink model for the design}
  \label{fig45}
\end{figure}

Figure \ref{fig44} shows the block diagram for the corresponding controller. The green block in the figure demonstrates the direct dynamics of the underlying robot. In other words this block represents the physical system. It takes in the torque vector and its output state is determined by the joint position and velocities, at each time instance. The yellow box, on the other hand, denotes the inverse dynamics of the manipulator. The dynamic matrices are found by solving equations given in previous chapters and then the terms $\alpha$, $\beta$ associated to the model-based portion of the controller are generated. Finally the servo portion serves as a PD controller to diminish the error between the real and desired position of the manipulator's joint variables. Note that there is also another unit to perform inverse kinematics calculations which is not portrayed in this figure. The role of the inverse kinematic unit is to translate the desired trajectories given in the Cartesian space into the joint space.
Figure \ref{fig45} depicts the Simulink model of the complete system. Like the previous plots, the green box in this figure indicates the manipulator which is equivalent to the direct dynamics of the manipulator modeled by the NOC method. This is elaborated with more details in the next figure.
 \begin{figure}[htp]
  \begin{center} 
	\includegraphics[width=13cm, height=7cm]{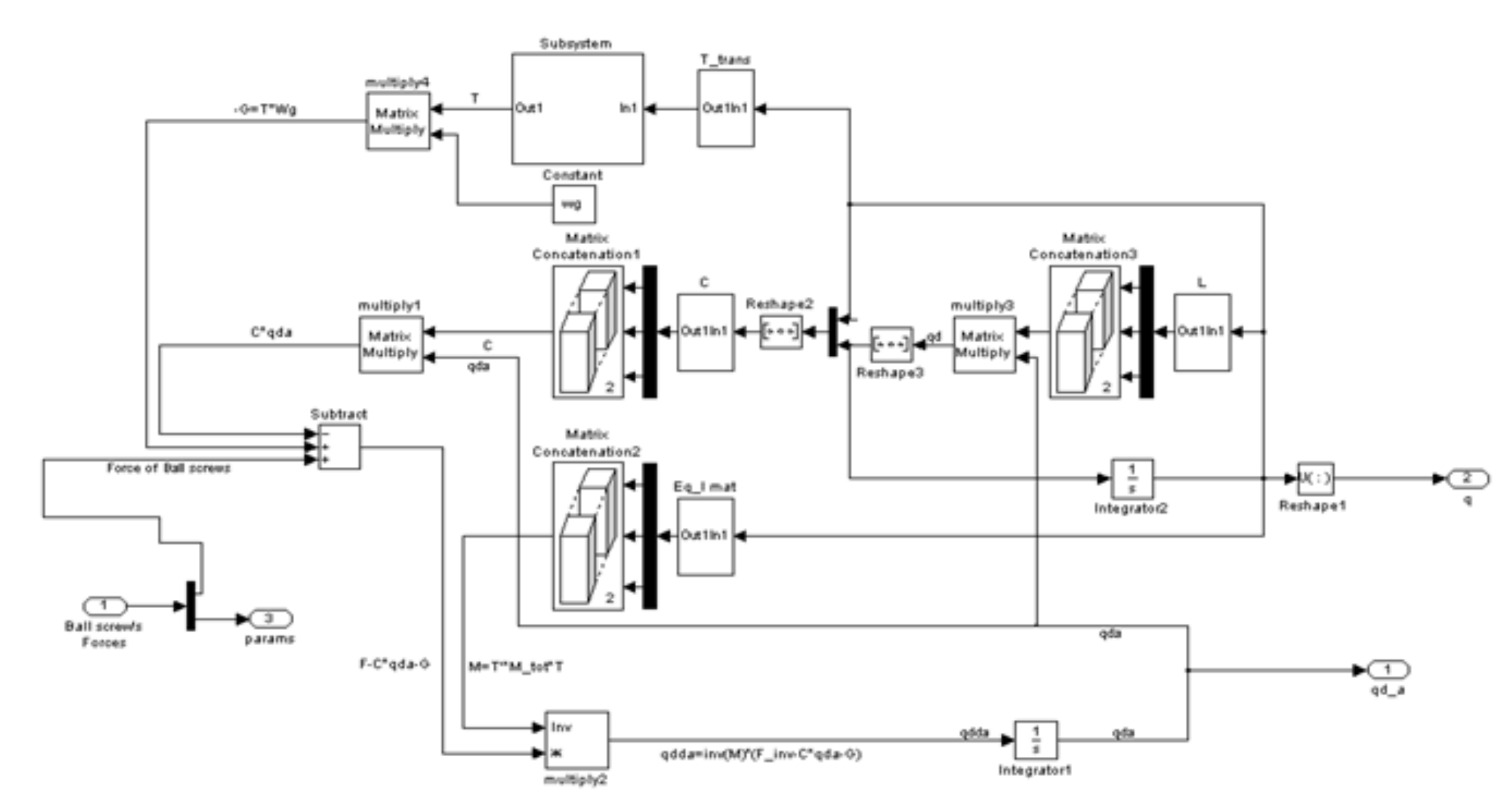}
  \end{center}
  \caption  {Simulink model for manipulator's direct dynamics}
  \label{fig46}
\end{figure}
\section{\textbf{Architecture of New Controller}}
\begin{figure}[htp]
  \begin{center} 
    \subfigure [ ]{\label{fig47-a}\includegraphics[width=13cm, height=7cm] {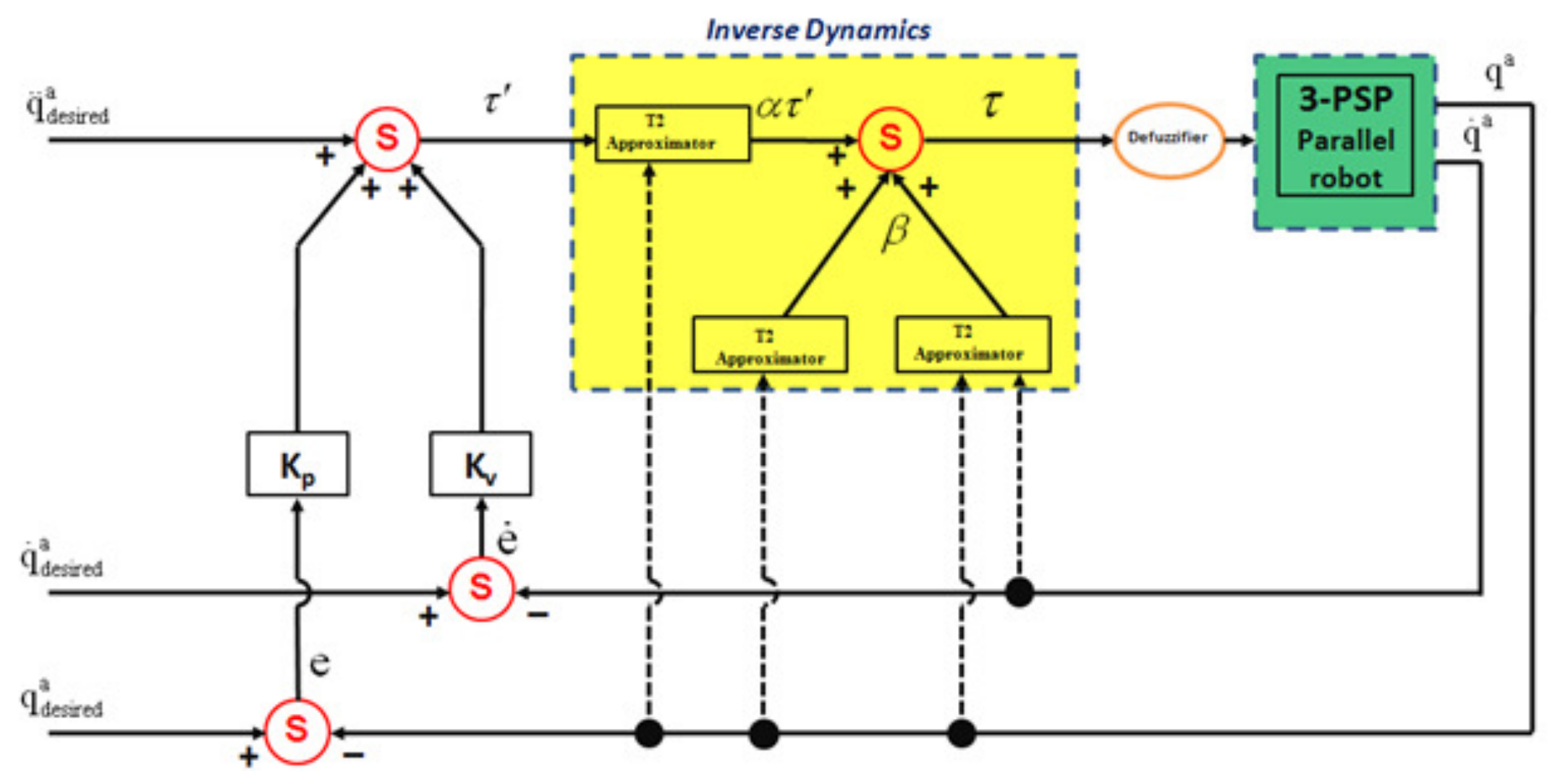}}\\
    \subfigure [ ]{\label{fig47-b}\includegraphics[width=10cm, height=5cm] {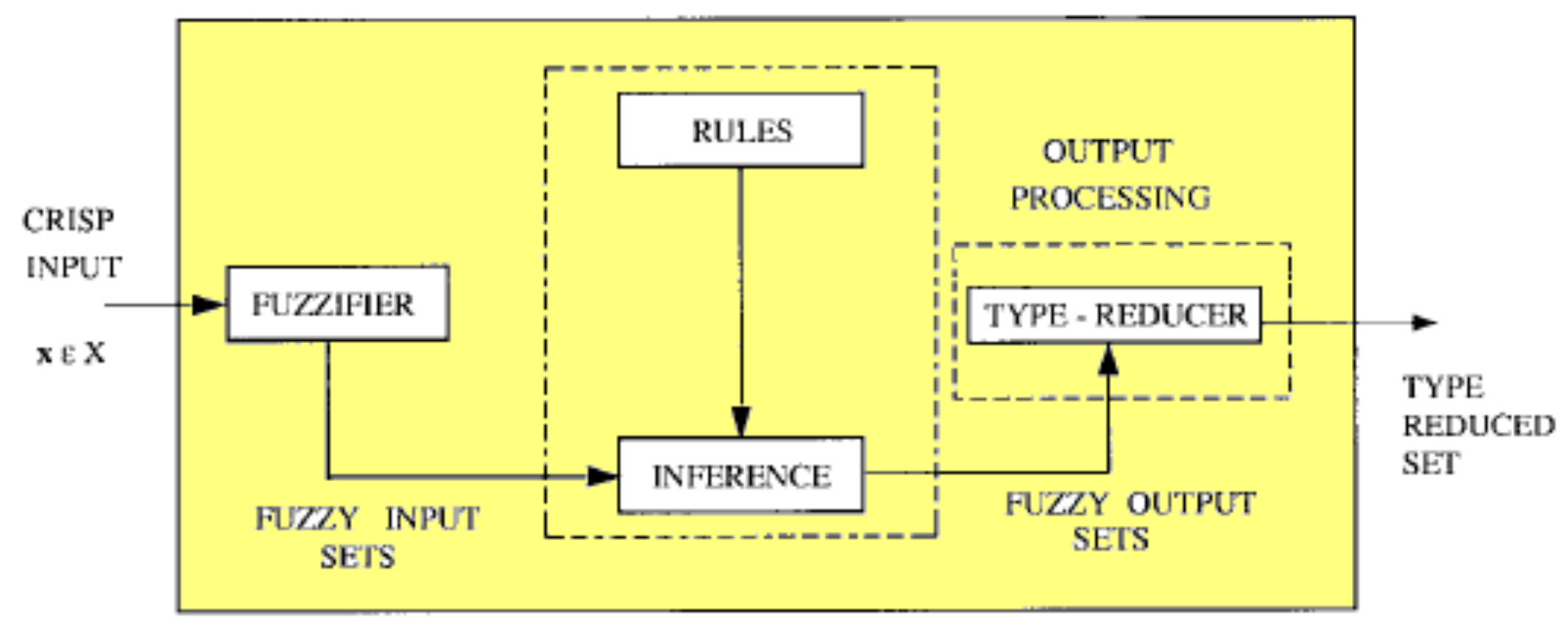}}\\
  \end{center}
  \caption{(a) The proposed model (b) Inside of the T2 approximators}
  \label{fig47}
\end{figure}

The problem with the aforementioned control method is that even if we manage to determine the foregoing matrices accurately we are yet to recalculate each matrix at each time step. This imposes a high amount of computational burden. Therefore to overcome this demanding task, we should find a closed form formula for each matrix in order not to perform intense computations that eventually leads to computation of these matrices again and again. One way as a remedy is to do approximation using classical fuzzy function approximators. The drawback is that we are not certain about he estimated parameters and as we will show, we cannot trust on the outputs of such systems. They might even entail unstable situations. Accordingly, a Type-II fuzzy approximator may probably perform better although it adds up more complexity. To sum up, Figures \ref{fig47-a}, \ref{fig47-b} illustrate the proposed architecture for control of dynamic model of the 3PSP robot.
A word under notation is that, for the special case of interval Type-II fuzzy logic systems (IT2FLS) it can be justified (by Theorem \ref{affineIT1comb}) that the defuzzifier units can be moved back to the approximator units and hence we don't need to manipulate fuzzy numbers to later defuzzify the results. This considerably reduces the computational burden. However it should be emphasized that the simplification is only applicable for the IT2FLS and not any other type of T2FLS (e.x. Gaussian T2FLSs).
\begin{figure}[htp]
  \begin{center} 
    \subfigure [ ]{\label{fig48-a}\includegraphics[width=13cm, height=7cm] {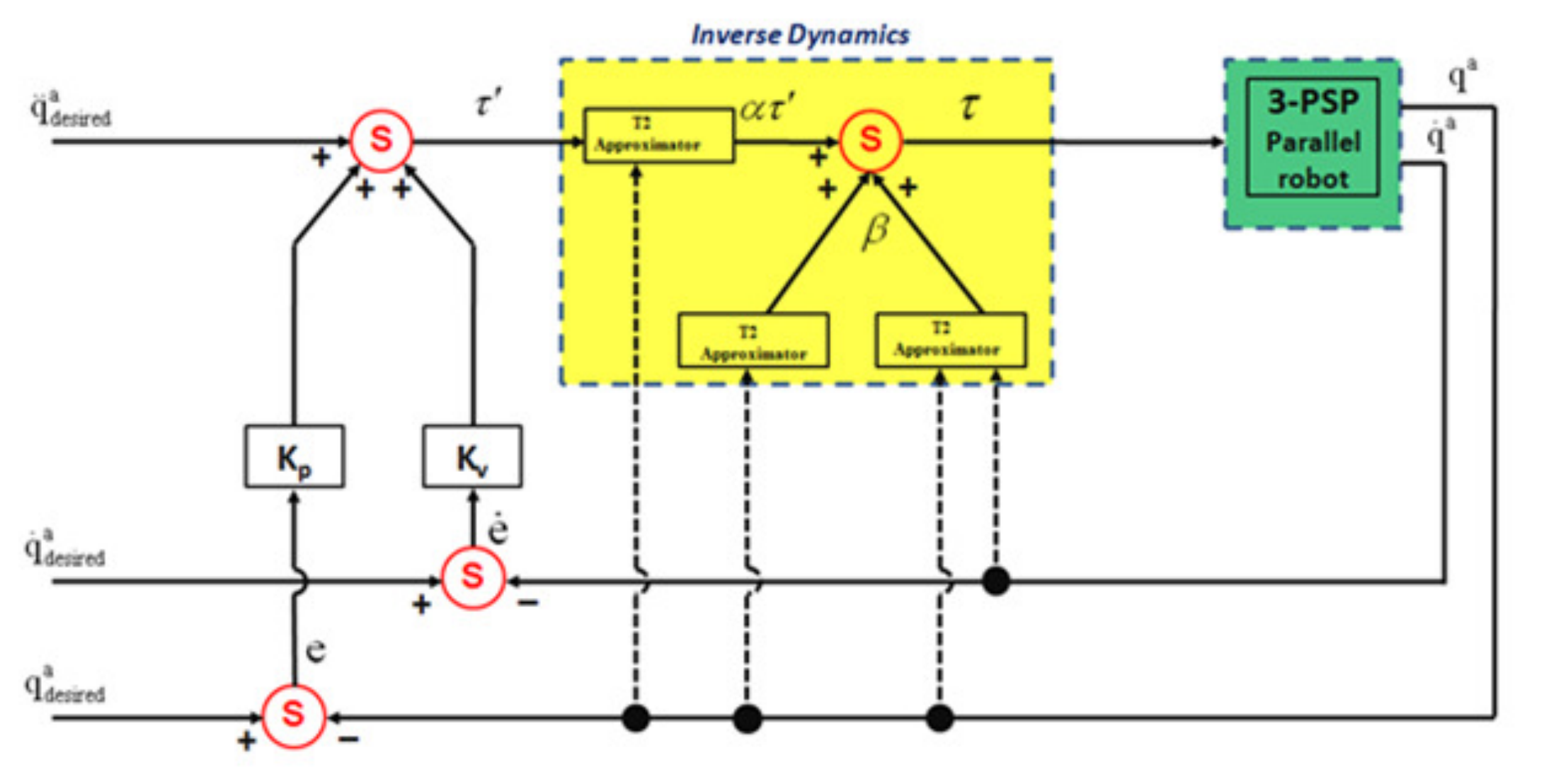}}\\
    \subfigure [ ]{\label{fig48-b}\includegraphics[width=13cm, height=7cm] {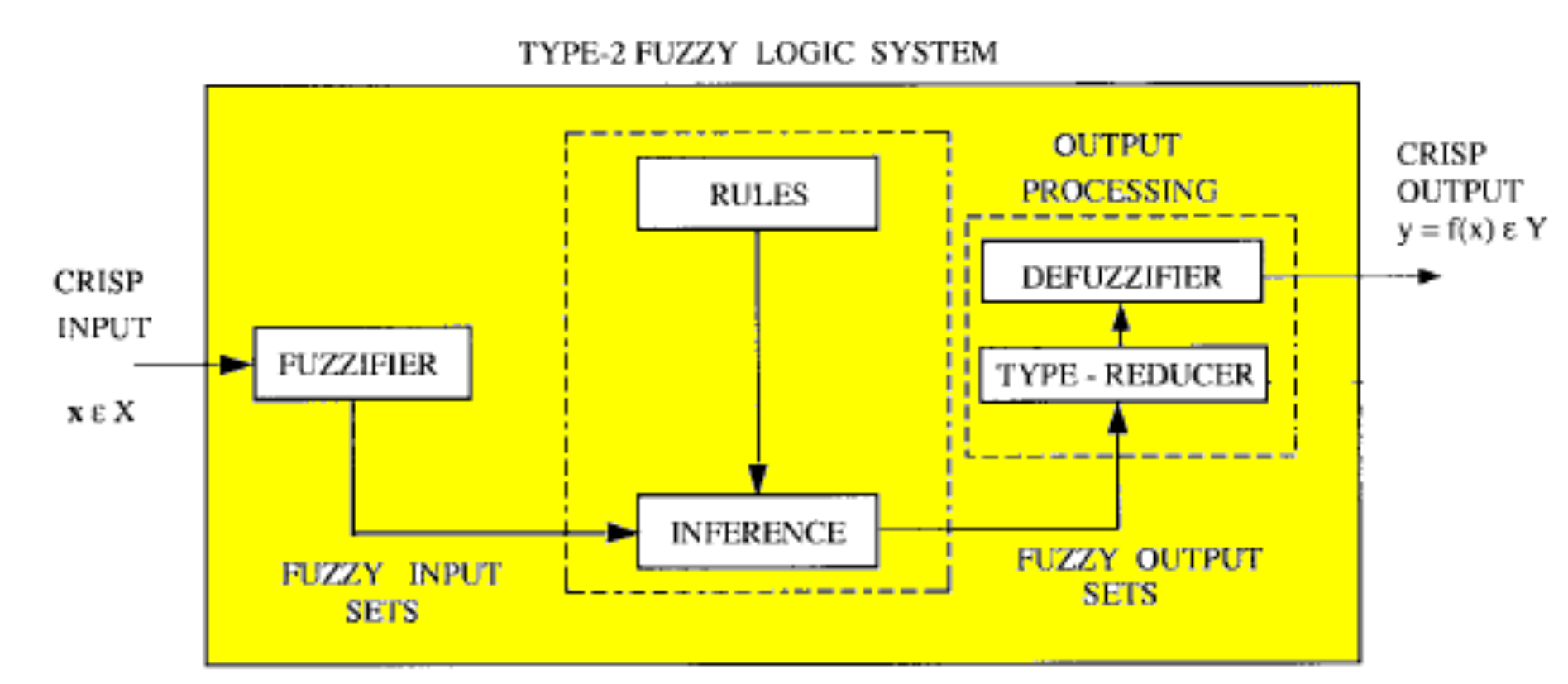}}\\
  \end{center}
  \caption{The proposed model for IT2FL MFs (b) inside of the T2 approximators}
  \label{fig48}
\end{figure} 
One of the main novelties of this work is the new architecture of the proposed controller which is based on Type-II fuzzy logic and manipulation of Type-I fuzzy numbers. These numbers are in fact the type reduced outputs of Type-II fuzzy systems. More specifically, we don't perform the defuzzification until the last step, right behind the manipulator block. In other words, each of the elements of the matrices $M$, $C$ and $G$ pertaining to the dynamics are estimated by fuzzy numbers (i.e. T1 fuzzy numbers) via Type-II fuzzy estimators. Using fuzzy algebra introduced in chapter 1, we perform algebraic operations. The results of these operations lead to fuzzy $\alpha$ and $\beta$ . These fuzzy numbers are then manipulated together to come to the last fuzzy number which is then defuzzified to produce the torque required for the physical (or simulated) system.

In this work, the type reducers are all of Center of Sets types and the Type-II fuzzy variables are of interval Type-II Gaussian primary membership functions with uncertain means (see chapter 1 for details). The mean variation might be roughly dependent on the noise level (signal to noise ratio). That is, having a sufficient knowledge of the SNR, the membership functions are created. To generate the required rules the following procedure is carried out:\\
At first step we have to produce a desired trajectory. This is done by help of the inverse kinematic unit. In so doing, the trajectory that we wish the tool tip to follow is created, after which the samples are given to the inverse kinematic unit whose job is to translate the given values into the joint space. The result is a set of vectors each containing 3 elements associated to the ball-screw lengths (the independent actuated joint variables; consult chapter 3 for details). Having found the desired set of joint variables, the non-fuzzy CTC controller is adopted to track the given points. Note that the CTC is not implementable in real-time. This becomes important if we wish to realize the controller in a real world environment. During the course of simulation with the pure CTC controller, we record the joints angles and rotational speeds (in fact the ball-screws' lengths and their speed). Notice that each of the parameters that we are going to estimate are not decoupled, in the sense that they are relevant to speed and position of the other two actuators, as well. At the first look, this amounts to a total of six variables (3 for speed and 3 for position), however, as we will soon show, four is enough. For generating the rules we made use of a well-known trajectory called the Helix.

Taking a closer look into the problem at hand reveals that the fuzzy approximators to be designed should not directly depend on these 6 variables. It looks as if there is something redundant. This becomes more evident by considering table \ref{tbl5.1}. This table shows the manipulator parameters (i.e. $C$, $M$, $G$) for 3 different manipulator configurations. As easily implied from the table, although the amount of ball-screw speeds and lengths differ considerably, the results are the same. This declares that there is certainly redundant data inherent in the foregoing set of variables which should be omitted from the fuzzy systems. Let's have a look to the foregoing variables from a different perspective to disclose the secrete behind the scene. In this respect, table \ref{tbl5.2} is rather similar to the previous with the exception that instead of the speeds and positions of the ball-screws, this time the difference values between them are denoted. Form this table it's clear that all the specifications associated with the noted configurations are the same relative to each other. That demonstrates that the amounts of the elements of the CTC matrices are not directly dependent to the 6 previous variables, rather, they are a function of their differences.

\begin{table}
\centering
\caption[Matrices $M$, $C$, $G$ for 3 different configurations, \dots]{Matrices $M$, $C$, $G$ for 3 different configurations, the "underline" sign indicates the actuated variable}
\begin{tabular}{p{4.4cm} p{3cm} p{3cm} p{3cm}}
\centering
& $1^{st}$ configuration & $2^{nd}$ configuration & $3^{rd}$ configuration\\
\hline\hline
$1^{st}$ ball-screw Length, $\underbar q_1$ & 0.1	&0.15	&0.05\\ \hline
$1^{st}$ ball-screw Speed, $\dot{\underbar{q}}_1$ & 0.15	& 0.25 & 0.25\\ \hline
$2^{nd}$ ball-screw Length, $\underbar q_2$ & 0.2&0.25 &0.15\\ \hline 
$2^{nd}$ ball-screw Speed, $\dot{\underbar{q}}_2$ & 0.12&0.22	&0.22\\ \hline
$3^{rd}$ ball-screw Length, $\underbar q_3$ & 0.05&0.15 &0.05\\ \hline 
$3^{rd}$ ball-screw Speed, $\dot{\underbar{q}}_3$&0.1&0.2	&0.2\\ \hline
& & & \\
M & \multicolumn{3}{c}{$\left[ \begin{array}{l l l }
  10.0161 & -0.5288 & -0.6026\\
  -0.528&9.2516&0.0091\\
  -0.6026 & 0.0091 & 9.4782\\
  \end{array}  \right]$} \\&&&\\ \hline 
  & & &\\
C & \multicolumn{3}{c}{$\left[ \begin{array}{l l l }
  10.0159 & 0.0004 & -0.0163\\
  0.0481 &-0.0247 &-0.0234\\
  0.0346 &0.0124& -0.0560\\
  \end{array}  \right]$}\\ \hline
&&&\\
G & \multicolumn{3}{c}{$\left[ \begin{array}{l } 
87.1237 \\85.6256 \\87.1237\\
  \end{array}  \right]$}
\label{tbl5.1}
\end{tabular}
\end{table}
Now consider that $\Delta \underbar q_{12}$, $\Delta \underbar q_{13}$, $\Delta \underbar q_{23}$ ($\Delta \dot{ \underbar q}_{12}$, $\Delta \dot{ \underbar q}_{13}$, $\Delta \dot{  \underbar q}_{23}$ ) are not independent of each other as $\Delta \underbar q_{13}-\Delta \underbar q_{12}=\Delta \underbar q_{23}$ ($\Delta \dot{ \underbar q}_{13}-\Delta \dot{\underbar q}_{12}=\Delta \dot{\underbar q}_{23}$) and among each three-tuple one is extra. In summary, we come up with the fact that, as opposed to intuition, the best set of inputs for the fuzzy approximators is $\{ \Delta \underbar{q} _{12} ,\Delta\underbar{q} _{13} ,\Delta \dot{\underbar{q}} _{12} ,\Delta \dot{\underbar{q}} _{13} \} _{}$ and not $\{ \underbar{q} _1 ,\underbar{q} _2 ,\underbar{q} _3 ,\dot{ \underbar{q}} _1 ,\dot{ \underbar{q}} _2 ,\dot{\underbar{q}}_3\}$. This not only increases the performance of the fuzzy systems but also considerably decreases the computational complexity of the resulting fuzzy systems due to huge saving in number of rules (as we are always doomed to the curse of dimensionality). This means that a linear decrease in the number of input variables amounts to an exponential computational saving.

We can also look into the point from a different perspective. Consider the aforementioned mixed degree of freedom platform. It has totally 3 D.O.F which is comprised of one translational freedom (along the $Z$-axis) and two rotational movements (about the $X$ and $Y$-axes). Clearly the system's dynamic does not incur any change if the system state merely undergoes a translation along the $Z$ axis as opposed to the rotational movements. Hence, the dynamics of the system only depends on two factors and each factor itself has a position and a speed which adds up to 4 independent variables.
  
\begin{figure}[htp]
  \begin{center} 
    \subfigure [ ]{\label{fig49-a}\includegraphics[width=5.5cm, height=5cm]{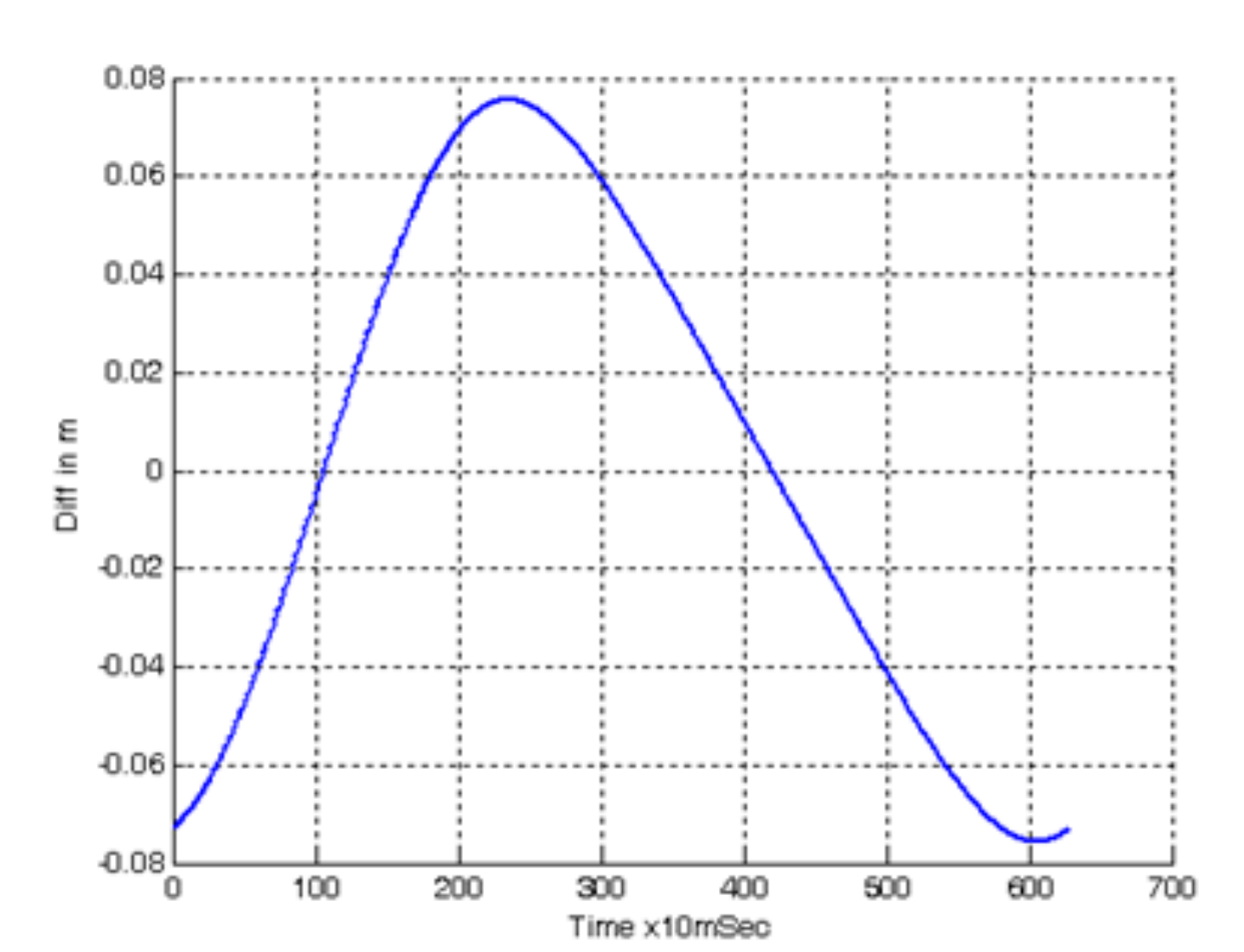}} 
    \subfigure [ ]{\label{fig49-b}\includegraphics[width=5.5cm, height=5cm]{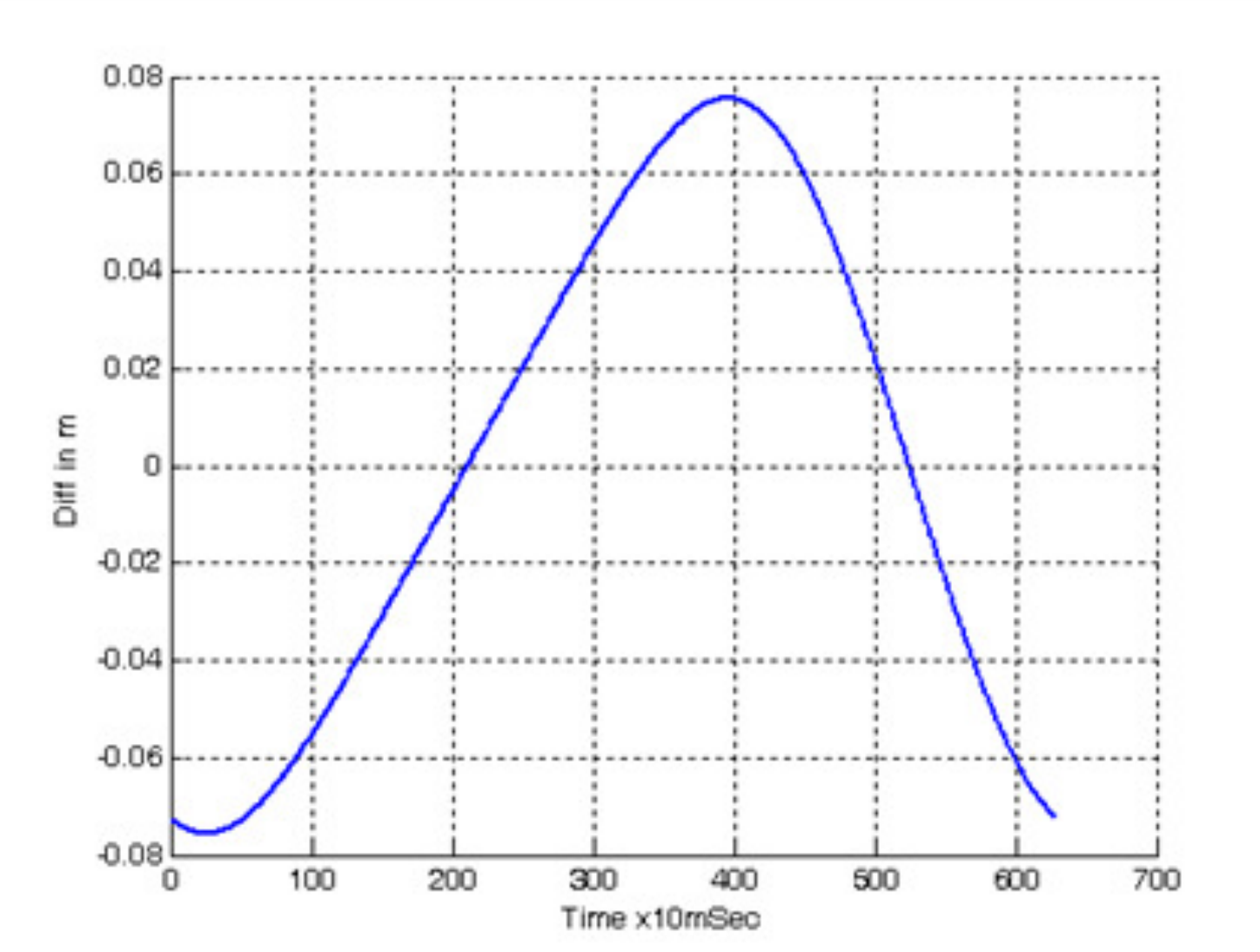}}\\
    \subfigure [ ]{\label{fig49-c}\includegraphics[width=5.5cm, height=5cm]{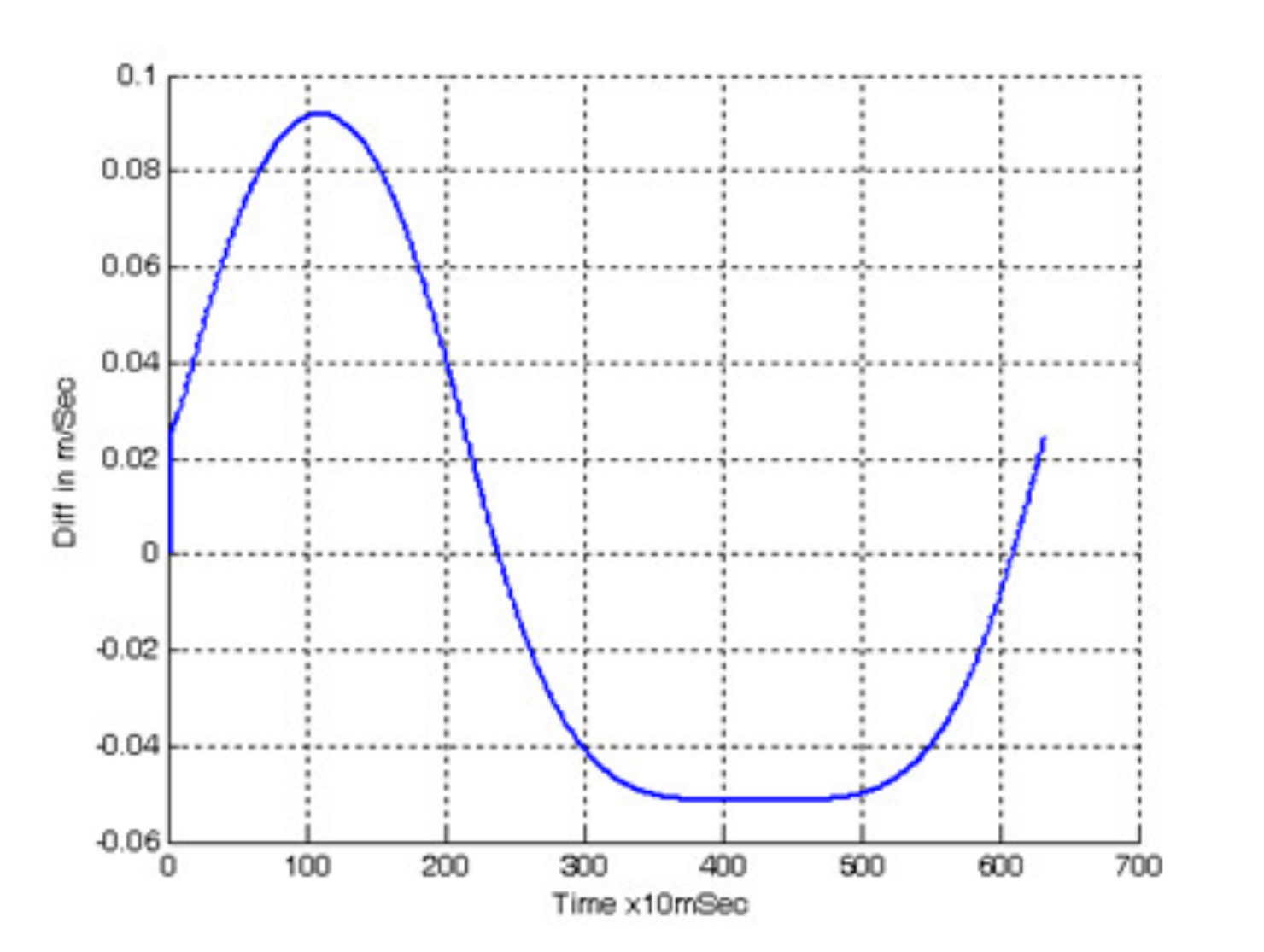}}
    \subfigure [ ]{\label{fig49-d}\includegraphics[width=5.5cm, height=5cm]{Images/deltaqdot12}}
  \end{center}
  \caption{Variations of (a) $\Delta q_{12}$ (b) $\Delta q_{13}$ (c) $\Delta \dot q_{12}$ (d) $\Delta \dot q_{13}$ }
  \label{fig49}
\end{figure}

\begin{table}
\centering
\caption{Differences between the 6 independent variables}
\begin{tabular}{p{4.2cm} p{3cm} p{3cm} p{3cm}}
& $1^{st}$ configuration & $2^{nd}$ configuration & $3^{rd}$ configuration\\
\hline\hline
Difference between $1^{st}$ and $2^{nd}$ ball-screw Lengths $\Delta \underbar q_{12}$ & -0.1 & -0.1 &-0.1 \\ \hline
Difference between $1^{st}$ and $2^{nd}$ ball-screw Speeds  $\Delta \dot {\underbar q}_{12}$ & 0.03 & 0.03 & 0.03 \\ \hline
Difference between $1^{st}$ and $3^{rd}$ ball-screw Lengths $\Delta \underbar q_{13}$ & 0.05 & 0.05&0.05\\ \hline
Difference between $1^{st}$ and $3^{rd}$ ball-screw Speeds  $\Delta \dot {\underbar q}_{13}$ & 0.05 & 0.05&0.05 \\ \hline
Difference between $2^{nd}$ and $3^{rd}$ ball-screw Lengths $\Delta \underbar q_{23}$ & 0.15 & 0.15&0.15\\ \hline
Difference between $2^{nd}$ and $3^{rd}$ ball-screw Speeds  $\Delta \dot {\underbar q}_{23}$ & 0.02 & 0.02&0.02 \\ \hline

\label{tbl5.2}
\end{tabular}
\end{table} 
The second step is to find the upper and lower bounds for $\Delta \underbar q_{12}$, $\Delta \underbar q_{13}$, $\Delta \dot {\underbar q}_{12}$, $\Delta \dot {\underbar q}_{13}$ and extract some samples from the trajectory already traversed such that they both cover all the feature space and be distributed uniformly. Figures \ref{fig49-a}-\ref{fig49-d}, depict the variations of each of the four introduced variables with respect to the time when a Helix trajectory is traversed.

As the last step, we follow the \emph{table lookup} procedure \cite{Wan97}, to find the rules. 
\section{\textbf{Type-I Equivalent System}}
\begin{figure}[htp]
  \begin{center} 
	\includegraphics[width=12.5cm, height=7cm]{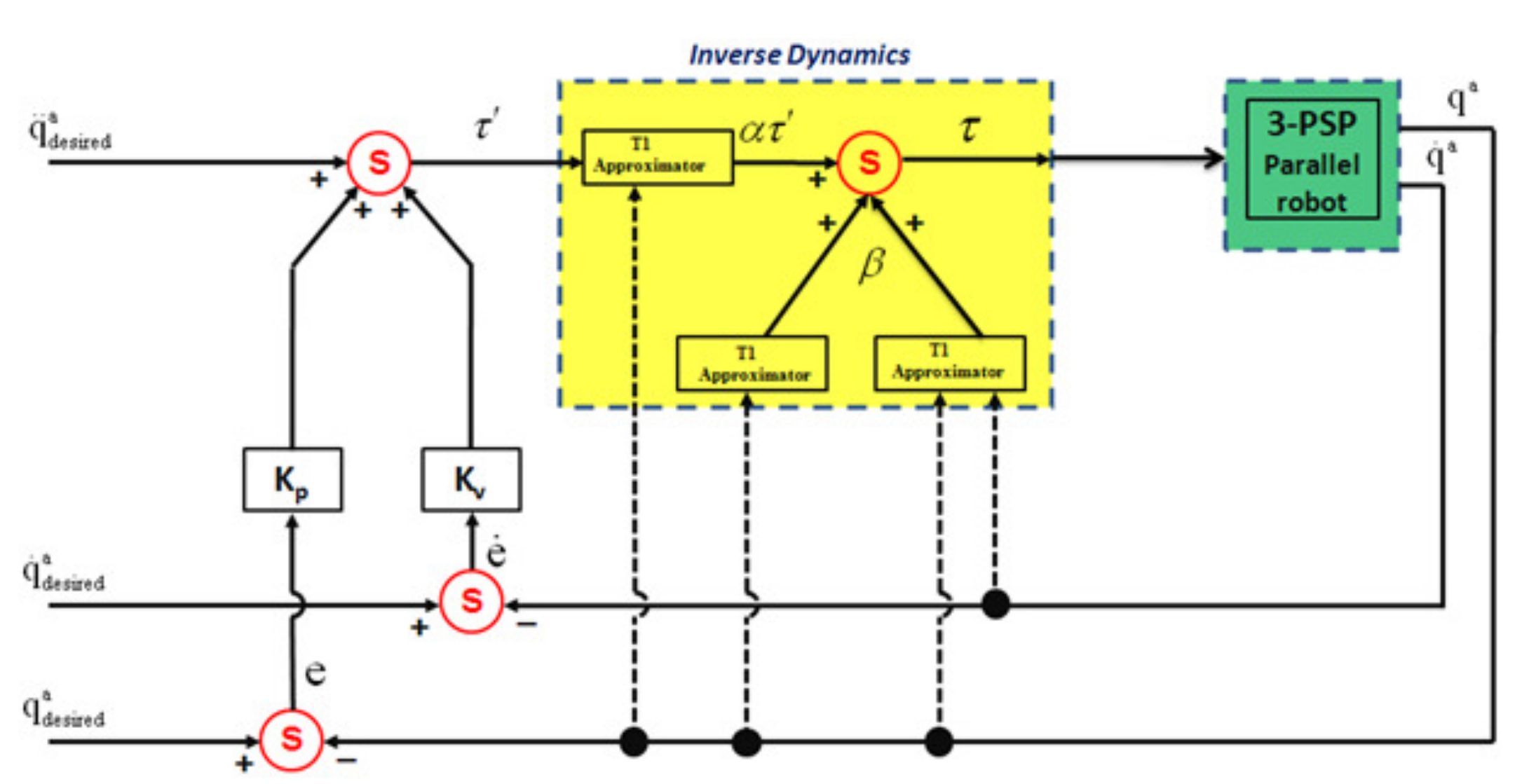}
  \end{center}
  \caption {Type-I downgraded model}
  \label{fig50}
\end{figure}

\begin{figure}[htp]
  \begin{center} 
    \subfigure [ ]{\label{fig51-a}\includegraphics[width=13cm, height=8cm] {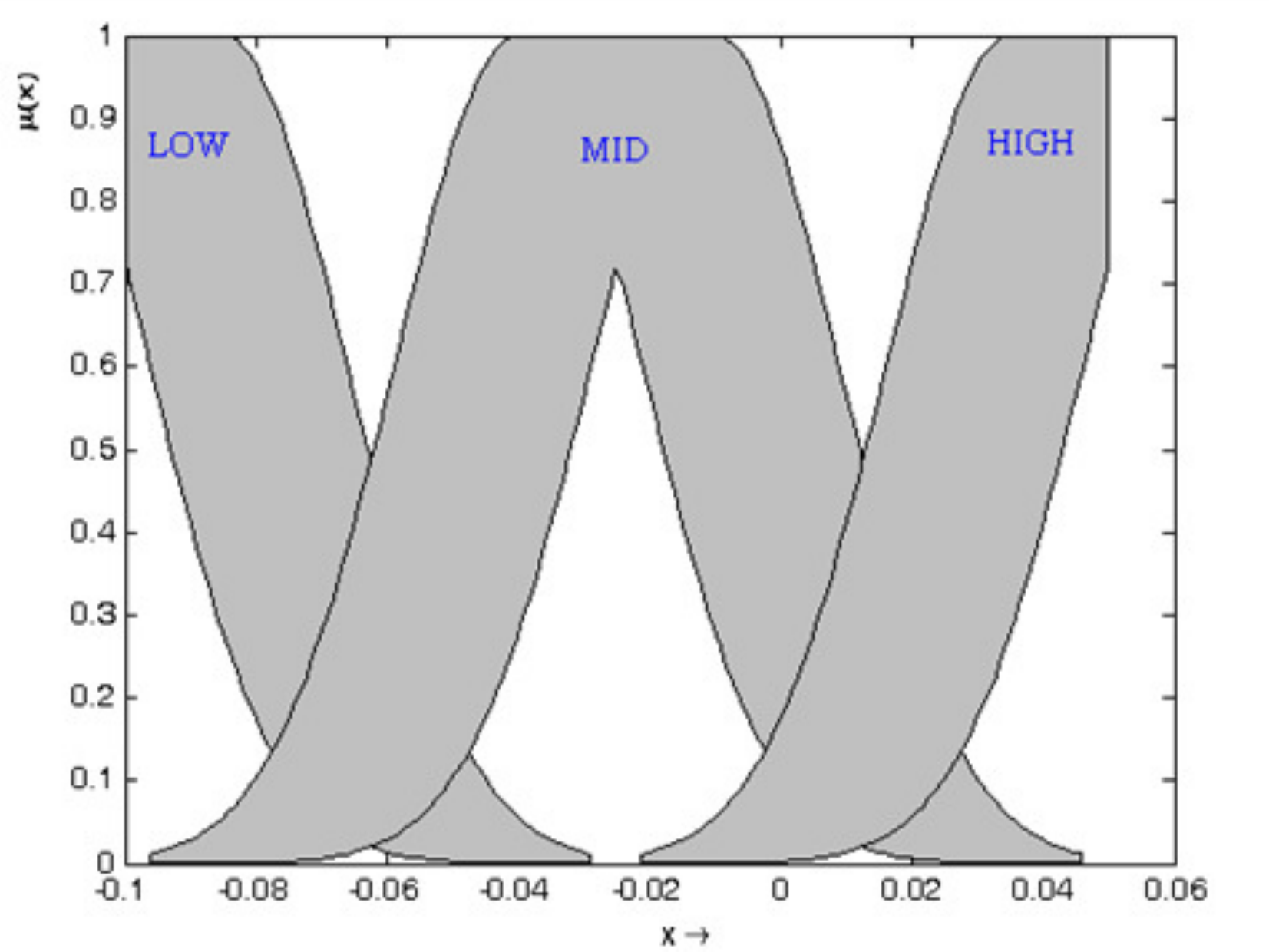}}\\
    \subfigure [ ]{\label{fig51-b}\includegraphics[width=13cm, height=8cm] {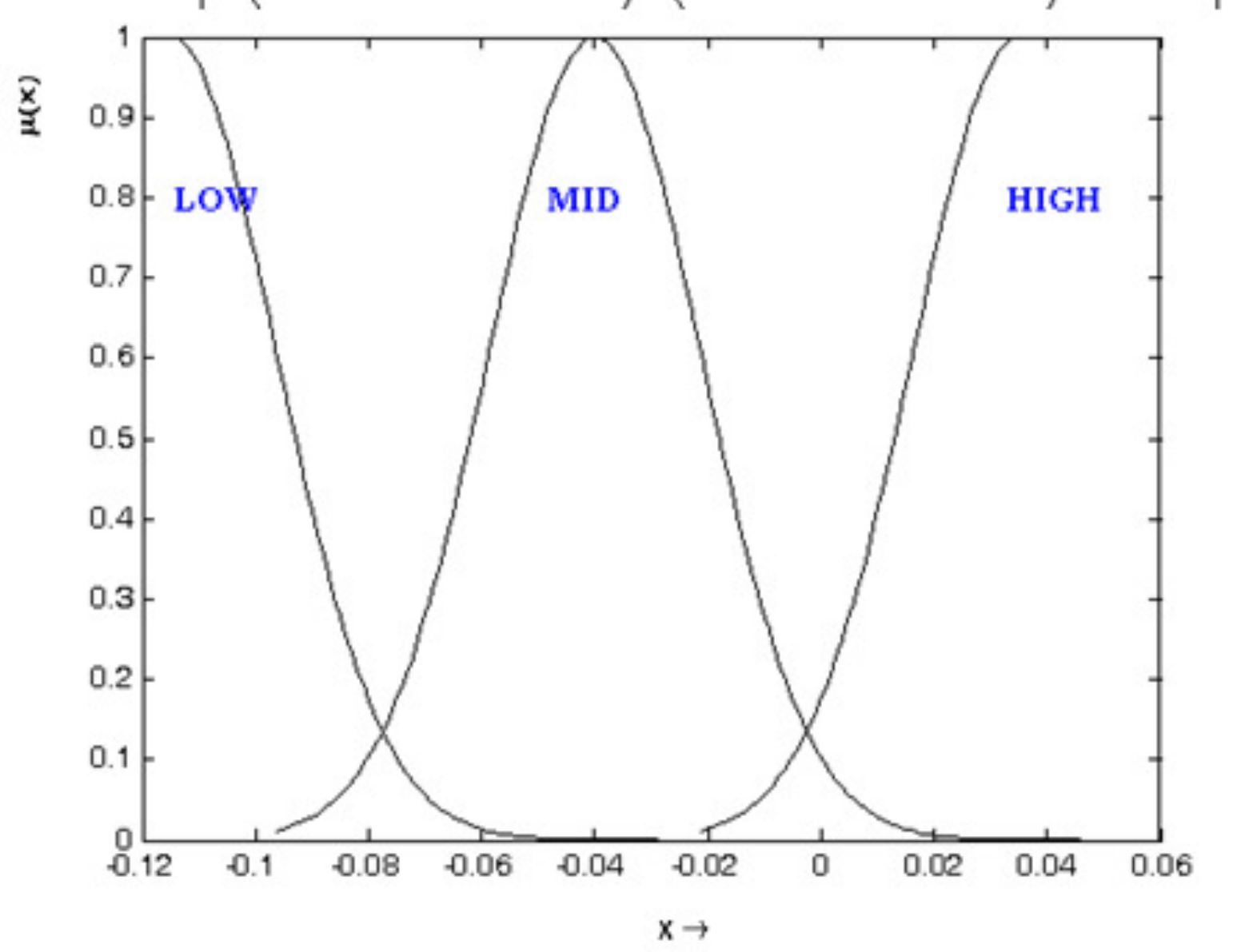}}\\
  \end{center}
  \caption{(a) The FOU for membership functions associated to    (b) the T1 reduced membership functions corresponding to (a)}
  \label{fig51}
\end{figure}

Like a wide variety of papers published yet, in this work our main focus is to consider the behaviour of the proposed controller for different signal to noise ratios. Accordingly, to compare the results, the best counterpart is the Type-I reduced controller. That is to say, we are interested to see how the new types of fuzzy systems can do the job with better performances for us. Usually this is done by downgrading the Type-II fuzzy system to a Type-I simply through reducing the Type-II membership functions to Type-I ones. In our method this is easily done by setting the mean variations to zero (this implies that there is not any uncertainty available in description of fuzzy variables). The noteworthy point 
is that in this thesis, we only consider the effects of \emph{numerical} uncertainty. Note also that, the 
\emph{numerical} uncertainty leads to uncertain inputs; contrary to \emph{linguistic} uncertainty which 
leads to uncertain rules. Figure \ref{fig50} depicts the downgraded architecture associated to its Type-II counter part (Fig. \ref{fig47}). Since the output of a Type-I approximator is a crisp number there is no need to any kind of algebraic operation on Type-I fuzzy numbers.
Finally, Figure \ref{fig51-a} depicts all the Type-II MFs associated with $\Delta \dot q_{13}$ (to give the reader a sense of what's going on) when 3 membership functions (\emph{Low, Mid} and \emph{High}) are defined over each variable in the presence of a 10\% noise in the system parameters and Figure \ref{fig51-b} depicts the corresponding downgraded Type-I fuzzy sets for the same variable (more on this in later chapter). Notice that for the former (Type-II membership functions) the footprint of uncertainty (FOU) is plotted.
In the next chapter we will show how the structure of fuzzy sets similar to Figure \ref{fig51-a} enhances the controllers with the ability to handle noises especially for high signal to noise ratios.

%% file: chapter6.tex
\chapter{The Simulation Results}
\section{\textbf{Preface}}
In the previous chapter we introduced the architecture of the proposed controller. We mentioned that the control system was a variant of a feedback linearizing controller with a flavour of Type-II fuzzy logic. We also alluded that due to vast computational burden of the CTC,
it is not implementable in real-time for complex dynamical systems. Thus, this type of controller is not suitable for the application of this thesis. In this chapter we show
that the presented controller not only handles the uncertainty in the system but also enhances the time taken for one loop execution (and hence making it appropriate for real-time realization).

In this chapter, we perform different trials and compare the results of a PD controller, a CTC controller, a Type-I fuzzy based CTC controller and a Type-II fuzzy based CTC controller with each other both in terms of 
uncertainty and execution time and interpret the results of the simulation scenarios. The first two sections give the results of the aforementioned classical controllers (PD \& CTC) and the last two fall in the category of non-classic controllers. The Matlab and C++ codes along with the stored simulation results are all included in the CD accompanied by this thesis.
Table \ref{tbl6.1} demonstrates all parameters set up for the rest of simulations for each method.
\begin{table}
\centering
\caption{Different parameters for each trial}
\begin{tabular}{p{4cm} p{1.5cm} p{1.5cm} p{2.5cm} p{1.5cm}}
\hline\hline
PD &1000 &100&-&-\\\hline
CTC&10&1&-&-\\ \hline
T1 fuzzy CTC&10&1&10, 15, 20, $\infty$&2\\\hline
T2 fuzzy CTC&10&1&10, 15, 20, $\infty$&2\\\hline
\label{tbl6.1}
\end{tabular}
\end{table} 
\pagebreak
\section{\textbf{PD}}
As a widespread classical controller of the PID family, we have implemented a PD controller according to the feedback gains demonstrated in table \ref{tbl6.1}. The gain with this type of controller is the ease of implementation and the fact that it is realizable in real-time and hence is useful for online applications. The pain, on the other hand, is that the controllers of this family are only suitable for linear plants; a problem which becomes more severe when the degree of 
non-linearity increases or when the robot is designed to work in high speeds or precisions. This is also easily understood by consulting table \ref{tbl6.1}. According to the table, the proportionate and derivative gains for the PD controller is by far greater than the ones associated to the other three or the controller becomes unstable otherwise. It should be underlined that
 in this we don't take into account the \emph{integral} gain to make the PD system more comparative to the 
 CTC controller presented earlier. Note that we only consider 2 MFs for each variable. The results for higher numbers of MFs do not offer significant improvements and hence disregarded in this thesis.

In all subsequent figures, the solid and dashed curves indicate the real and desired trajectories respectively.
The \emph{sum of squared} errors corresponding to all trials are gathered and demonstrated in table \ref{tbl6.2}. 

\begin{figure}[htp]
  \begin{center} 
    \subfigure [ ]{\label{fig52-a}\includegraphics[width=6cm, height=5cm] {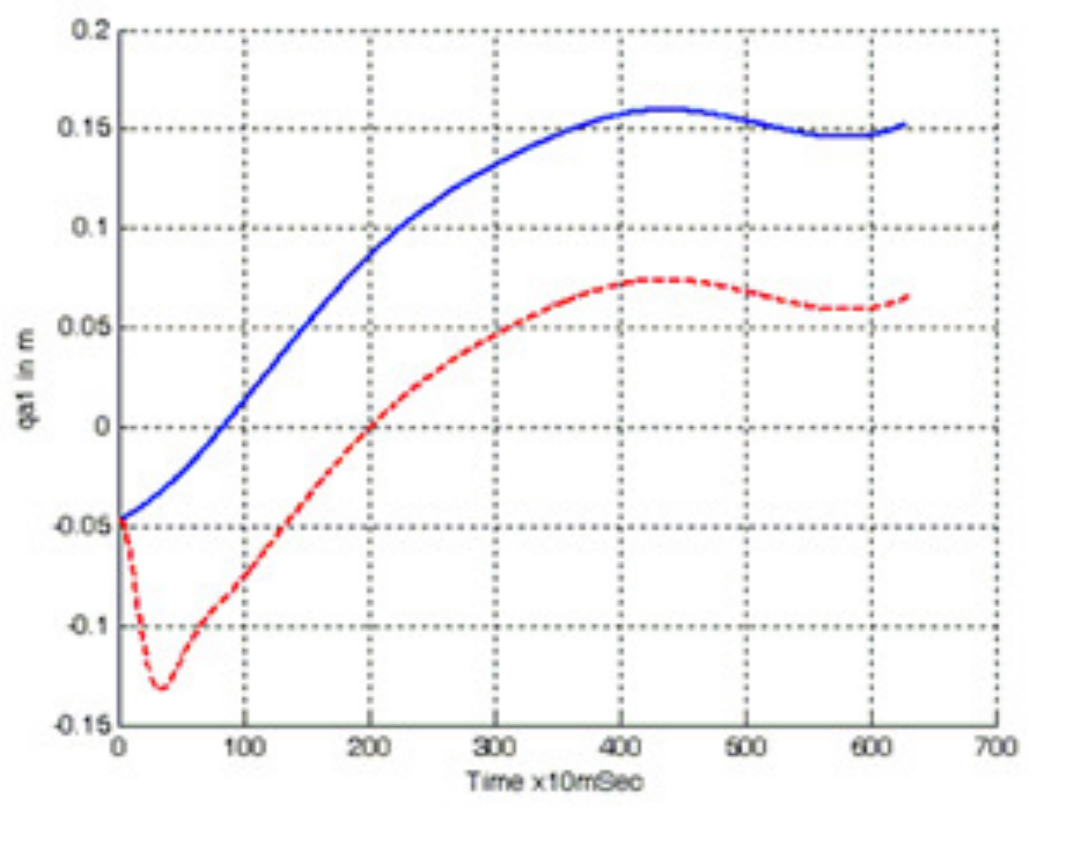}}
    \subfigure [ ]{\label{fig52-b}\includegraphics[width=6cm, height=5cm] {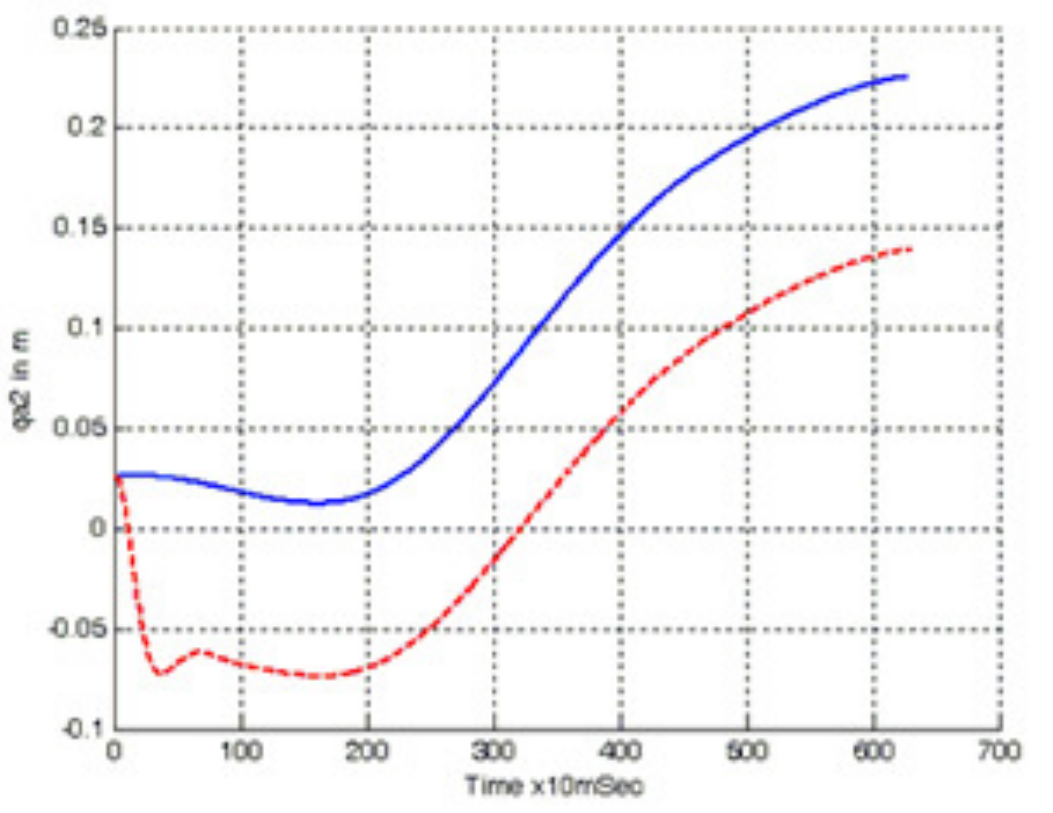}}\\
    \subfigure [ ]{\label{fig52-c}\includegraphics[width=6cm, height=5cm] {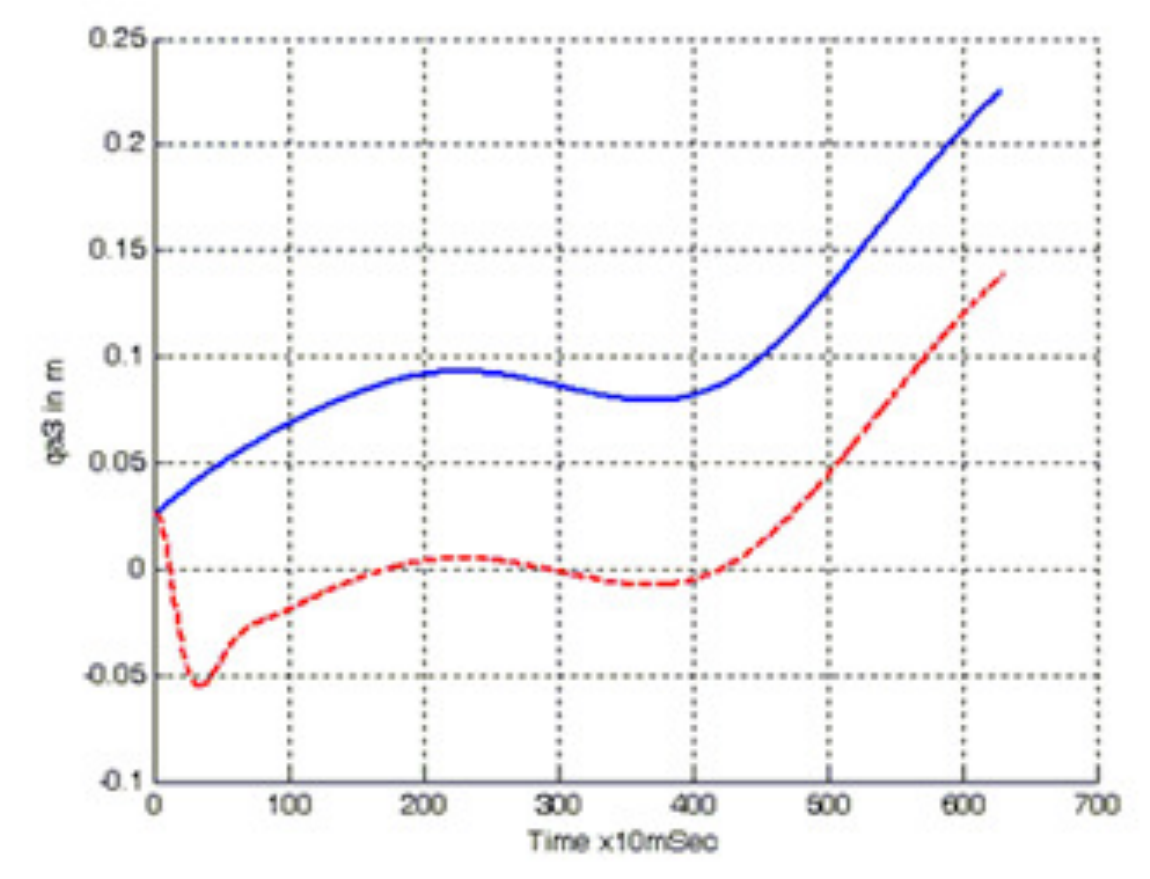}}
    \subfigure [ ]{\label{fig52-d}\includegraphics[width=6cm, height=5cm] {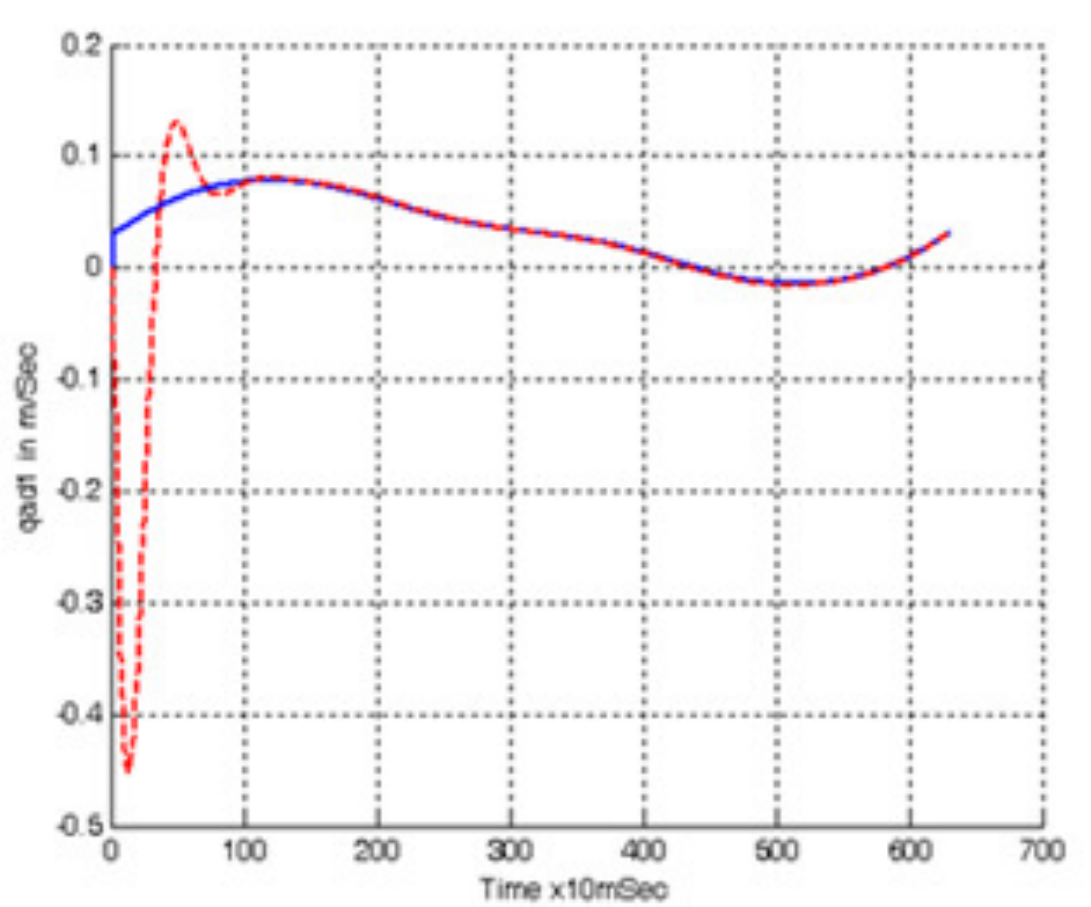}}\\
  \end{center}
   \label{fig52}
  
\end{figure}

\begin{figure}[htp]
  \begin{center} 
    \subfigure [ ]{\label{fig52-e}\includegraphics[width=6cm, height=5cm] {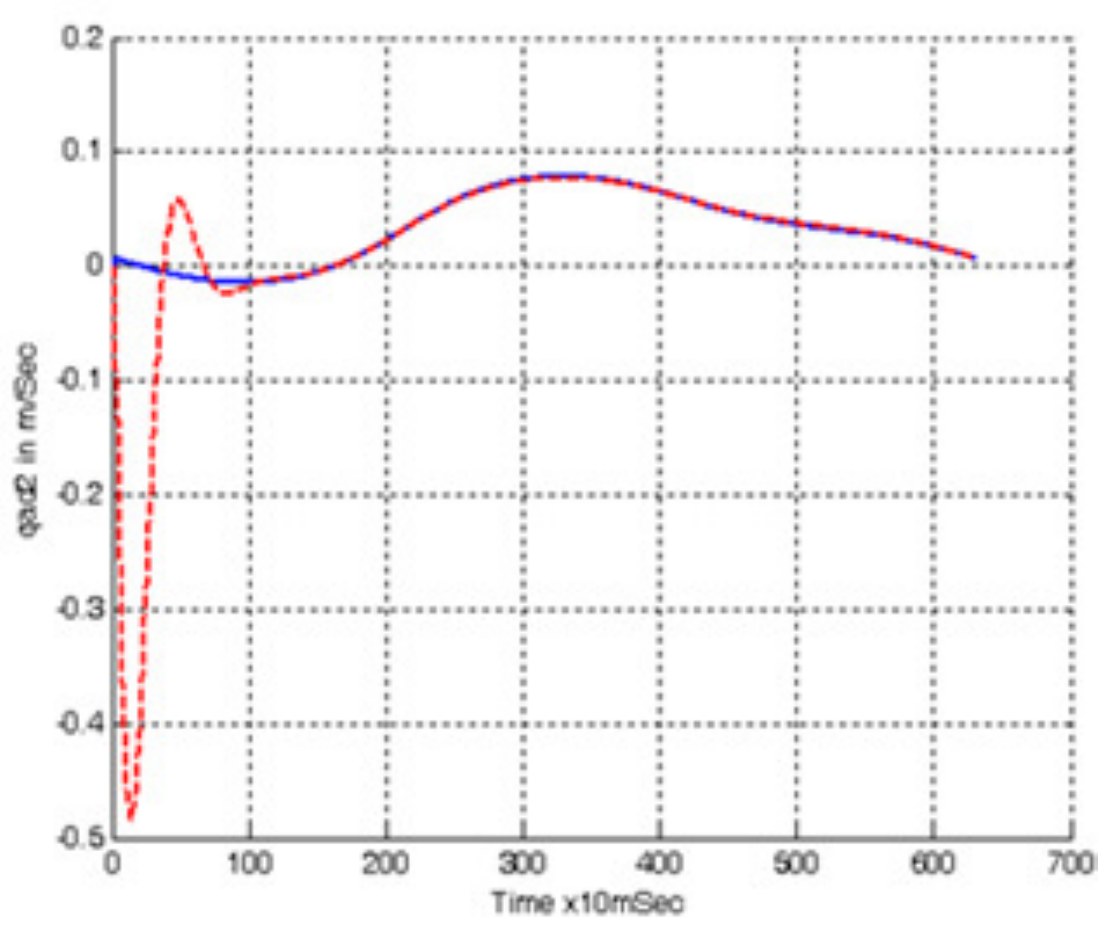}}
    \subfigure [ ]{\label{fig52-f}\includegraphics[width=6cm, height=5cm] {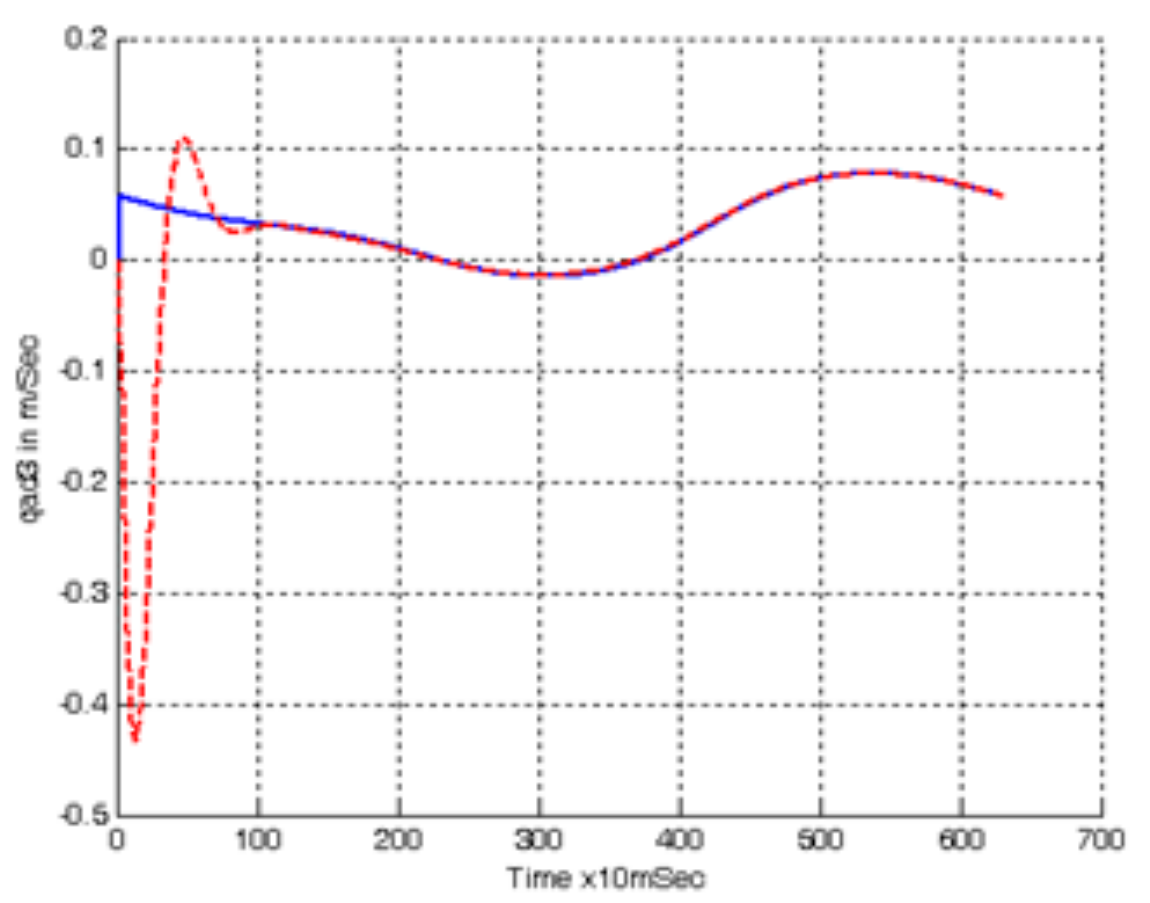}}\\
    \subfigure [ ]{\label{fig52-g}\includegraphics[width=7cm, height=6cm] {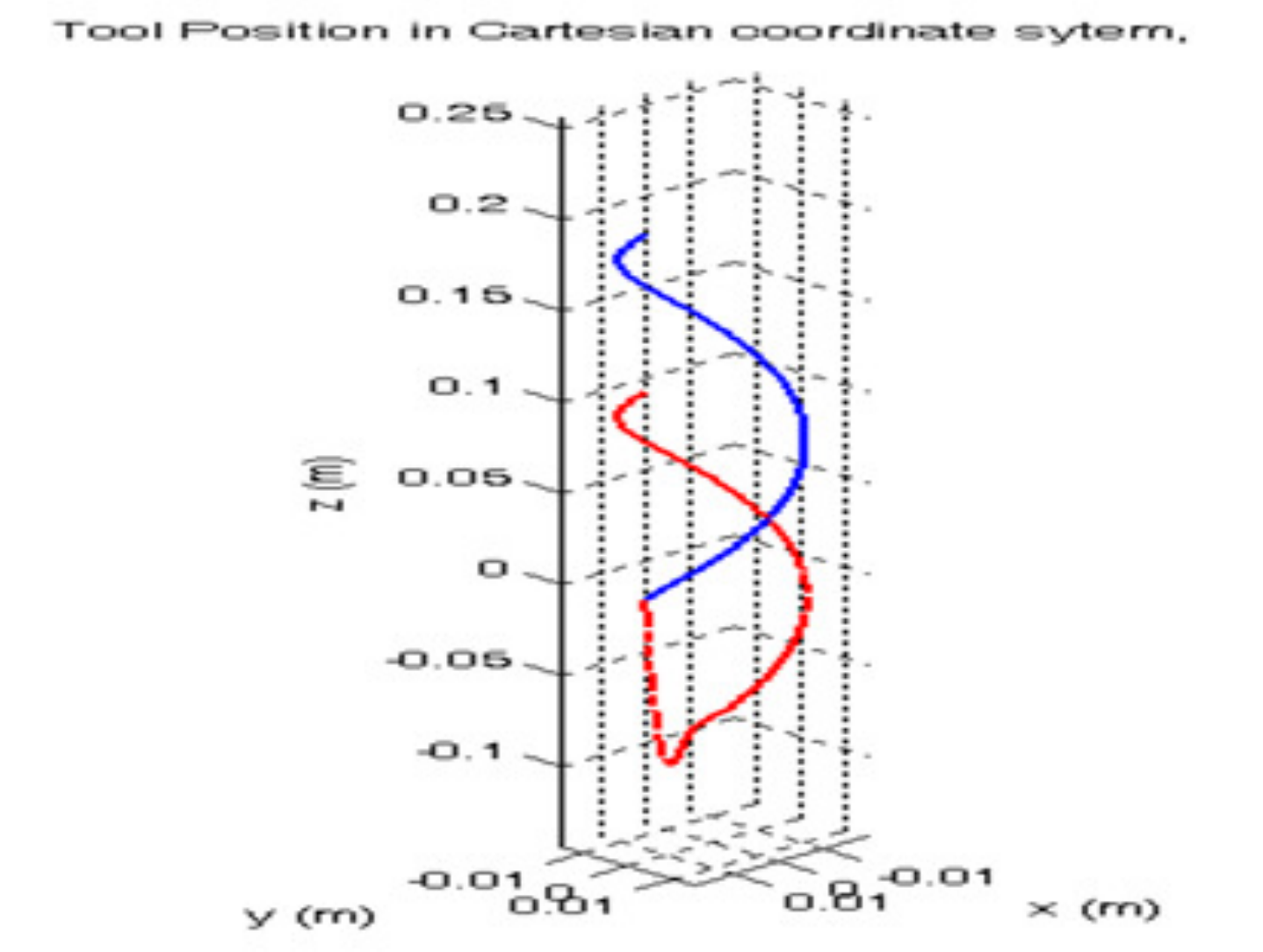}}\\
	\end{center}
  \caption[PD controller:(a)-(f)...]{PD controller:(f) Manipulator's desired and actual outputs $\{ \underbar q_1 ,\underbar q_2 ,\underbar q_3 ,\dot {\underbar q}_1 ,\dot{\underbar q} _2 ,\dot{\underbar q} _3 \}$ during the simulation time  (g) The followed desired and actual trajectory}
  \label{fig52}
\end{figure}

Figures \ref{fig52-a}-\ref{fig52-f} show each of the system outputs, $\{\underbar q_1 ,\underbar q_2 ,\underbar q_3 , \dot {\underbar{q}}_1 ,\dot {\underbar{q}} _2 ,\dot{\underbar{ q}} _3 \}$, both actual and desired during the simulation time. Notice how the positions lag behind the desired trajectories (although the actual speeds meet our expectations) in spite of the high PD gains 
(of course this could be mitigated if we had exploited the integral gain).  Figure \ref{fig52-g} depicts the final trajectory followed by the manipulator's tool-tip.

\section{\textbf{CTC}}
\begin{figure}[htp]
  \begin{center} 
    \subfigure [ ]{\label{fig53-a}\includegraphics[width=6cm, height=5cm] {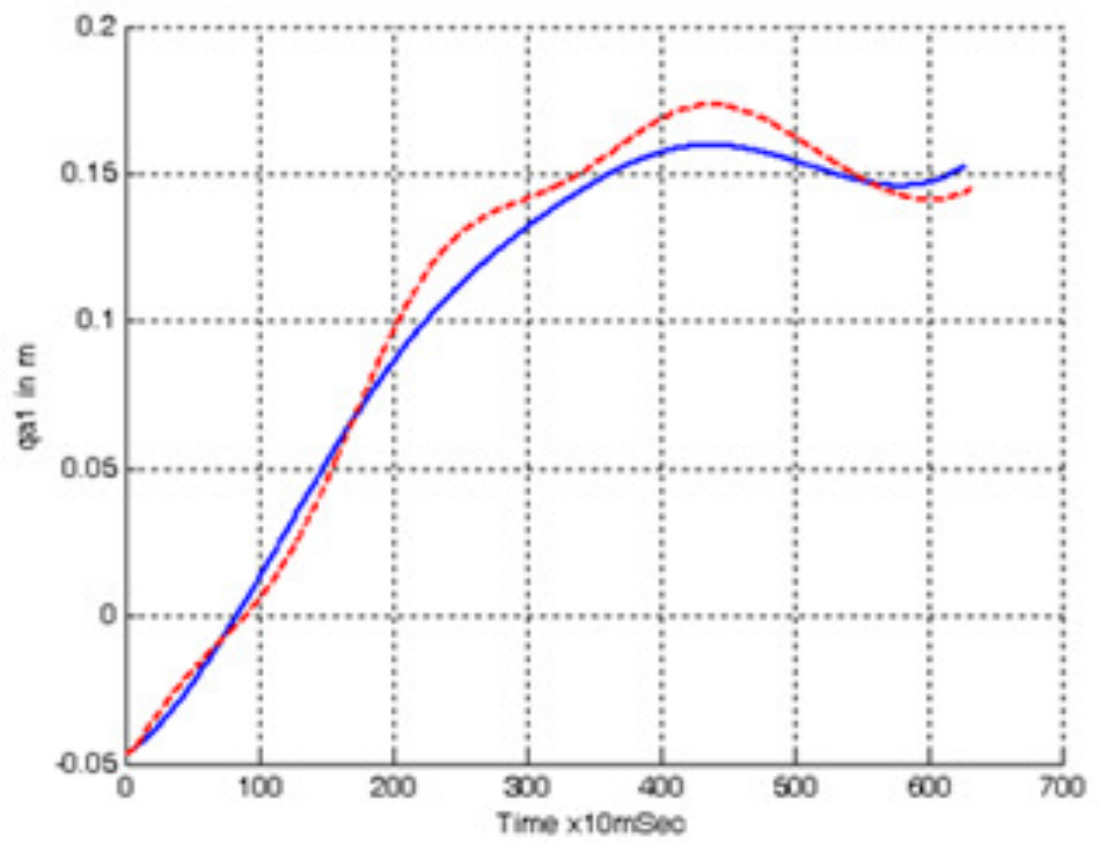}}
    \subfigure [ ]{\label{fig53-b}\includegraphics[width=6cm, height=5cm] {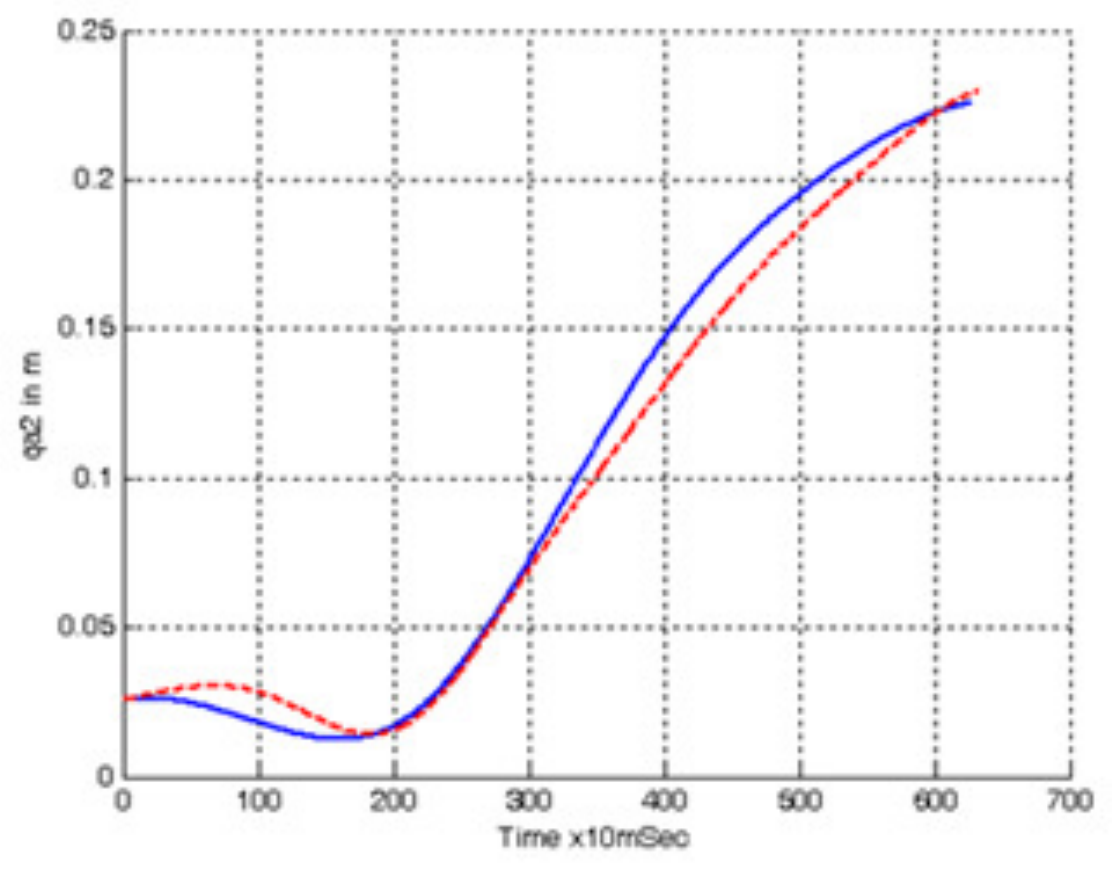}}\\
    \subfigure [ ]{\label{fig53-c}\includegraphics[width=6cm, height=5cm] {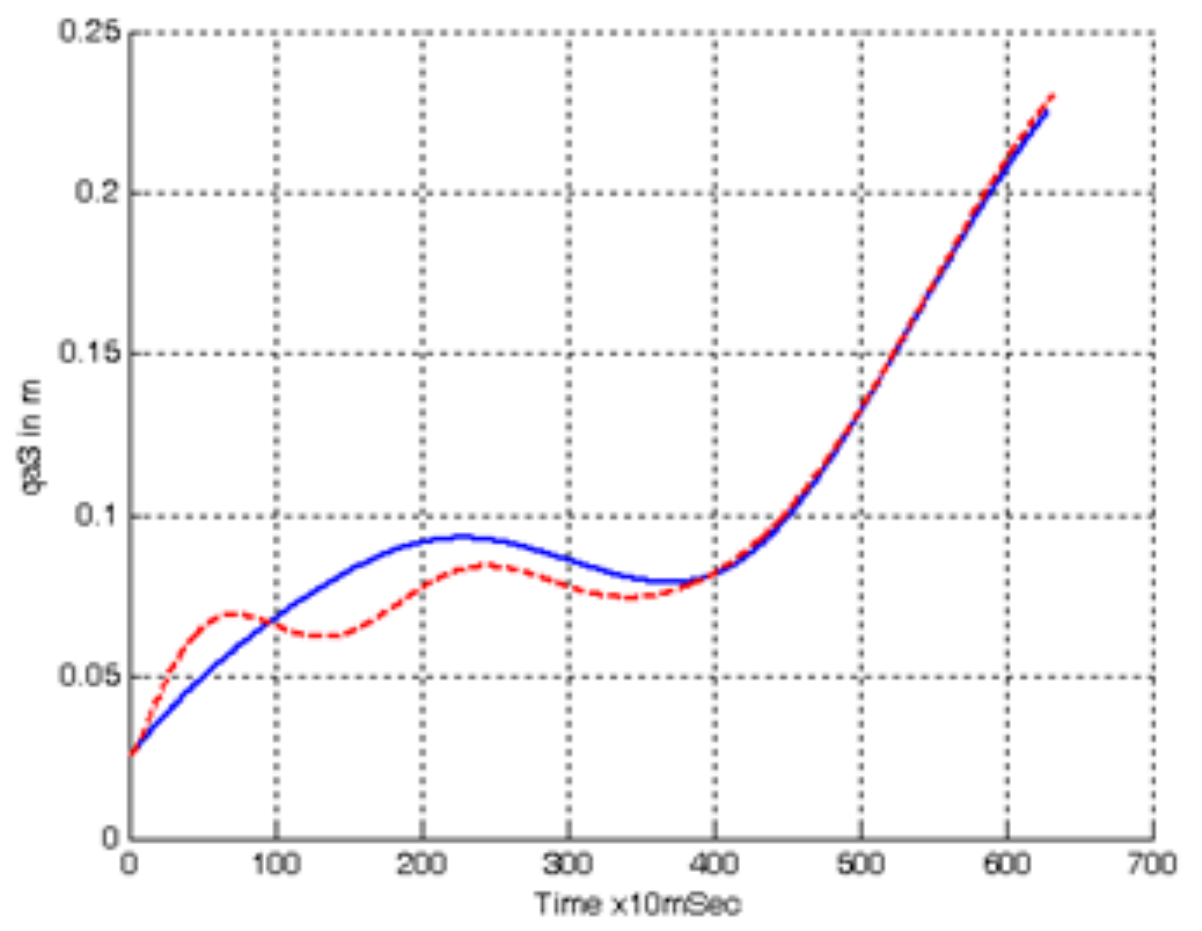}}
    \subfigure [ ]{\label{fig53-d}\includegraphics[width=6cm, height=5cm] {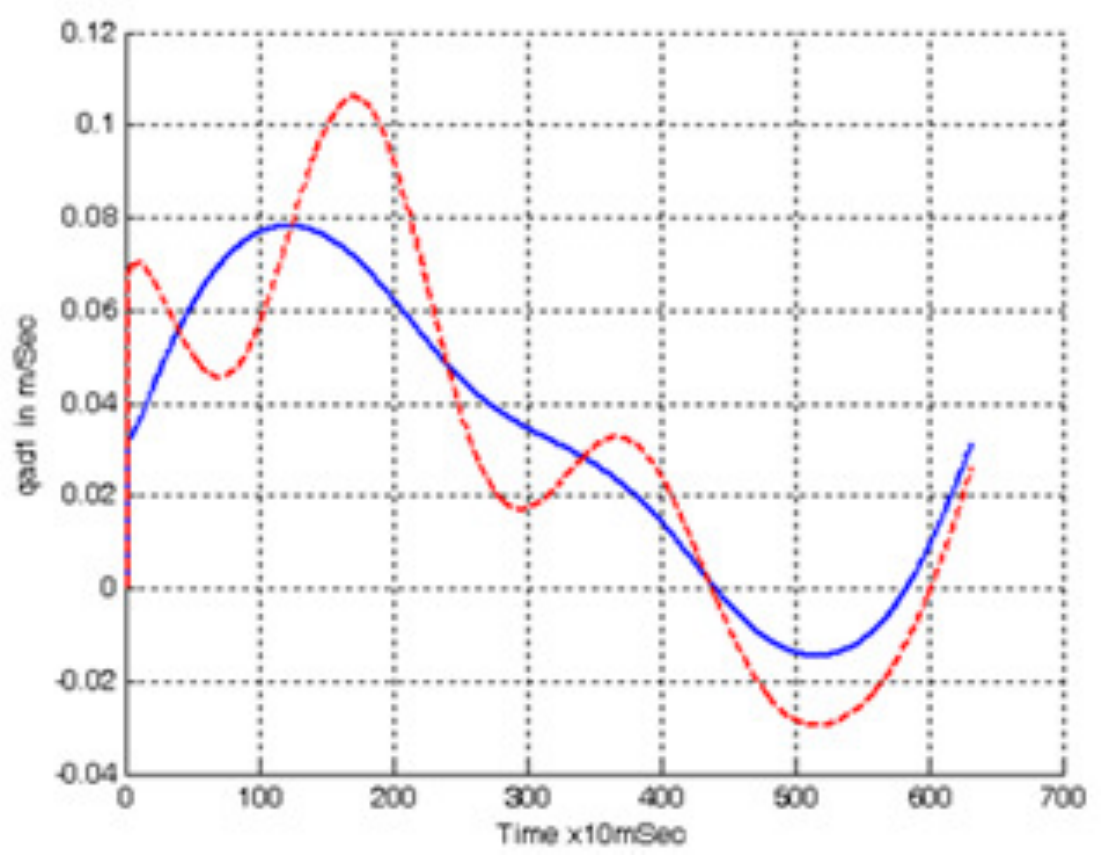}}\\
\end{center}
\label{fig53}
\end{figure}

\begin{figure}[htp]
  \begin{center} 
    \subfigure [ ]{\label{fig53-e}\includegraphics[width=6cm, height=5cm] {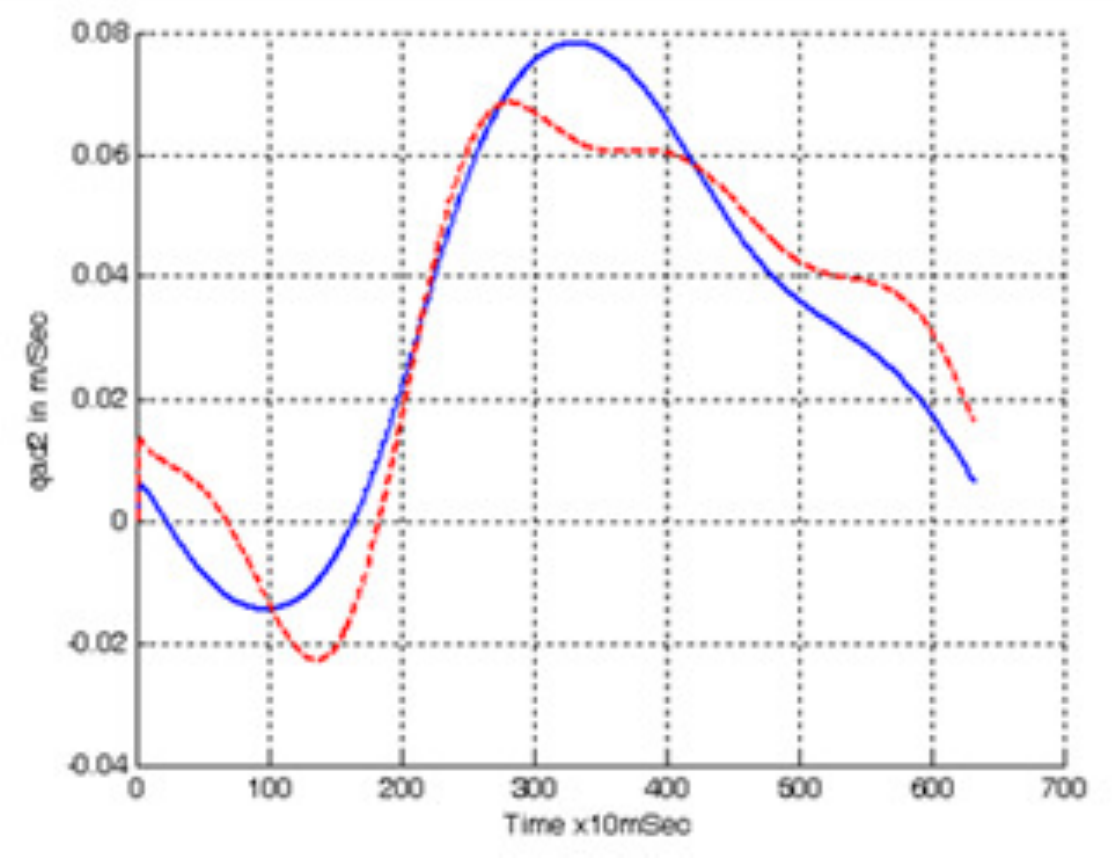}}
    \subfigure [ ]{\label{fig53-f}\includegraphics[width=6cm, height=5cm] {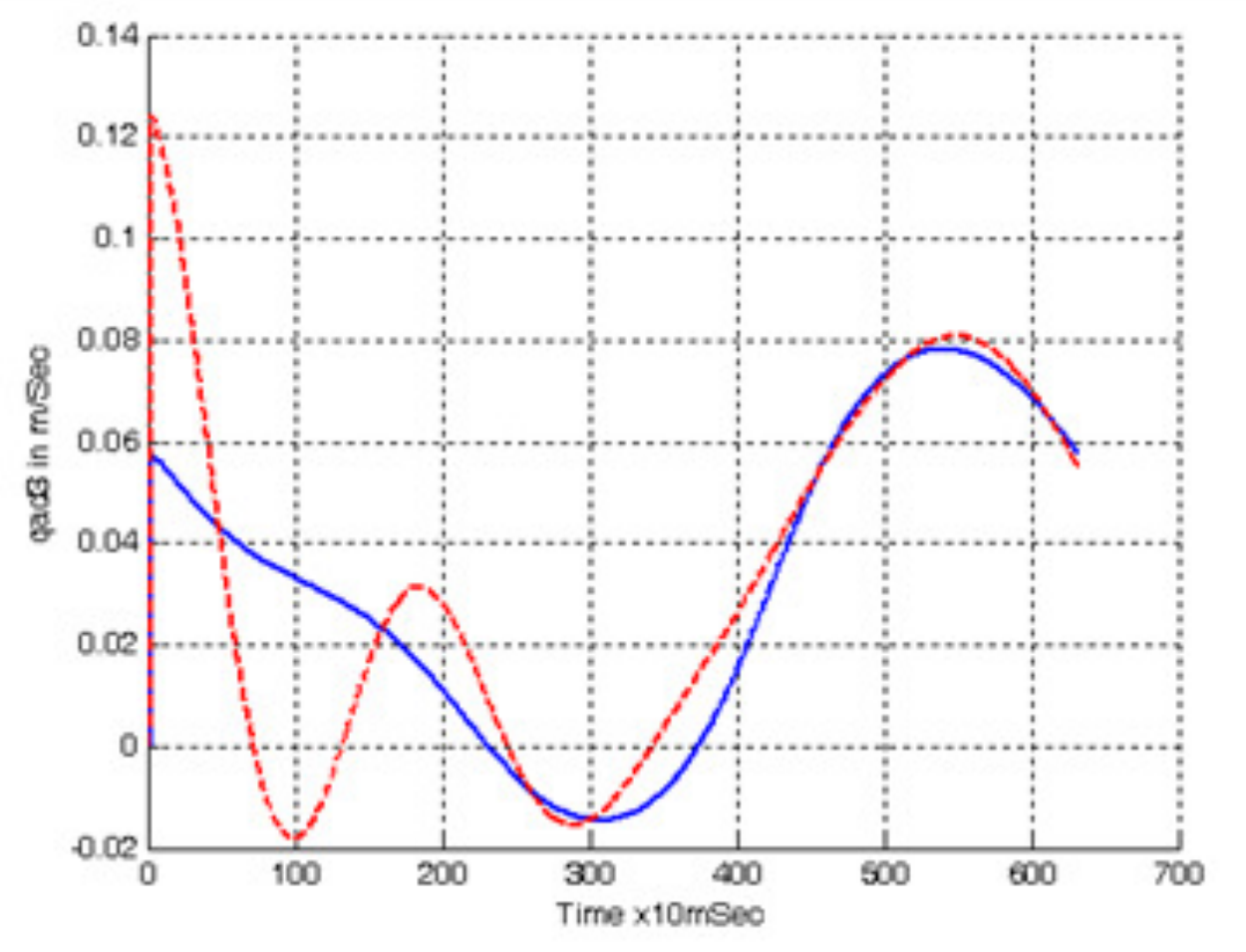}}\\
    \subfigure [ ]{\label{fig53-g}\includegraphics[width=7cm, height=6cm] {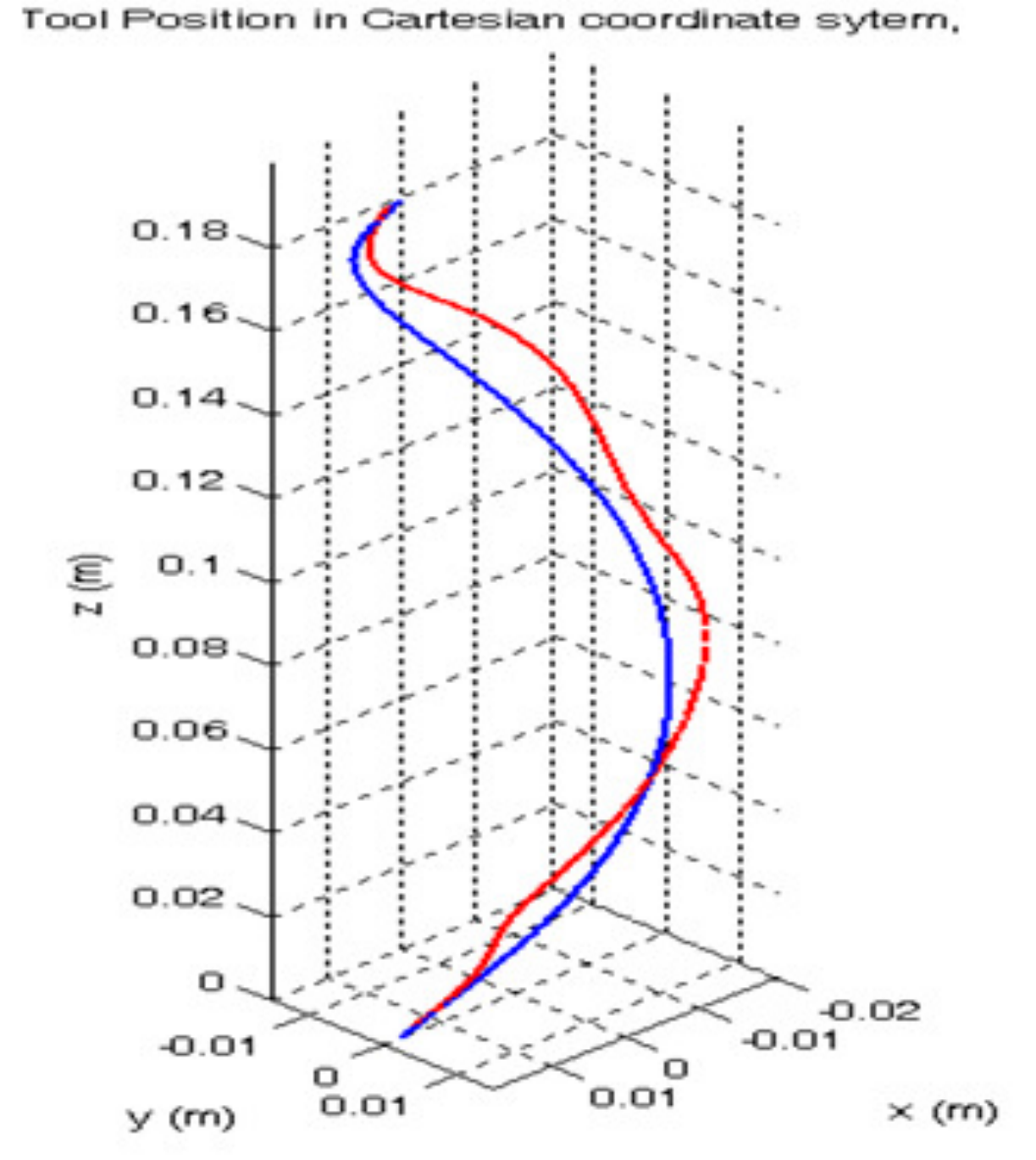}}\\
\end{center}
  \caption[CTC controller: (a)-(f)...]{CTC controller: (a)-(f) Manipulator's desired and actual outputs, $\{ \underbar q_1 ,\underbar q_2 ,\underbar q_3 ,\dot {\underbar q}_1 ,\dot{\underbar q} _2 ,\dot{\underbar q} _3 \}$ during the simulation time  (g) The followed desired and actual trajectory}
  \label{fig53}
\end{figure}

This trial is similar to the previous but this time a CTC controller is employed instead of the PD controller. Figures \ref{fig53-a}-\ref{fig53-f} show each of the system actual and desired outputs, $\{ \underbar q_1 ,\underbar q_2 ,\underbar q_3 ,\dot {\underbar q}_1 ,\dot{\underbar q} _2 ,\dot{\underbar q} _3 \}$ , both actual and desired during the simulation time. Notice from table \ref{tbl6.3} and this figure how the results outperform the ones corresponding to the previous method and as will be shown later, the CTC has the best performance provided that the underlying plant is completely determined. The main drawbacks are however the high computational burden (so much so it is not realizable in real-time applications) and the fact that there is not any possibility to improvise the uncertainty handling mechanism. Finally, Figure \ref{fig53-g} depicts the final trajectory followed by the manipulator's tool-tip.

\section{\textbf{Type-I Fuzzy Based CTC}}
\begin{figure}[htp]
  \begin{center} 
    \subfigure [ ]{\label{fig54-a}\includegraphics[width=6cm, height=5cm] {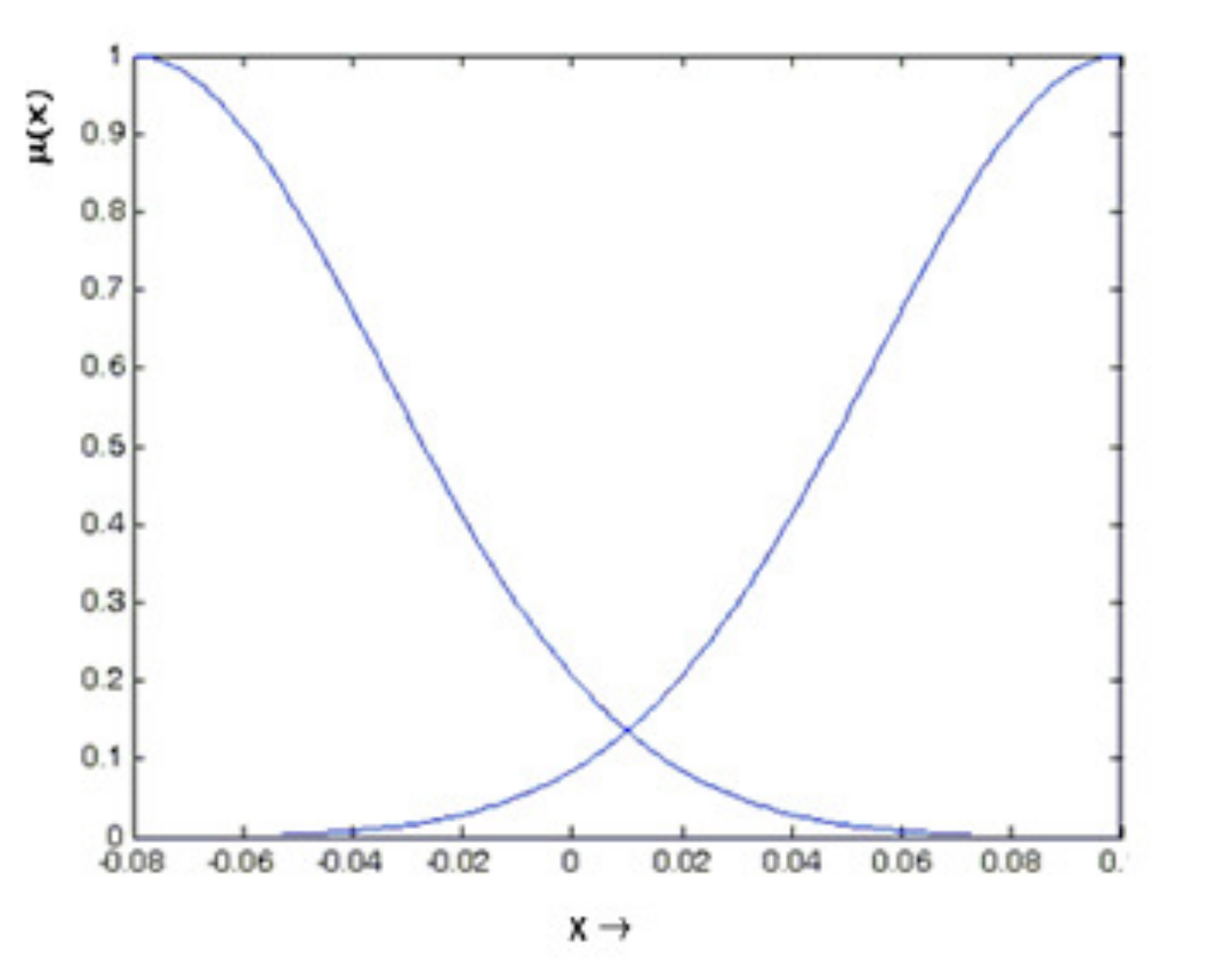}}
    \subfigure [ ]{\label{fig54-b}\includegraphics[width=6cm, height=5cm] {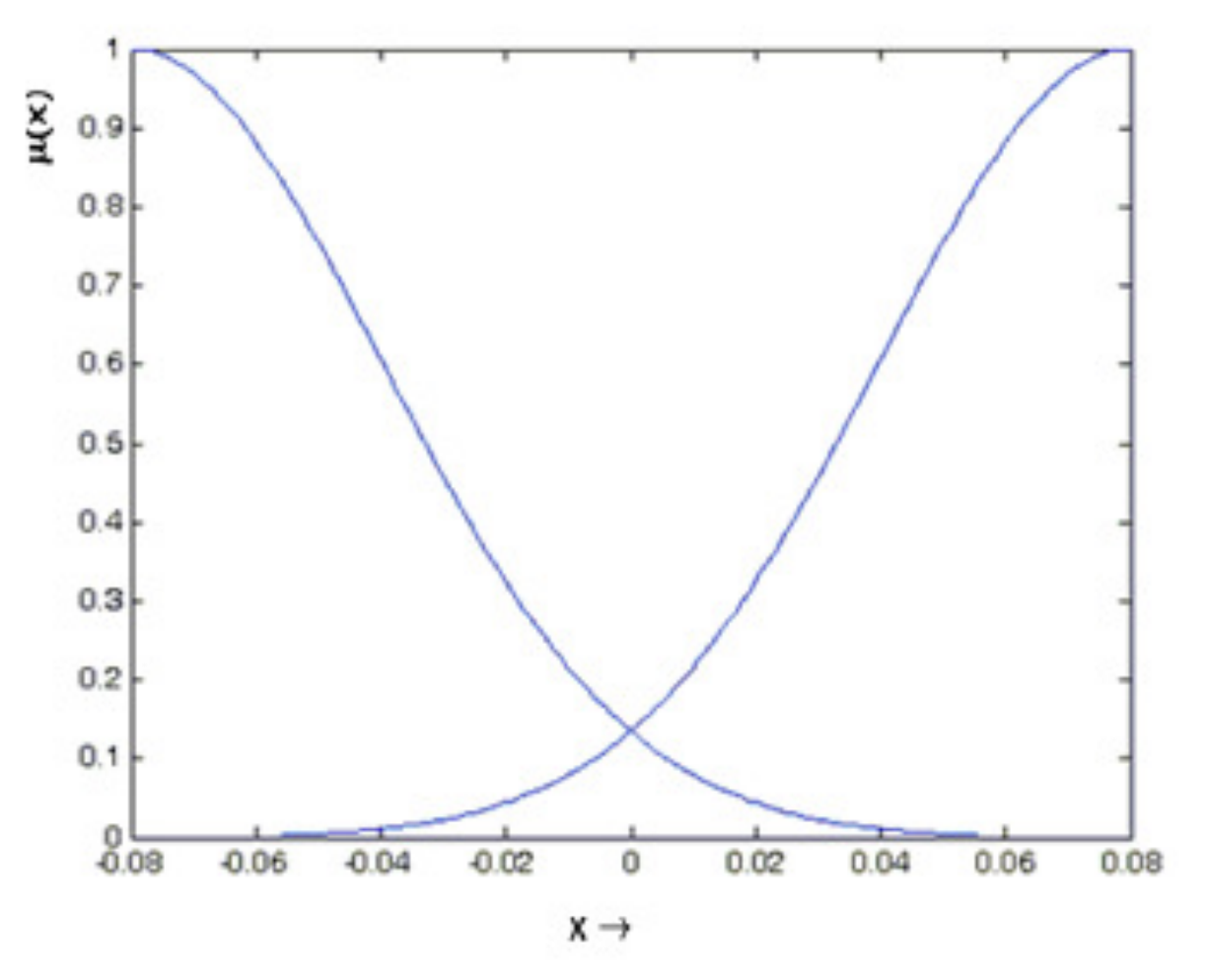}}\\
    \subfigure [ ]{\label{fig54-c}\includegraphics[width=6cm, height=5cm] {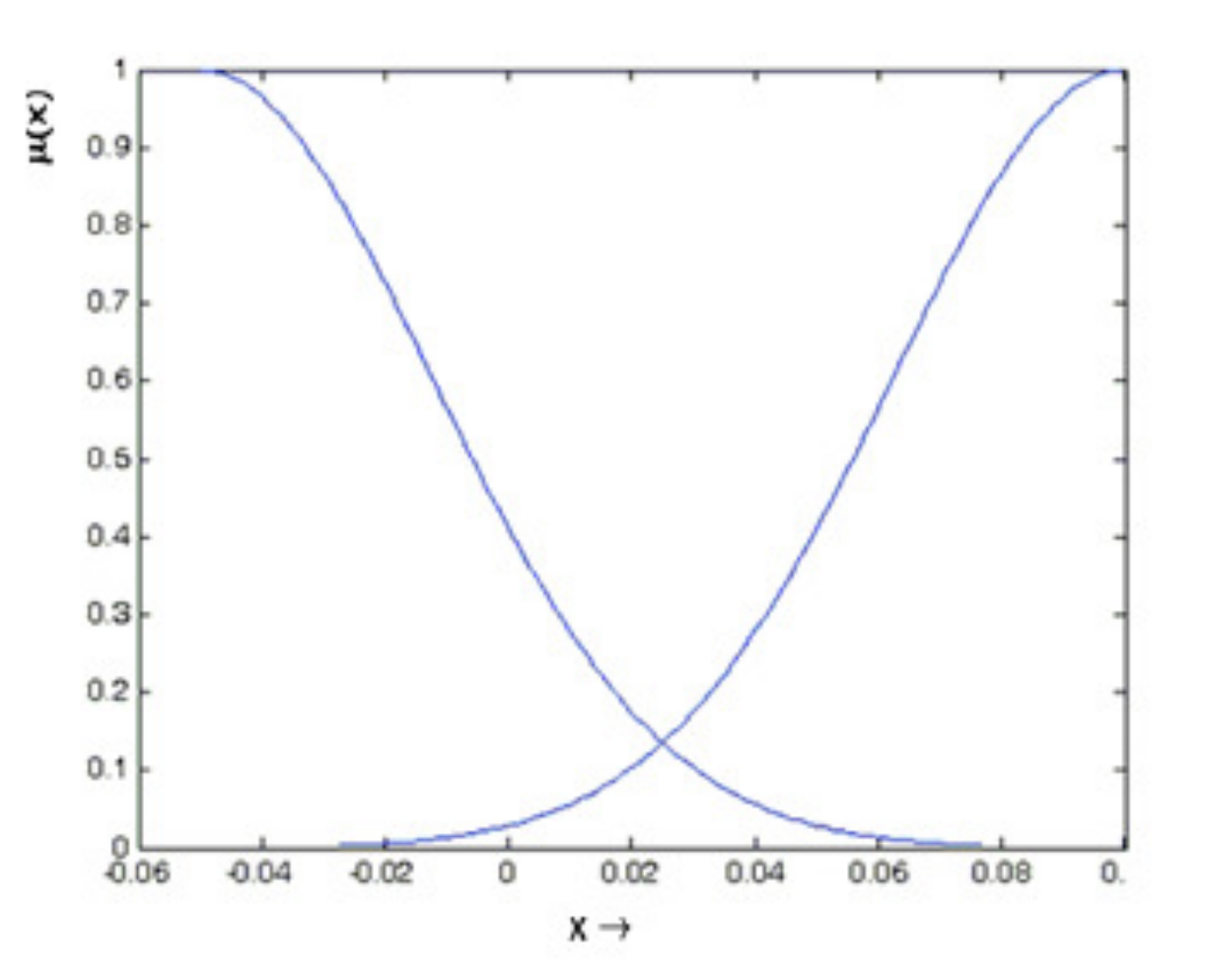}}
    \subfigure [ ]{\label{fig54-d}\includegraphics[width=6cm, height=5cm] {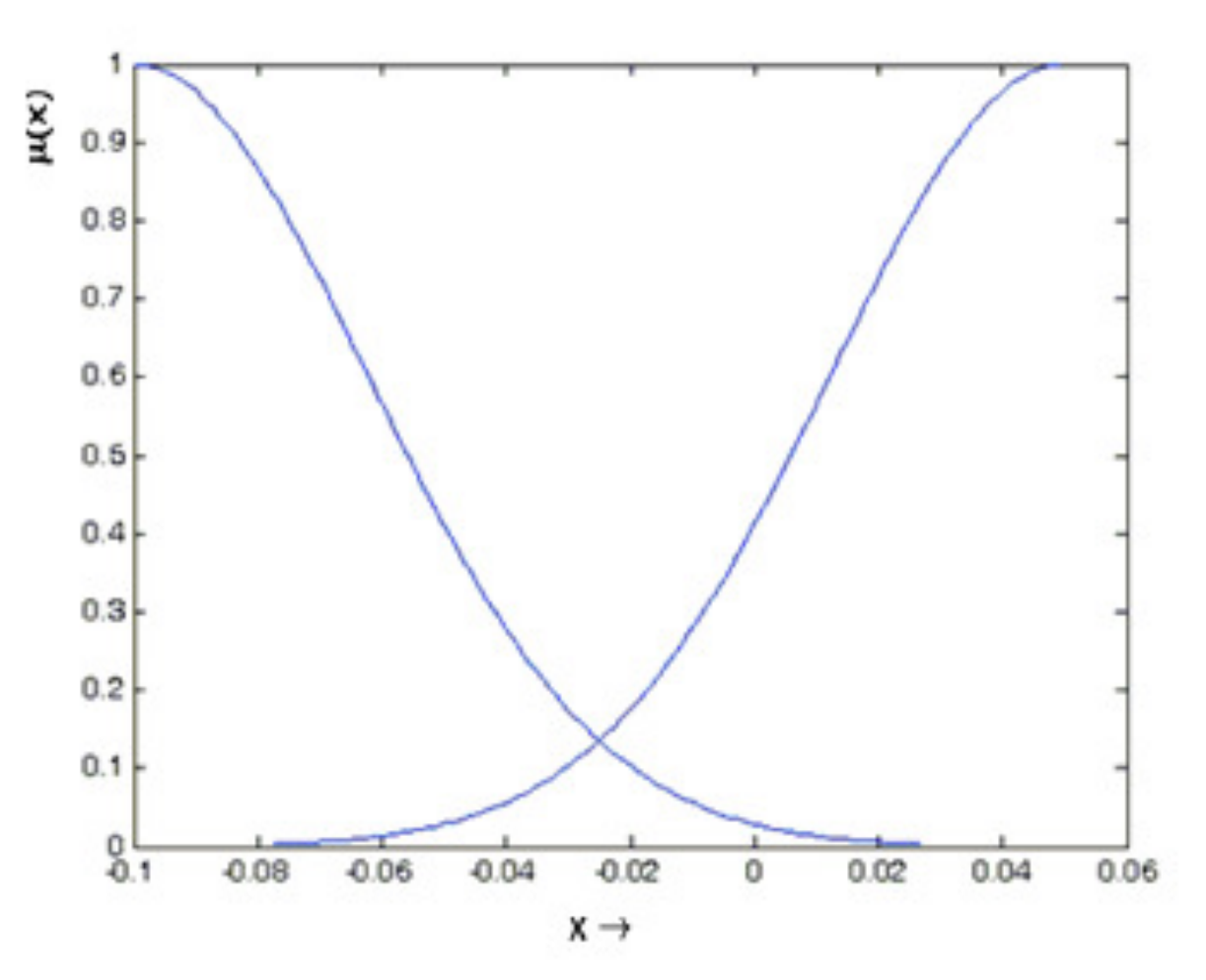}}\\
\end{center}
  \caption{Membership functions corresponding to the inputs space}
  \label{fig54}
\end{figure}

In the next scenarios a Helix trajectory is given to the downgraded T1 controller and the additive noises with different noise levels are injected into the system parameters (recall that 
we named this type of uncertainty, the \emph{numerical} uncertainty. Four different trials have been tested each associated to a specific signal to noise ratio (SNR). For each trial, 5 figures are depicted the first four pertaining to real and desired values of $\Delta \underbar{q}_{12}$,$\Delta \underbar{q}_{13}$, $\Delta \dot{ \underbar{q}}_{12}$, $\Delta \dot{ \underbar{q}}_{13}$ during the simulation time and the last shows the traversed trajectory. \\
Figure \ref{fig54} depicts the membership functions corresponding to the input space of the T1FLC.
Figure \ref{fig55} depicts the result of applying the aforementioned T1FLS to the plant for a Helix trajectory when no noise is present.  The first four plots show the desired and actual values associated to \{$\Delta \underbar{q}_{12}$,$\Delta \underbar{q}_{13}$, $\Delta \dot{ \underbar{q}}_{12}$, $\Delta \dot{ \underbar{q}}_{13}$\} respectively and the last plot shows the traversed trajectory and the desired one. 

Figures \ref{fig56} through \ref{fig58} show the same results when the signal to noise ratio (SNR) is 20db, 15db and 10db respectively. Notice how the performance of the controller degrades remarkably when the SNR decreases. See table \ref{tbl6.3} for a quantitative measure of the error.

\begin{figure}[htp]
  \begin{center} 
    \subfigure [ ]{\label{fig55-a}\includegraphics[width=6cm, height=5cm] {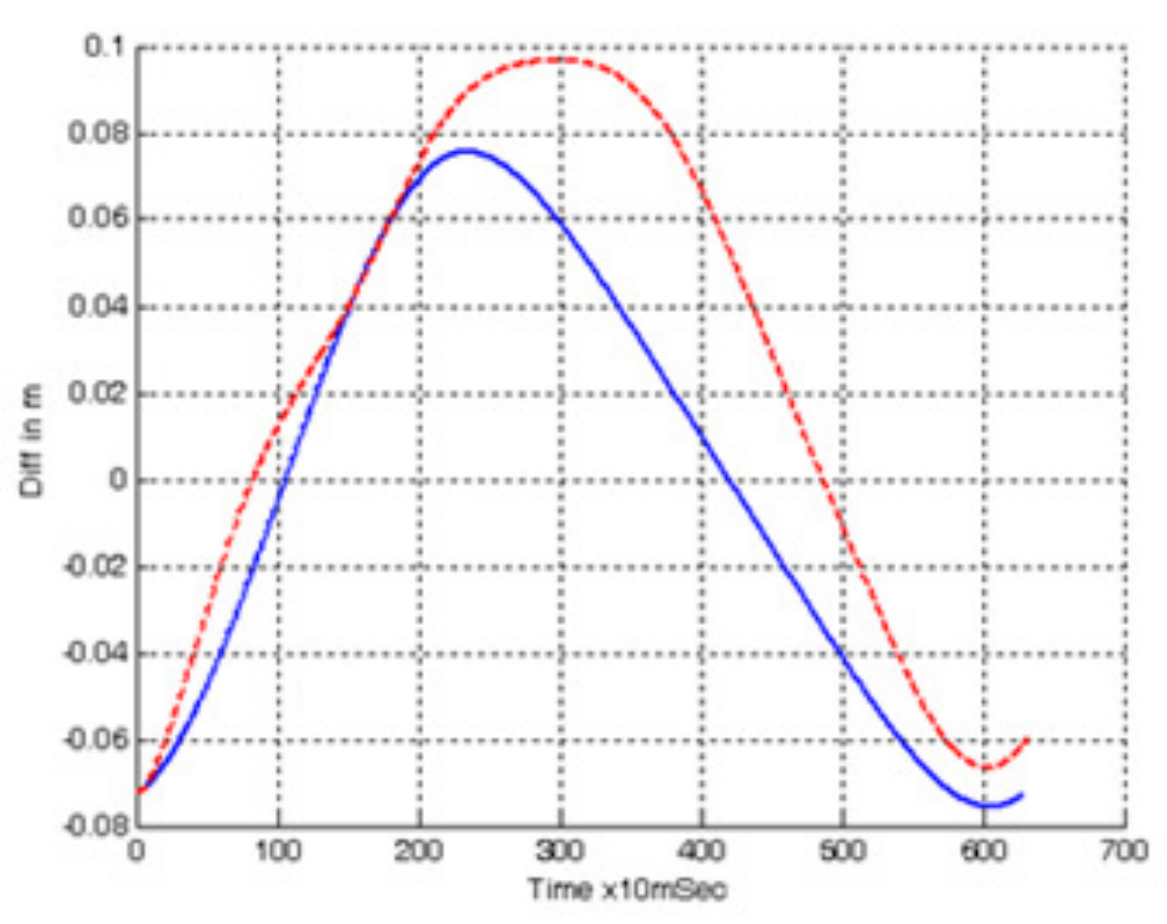}}
    \subfigure [ ]{\label{fig55-b}\includegraphics[width=6cm, height=5cm] {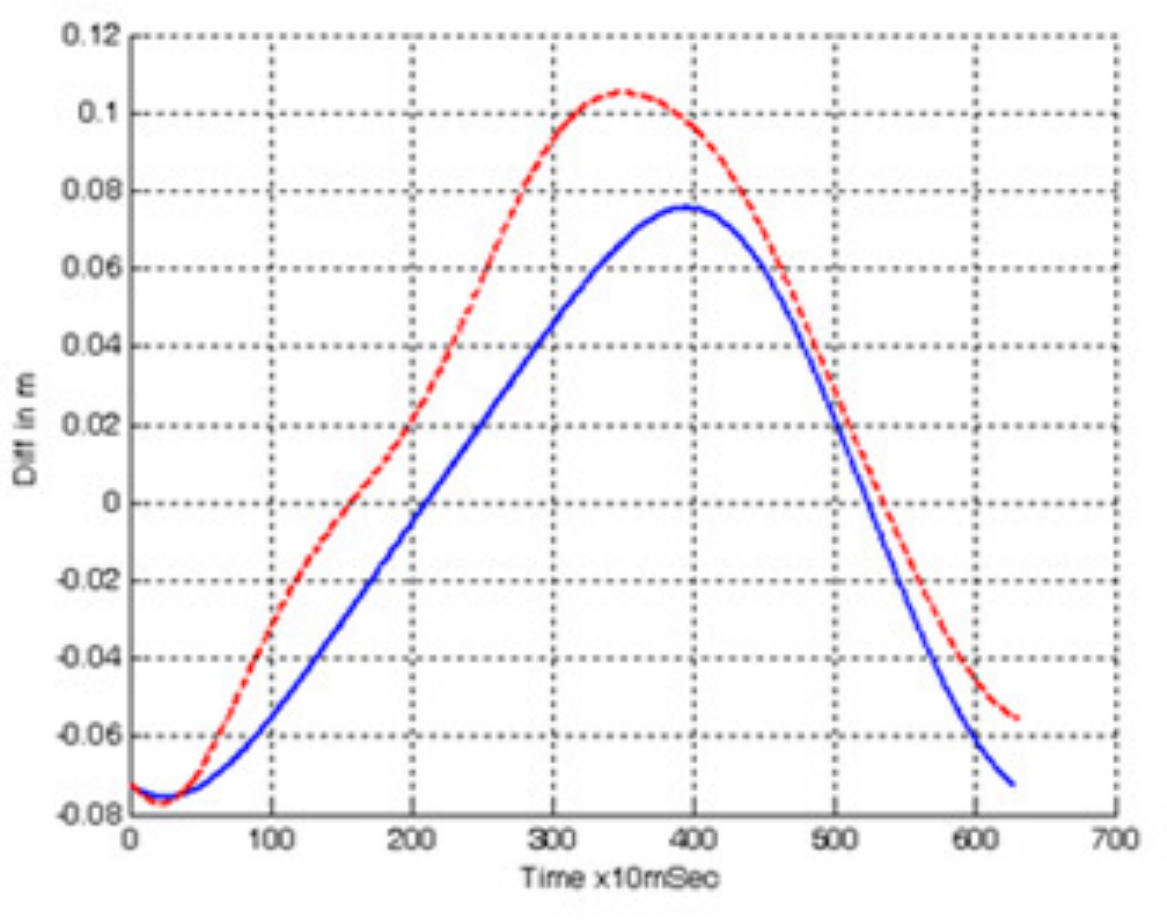}}\\
    \subfigure [ ]{\label{fig55-c}\includegraphics[width=6cm, height=5cm] {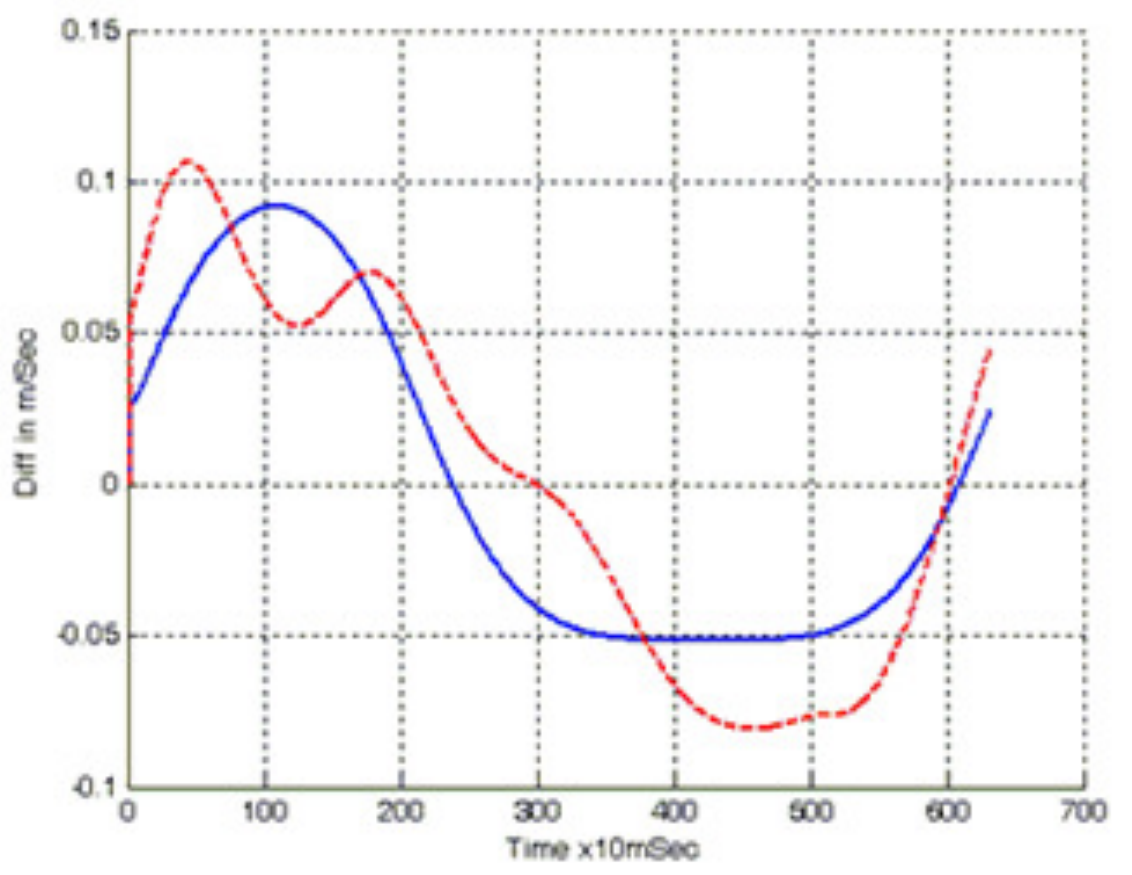}}
    \subfigure [ ]{\label{fig55-d}\includegraphics[width=6cm, height=5cm] {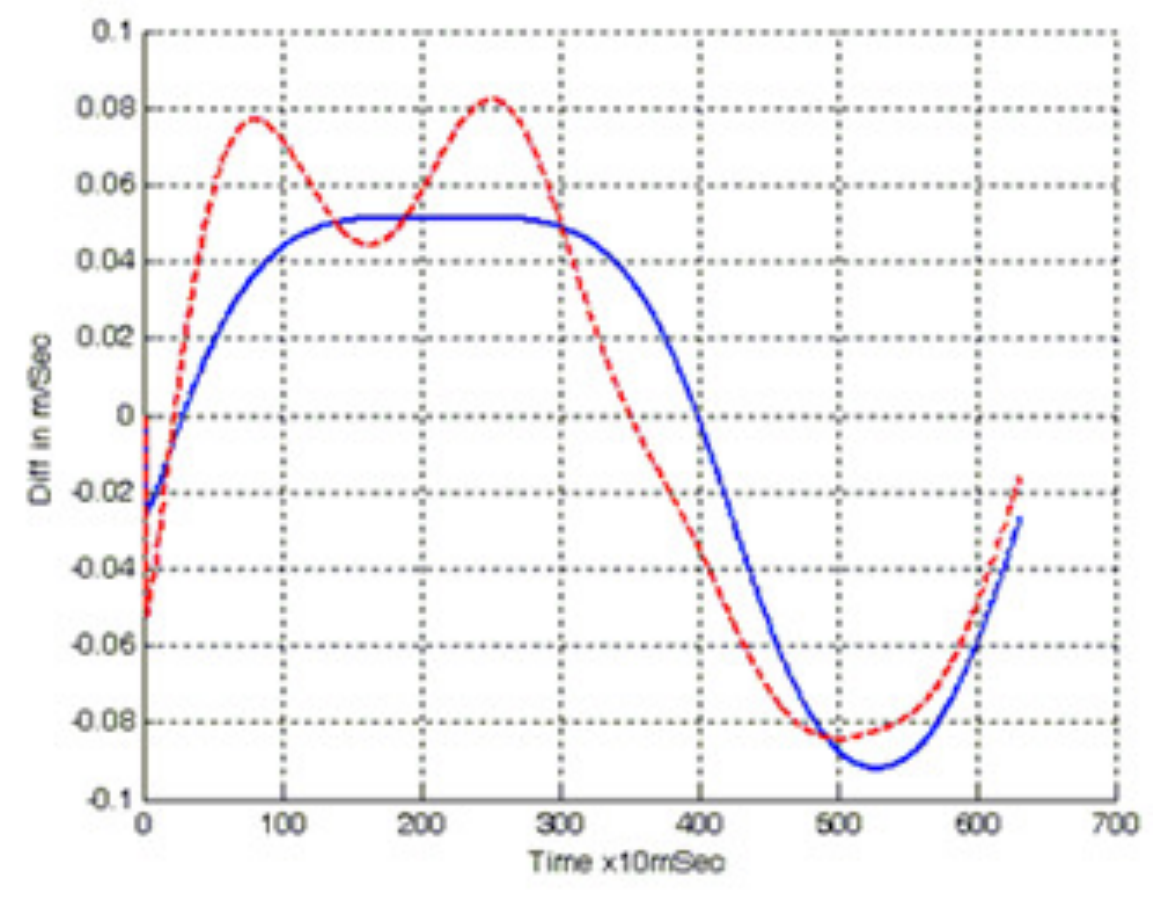}}\\
\end{center}
  \label{fig55}
\end{figure} 

\begin{figure}[htp]
  \begin{center} 
    \subfigure [ ]{\label{fig55-e}\includegraphics[width=7cm, height=6cm] {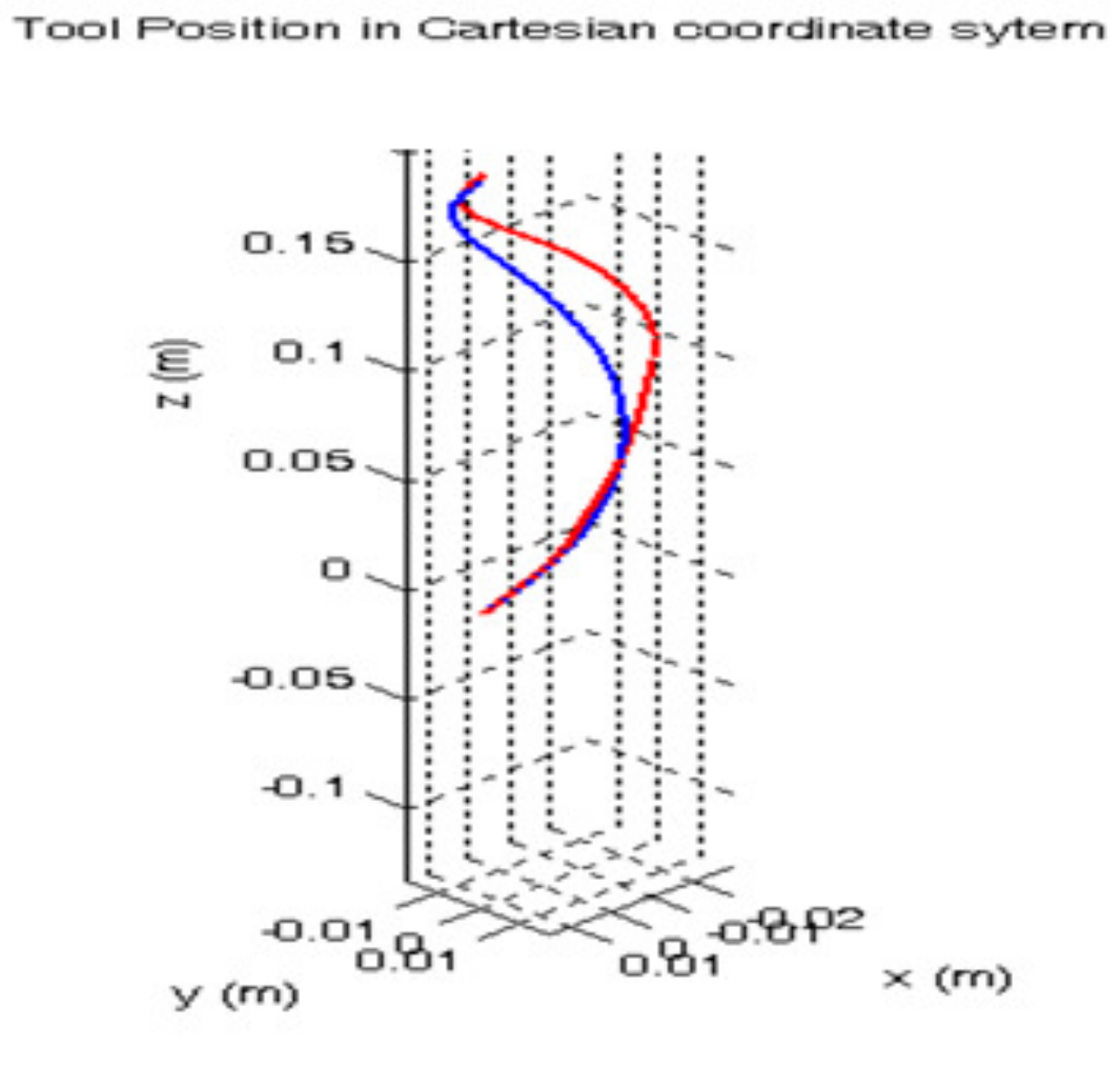}}\\
\end{center}
  \caption[T1 Fuzzy controller, SNR=$\infty$:...]{T1 Fuzzy controller, SNR=$\infty$: (a)-(f) Manipulator's desired and actual outputs,  \{$\Delta \underbar{q}_{12}$,$\Delta \underbar{q}_{13}$, $\Delta \dot{ \underbar{q}}_{12}$, $\Delta \dot{ \underbar{q}}_{13}$\} during the simulation time  (g) The followed desired and actual trajectory}
  \label{fig55}
\end{figure} 
\begin{figure}[htp]
  \begin{center} 
   \subfigure [ ]{\label{fig56-a}\includegraphics[width=6cm, height=5cm] {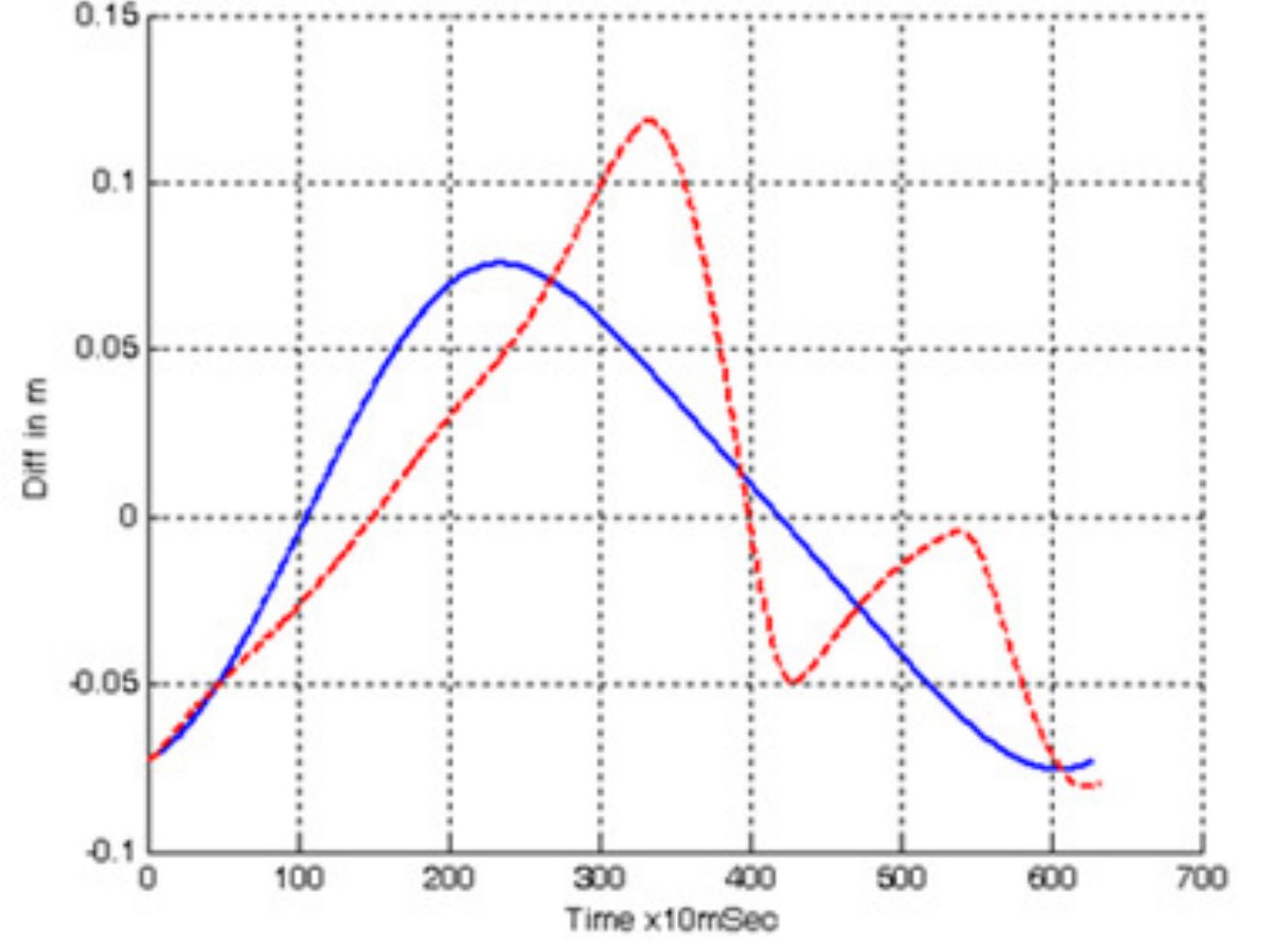}}
    \subfigure [ ]{\label{fig56-b}\includegraphics[width=6cm, height=5cm] {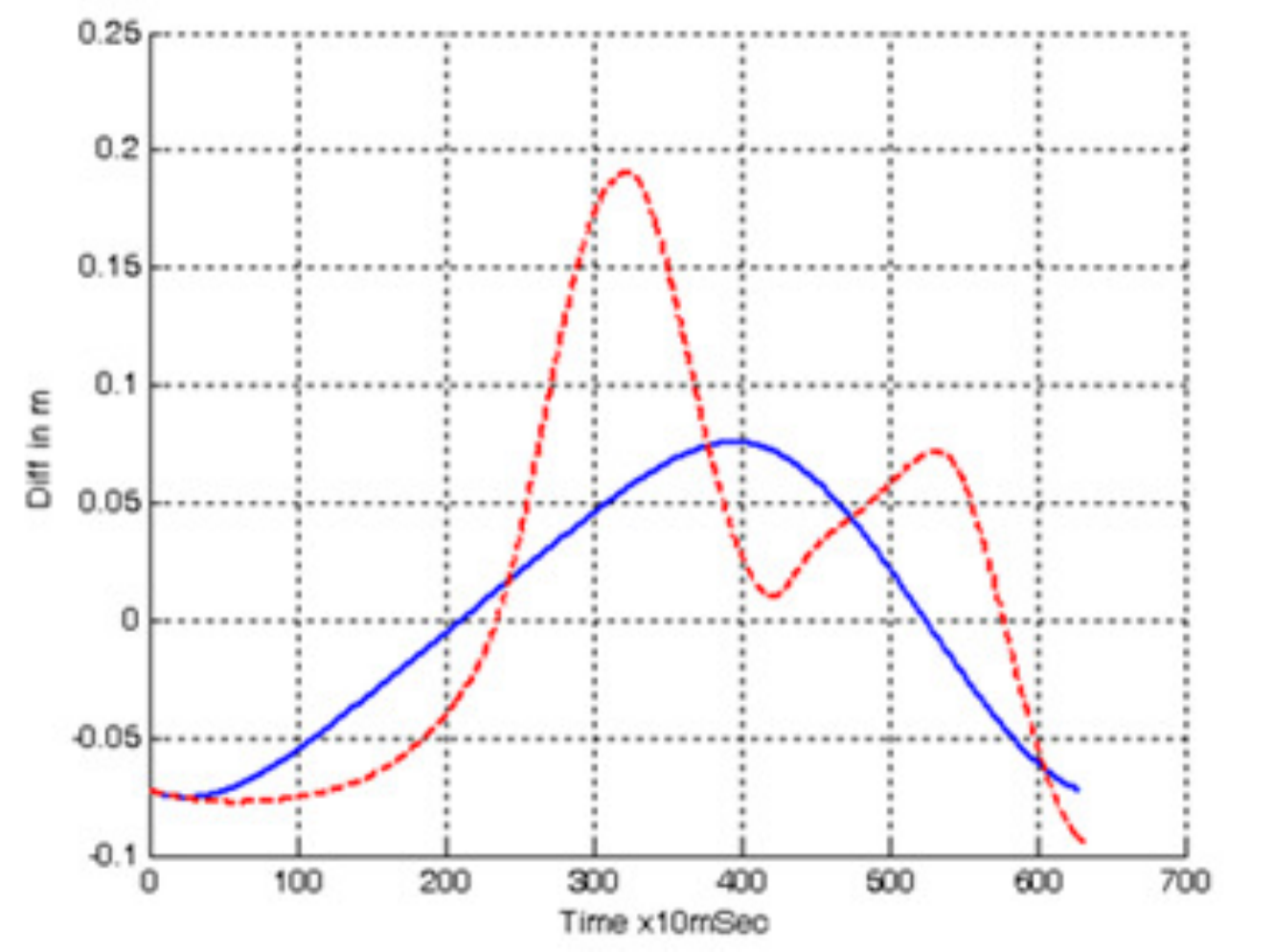}}\\
    \subfigure [ ]{\label{fig56-c}\includegraphics[width=6cm, height=5cm] {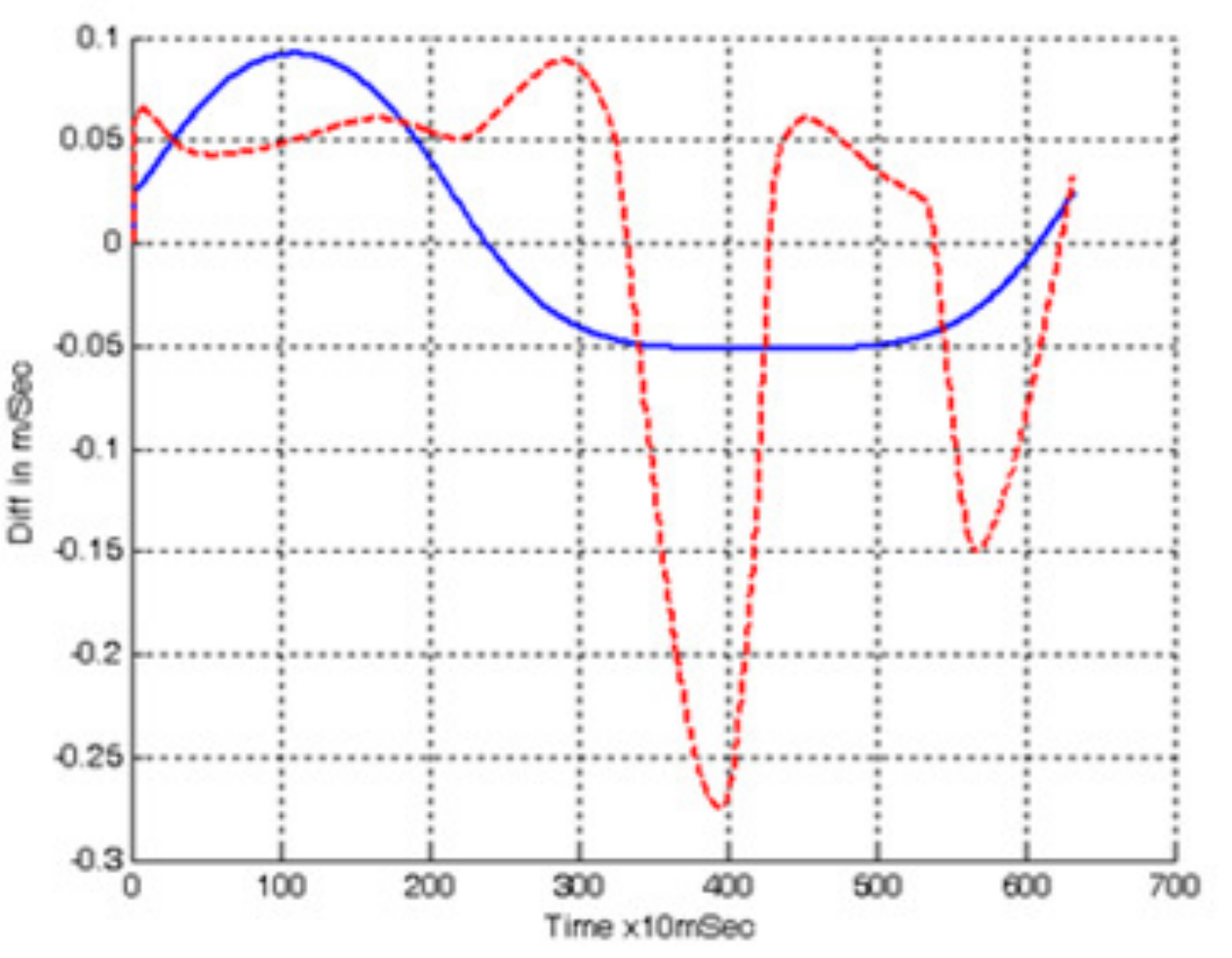}}
    \subfigure [ ]{\label{fig56-d}\includegraphics[width=6cm, height=5cm] {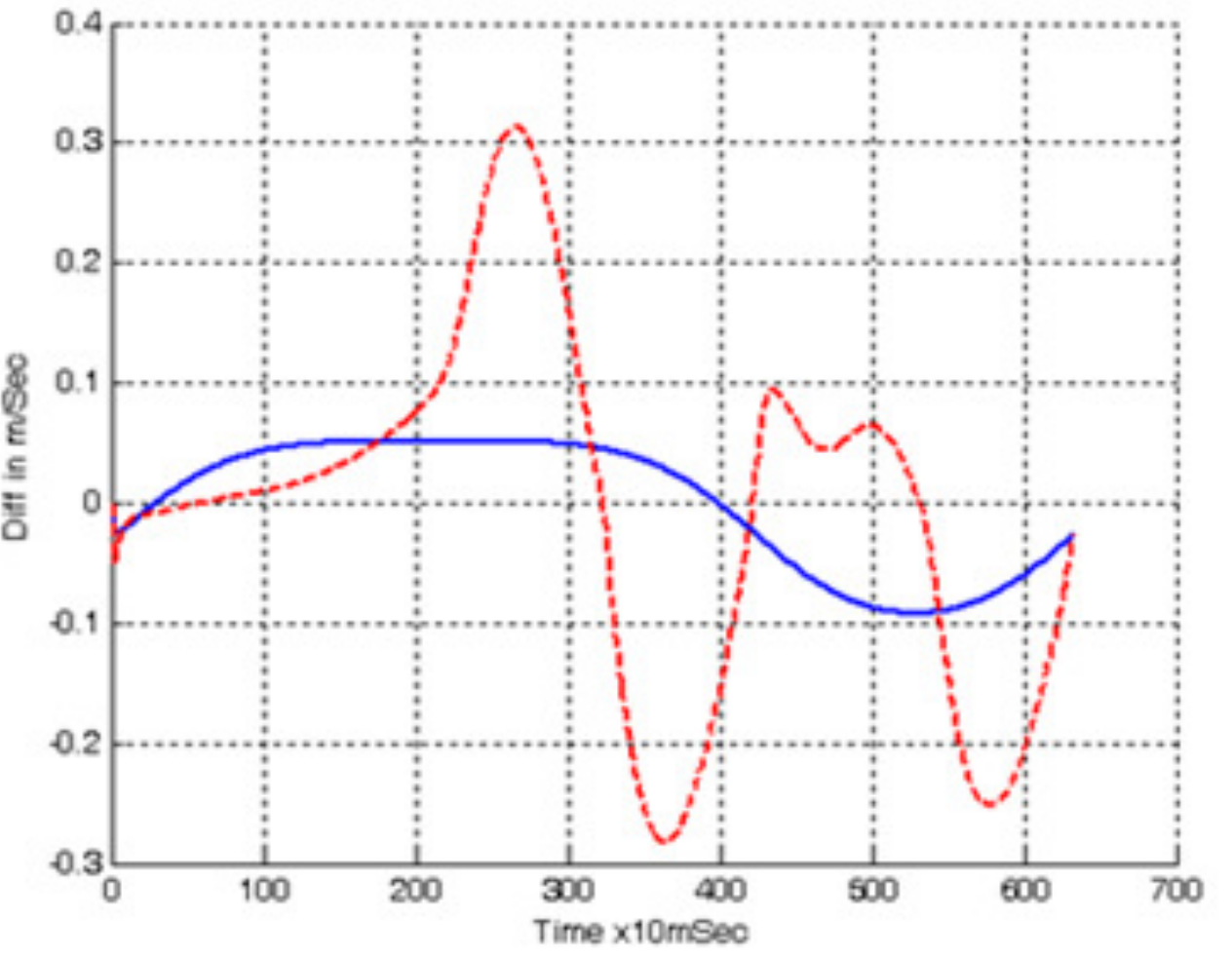}}\\
    \end{center}
    \label{fig56}
\end{figure} 
\begin{figure}[htp]
  \begin{center} 
   \subfigure [ ]{\label{fig56-e}\includegraphics[width=7cm, height=6cm] {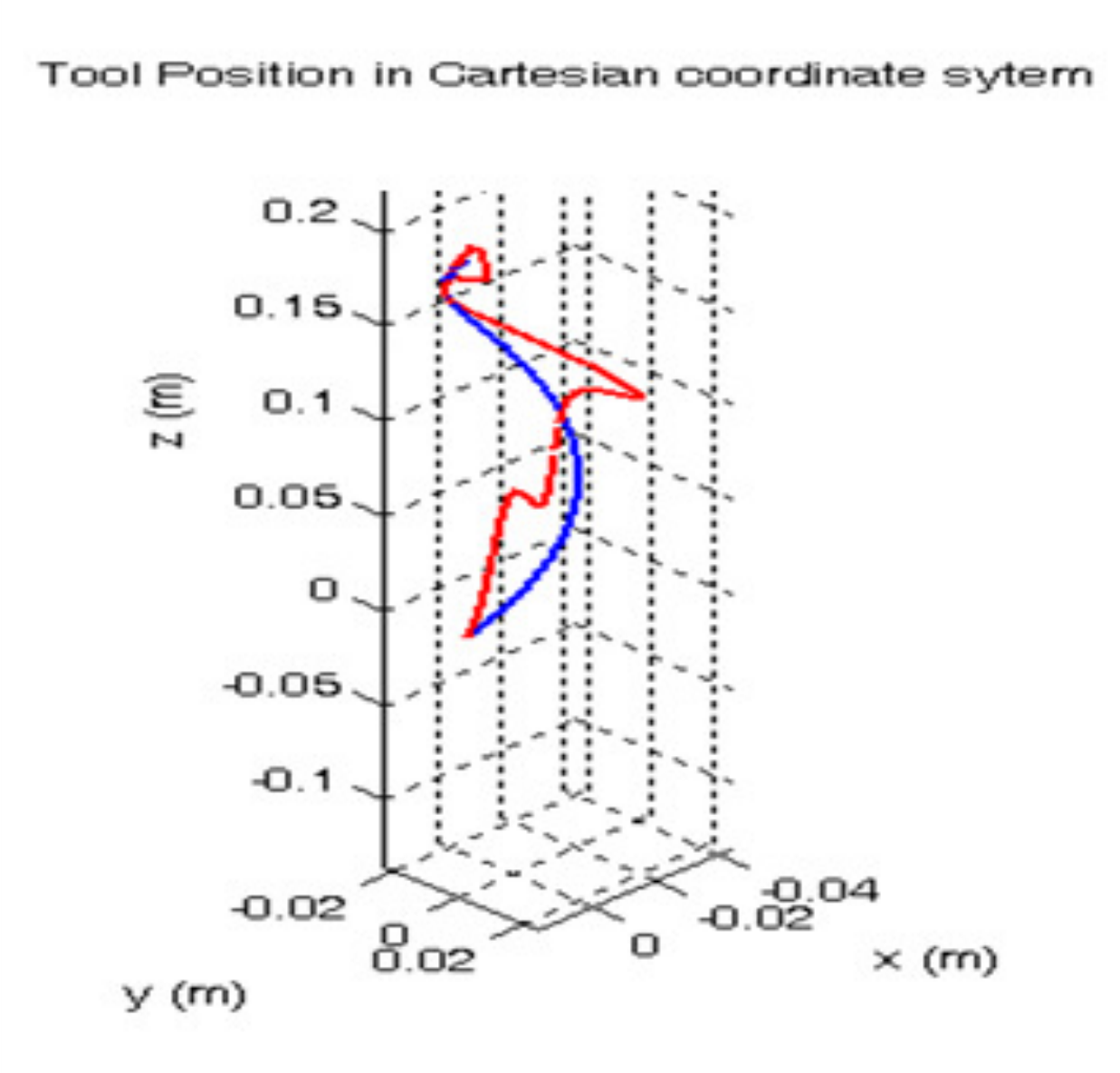}}\\
  \end{center}
  \caption[T1 Fuzzy controller, SNR=20dB: (a)-(f)...]{T1 Fuzzy controller, SNR=20dB: (a)-(f) Manipulator's desired and actual outputs,  \{$\Delta \underbar{q}_{12}$,$\Delta \underbar{q}_{13}$, $\Delta \dot{ \underbar{q}}_{12}$, $\Delta \dot{ \underbar{q}}_{13}$\} during the simulation time  (g) The followed desired and actual trajectory}
  \label{fig56}
\end{figure} 

\begin{figure}[htp]
  \begin{center} 
   \subfigure [ ]{\label{fig57-a}\includegraphics[width=6cm, height=5cm] {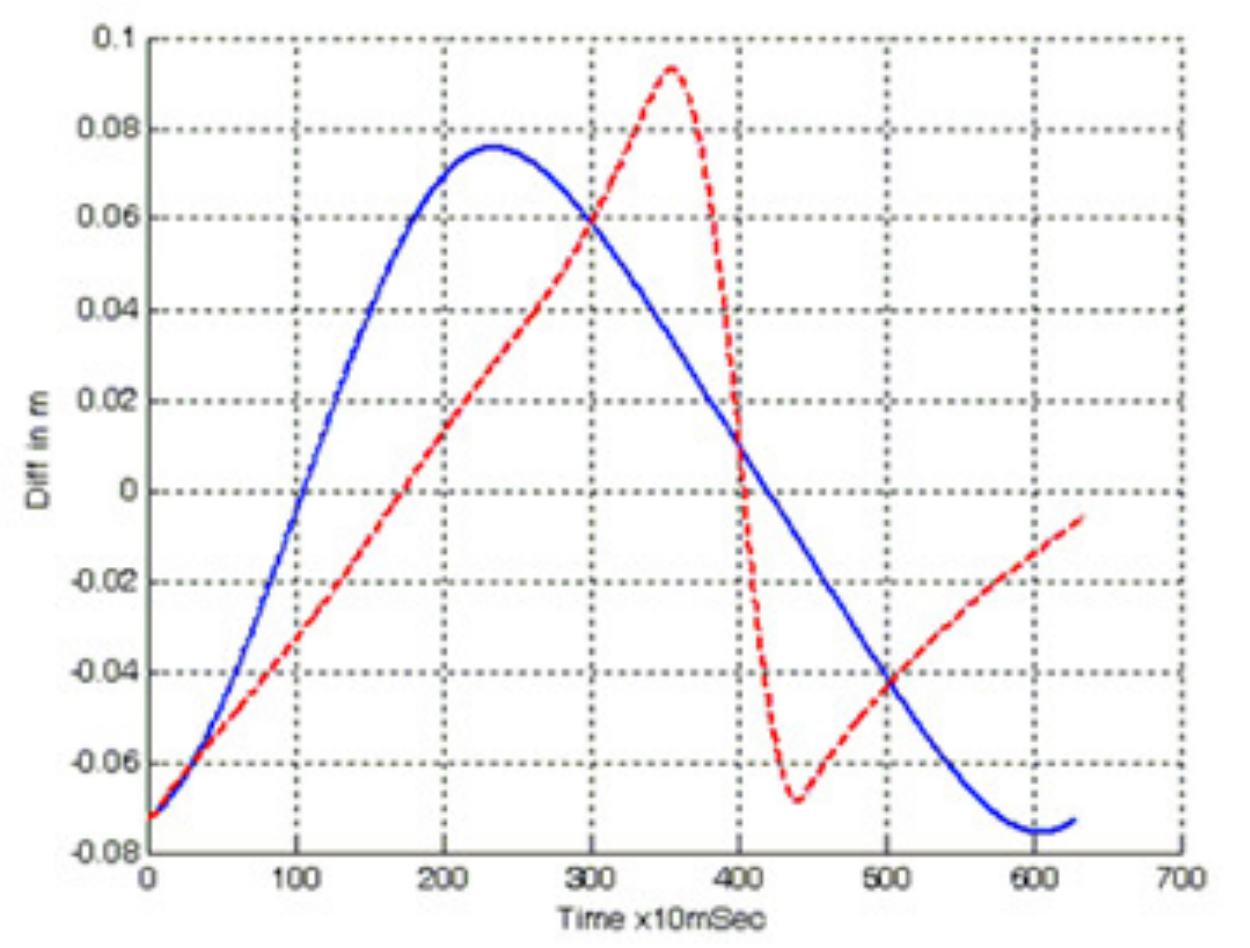}}
    \subfigure [ ]{\label{fig57-b}\includegraphics[width=6cm, height=5cm] {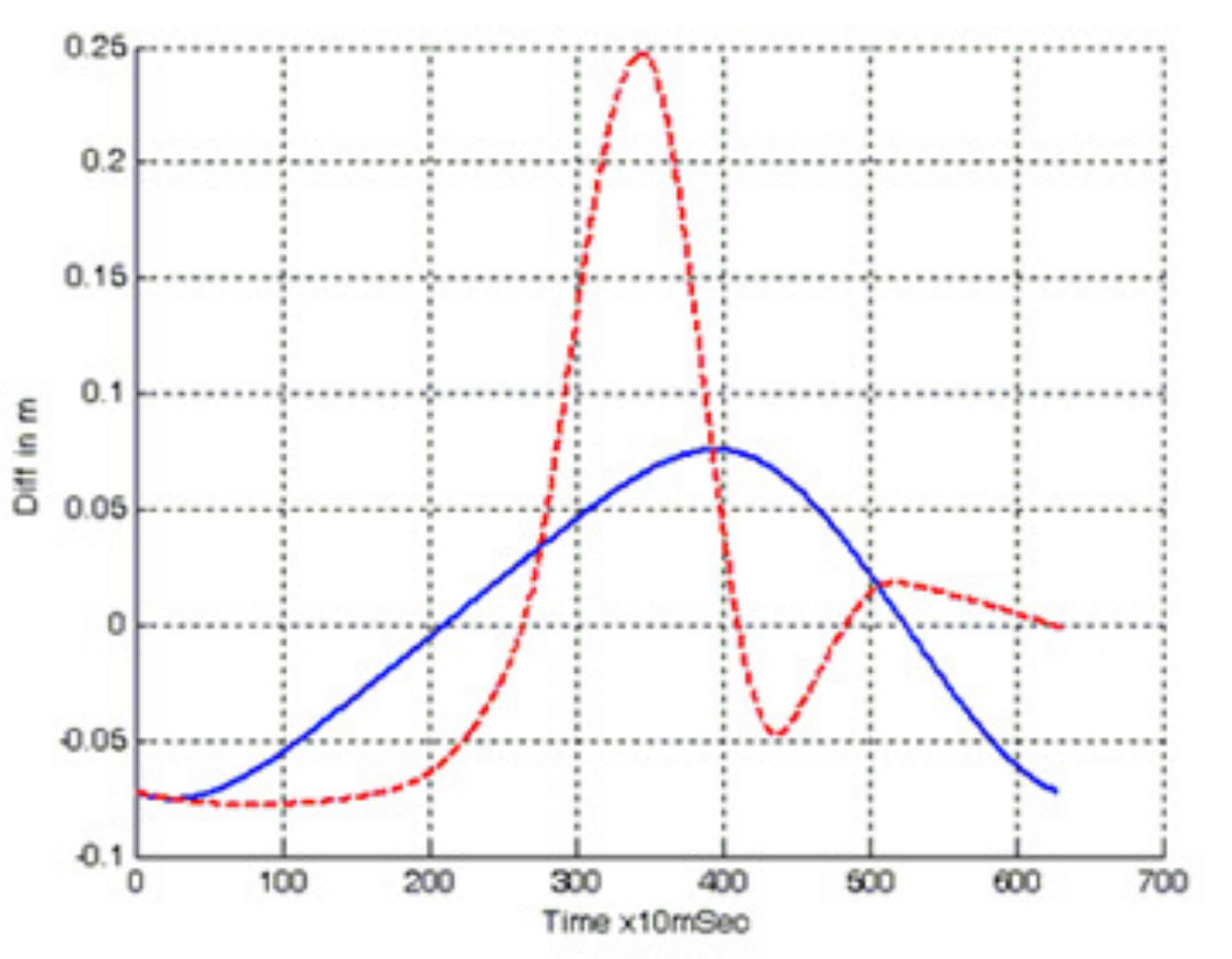}}\\
    \subfigure [ ]{\label{fig57-c}\includegraphics[width=6cm, height=5cm] {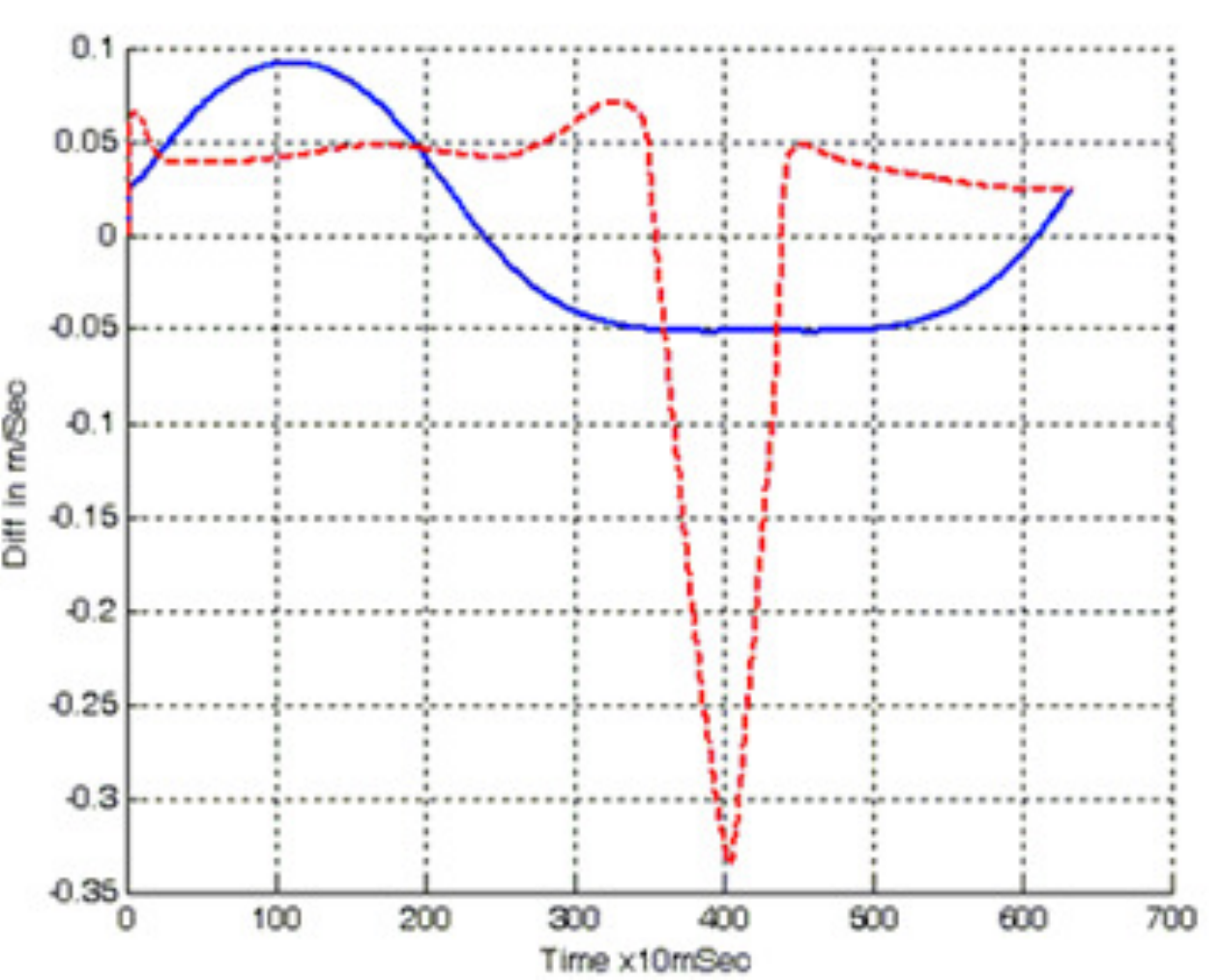}}
    \subfigure [ ]{\label{fig57-d}\includegraphics[width=6cm, height=5cm] {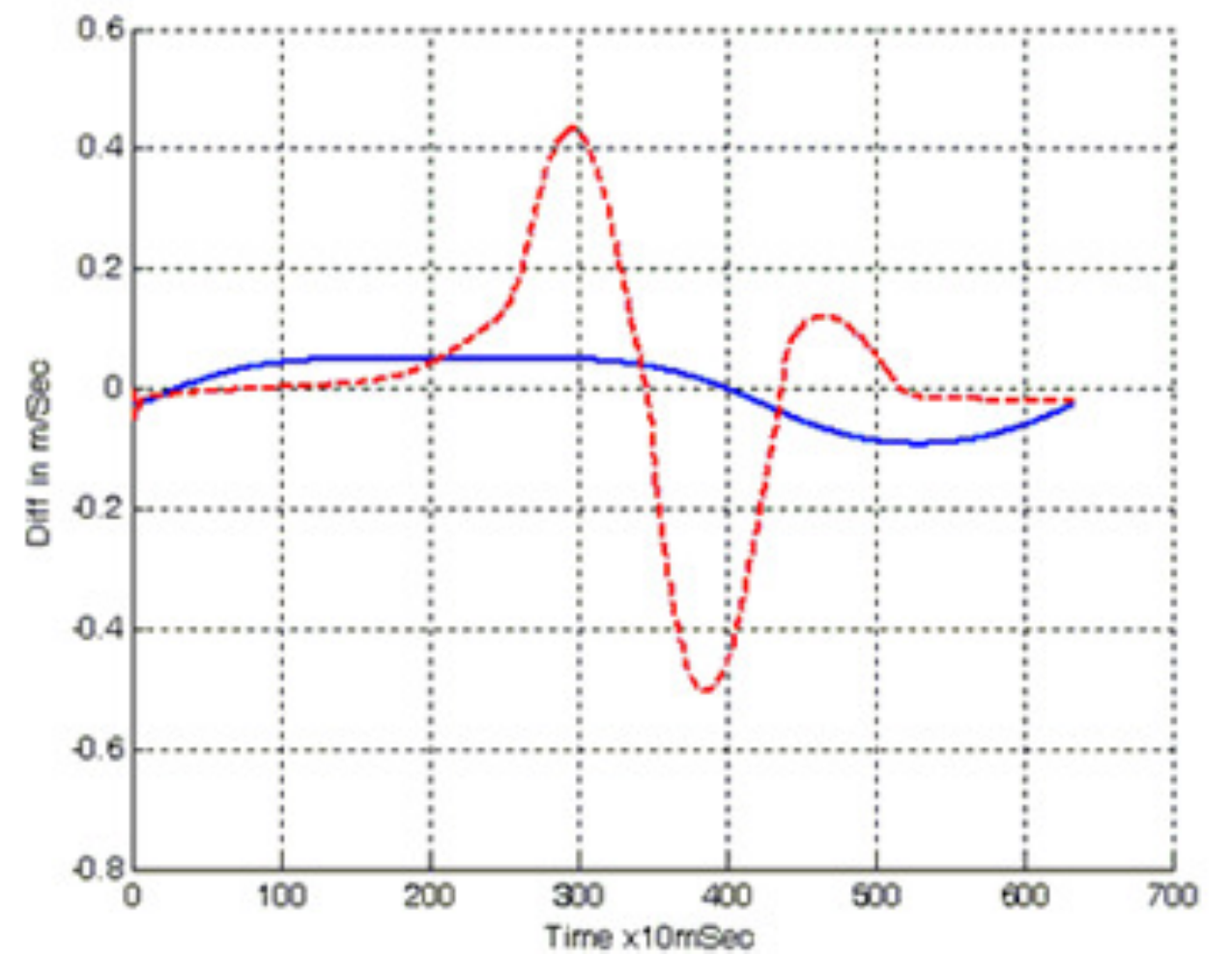}}\\
\end{center}
  \label{fig57}
\end{figure} 

\begin{figure}[htp]
  \begin{center}
    \subfigure [ ]{\label{fig57-e}\includegraphics[width=7cm, height=6cm] {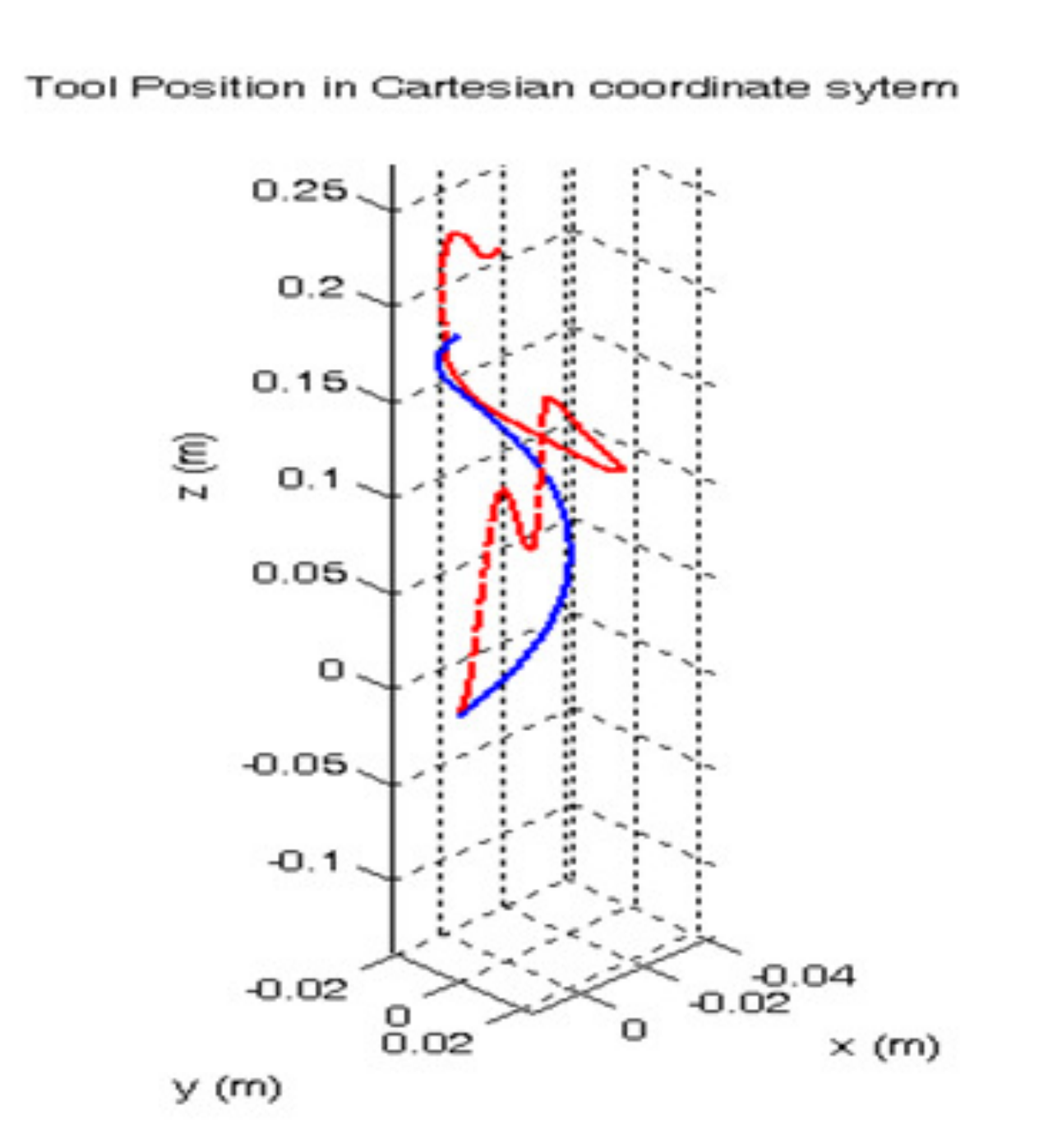}}\\
\end{center}
  \caption[T1 Fuzzy controller, SNR=15dB: (a)-(f)...]{T1 Fuzzy controller, SNR=15dB: (a)-(f) Manipulator's desired and actual outputs,  \{$\Delta \underbar{q}_{12}$,$\Delta \underbar{q}_{13}$, $\Delta \dot{ \underbar{q}}_{12}$, $\Delta \dot{ \underbar{q}}_{13}$\} during the simulation time  (g) The followed desired and actual trajectory}
  \label{fig57}
\end{figure}

\begin{figure}[htp]
  \begin{center} 
   \subfigure [ ]{\label{fig58-a}\includegraphics[width=6cm, height=5cm] {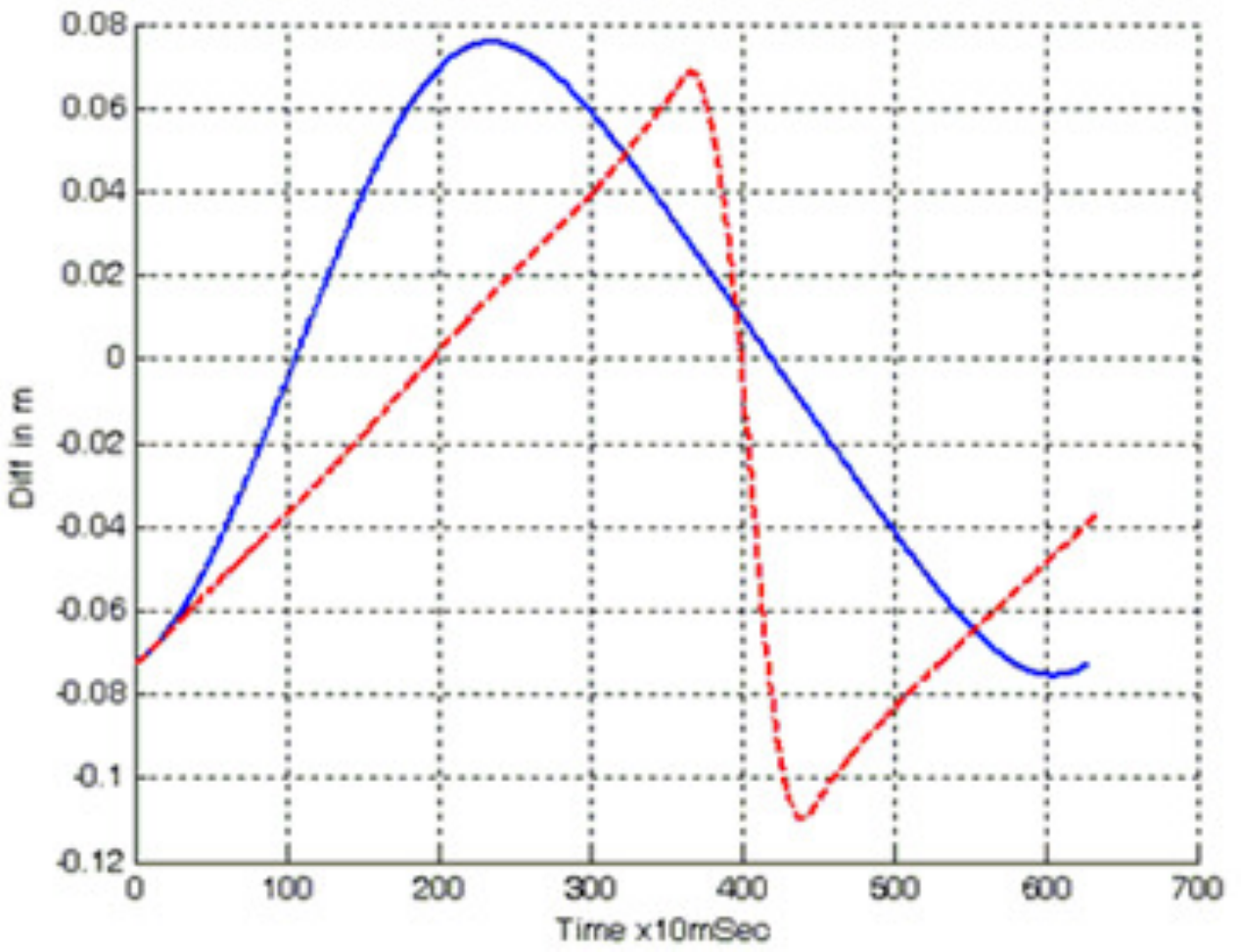}}
    \subfigure [ ]{\label{fig58-b}\includegraphics[width=6cm, height=5cm] {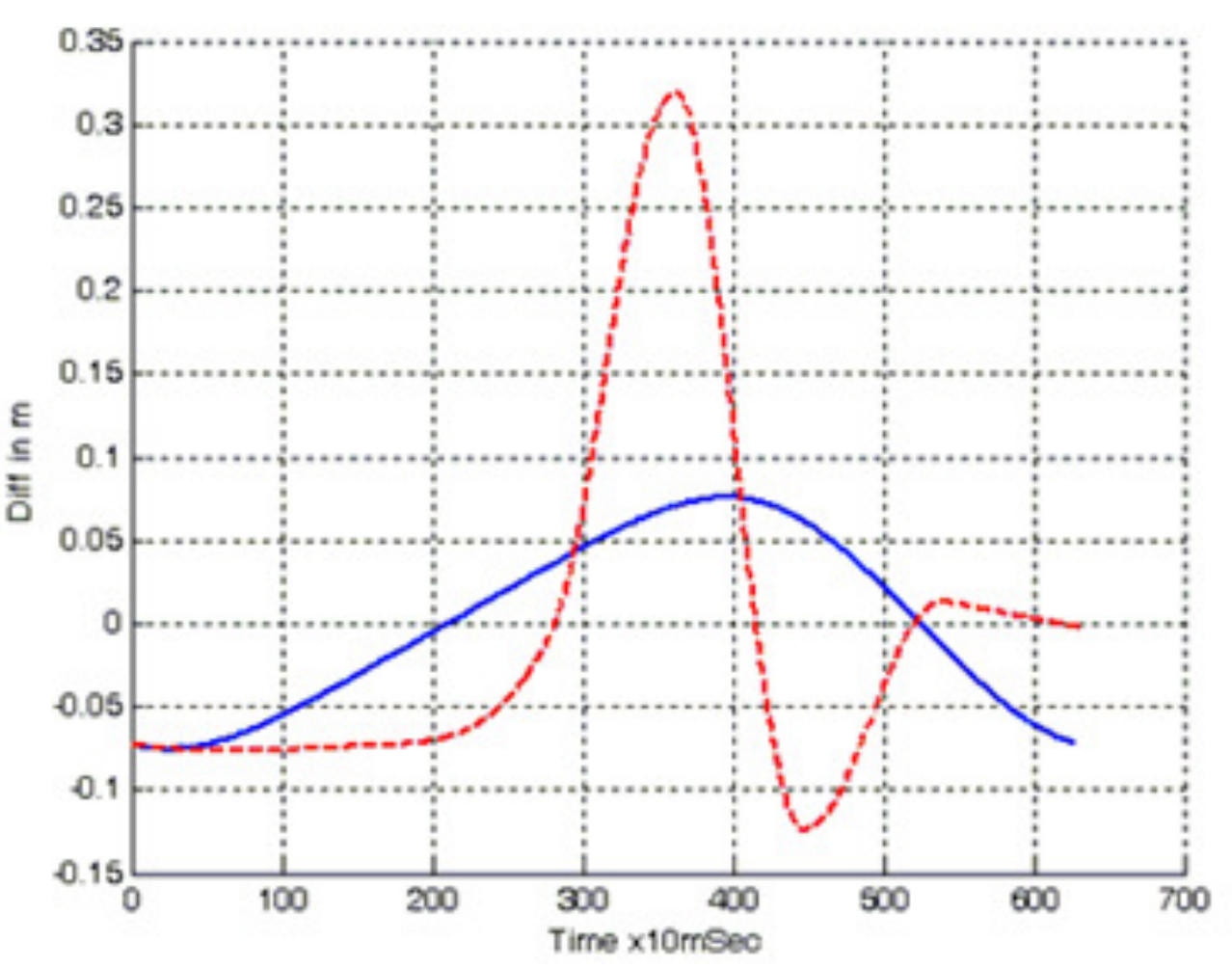}}\\
    \subfigure [ ]{\label{fig58-c}\includegraphics[width=6cm, height=5cm] {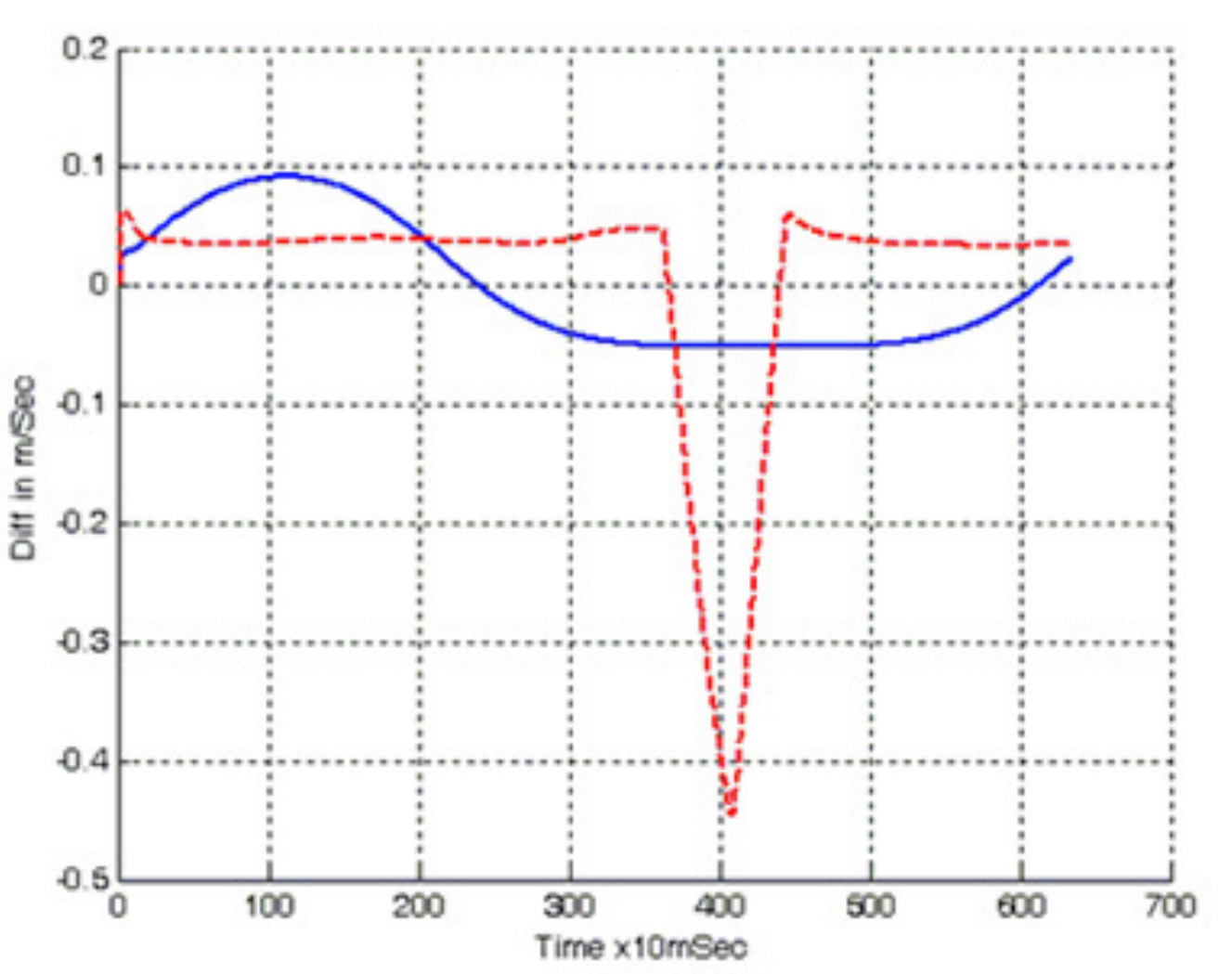}}
    \subfigure [ ]{\label{fig58-d}\includegraphics[width=6cm, height=5cm] {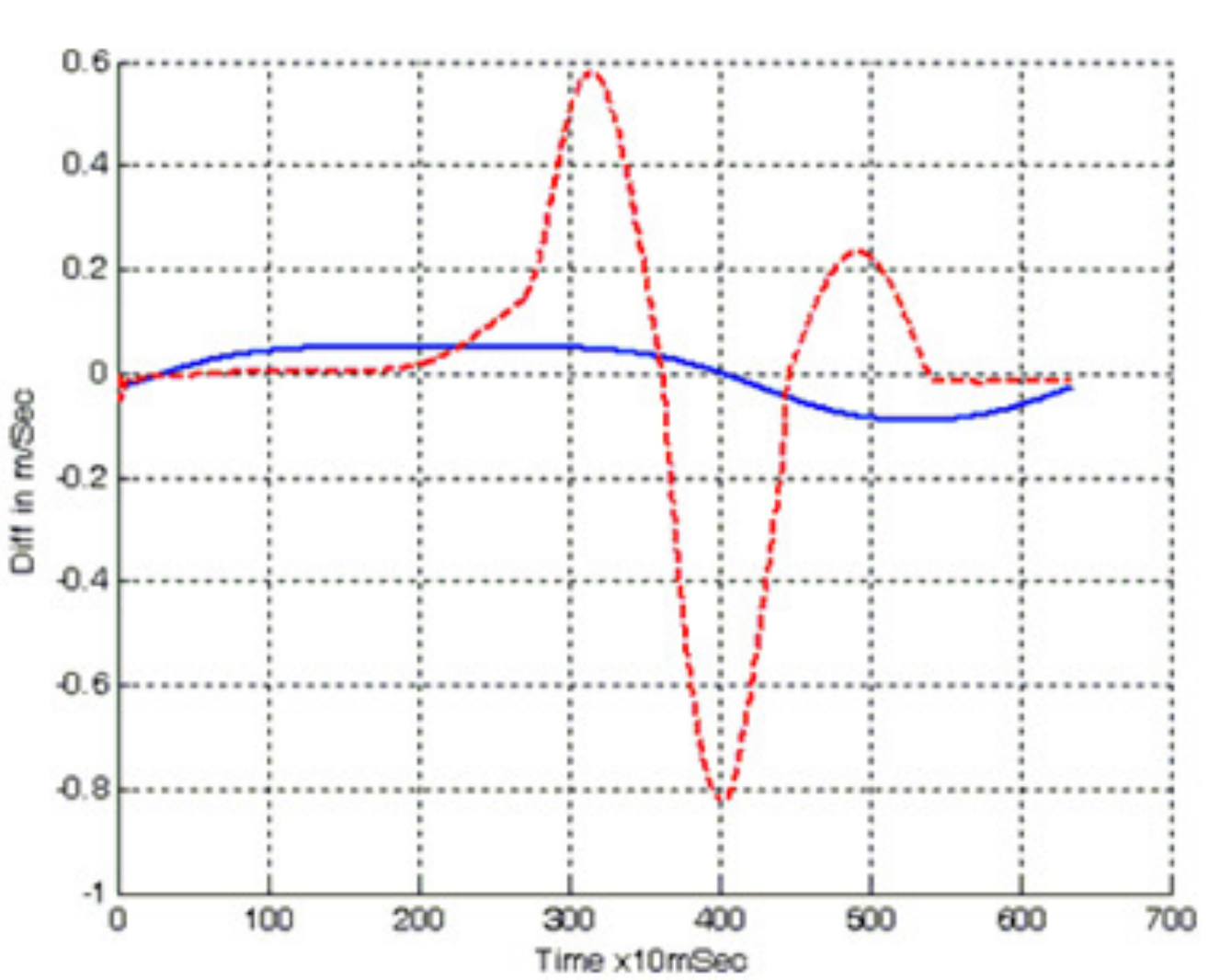}}\\
\end{center}
  \label{fig58}
\end{figure} 

\begin{figure}[htp]
  \begin{center} 
    \subfigure [ ]{\label{fig58-e}\includegraphics[width=7cm, height=6cm] {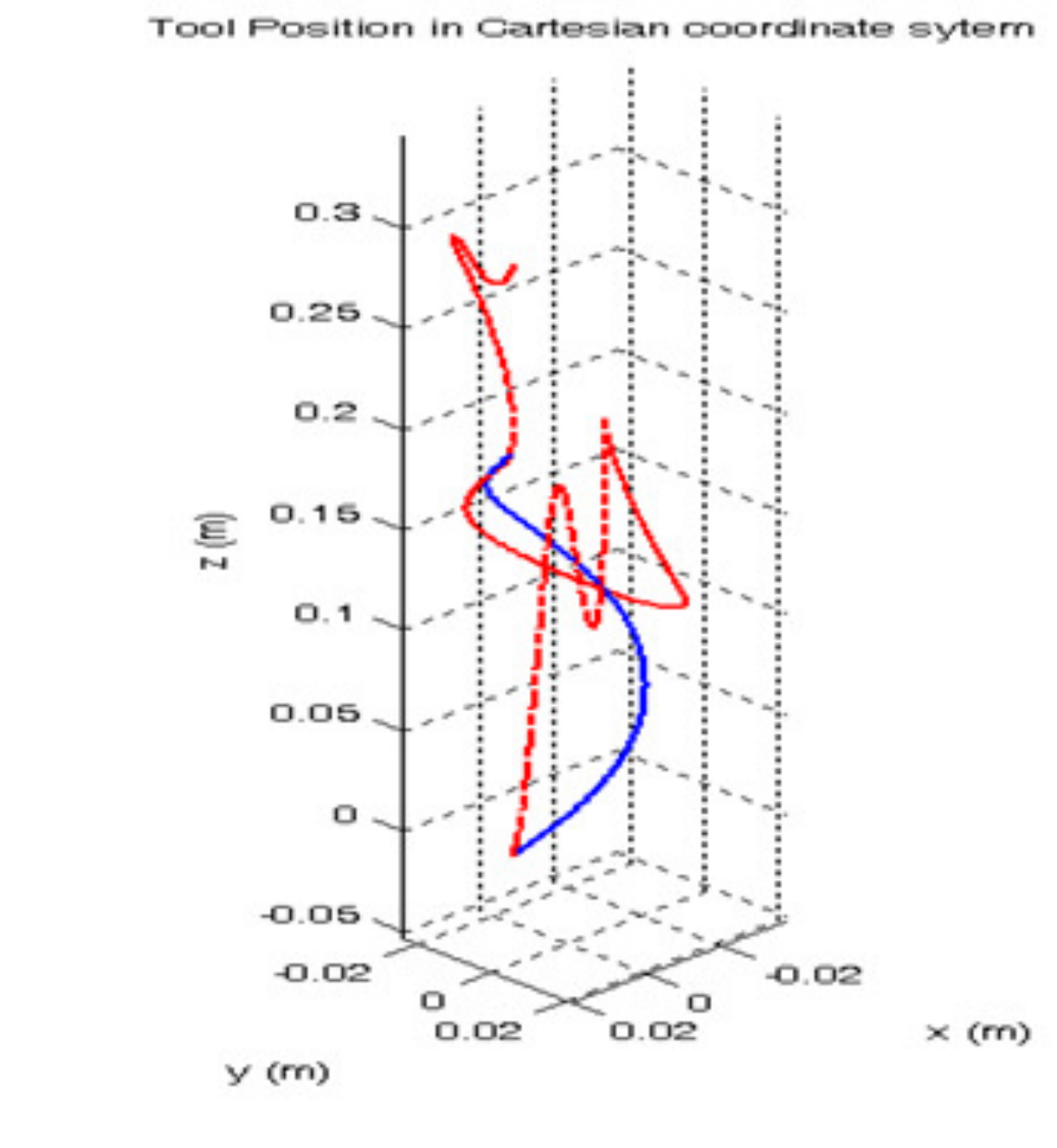}}\\
\end{center}
  \caption[T1 Fuzzy controller, SNR=10dB: (a)-(f)...]{T1 Fuzzy controller, SNR=10dB: (a)-(f) Manipulator's desired and actual outputs,  \{$\Delta \underbar{q}_{12}$,$\Delta \underbar{q}_{13}$, $\Delta \dot{ \underbar{q}}_{12}$, $\Delta \dot{ \underbar{q}}_{13}$\} during the simulation time  (g) The followed desired and actual trajectory}
  \label{fig58}
\end{figure} 

\section{\textbf{Type-II Fuzzy Based CTC}}
\begin{figure}[htp]
  \begin{center} 
    \subfigure [ ]{\label{fig59-a}\includegraphics[width=6cm, height=5cm] {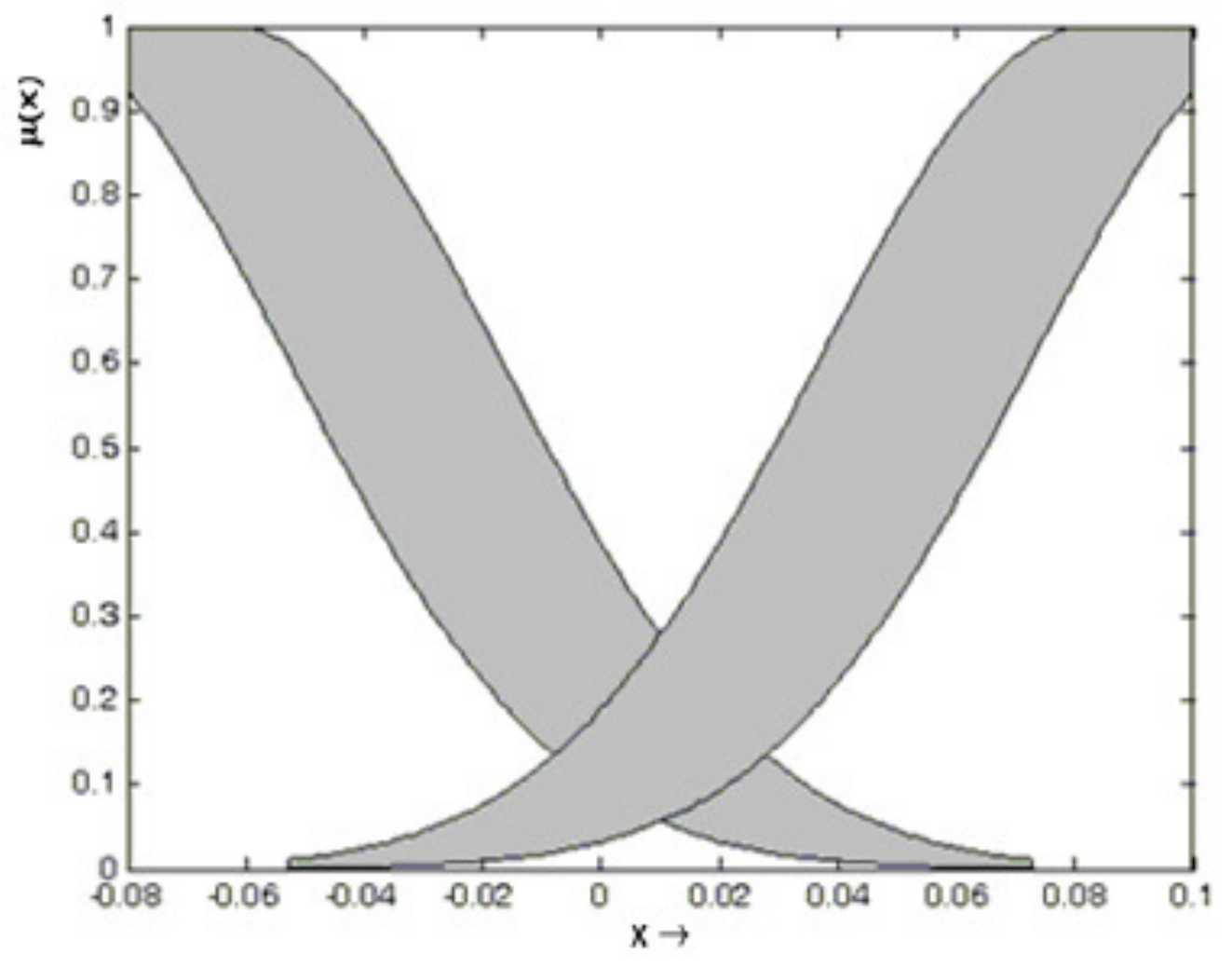}}
    \subfigure [ ]{\label{fig59-b}\includegraphics[width=6cm, height=5cm] {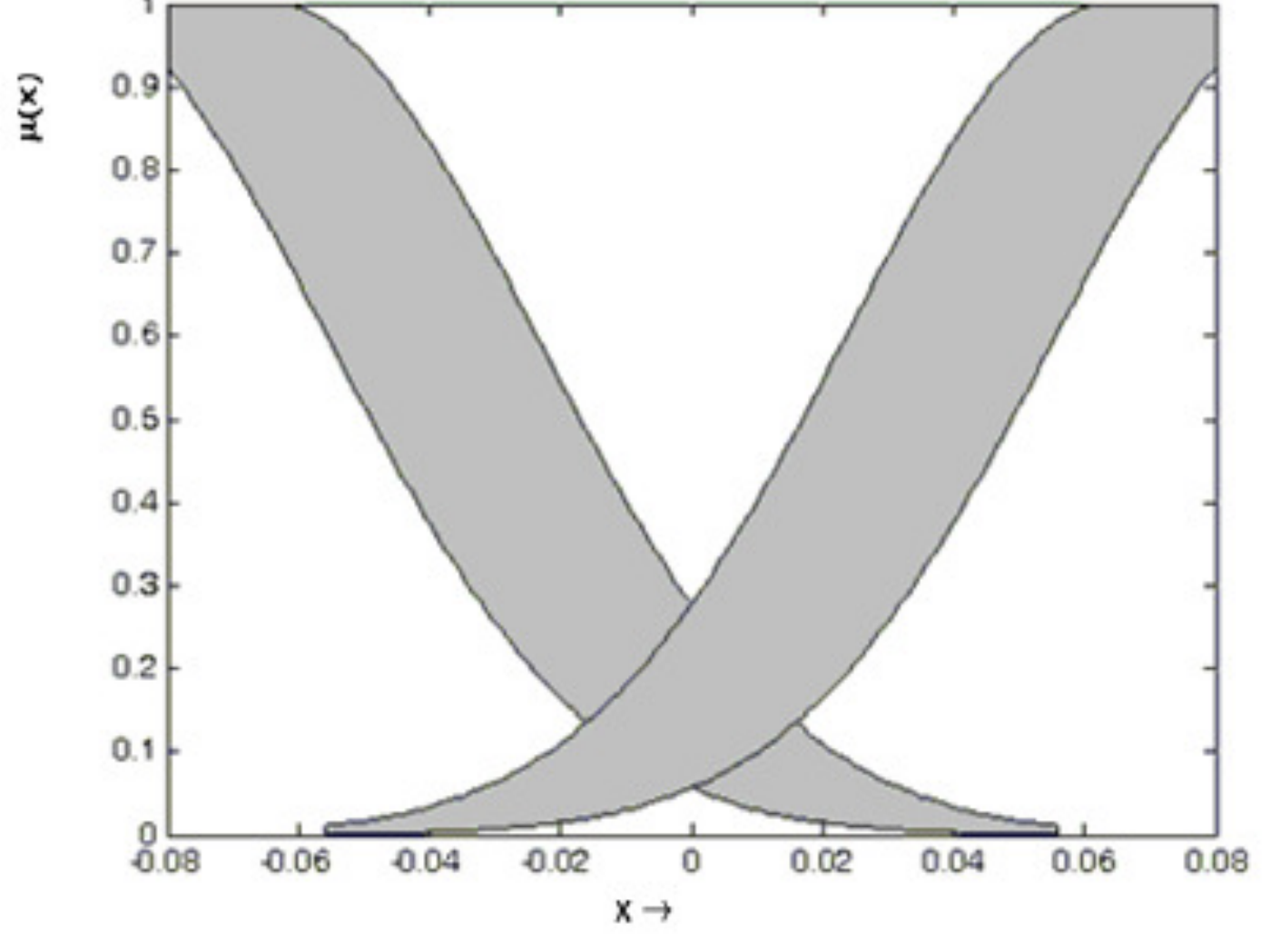}}\\
    \subfigure [ ]{\label{fig59-c}\includegraphics[width=6cm, height=5cm] {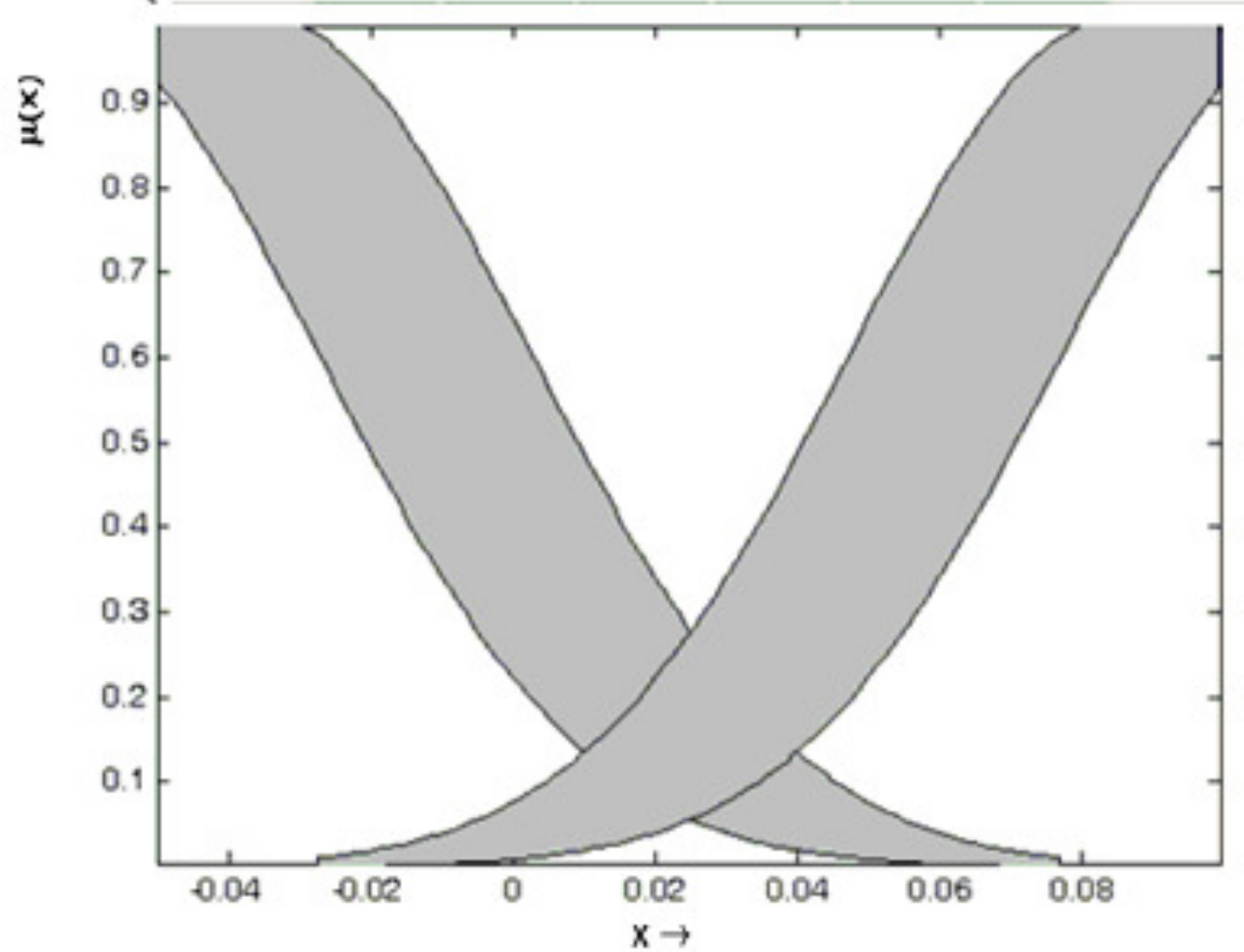}}
    \subfigure [ ]{\label{fig59-d}\includegraphics[width=6cm, height=5cm] {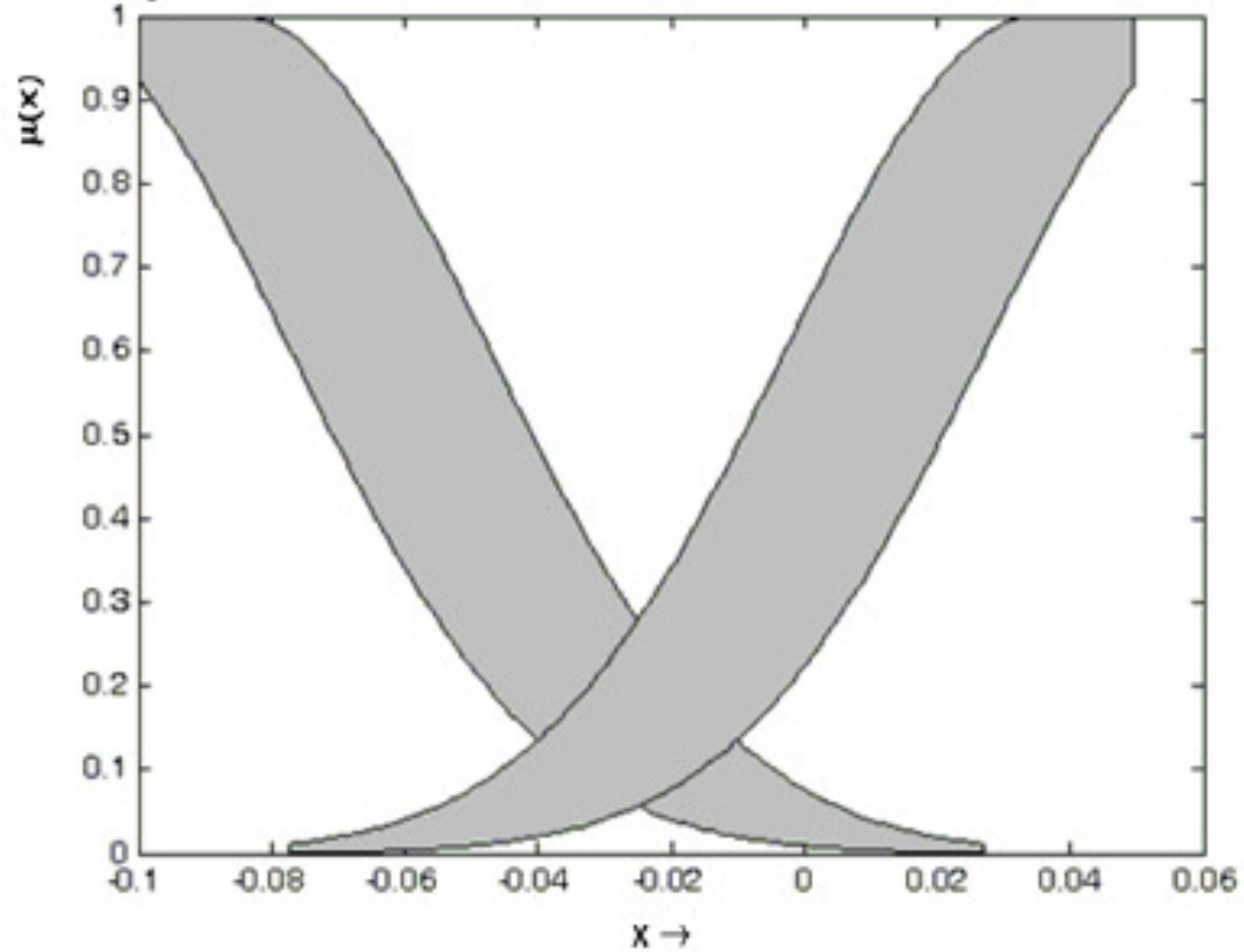}}\\
\end{center}
  \caption{Membership functions corresponding to the inputs space}
  \label{fig59}
\end{figure}

In this section we test our proposed controller in the same way we did for the previous section (Type-I Fuzzy Based CTC). We show that our method is remarkably more robust to noise. \\
In the next scenarios a Helix trajectory is given to the newly introduced controller and the additive noises with different noise levels are injected into the system parameters. Four different trials have been tested each associated to a specific signal to noise ratio (SNR). For each trial, 5 figures are depicted the first four pertaining to real and desired values of $\Delta \underbar{q}_{12}$,$\Delta \underbar{q}_{13}$, $\Delta \dot{ \underbar{q}}_{12}$, $\Delta \dot{ \underbar{q}}_{13}$ during the simulation time and the last shows the traversed trajectory. 
Figure \ref{fig59} depicts the membership functions corresponding to the input space of the T1FLC. As explained earlier, these MFs are in fact obtained by downgrading the T2 MFs of the T2FLS given later.

\begin{figure}[htp]
  \begin{center} 
    \subfigure [ ]{\label{fig60-a}\includegraphics[width=6cm, height=5cm] {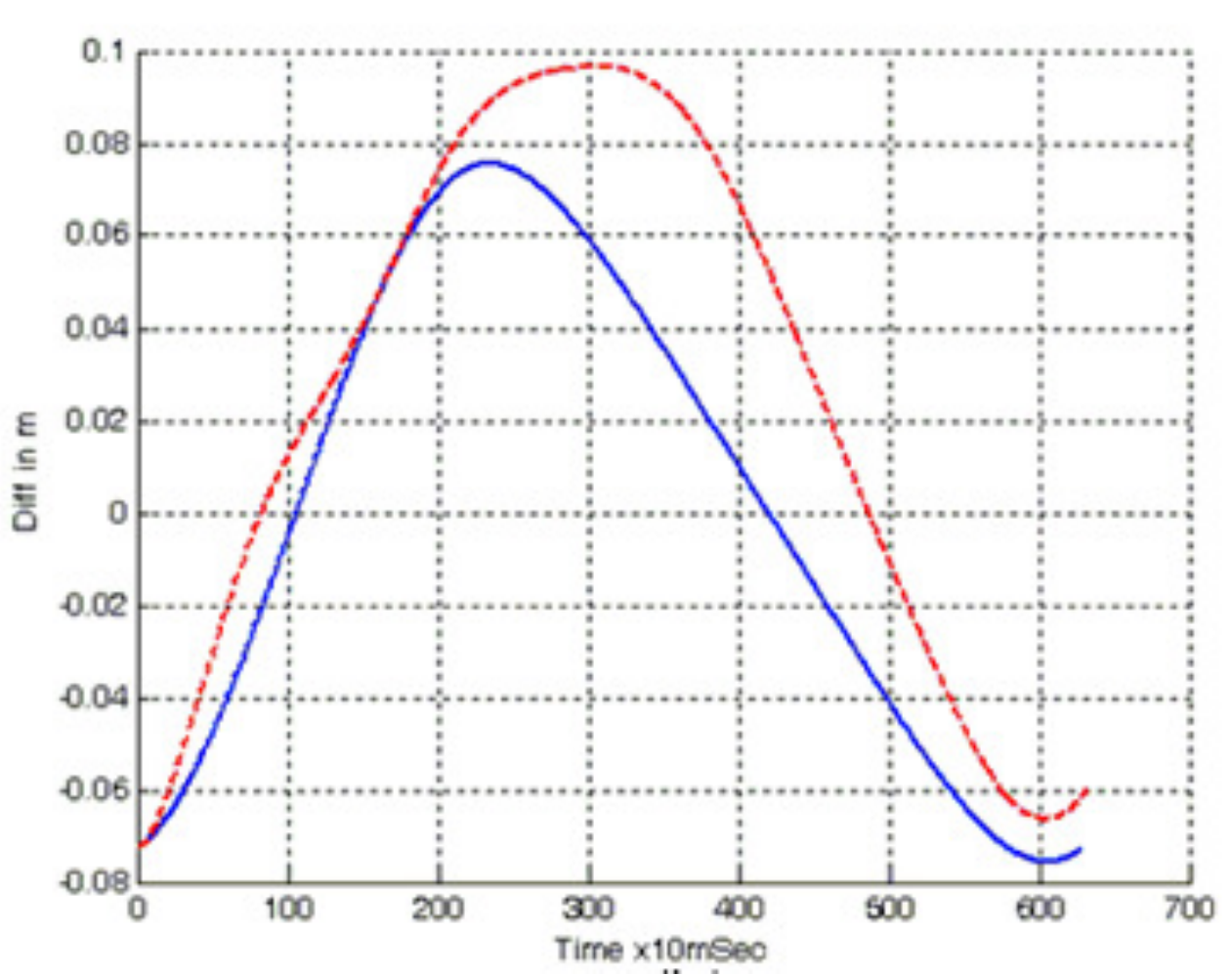}}
    \subfigure [ ]{\label{fig60-b}\includegraphics[width=6cm, height=5cm] {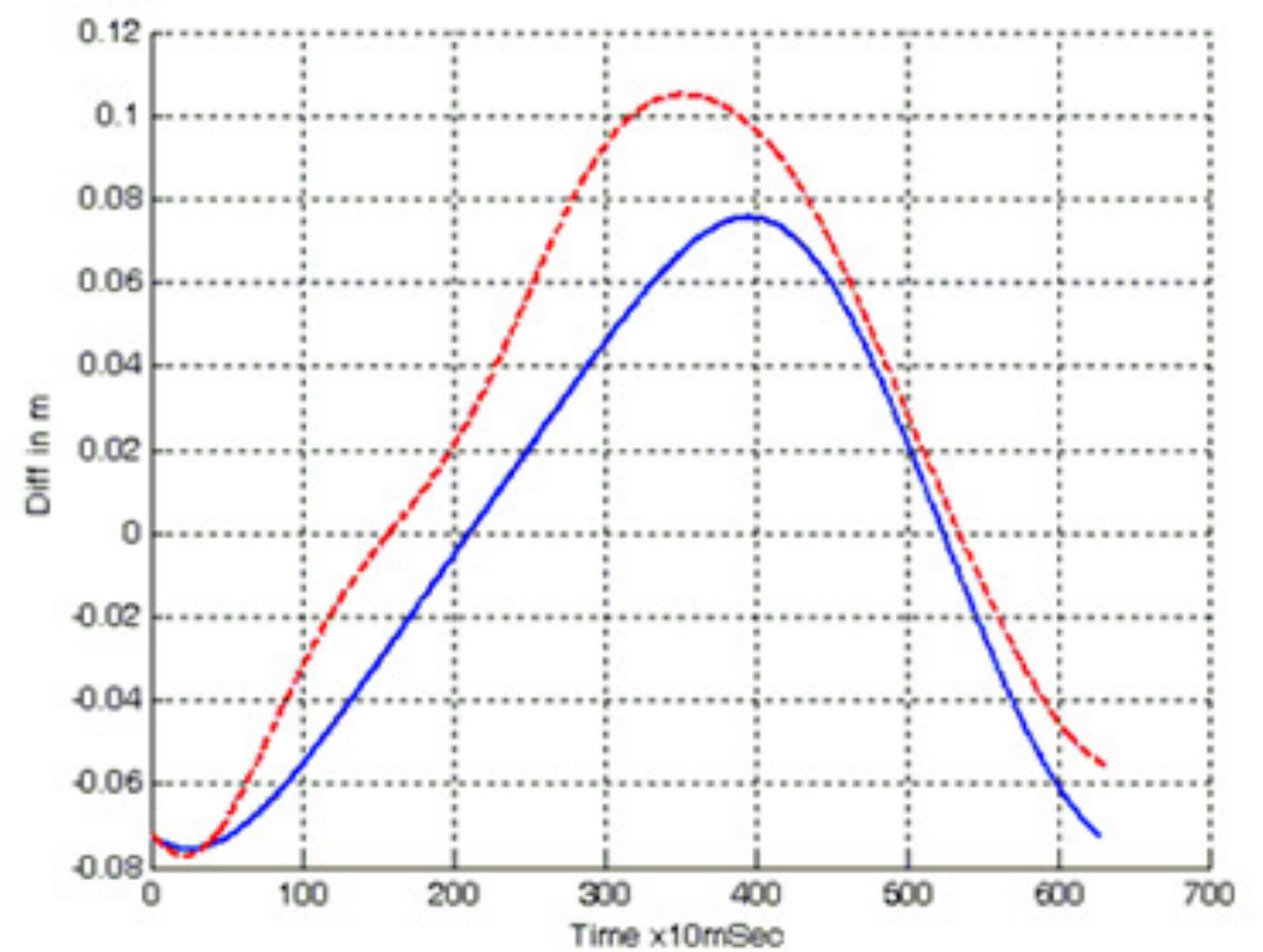}}\\
    \subfigure [ ]{\label{fig60-c}\includegraphics[width=6cm, height=5cm] {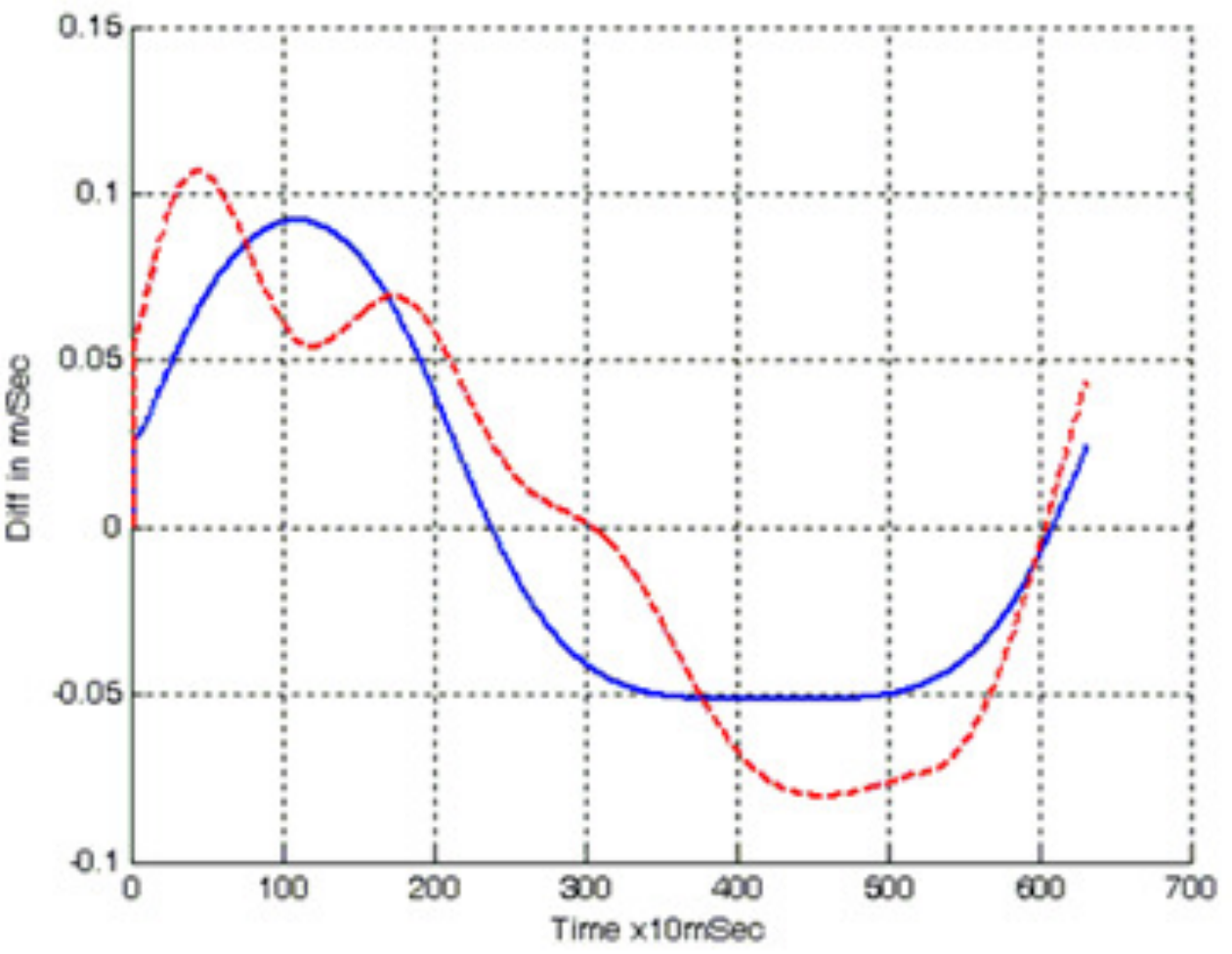}}
    \subfigure [ ]{\label{fig60-d}\includegraphics[width=6cm, height=5cm] {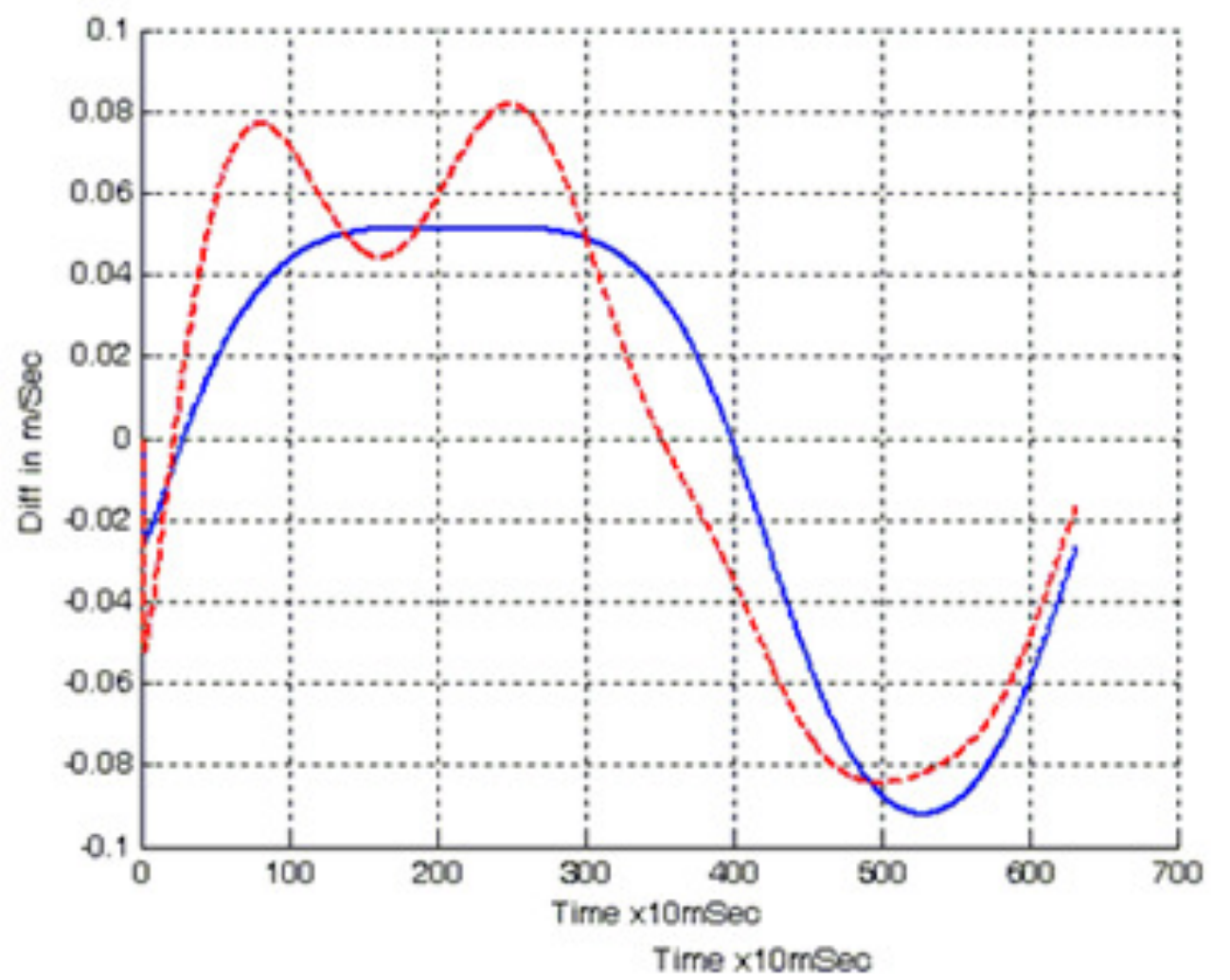}}\\
\end{center}
  \label{fig60}
\end{figure} 

\begin{figure}[htp]
  \begin{center} 
    \subfigure [ ]{\label{fig60-e}\includegraphics[width=7cm, height=6cm] {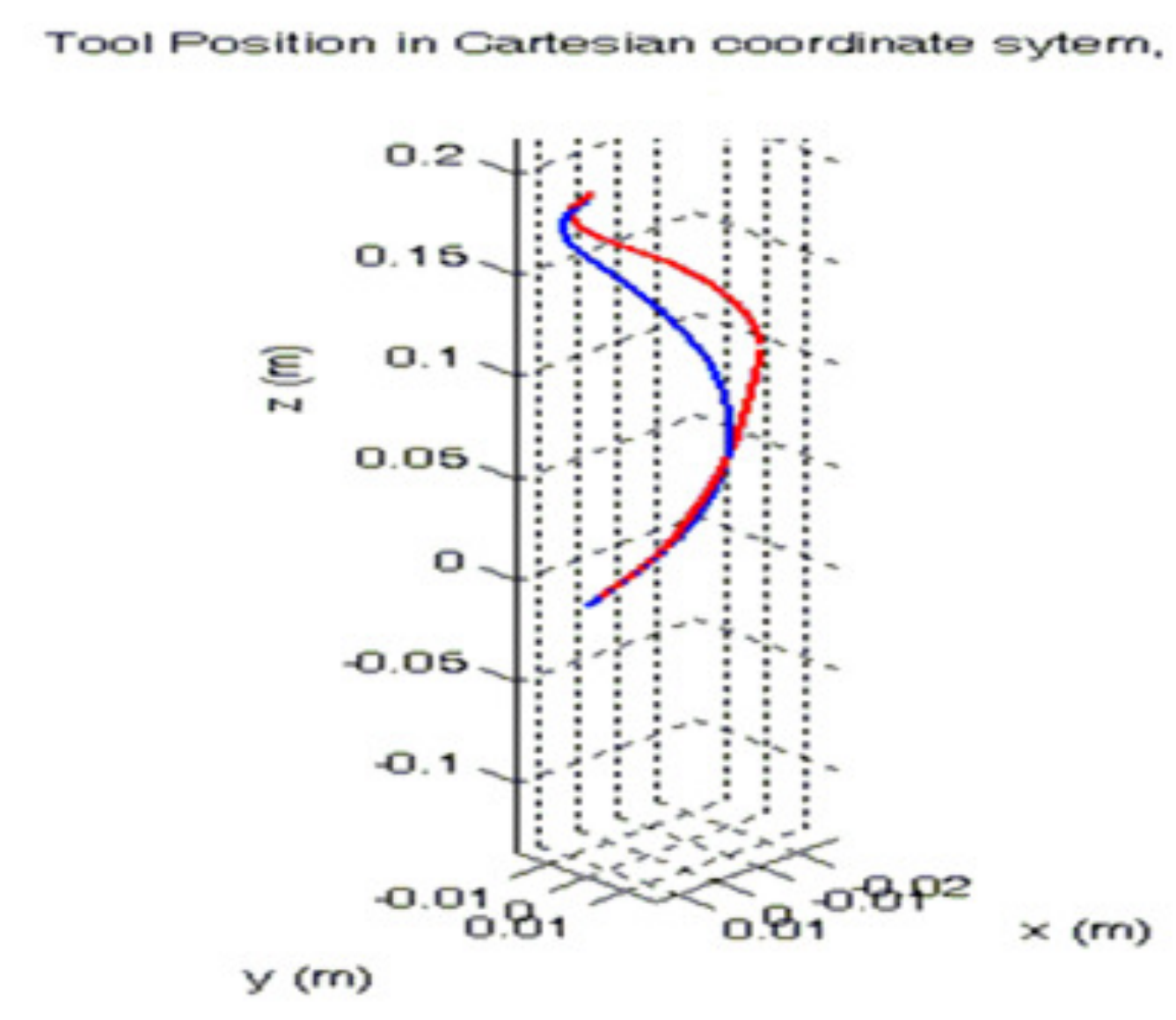}}\\
\end{center}
  \caption[T2 Fuzzy controller, SNR=$\infty$: (a)-(f)...]{T2 Fuzzy controller, SNR=$\infty$: (a)-(f) Manipulator's desired and actual outputs,  \{$\Delta \underbar{q}_{12}$,$\Delta \underbar{q}_{13}$, $\Delta \dot{ \underbar{q}}_{12}$, $\Delta \dot{ \underbar{q}}_{13}$\} during the simulation time  (g) The followed desired and actual trajectory}
  \label{fig60}
\end{figure} 

\begin{figure}[htp]
  \begin{center} 
   \subfigure [ ]{\label{fig61-a}\includegraphics[width=6cm, height=5cm] {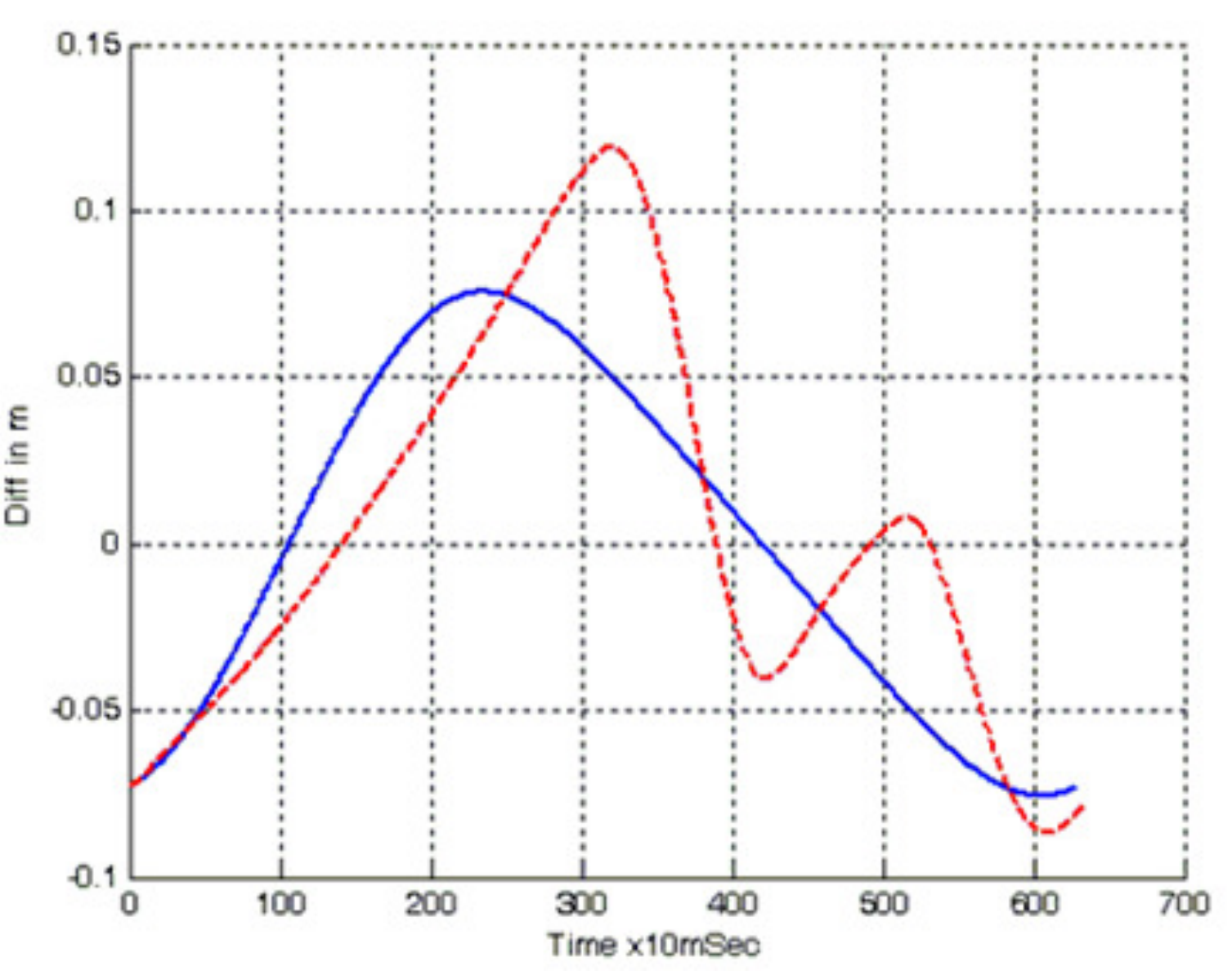}}
    \subfigure [ ]{\label{fig61-b}\includegraphics[width=6cm, height=5cm] {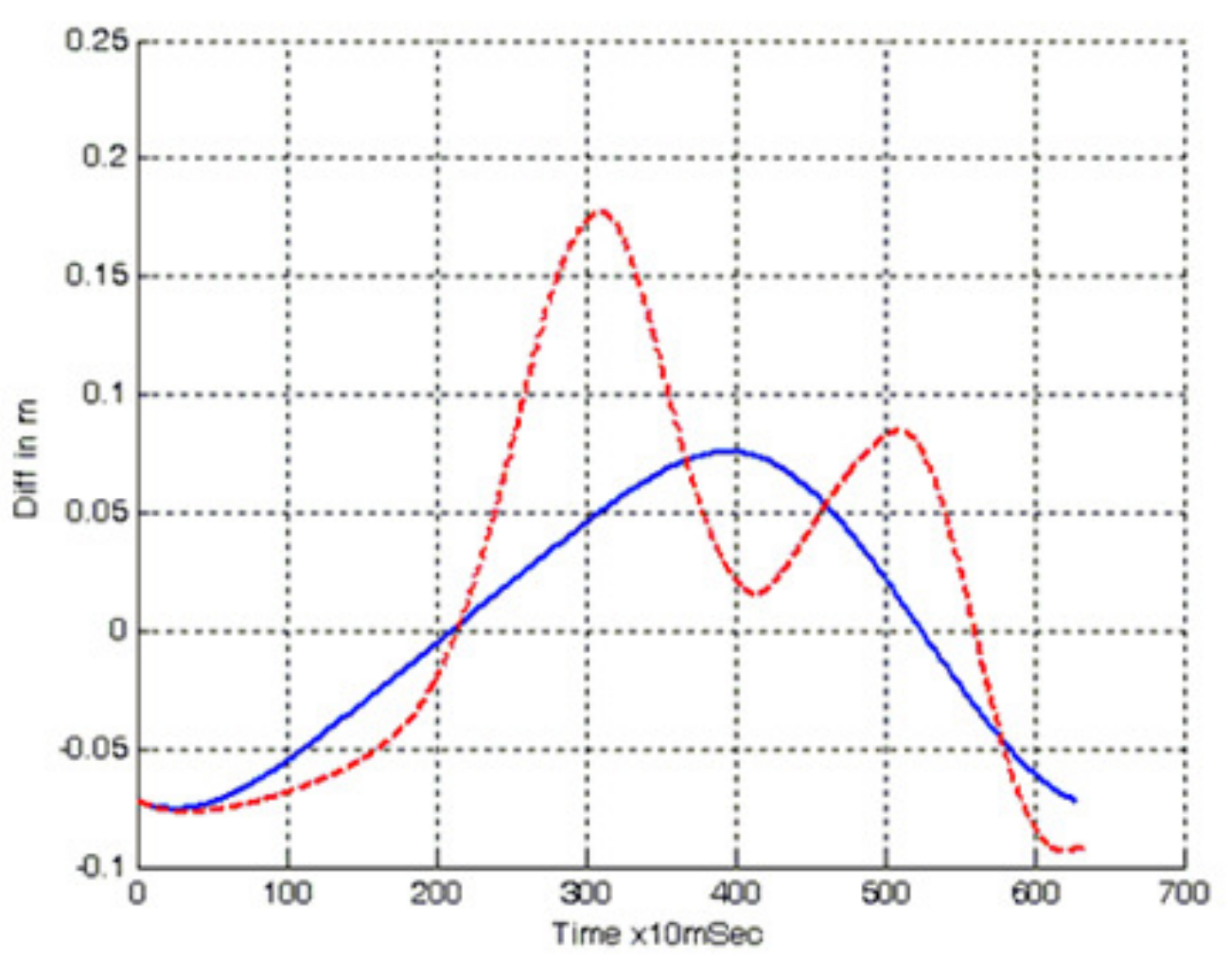}}\\
    \subfigure [ ]{\label{fig61-c}\includegraphics[width=6cm, height=5cm] {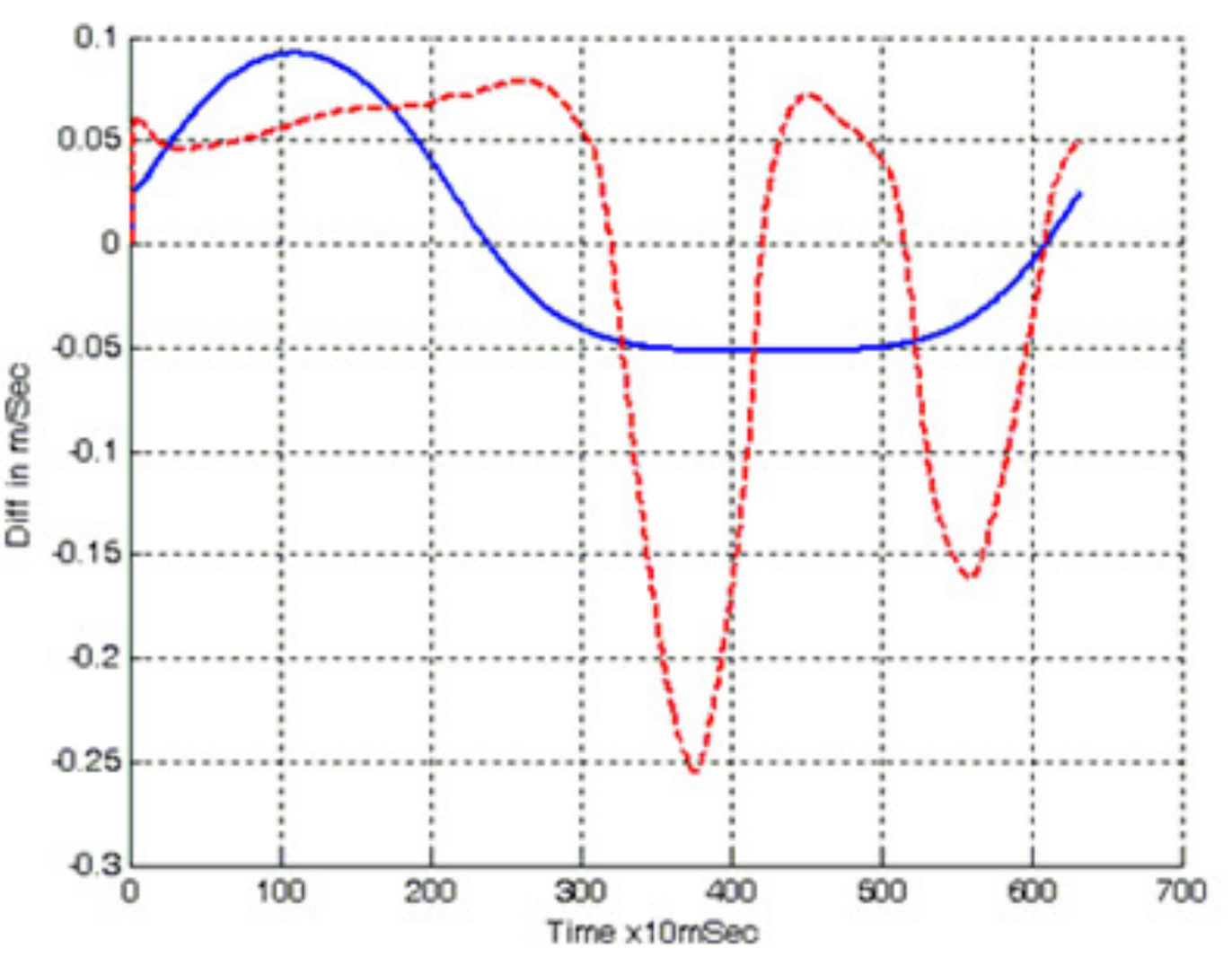}}
    \subfigure [ ]{\label{fig61-d}\includegraphics[width=6cm, height=5cm] {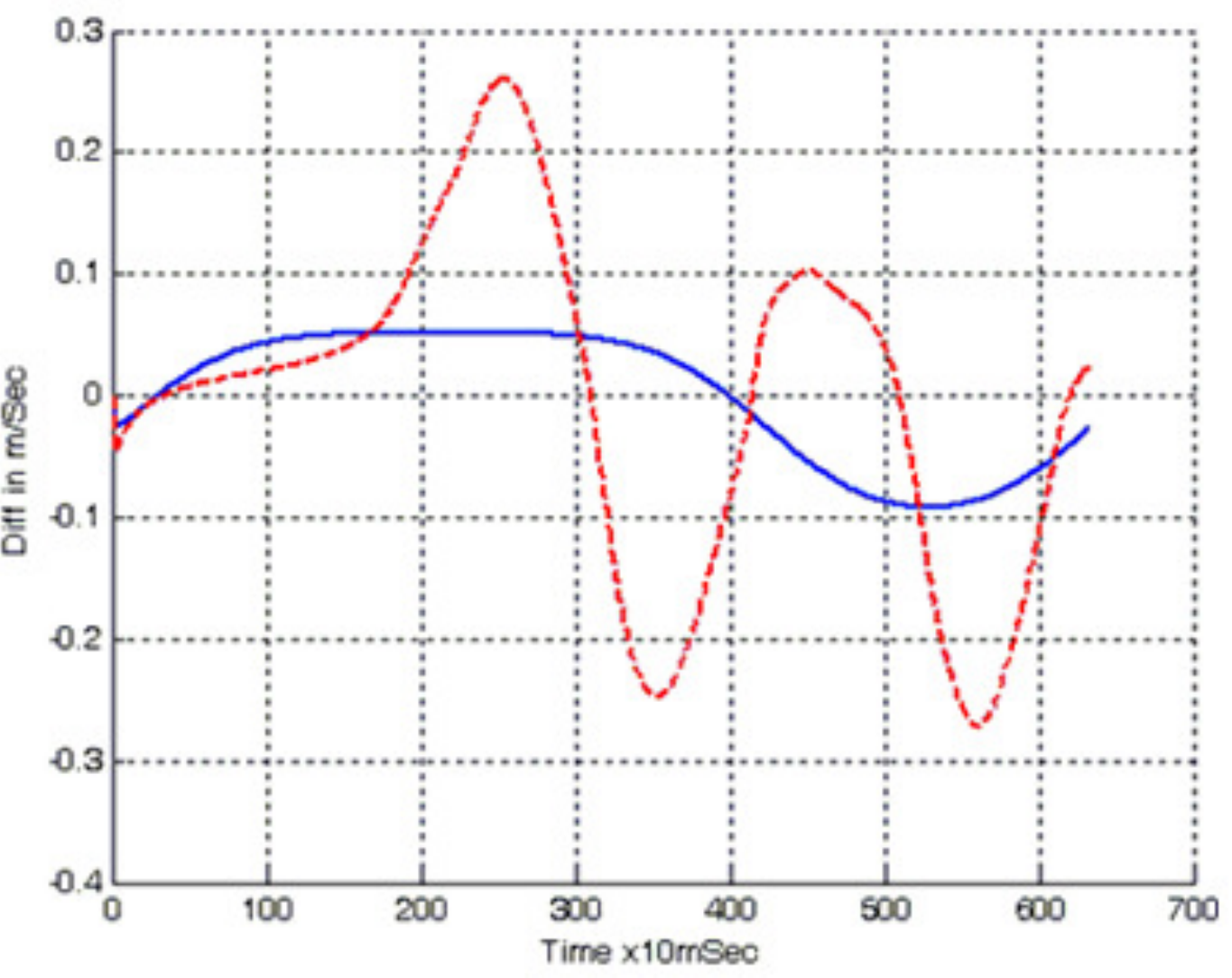}}\\
\end{center}
  \label{fig61}
\end{figure} 

\begin{figure}[htp]
  \begin{center} 
    \subfigure [ ]{\label{fig61-e}\includegraphics[width=7cm, height=6cm] {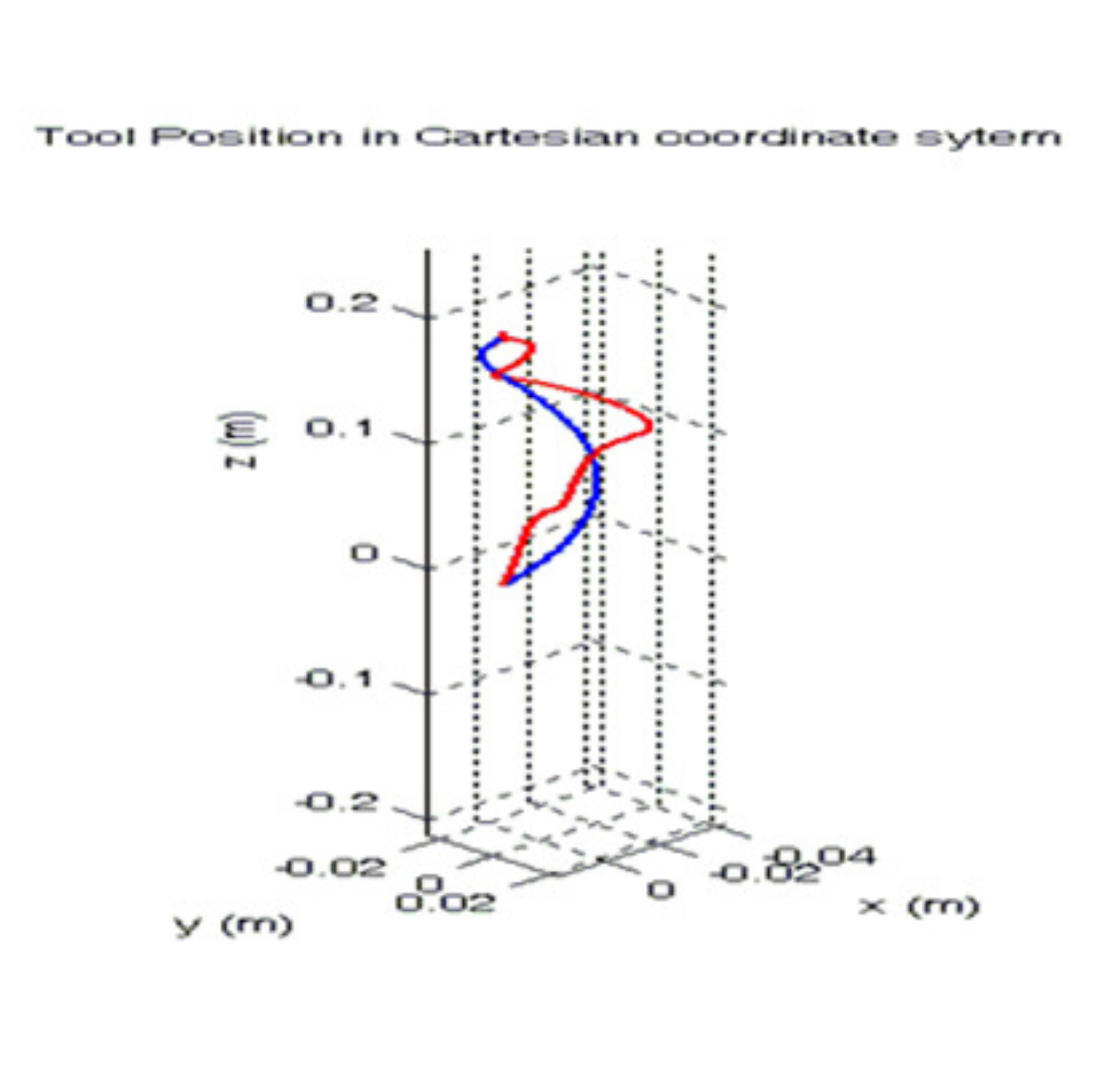}}\\
\end{center}
  \caption[T2 Fuzzy controller, SNR=20dB: (a)-(f)...]{T2 Fuzzy controller, SNR=20dB: (a)-(f) Manipulator's desired and actual outputs,  \{$\Delta \underbar{q}_{12}$,$\Delta \underbar{q}_{13}$, $\Delta \dot{ \underbar{q}}_{12}$, $\Delta \dot{ \underbar{q}}_{13}$\} during the simulation time  (g) The followed desired and actual trajectory}
  \label{fig61}
\end{figure}

\begin{figure}[htp]
  \begin{center} 
   \subfigure [ ]{\label{fig62-a}\includegraphics[width=6cm, height=5cm] {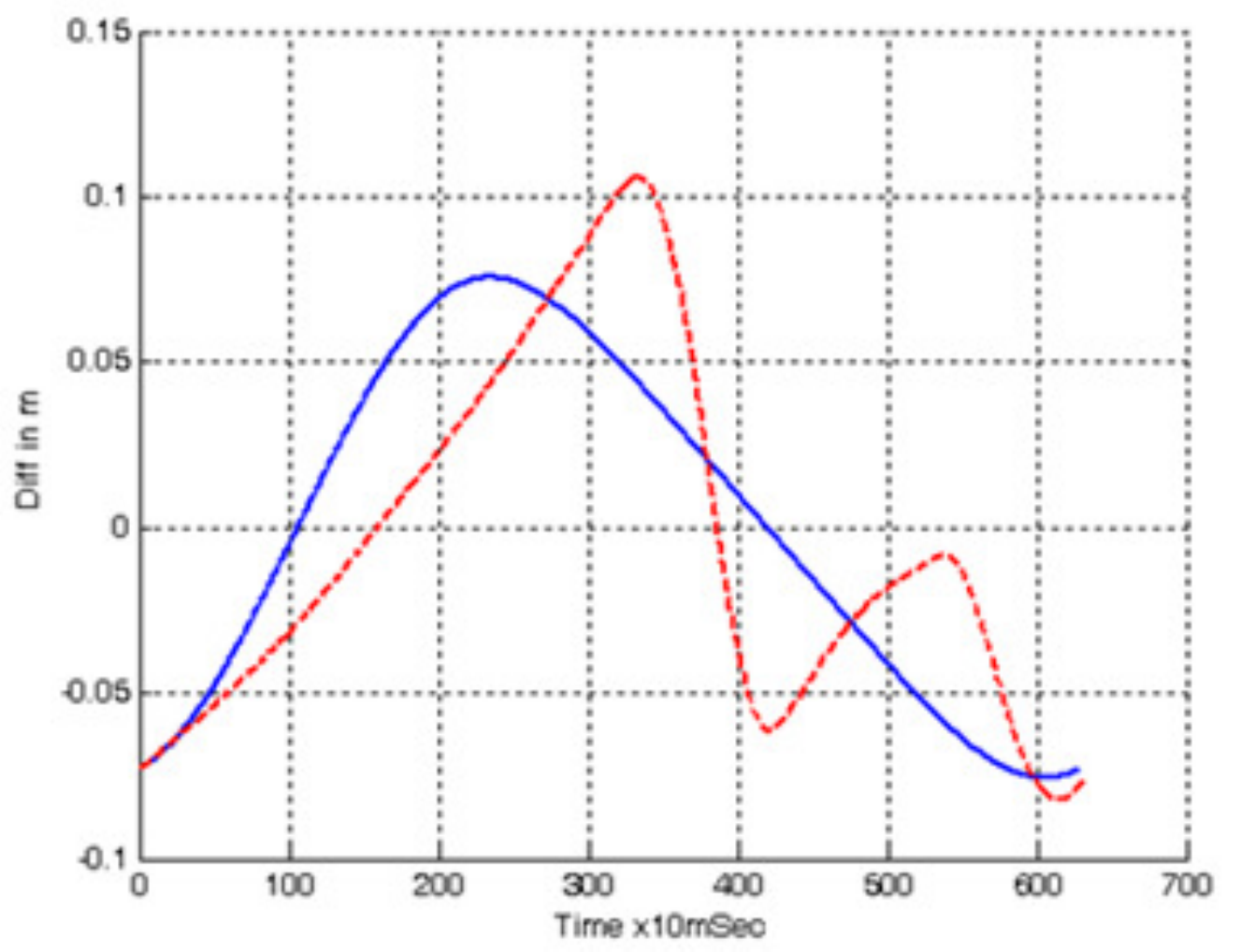}}
    \subfigure [ ]{\label{fig62-b}\includegraphics[width=6cm, height=5cm] {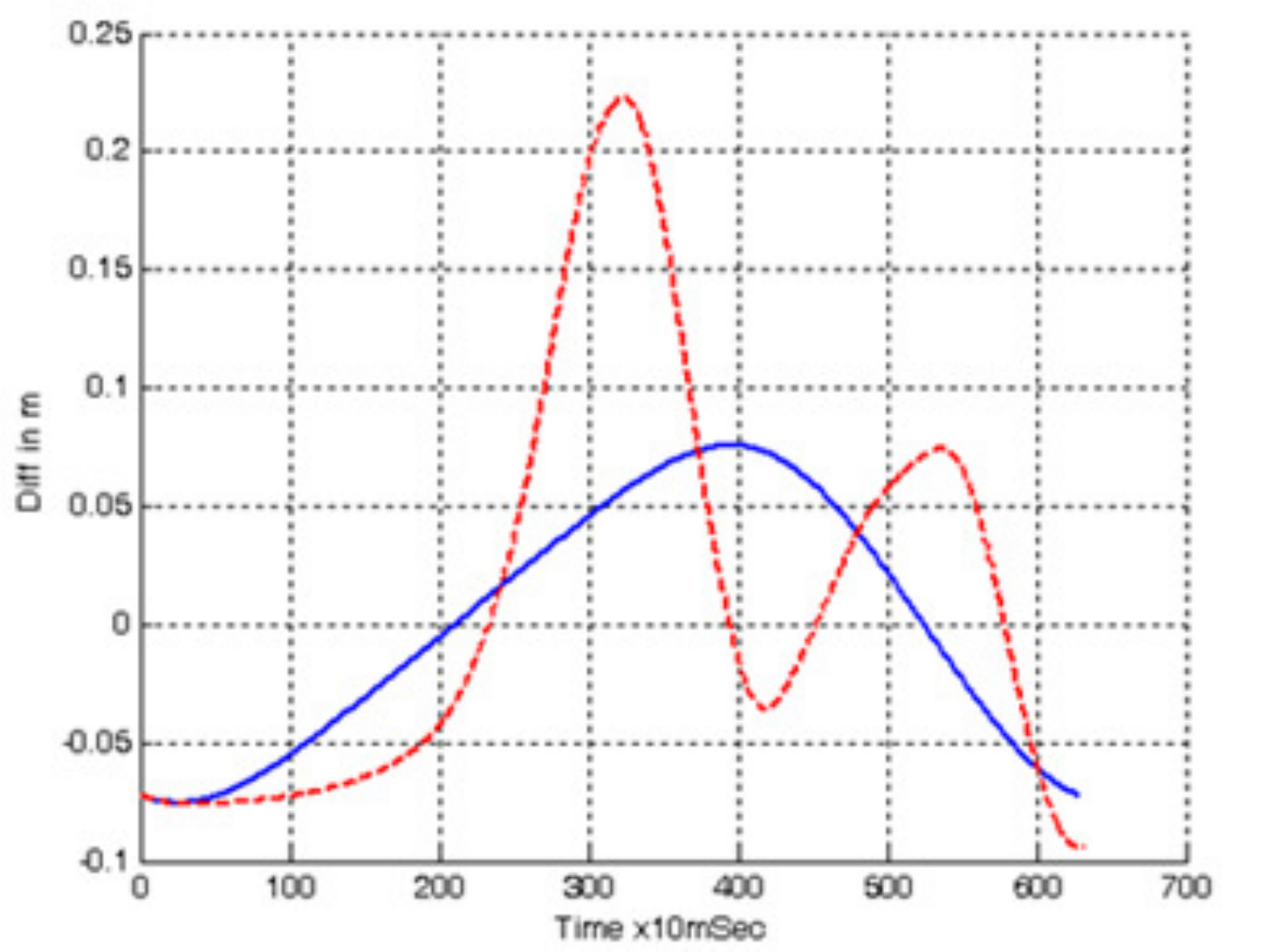}}\\
    \subfigure [ ]{\label{fig62-c}\includegraphics[width=6cm, height=5cm] {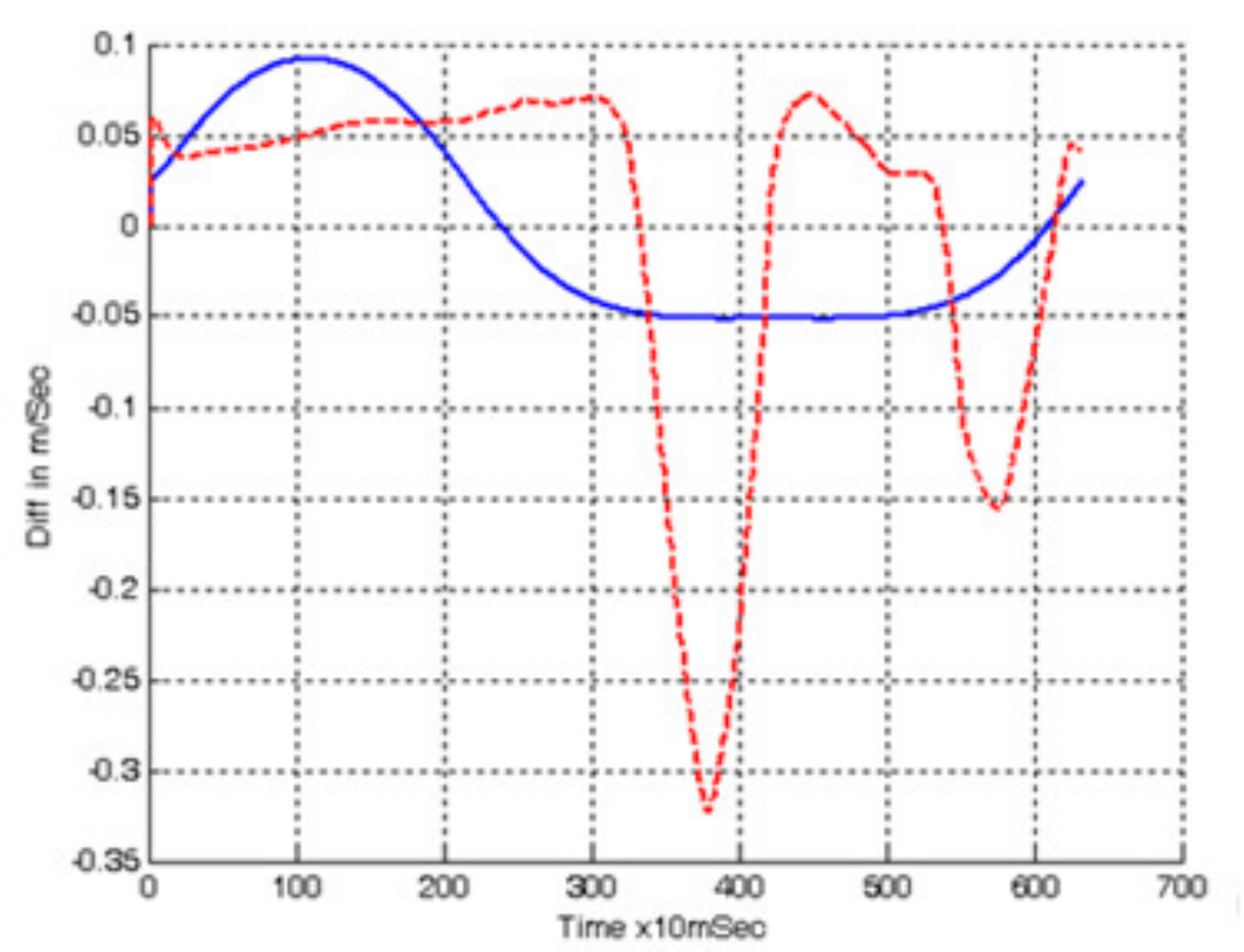}}
    \subfigure [ ]{\label{fig62-d}\includegraphics[width=6cm, height=5cm] {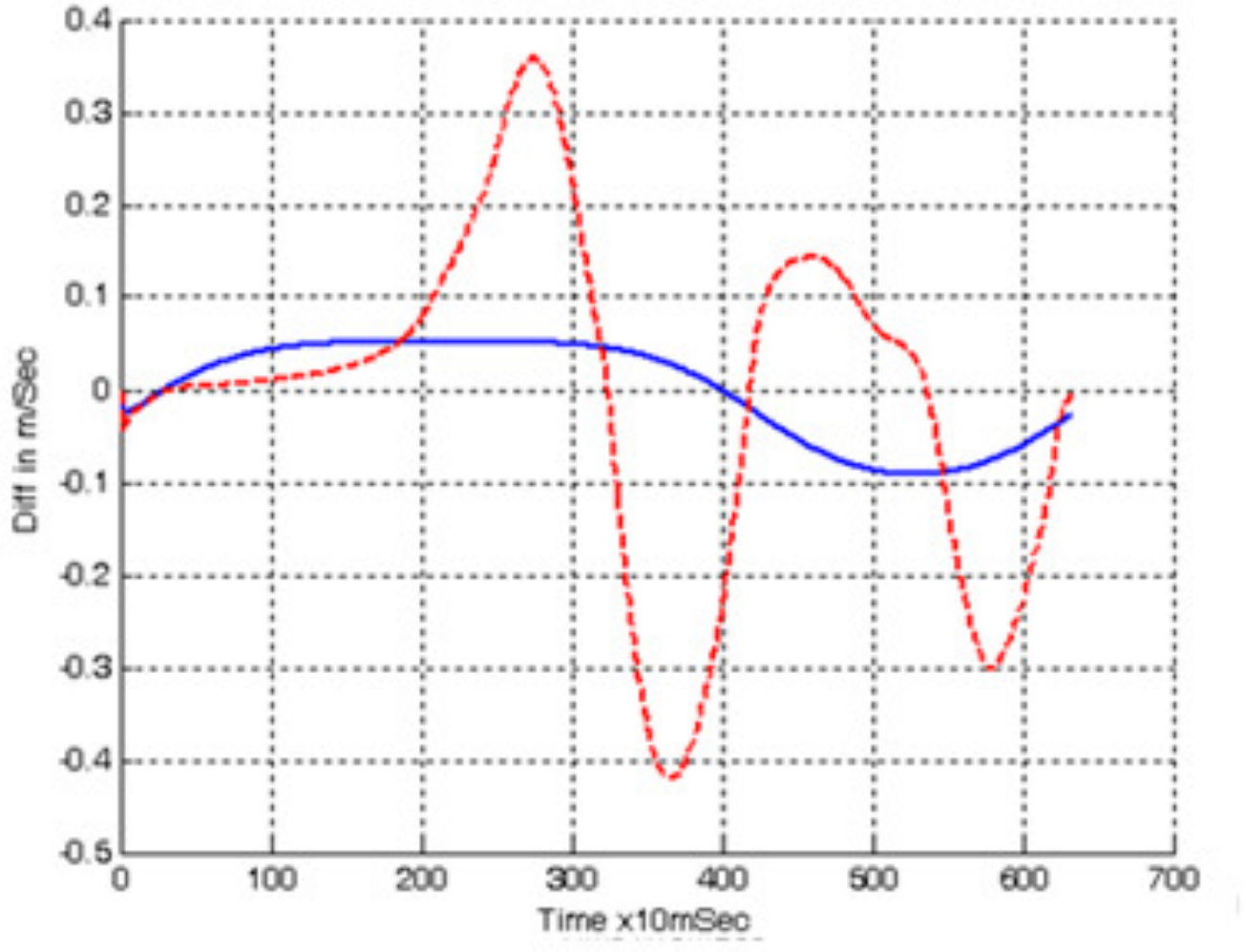}}\\
 \end{center}
 \label{fig62}
\end{figure} 

\begin{figure}[htp]
  \begin{center} 
   \subfigure [ ]{\label{fig62-e}\includegraphics[width=7cm, height=6cm] {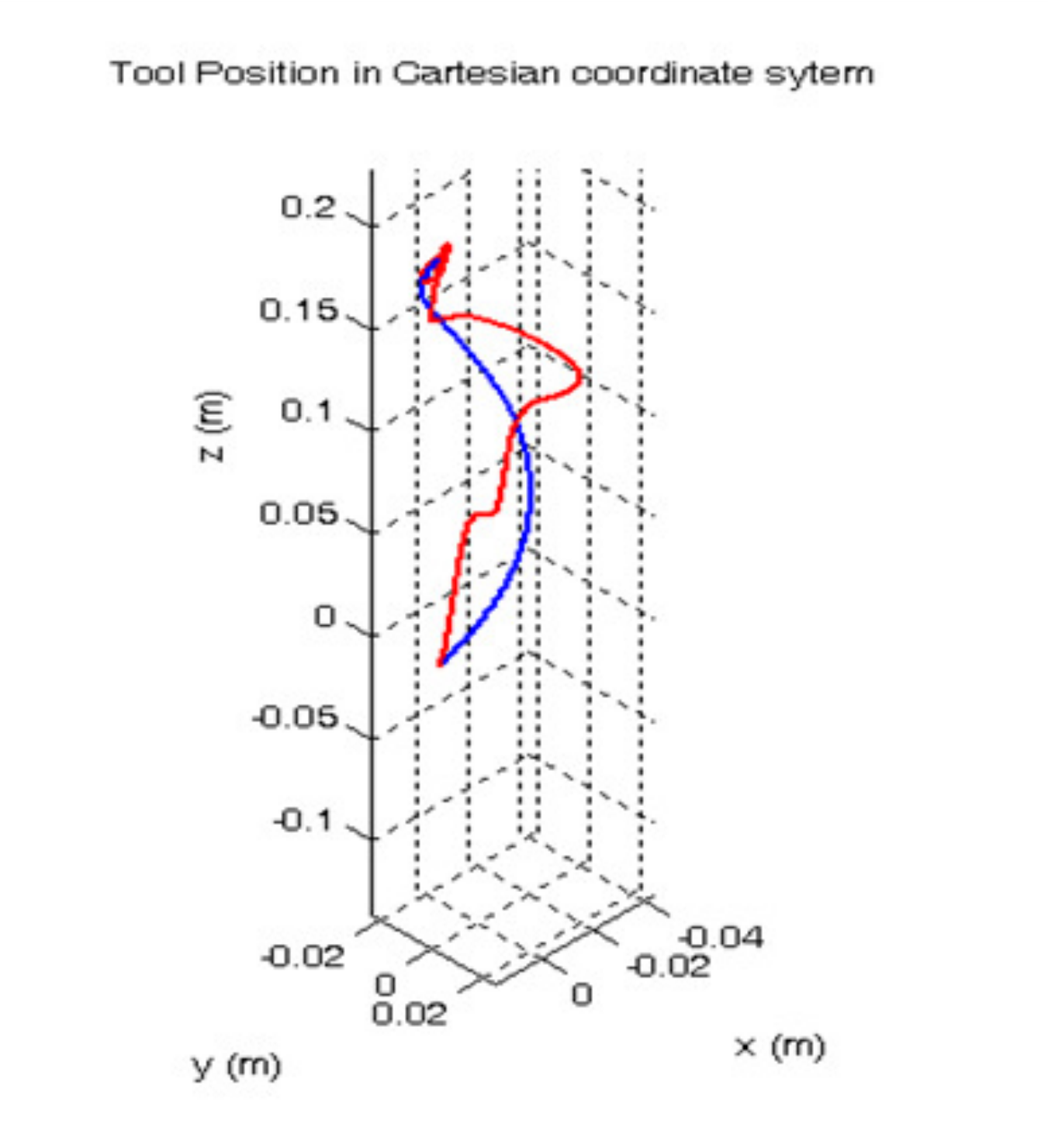}}\\
\end{center}
 \caption[T2 Fuzzy controller, SNR=15dB: (a)-(f)...]{T2 Fuzzy controller, SNR=15dB: (a)-(f) Manipulator's desired and actual outputs,  \{$\Delta \underbar{q}_{12}$,$\Delta \underbar{q}_{13}$, $\Delta \dot{ \underbar{q}}_{12}$, $\Delta \dot{ \underbar{q}}_{13}$\} during the simulation time  (g) The followed desired and actual trajectory}
  \label{fig62}
\end{figure} 

\begin{figure}[htp]
  \begin{center} 
   \subfigure [ ]{\label{fig63-a}\includegraphics[width=6cm, height=5cm] {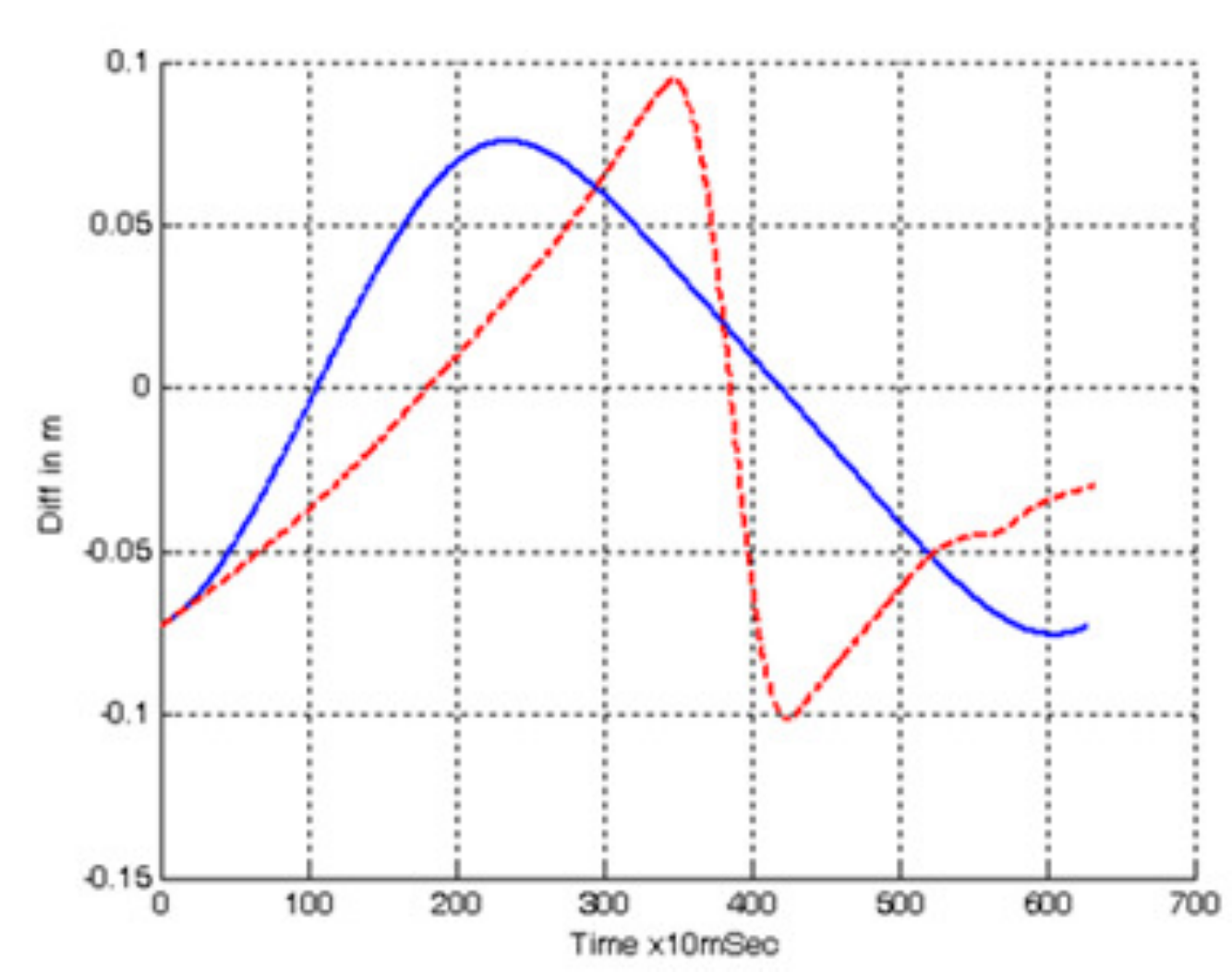}}
    \subfigure [ ]{\label{fig63-b}\includegraphics[width=6cm, height=5cm] {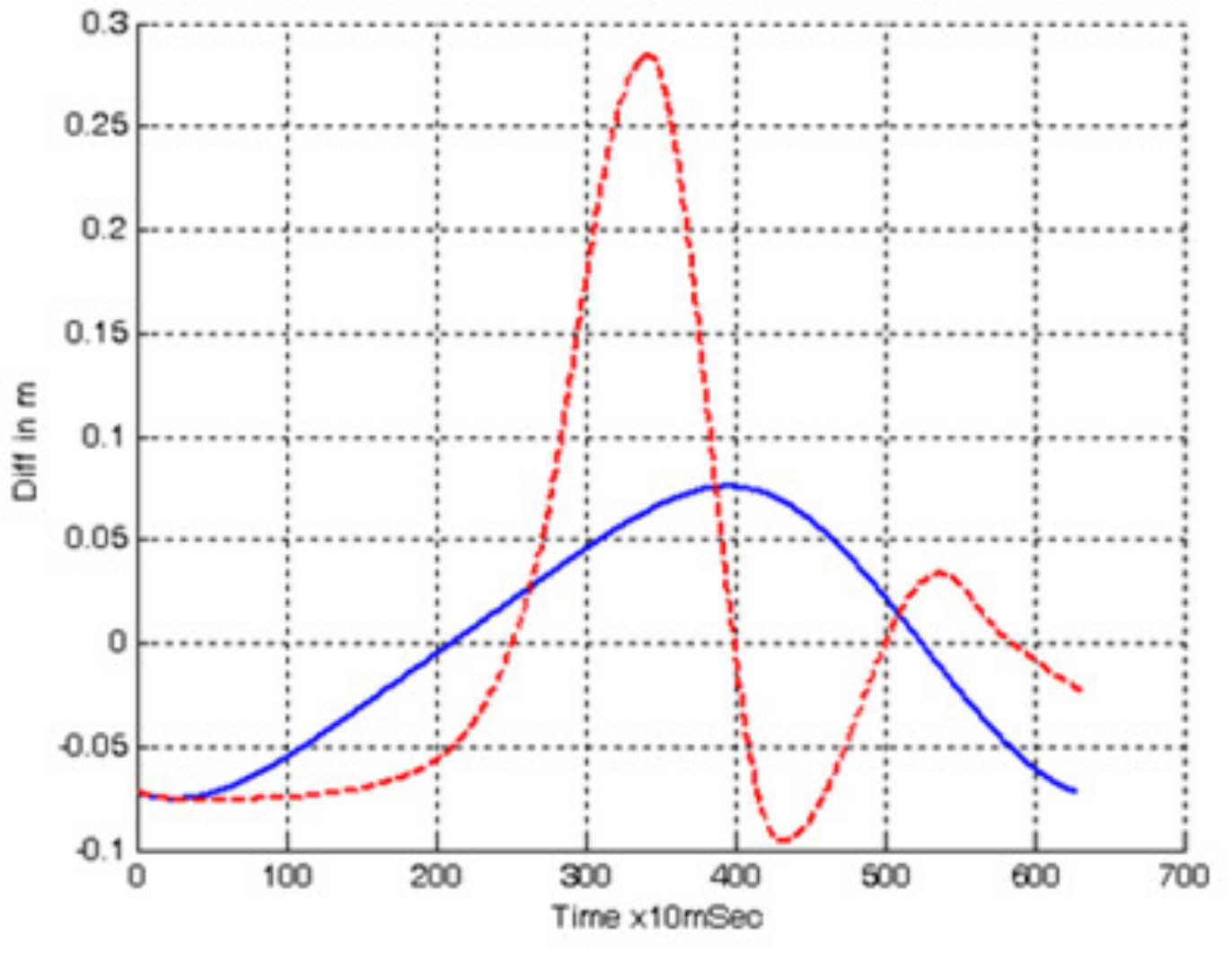}}\\
    \subfigure [ ]{\label{fig63-c}\includegraphics[width=6cm, height=5cm] {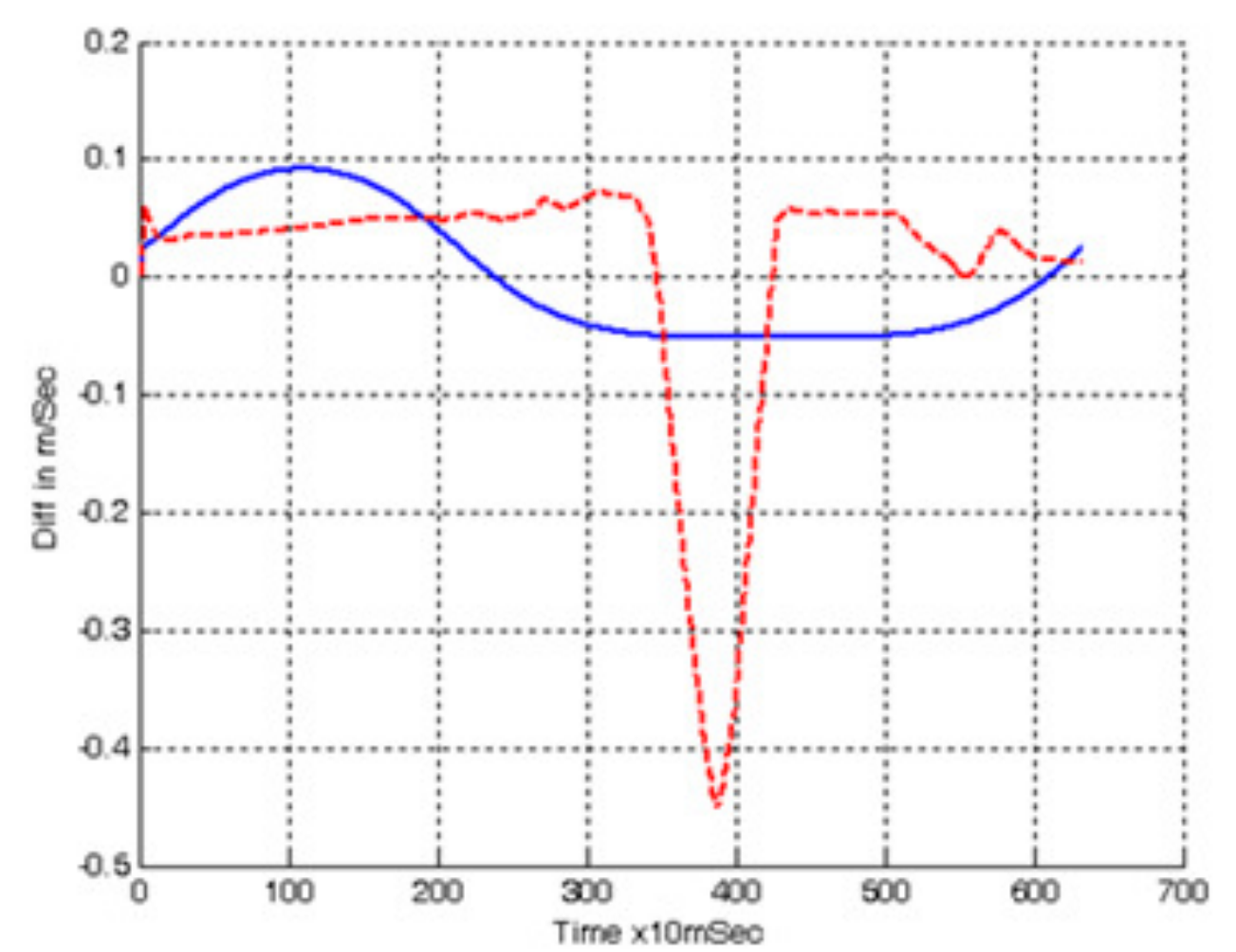}}
    \subfigure [ ]{\label{fig63-d}\includegraphics[width=6cm, height=5cm] {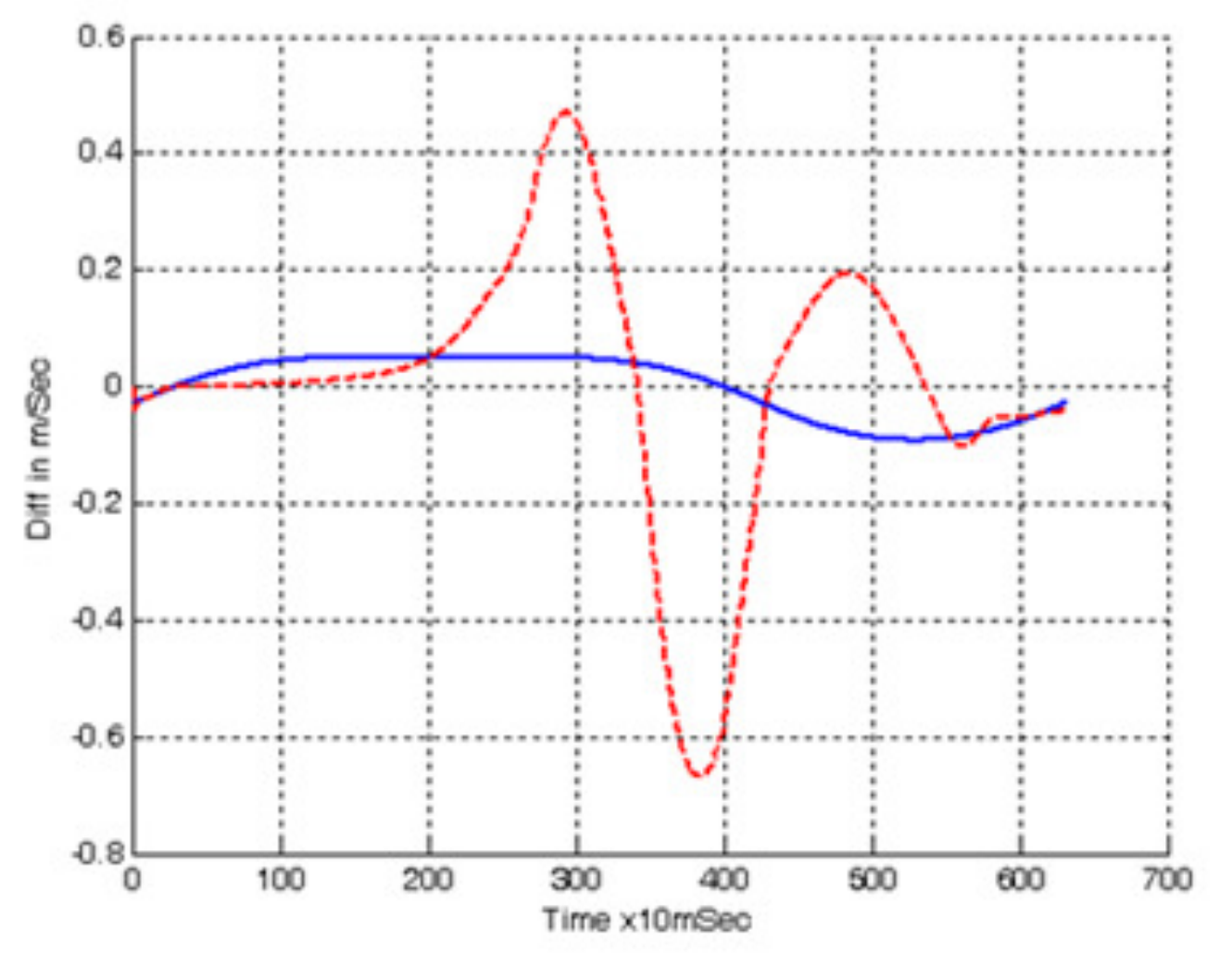}}\\  
\end{center}
  \label{fig63}
\end{figure}

\begin{figure}[htp]
  \begin{center} 
    \subfigure [ ]{\label{fig63-e}\includegraphics[width=7cm, height=6cm] {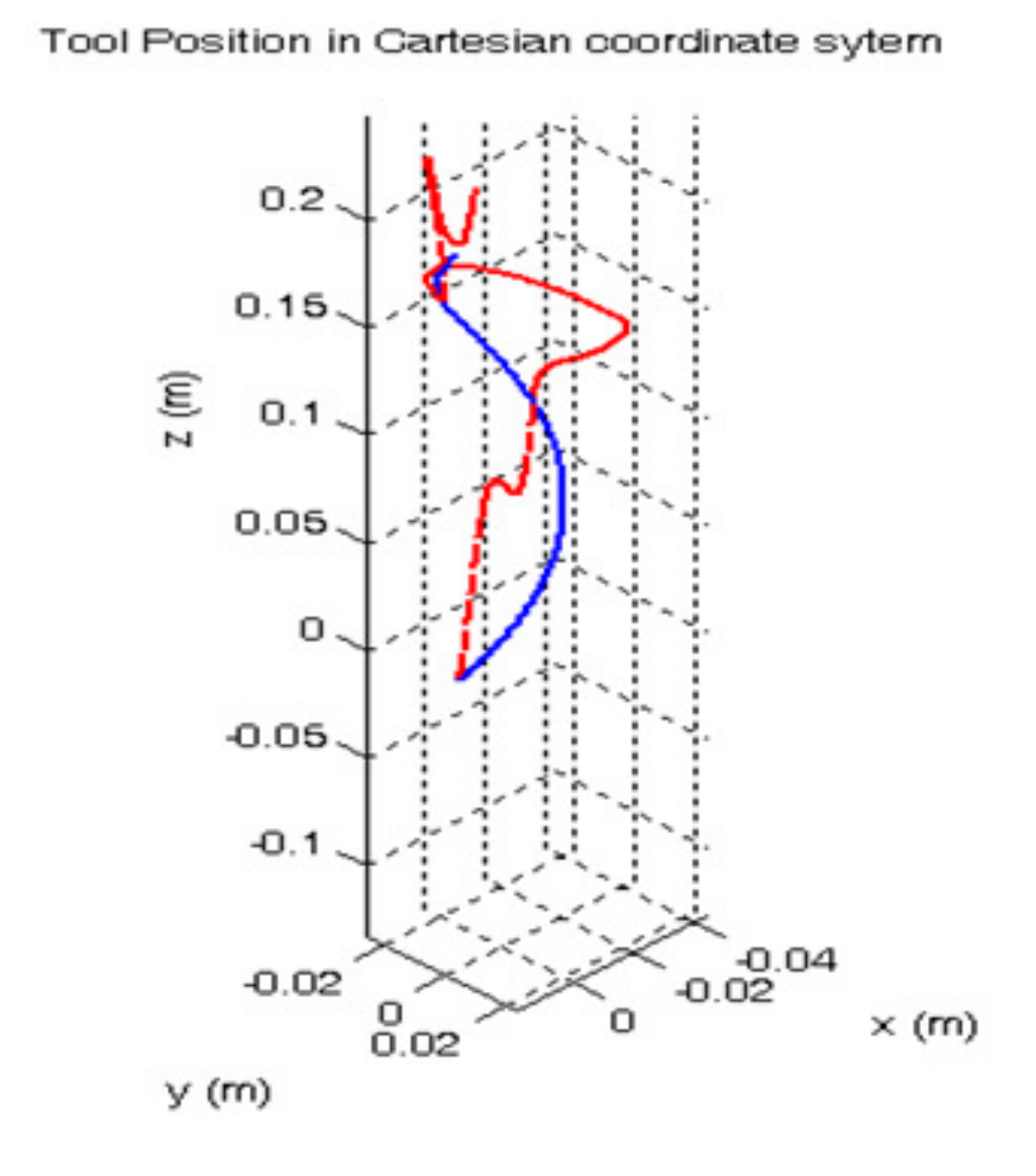}}\\
\end{center}
  \caption[T2 Fuzzy controller, SNR=10dB: (a)-(f)...]{T2 Fuzzy controller, SNR=10dB: (a)-(f) Manipulator's desired and actual outputs,  \{$\Delta \underbar{q}_{12}$,$\Delta \underbar{q}_{13}$, $\Delta \dot{ \underbar{q}}_{12}$, $\Delta \dot{ \underbar{q}}_{13}$\} during the simulation time (g) The followed desired and actual trajectory}
  \label{fig63}
\end{figure} 

\section{\textbf{Comparisons}}
Table \ref{tbl6.2}, demonstrates the time taken to execute all four methods in a computer with a 3.0GHz Pentium 4 processor. As easily implied by the table, the PD controller is the fastest but according to the discussion presented earlier, it has a poor performance for non-linear systems. The situation gets even worse if the controller is to make the system follow a trajectory in high speeds. In fact it is in such circumstances that the effect of dynamic terms emerge.
  
The CTC, on the other hand, has the highest performance among the other three, but at cost of high computational overhead; to the extent that it's not realizable in real-time if the aforementioned processor is employed. To yield an acceptable performance in realization of any controller including the CTC, the response time should not exceeds 2 mili-seconds.
  
Finally, both the T1 and T2 fuzzy based CTC meet our timing expectations (although the former is faster) and thus can be tested in a real experiment (as suggested in the \emph{future works} section.)

\begin{table}
\centering
\caption{\emph{Sum of squared} error for each trial}
\begin{tabular}{p{5cm} p{6cm}}
\hline\hline
PD &less than 0.1 mSec\\\hline
CTC&255 mSec\\ \hline
T1 fuzzy CTC&0.15 mSec\\\hline
T2 fuzzy CTC&1.5 mSec \\
\label{tbl6.2}
\end{tabular}
\end{table} 
Table \ref{tbl6.3} shows the simulation results corresponding to the four controllers. According to the table several points become evident:
Firstly, Notice that when SNR factor is infinite (i.e. when no noise is injected into the system) the 
performance of both T1 and T2 systems are the same which makes sense of course.

Secondly, when noises (with different noise levels) are fed into the system, the T2 controller outperforms the T1 reduced controller.
 
\begin{figure}[htp]
  \begin{center} 
    \includegraphics[width=9cm, height=7cm] {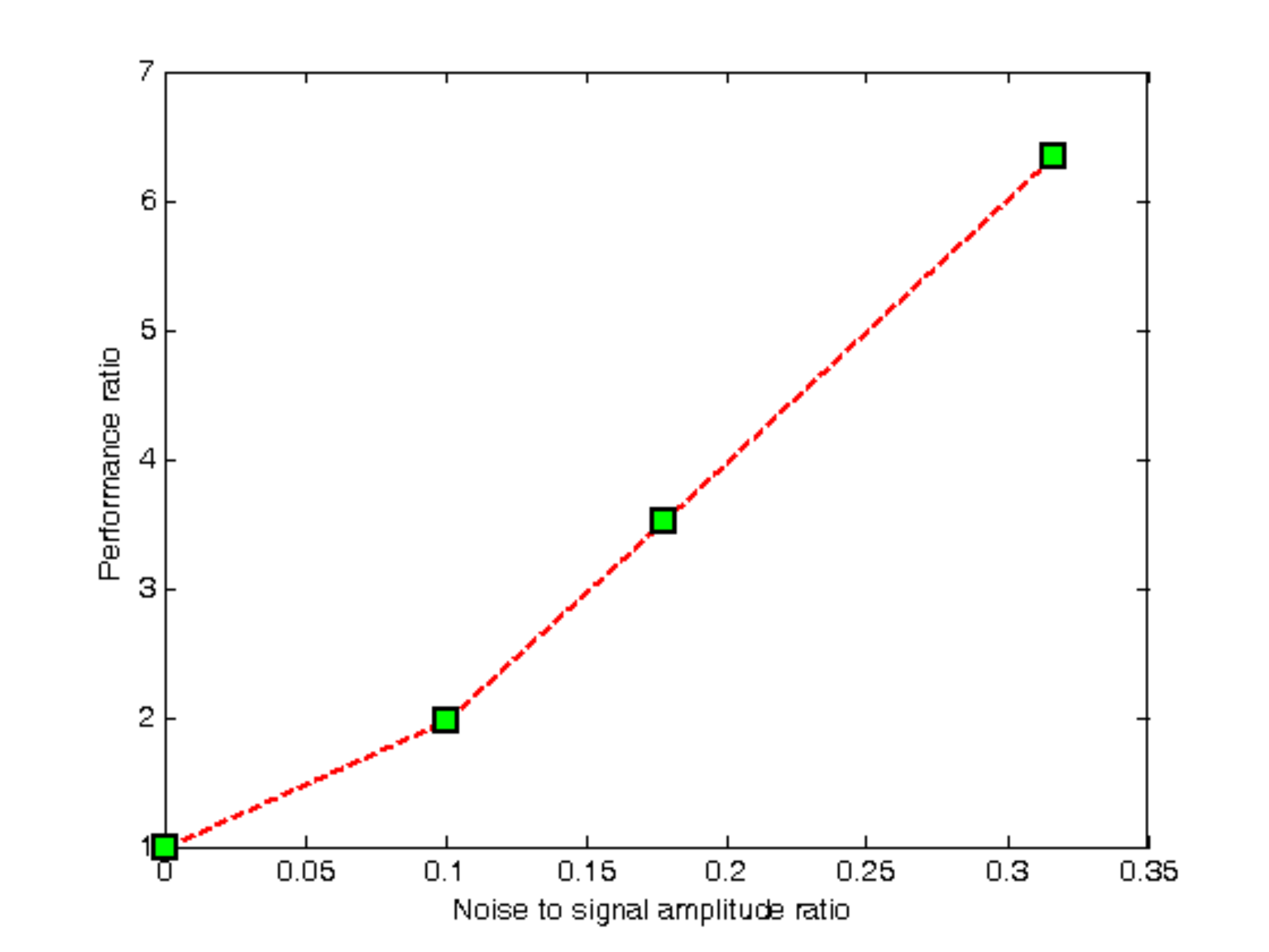}
  \end{center}
  \caption{T2 to T1 performance ratio vs. noise level.} 
  \label{fig64}
\end{figure}

Finally, as the noise level increases the performance of the T1 system deteriorates in a higher pace than that of T2 system. To clarify this, figure \ref{fig64} depicts the ratio of sum of squared error in the T2 system to the corresponding error in the T1 coutnerpart for different noise levels. Notice how this ratio monotonically increases in hormony with noise level increase. This verifies that in the presence of noise, a T2 fuzzy based system do the job better (recall the paradox of uncertainty presented in chapter 1.)

\begin{table}
\centering
\caption{Comparisons}
\begin{tabular}{p{2.5cm}| p{4cm} p{4cm}| }
\hline\hline
PD  &\multicolumn {2}{|c|}{4.6339}\\
CTC &\multicolumn{2}{|c|}{0.0241}\\
& \centering \bf{Type-I} &{\bf{Type-II}} \\ 
& & \\
SNR = $\infty$ & \centering {0.0684}&{0.0684}\\
SNR = 20 & \centering {0.5543} & {0.2816} \\
SNR = 15 & \centering {1.9421}&{0.5530}\\
SNR = 10 & \centering {5.6517}&{1.3930}\\
\label{tbl6.3}
\end{tabular}
\end{table} 

\section{\textbf{Conclusion}}
In this thesis we proposed a new T2FL based method for control of a complex non-linear dynamic system subject to different noise levels. We compared the introduced method with two classical controllers as well as the type reduced controller with similar architecture. In fact we enhanced a non-linear classical controller to cover two important issues, the uncertainty and the real-time realization. We then compared the results in terms of these two criteria to show that the newly introduced method is a promising one and hence can be adopted in other dynamic systems, too.
\section{\textbf{Future Works}}
Although in this work (and most of the T2FLCs proposed to date) the interval Type-II fuzzy MFs are used but this is not necessarily the best choice. The only advantage of the IT2FLSs is the computational costs. The pay off is that interval Type-II fuzzy sets are not the best ones describing linguist variables. In this respect, Gaussian T2FSs may be better options. 
Thus, as a future research one may investigate the role of these types of fuzzy sets in parameter estimation of CTC. In this case, the point at the center of concern will be the real-time issues.

Note also that, as is for every proposed controller, this work is aimed at controlling a real plant 
in the presence of noise and logically the next step will be to implement the proposed controller on 
our real platform (i.e. the 3PSP parallel manipulator developed at FUM Robotics Lab. check for details at \url{http://robotics.um.ac.ir})

%% file: proofs.tex
\chapter{Proofs}

\section{Proof of Theorem \ref{thr1.1}}
\label{prf1.1}
\begin{figure}[htp]
  \begin{center}
    \subfigure [Two fuzzy numbers, $f_1(\theta)$, $f_2(\theta)$]{\label{fig5-a}\includegraphics[width=7cm, height=3cm]{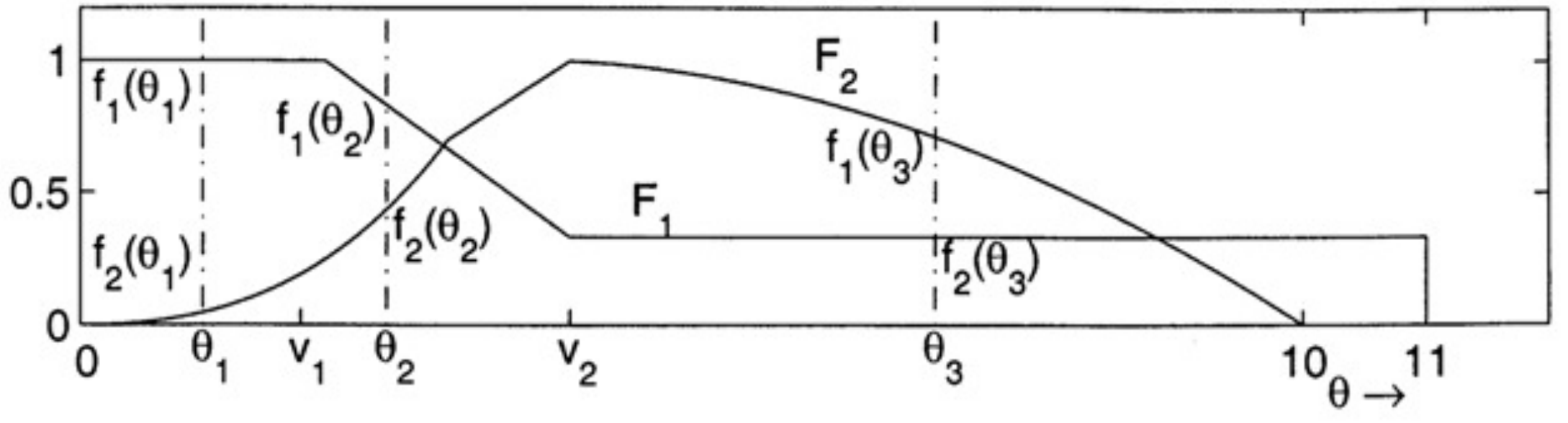}}\\
    \subfigure [The result of join operation]{\label{fig5-b}\includegraphics[width=7cm, height=3cm]{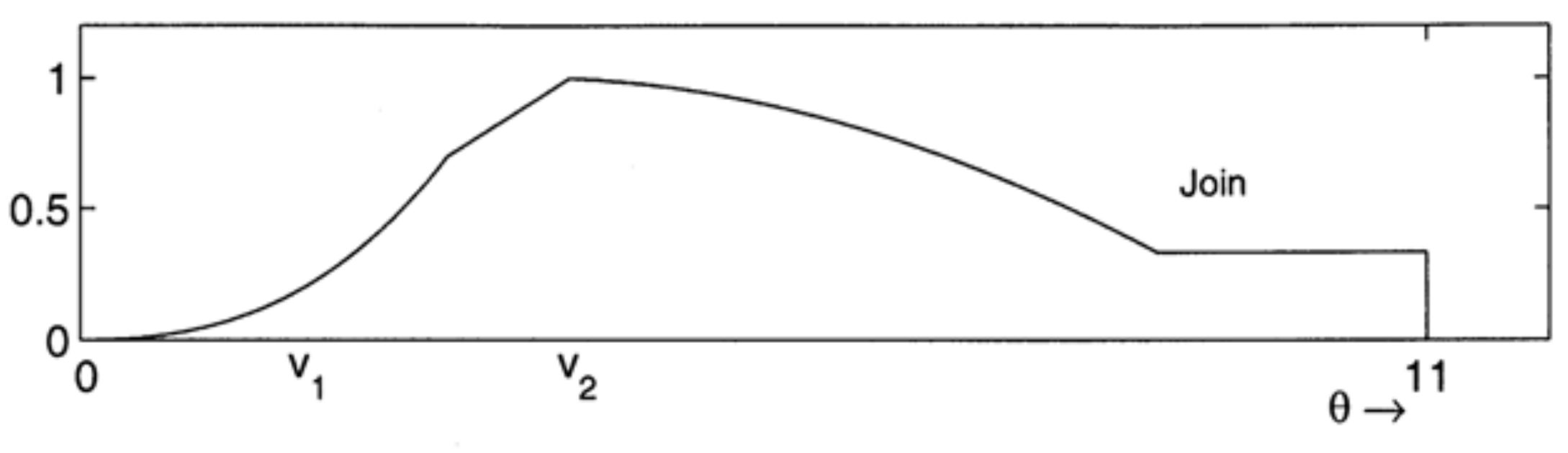}} \\
    \subfigure [The result of meet operation ]{\label{fig5-c}\includegraphics[width=7cm, height=3cm]{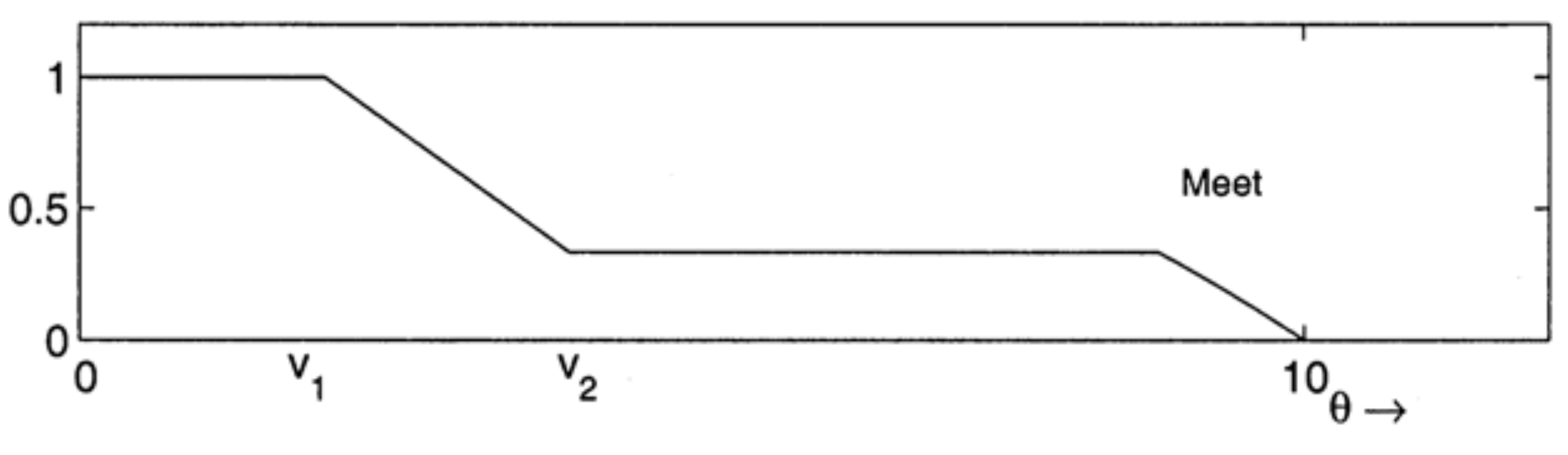}} \\
  \end{center}
  \caption{Meet and join operations for two convex, normal fuzzy numbers under minimum t-norm and maximum t-conorm. \cite{Kar01b}}
  \label{fig5}
\end{figure}

We prove the former formula and leave the rest to the reader. The proof of the latter is quite similar to the former with slight changes due to replacement of the meet operation.
Let $\int_vf_i(v)/v$ $(i=1,2,\dots,n)$ where $v$ assumed to take any real number. We firstly prove the theorem for join operation (first part) and then give the proof for meet operation.  
Consider the case where n=2. The \emph{join} operation between $F_1$ and $F_2$ can thus be expressed as 
\begin{equation}
F_1 \coprod F_2 = \int_v {\int_w {[f_1 (v) \wedge f_2 (w)/(v \vee w)} } 
\label{1.16}
\end{equation}
The process we follow to find the result of \eqref{1.16} is as follows; for each $\theta \in F_1 \coprod F_2$  we should find every possible pairs $\{v,w\}$ where $v \in F_1$ and $w \in F_2$ such that $v\vee w = \theta$  and then find the minimum of the membership grades corresponding to each and finally find the maximum of all the resulting membership grades and ascribe it to the $\mu_{F_1\coprod F_2}(\theta)$  . Now, assuming that  $\theta \in F_1 \coprod F_2$ the possible pairs $\{v,w\}$ that can yield  obtained by the result of maximums between ${\theta ,w}$ where $w \in [-\infty,\theta]$ and ${v,\theta }$ where  $v \in [-\infty,\theta]$ . In so doing, we can break the process to find 
the associated membership grade as follows:\\
\begin{enumerate}
\item Find the minima between memberships of all possible pairs in the form $\{\theta ,w\}$ where $w \in [-\infty,\theta]$.
\item Find the minima between memberships of all possible pairs in the form $\{v,\theta \}$ where $v \in [-\infty,\theta]$.
\item Find the supremum of the above two membership values. That is,
\begin{equation}
\mu _{F_1 \coprod F_2 } (\theta ) = \phi _1 (\theta ) \vee \phi _2 (\theta )
\label{1.17}
\end{equation}\\
where $\phi _1 (\theta ) = \mathop {\sup }\limits_{w \in [ - \infty ,\theta ]} \{ f_1 (\theta ) \wedge f_2 (w)\}  = f_1 (\theta ) \wedge \mathop {\sup }\limits_{w \in [ - \infty ,\theta ]} f_2 (w)
$ and $\phi _2 (\theta ) = \mathop {\sup }\limits_{v \in [ - \infty ,\theta ]} \{ f_1 (v) \wedge f_2 (\theta )\}  = f_2 (\theta ) \wedge \mathop {\sup }\limits_{v \in [ - \infty ,\theta ]} f_1 (v)$
\end{enumerate}
Notice that due to conditions mentioned in the problem description, $f_i(v_i)=1\quad (i=1,2,\dots,n)$ and that  is monotonic non-decreasing for $v\leq v_i$ and is monotonic non-increasing for $v\geq v_i$ .\\  
Let setting apart $\theta$ into three ranges, as follows:	 
\begin{itemize}
\item $\theta=\theta_1<v_1$\\
Since both $f_1$ and $f_2$ are non-decreasing in this domain, $\varphi_1(\theta) $ and $\varphi_2(\theta) $ become,
\begin{equation}
\phi _1 (\theta ) = \mathop {\sup }\limits_{w \in [ - \infty ,\theta ]} \{ f_1 (\theta ) \wedge f_2 (w)\}  = f_1 (\theta ) \wedge \mathop {\sup }\limits_{w \in [ - \infty ,\theta ]} f_2 (w) = f_1 (\theta ) \wedge f_2 (\theta )
\label{1.18a}
\end{equation}\\
and
\begin{equation}
\phi _2 (\theta ) = \mathop {\sup \,}\limits_{v \in [ - \infty ,\theta ]} \{ f_1 (v) \wedge f_2 (\theta )\}  = f_2 (\theta ) \wedge \mathop {\sup }\limits_{v \in [ - \infty ,\theta ]} f_1 (v) = f_2 (\theta ) \wedge f_1 (\theta )
\\
\label{1.18b}
\end{equation}\\  
which follows that: 
\begin{equation}
\mu _{F_1 \coprod F_2 } (\theta ) = \phi _1 (\theta ) \vee \phi _2 (\theta ) = f_1 (\theta ) \vee f_2 (\theta )\qquad theta  < v_1 
\label{1.19}
\end{equation}
\item $v_1  \le \theta  = \theta _2  < v_2 $\\
Recall that $f_1(v)=1$ and that $f_2(w)$ is non-decreasing in this domain, hence,
\begin{equation}
\phi _1 (\theta ) = \mathop {\sup }\limits_{w \in [ - \infty ,\theta ]} \{ f_1 (\theta ) \wedge f_2 (w)\}  = f_1 (\theta ) \wedge \mathop {\sup }\limits_{w \in [ - \infty ,\theta ]} f_2 (w) = f_1 (\theta ) \wedge f_2 (\theta )
\label{1.20a}
\end{equation}
and
\begin{equation}
\phi _2 (\theta ) = \mathop {\sup }\limits_{v \in [ - \infty ,\theta ]} \{ f_1 (v) \wedge f_2 (\theta )\}  = f_2 (\theta ) \wedge \mathop {\sup }\limits_{v \in [ - \infty ,\theta ]} f_1 (v) = f_2 (\theta )
\label{1.20b}
\end{equation}      
which then follows,
\begin{equation}
\mu _{F_1 \coprod F_2 } (\theta ) = f_2 (\theta ) \vee [f_1 (\theta ) \wedge f_2 (\theta )] = f_2 (\theta )\,\,\,\,v_1  \le \theta  = \theta _2  < v_2 
\label{1.21}\end{equation}
\item $\theta=\theta_3\geq v_3$\\
Since both curves have already passed their extremum at some points inside $[-\infty,\theta]$  it therefore follows,
\begin{equation}
\phi _1 (\theta ) = \mathop {\sup }\limits_{w \in [ - \infty ,\theta ]} \{ f_1 (\theta ) \wedge f_2 (w)\}  = f_1 (\theta ) \wedge \mathop {\sup }\limits_{w \in [ - \infty ,\theta ]} f_2 (w) = f_1 (\theta )   
\label{1.22a}
\end{equation}
and
\begin{equation}
\phi _2 (\theta ) = \mathop {\sup }\limits_{v \in [ - \infty ,\theta ]} \{ f_1 (v) \wedge f_2 (\theta )\}  = f_2 (\theta ) \wedge \mathop {\sup }\limits_{v \in [ - \infty ,\theta ]} f_1 (v) = f_2 (\theta ) 
\label{1.22b}
\end{equation}
Which yields,
\begin{equation}
\mu _{F_1 \coprod F_2 } (\theta ) = f_1 (\theta ) \vee f_2 (\theta )\,\,\theta  \ge v_2 
\label{1.23}
\end{equation}
\end{itemize} 	 
Figure \ref{fig5} portrays all the three cases visually for two sample normal, convex curves $f_1(v)$ 
and $f_2(v)$, respectively and the result of \emph{meet} and \emph{join} operations are depicted.

To sum up, we can derive a closed form formula for the join operation as follows,
\begin{equation}
\mu _{F_1 \coprod F_2 } (\theta ) = \left\{ \begin{array}{c l}
   f_1 (\theta) \wedge f_2 (\theta ) & \theta  < v_1   \\ 
   f_2 (\theta) & v_1  \le \theta  < v_2 \\
   f_1 (\theta ) \vee f_2 (\theta ) & \theta  \ge v_2 
	\end{array}  \right.
\label{1.24}
\end{equation}
Observe from the figure and \eqref{1.24} that
\begin{itemize}
\item For $\theta<v_1$ both $f_1$ and $f_2$ are non-decreasing, thus,$f_1(\theta)\wedge f_2(\theta)$ is also non-decreasing in this region.
\item For $v_1\leq \theta < v_2$, $f_2$ is non-decreasing.
\item For $v_2\leq \theta$ both $f_1$ and $f_2$ are non-increasing, hence, $f_1(\theta) \vee f_2(\theta)$ is also non-increasing.
\end{itemize}
Accordingly $\mu_{F_1 \coprod F_2}(\theta)$ will be a convex curve with maximum achieved at 
$\theta = v_2$. This is easily implied from the foregoing figure. Note also that the joined set is normal as well since $f_1(\theta) \vee f_2(\theta)=1$ .

Now let's prove the theorem for $n>2$ by taking advantage of properties of fuzzy sets \cite{Wan08}. 
Using the associativity law for join operator, we have $F_1\prod F_2 \prod F_3=(F_1\prod F_2) \prod F_3$ where by the proof for $n=2$ , $(F_1\coprod F_2)$ is another convex, normal fuzzy set with its maximum located at $v_2$. 
Assuming that $F_3(v_3)=1$, we can generalize \eqref{1.24} to three fuzzy numbers as follows:
\begin{equation}
\mu _{F_1 \coprod F_2 \coprod F_3 } (\theta ) = \left\{ 
\begin{array}{c l}
   \mu _{F_1 \coprod F_2 } (\theta ) \wedge f_3 (\theta ) & \theta  < v_2   \\
   f_3 (\theta ) & v_2  \le \theta  < v_3   \\
   \mu _{F_1 \coprod F_2} (\theta) \vee f_3 (\theta ) &\theta  \ge v_3  \\
\end{array} \right.
\label{1.25}
\end{equation}
Now as  we can rewrite \eqref{1.25} in the following sense,
\begin{equation}
\mu _{F_1 \coprod F_2 } (\theta ) = \left\{\begin{array}{c l}
   f_1 (\theta ) \wedge f_2 (\theta ) & \theta  < v_1  \\ 
   f_2 (\theta ) &  v_1  \le \theta  < v_2   \\ 
   f_1 (\theta ) \vee f_2 (\theta ) & v_2  \le \theta  < v_3  \hfill \\
   f_1 (\theta ) \vee f_2 (\theta ) & v_3  \le \theta  \hfill \\
\end{array}
\right.
\label{1.26}
\end{equation}
Finally substituting \eqref{1.26} into \eqref{1.25}, gives
\begin{equation}
\begin{array}{l}  \mu _{F_1 \coprod F_2 \coprod F_3 } (\theta ) = \left\{ \begin{array}{c l}
   f_1 (\theta ) \wedge f_2 (\theta ) \wedge f_3 (\theta ) & \theta  < v_1   \\
   f_2 (\theta ) \wedge f_3 (\theta ) & v_1  \le \theta  < v_2  \hfill \\ 
	f_3 (\theta ) & v_2  \le \theta  < v_3  \hfill   \\ 
   f_1 (\theta ) \vee f_2 (\theta ) \vee f_3 (\theta ) & \theta  \ge v_3   \\
 	\end{array} \right. = \\ \\
  \left\{ \begin{array}{c l}
  \mathop  \wedge \nolimits_{i = 1}^3 f_i (\theta ) & \theta  < v_1  \hfill \\
  \mathop  \wedge \nolimits_{i = k + 1}^3 f_i (\theta ) & v_k  \le \theta  < v_{k + 1} \,,\,1 \le k \le 2 \hfill \\ 
  \mathop  \vee \nolimits_{i = 1}^3 f_i (\theta ) & v_3  \le \theta  \hfill \\  
	\end{array} \right. \end{array}  
\label{1.27}
\end{equation}
It's rather straight forward to show that \eqref{1.27} is convex and normal. Using the same line of thought we can do the same for $ n = 4 $ $ ( \mu_ {F_1 \coprod F_2 \coprod F_3 \coprod F_4}(\theta))$ and eventually for any desired $n$.

\section{Proof of Theorem \ref{thr1.3}}
\label{prf1.3}
Before we present the proof, it should be noted that unlike the case of IT1FSs, for Gaussian fuzzy sets there is not an exact closed form formula which calculates the meet operation
under product t-norm. However we can approximately find the result for this special case expressed in theorem description. 
Let $F_1$ and $F_2$ be two normal fuzzy number (notice that any Gaussian fuzzy number is normal, too) that are characterized by membership functions $f_1$ and $f_1$ with means and standard deviations $m_1, m_2$ and $\sigma_1, \sigma_2$ respectively. Hence, the \emph{meet} operation under \emph{produc}t t-norm can be found by,
\begin{equation}
F_1 \prod F_2  = \int_v {\int_w {[f_1 (v)f_2 (w)/vw} } 
\label{1.31}
\end{equation}
Assuming that $\theta$ is an element of $F_1 \prod F_2$ the associated membership value can be calculated by finding all admissible pairs $\{v,w\}$ such that $\theta=vw, v\in F_1, w \in F_2$, multiplying membership grades corresponding to $\{v,w\}$ and finally 
finding the maximum value over all possible results. Consider the following cases:
\begin{itemize}
\item $\theta=0$
In this case either $v=0$ or $w=0$ (or both). Thus, the result of meet operation is:
\begin{equation}
\begin{array}{l}
\mu _{F_1 \prod F_2 } (0) = \left[ {\mathop {\sup }\limits_{v \in \Re} \,\,f_1 (v)f_2 (0)} \right] \vee \left[ {\mathop {\sup }\limits_{w \in \Re} \,\,f_1 (0)f_2 (w)} \right] \\ = \left[ {f_2 (0)\mathop {\sup }\limits_{v \in \Re} \,\,f_1 (v)} \right] \vee \left[ {f_1 (0)\mathop {\sup }\limits_{w \in \Re} \,\,f_2 (w)} \right]
\end{array}
\label{1.32}
\end{equation}
Since $F_1$ and $F_2$ are normal fuzzy numbers, equation \eqref{1.32} reduces to,
\begin{equation}
\mu _{F_1 \prod F_2 } (0) = f_2 (0) \vee f_1 (0)
\label{1.33}
\end{equation}
\item $\theta \neq 0$
Let $w = {\theta  \over v}$, thus we get,
\begin{equation}
\mu _{F_1 \prod F_2 } (\theta ) = \mathop {\sup }\limits_{v \in \Re,v \ne 0} \,\,f_1 (v)f_2 ({\theta  \over v})\,\,\,\theta  \in ,\theta  \ne 0
\label{1.34}
\end{equation}
\end{itemize}
Considering \eqref{1.33} and \eqref{1.34}, it becomes obvious why the aforementioned equations can not be reduced to a closed form expression for general fuzzy numbers under product t-norm operation.
Now, let $F_1$ and $F_2$ be Gaussian, too. For the rest of the proof we make an approximation and assume that $F_1$ and $F_2$ are not clipped by their ends. More specifically, 
since we are manipulating secondary membership function of T2FSs, we know that the support associated 
to each of the T1FSs must lie within the range of $[0,1]$. Accordingly, the Gaussian functions should be clipped when they meet the borders. However, for simplicity we take this clipping for granted and consider each T1 fuzzy number as a well shaped Gaussian, regardless of its domain interval. \\
Applying \eqref{1.33} and \eqref{1.34} for Gaussian membership functions, we have 
\begin{itemize}
\item $\theta=0$
\begin{equation}
\mu _{F_1 \prod F_2 } (0) = \exp \left\{ { - {1 \over 2}\left( {{{m_1 } \over {\sigma _1 }}} \right)^2 } \right\} \vee \exp \left\{ { - {1 \over 2}\left( {{{m_2 } \over {\sigma _2 }}} \right)^2 } \right\}
\label{1.35}
\end{equation}
\item $\theta \neq 0$
\begin{equation}
\begin{array}{l}
\mu _{F_1 \prod F_2 } (\theta ) = \mathop {\sup }\limits_{v \in \Re,v \ne 0} \,\,\exp \left\{ { - {1 \over 2}\left[ {\left( {{{v - m_1 } \over {\sigma _1 }}} \right)^2  + \left( {{{\theta /v - m_2 } \over {\sigma _2 }}} \right)^2 } \right]} \right\}\\ \theta  \in \Re,\theta  \ne 0
\label{1.36}
\end{array}
\end{equation}
\end{itemize}
To get rid of the supremum, one should minimize
\begin{equation}
J(v) = ({{v - m_1 } \over {\sigma _1 }})^2  + ({{\theta /v - m_2 } \over {\sigma _2 }})^2  = ({{v - m_1 } \over {\sigma _1 }})^2  + ({{\theta  - m_2 v} \over {\sigma _2 v}})^2 
\label{1.37}
\end{equation}
Note that $J(v)$ is non-convex and hence has local extremums which makes it hard to find the global minimum. To eliminate this problem, we substitute a constant, k, for v in the denominator which leads to a modified objective function, as follows:
\begin{equation}
H(v) = ({{v - m_1 } \over {\sigma _1 }})^2  + ({{\theta  - m_2 v} \over {k\sigma _2 }})^2 
\label{1.38}
\end{equation}
Observe that $H(v)$ is convex since,
\begin{equation}
H''(v) = {2 \over {\sigma _1 ^2 }} + {{2m_2 ^2 } \over {\sigma _2 ^2 }} > 0
\label{1.39}
\end{equation}
Therefore the optimum placed at $v^\star$ , where the first derivate becomes zero. Equating $H'(v)$ to zero gives,
\begin{equation}
H'(v) = 0 \to 2({{v^ *   - m_1 } \over {\sigma _1 }})({1 \over {\sigma _1 }}) + 2({{\theta  - m_2 v^ *  } \over {k\sigma _2 }})({{ - m_2 } \over {k\sigma _2 }}) = 0 \to v^ *   = {{\theta m_2 \sigma _1 ^2  + m_1 k^2 \sigma _2 ^2 } \over {m_2 ^2 \sigma _1 ^2  + k^2 \sigma _2 ^2 }}
\label{1.40}
\end{equation}
Finally substituting \eqref{1.40} into \eqref{1.38}, gives 
\begin{equation}
\mathop {\inf }\limits_v \,H(v) = \left( {{{\sigma  - m_1 m_2 } \over {\sqrt {m_2 ^2 \sigma _1 ^2  + m_1 ^2 \sigma _2 ^2 } }}} \right)^2 
\label{1.41}
\end{equation}
Notice that the constant, $k$, substituted for $v$¸ should lie within the range of $[0,1]$ if a good approximation is desired. This is plausible since $v$ is the primary membership for some point in the underlying T2FS which does not fall outside the foregoing range. Notice also that if we put $m_1$ in place of $v$, then \eqref{1.40} becomes commutative with respect to $\{m_1,\sigma_1\}$ and $\{m_2,\sigma_2\}$ which makes sense since the meet operation is inherently commutative. Thus by doing this final approximation we come to,
\begin{equation}
\mu _{F_1 \prod F_2 } (\theta ) = \exp \left\{ { - {1 \over 2}\left( {{{\theta  - m_1 m_2 } \over {\sqrt {m_2 ^2 \sigma _1 ^2  + m_1 ^2 \sigma _2 ^2 } }}} \right)^2 } \right\}
\label{1.42}
\end{equation}
It is now a straightforward task to show the generalized version of the previous formula to complete the proof. So, we leave it to the reader. 

\section{Proof of Theorem \ref{thr1.5}}
\label{prf1.5}
The proof is given in two parts, firstly, we show that, $\alpha _i F_i  + \beta 
$ is a Gaussian fuzzy number with corresponding mean and standard deviation of $
\alpha _i m_i  + \beta 
$ and $|\alpha _i |\sigma _i$. Secondly, we show that $\sum\limits_{i = 1}^n {F_i } 
$ is a Gaussian fuzzy number with mean $\sum\limits_{i = 1}^n {m_i } 
$ and standard deviation $\Sigma '' = \left\{ {\begin{array}{l}
   \sqrt {\sum\nolimits_{i = 1}^n {\sigma _i^2 } }   \\ 
   \sum\nolimits_{i = 1}^n {\sigma _i }   \\ 
	\end{array}
 }  \right.$. To prove the first part, consider $
F_i  = \int\limits_v {{{\exp \left\{ {{{ - 1} \over 2}({{v - m_i } \over {\sigma _i }})^2 } \right\}} \mathord {\left/
 {\vphantom {{\exp \left\{ {{{ - 1} \over 2}({{v - m_i } \over {\sigma _i }})^2 } \right\}} v}} \right.
 \kern-\nulldelimiterspace} v}} 
$. Multiplying $F_i$ by constant $\alpha _i ( = 1/\alpha _i )
$, and the result to $\beta ( = 1/\beta )$ gives,
\begin{equation}
\begin{array}{l l} \alpha _i F_i  + \beta  = \int\limits_v {{{\exp \left\{ {{{ - 1} \over 2}({{v - m_i } \over {\sigma _i }})^2 } \right\} * 1} \mathord{\left/
 {\vphantom {{\exp \left\{ {{{ - 1} \over 2}({{v - m_i } \over {\sigma _i }})^2 } \right\} * 1} {(\alpha _i v + \beta )}}} \right.
 \kern-\nulldelimiterspace} {(\alpha _i v + \beta )}}}  = \\
\int\limits_v {{{\exp \left\{ {{{ - 1} \over 2}({{v - m_i } \over {\sigma _i }})^2 } \right\}} \mathord{\left/
 {\vphantom {{\exp \left\{ {{{ - 1} \over 2}({{v - m_i } \over {\sigma _i }})^2 } \right\}} {(\alpha _i v + \beta )}}} \right.
 \kern-\nulldelimiterspace} {(\alpha _i v + \beta )}}} \end{array}
\label{1.49}
\end{equation}
Substituting $v'$ for $\alpha_i v +\beta$ yields,
\begin{equation}
\alpha _i F_i  + \beta  = \int\limits_v {{{\exp \left\{ {{{ - 1} \over 2}({{{{v' - \beta } \over {\alpha _i }} - m_i } \over {\sigma _i }})^2 } \right\}} \mathord{\left/
 {\vphantom {{\exp \left\{ {{{ - 1} \over 2}({{{{v' - \beta } \over {\alpha _i }} - m_i } \over {\sigma _i }})^2 } \right\}} {v'}}} \right.
 \kern-\nulldelimiterspace} {v'}}}  = \int\limits_{v'} {{{\exp \left\{ {{{ - 1} \over 2}({{v' - (\alpha _i m_i  + \beta )} \over {\alpha _i \sigma _i }})^2 } \right\}} \mathord{\left/
 {\vphantom {{\exp \left\{ {{{ - 1} \over 2}({{v' - (\alpha _i m_i  + \beta )} \over {\alpha _i \sigma _i }})^2 } \right\}} {v'}}} \right.
 \kern-\nulldelimiterspace} {v'}}} 
\label{1.50}
\end{equation}
which is a Gaussian fuzzy number with mean $(\alpha _i m_i  + \beta )$ and standard deviation $|\alpha _i |\sigma _i 
$. Notice that till now we have not restricted our proof to the choice of t-norm.
No consider $F_1$ and $F_2$ with respective means and standard deviation $m_1$, $\sigma_1$ and $m_2$, $\sigma_2$ . The sum of these two fuzzy numbers can be expressed as,
\begin{equation}
F_1  + F_2  = \int\limits_{v \in F_1 } {{{\int\limits_{w \in F_2 } {\exp \left\{ {{{ - 1} \over 2}({{v - m_1 } \over {\sigma _1 }})^2 } \right\}}  * \exp \left\{ {{{ - 1} \over 2}({{w - m_2 } \over {\sigma _2 }})^2 } \right\}} \mathord{\left/
 {\vphantom {{\int\limits_{w \in F_2 } {\exp \left\{ {{{ - 1} \over 2}({{v - m_1 } \over {\sigma _1 }})^2 } \right\}}  * \exp \left\{ {{{ - 1} \over 2}({{w - m_2 } \over {\sigma _2 }})^2 } \right\}} {(v + w)}}} \right.
 \kern-\nulldelimiterspace} {(v + w)}}} 
\label{1.51}
\end{equation}
where $*$ indicates the t-norm operation. We continue \eqref{1.51} considering two different t-norms, namely the product t-norm and the minimum t-norm (which are among the most well-known t-norms used to date).
\begin{itemize}
\item \emph{Product} t-norm: For product t-norm, \eqref{1.51} reduces to 
\begin{equation}
F_1  + F_2  = \int\limits_{v \in F_1 } {{{\int\limits_{w \in F_2 } {\exp \left\{ {{{ - 1} \over 2}({{v - m_1 } \over {\sigma _1 }})^2 } \right\}} \exp \left\{ {{{ - 1} \over 2}({{w - m_2 } \over {\sigma _2 }})^2 } \right\}} \mathord{\left/
 {\vphantom {{\int\limits_{w \in F_2 } {\exp \left\{ {{{ - 1} \over 2}({{v - m_1 } \over {\sigma _1 }})^2 } \right\}} \exp \left\{ {{{ - 1} \over 2}({{w - m_2 } \over {\sigma _2 }})^2 } \right\}} {(v + w)}}} \right.
 \kern-\nulldelimiterspace} {(v + w)}}} 
\label{1.52}
\end{equation}
Now Consider $\theta  \in F_1  + F_2 $, the membership grade associated with this point is obtained by enumerating all possible pairs $\{v,w\}$ where $\theta=v+w$ and $v \in F_1 ,\,\,w \in F_2 $, multiplying the membership functions associated to $\{v,w\}$ and finding the maximum over all resulting membership products. This all can be expressed as,
\begin{equation}
\mu _{F_1  + F_2 } (\theta ) = \mathop {\sup }\limits_v \exp \left\{ {{{ - 1} \over 2}\left[ {({{v - m_1 } \over {\sigma _1 }})^2  + ({{(\theta  - v) - m_2 } \over {\sigma _2 }})^2 } \right]} \right\}
\label{1.53}
\end{equation}
Let $v^*$ be optimum point we are seeking for, obviously it can be found by minimizing the bracketed term in right hand side of \eqref{1.53}. In other words:
\begin{equation}
v^ *   = \mathop {\arg \min }\limits_v J(v)
\label{1.54}
\end{equation}
with $ J(v) = ({{v - m_1 } \over {\sigma _1 }})^2  + ({{(\theta  - v) - m_2 } \over {\sigma _2 }})^2$.\\
Taking the second derivative of \eqref{1.54}, we get,
\begin{equation}
J''(v) = {1 \over {\sigma _1 ^2 }} + {1 \over {\sigma _2 ^2 }}
\label{1.55}
\end{equation}
which is a positive quantity meaning that $J(v)$ is a convex function of $v$. Thus the global minimum for this curve is located at the point with its first derivative be zero. Therefore equating the first derivative of $J(v)$ to zero follows:
\begin{equation}
\begin{array}{l}
J'(v) = 0 \to 2({{v^ *   - m_1 } \over {\sigma _1 }})({1 \over {\sigma _1 }}) + 2({{(\theta  - v^ *  ) - m_2 } \over {\sigma _2 }})( - {1 \over {\sigma _2 }}) = 0,  \\
v^ *  ({1 \over {\sigma _1 ^2 }} + {1 \over {\sigma _2 ^2 }}) = {{m_1 } \over {\sigma _1 ^2 }} + {{\theta  - m_2 } \over {\sigma _2 ^2 }},  \\
v^ *   = {{m_1 \sigma _2 ^2  + (\theta  - m_2 )\sigma _1 ^2 } \over {\sigma _1 ^2  + \sigma _2 ^2 }}\\
\end{array} 
\label{1.56}
\end{equation}
Substituting \eqref{1.56} into \eqref{1.53} we get
\begin{equation}
\mu _{F_1  + F_2 } (\theta ) = \exp \{ {{ - 1} \over 2}J(v^ *  )\}  = \exp \left\{ {{{ - 1} \over 2}\left[ {{{\theta  - (m_1  + m_2 )} \over {\sqrt {\sigma _1 ^2  + \sigma _2 ^2 } }}} \right]^2 } \right\}
\label{1.57}
\end{equation}
which is a Gaussian fuzzy number with mean $m_1+m_2$ and standard deviation $\sqrt {\sigma _1 ^2  + \sigma _2 ^2 } $. Now using the induction principle we can easily generalize \eqref{1.57} to the case of n Gaussian type-I sets, $F_1 , \cdots ,F_n $ to attain a Gaussian fuzzy set with mean $ \sum\limits_{i = 1}^n {m_i }$ and $ \sqrt {\sum\limits_{i = 1}^n {\sigma _i ^2 } } 
$.
\item \emph{Min} t-norm: In this case \eqref{1.51} reduces to,
\begin{equation}
F_1  + F_2  = \int\limits_{v \in F_1 } {{{\int\limits_{w \in F_2 } {\exp \left\{ {{{ - 1} \over 2}({{v - m_1 } \over {\sigma _1 }})^2 } \right\}}  \wedge \exp \left\{ {{{ - 1} \over 2}({{w - m_2 } \over {\sigma _2 }})^2 } \right\}} \mathord{\left/
 {\vphantom {{\int\limits_{w \in F_2 } {\exp \left\{ {{{ - 1} \over 2}({{v - m_1 } \over {\sigma _1 }})^2 } \right\}}  \wedge \exp \left\{ {{{ - 1} \over 2}({{w - m_2 } \over {\sigma _2 }})^2 } \right\}} {(v + w)}}} \right.
 \kern-\nulldelimiterspace} {(v + w)}}} 
\label{1.58}
\end{equation}
Similar to previous approach, consider $\theta  \in F_1  + F_2 $, the membership grade associated with this point is obtained by enumerating all possible pairs $\{v,w\}$, where $\theta=v+w$ and $v \in F_1 ,\,\,w \in F_2 $, calculating the minimum of membership functions associated to 
$\{v,w\}$ pairs and finding the maximum over all the resulting values. This all can be expressed as,
\begin{equation}
\mu _{F_1  + F_2 } (\theta ) = \mathop {\sup }\limits_v \left[ {\exp \left\{ {{{ - 1} \over 2}({{v - m_1 } \over {\sigma _1 }})^2 } \right\} \wedge \exp \left\{ {{{ - 1} \over 2}({{(\theta  - v) - m_2 } \over {\sigma _2 }})^2 } \right\}} \right]
\label{1.59}
\end{equation}
Now, since for two normal Gaussian fuzzy numbers, the supremum point of the minimum between the two membership functions is located at their intersection point, we equate the equation of the two Gaussians which follows,
\begin{equation}
\begin{array}{l}
\exp \left\{ {{{ - 1} \over 2}({{v^ *   - m_1 } \over {\sigma _1 }})^2 } \right\} = \exp \left\{ {{{ - 1} \over 2}({{(\theta  - v^ *  ) - m_2 } \over {\sigma _2 }})^2 } \right\} \to \\ v^ *   = {{m_1 \sigma _2  + (\theta  - m_2 )\sigma _1 } \over {\sigma _1  + \sigma _2 }}
\end{array}
\label{1.60}
\end{equation}
Obviously the membership grade at $v^*$ for both membership functions is the same.
\end{itemize}